\pdfoutput=1

\documentclass[11pt,twoside,a4paper,cmspaper,final,collab]{cms-tdr}
\def\svnVersion{428533P}\def\cmsCernNoTag{CERN-EP-2017-110}\def\cmsCernDate{\today}\def\cmsMessage{Published in the Journal of Instrumentation as \href{http://dx.doi.org/10.1088/1748-0221/12/10/P10003}{\doi{10.1088/1748-0221/12/10/P10003.}}}

\begin{document}\cmsNoteHeader{PRF-14-001}

\hyphenation{had-ron-i-za-tion}
\hyphenation{cal-or-i-me-ter}
\hyphenation{de-vices}
\RCS$Revision: 423248 $
\RCS$HeadURL: svn+ssh://svn.cern.ch/reps/tdr2/papers/PRF-14-001/trunk/PRF-14-001.tex $
\RCS$Id: PRF-14-001.tex 423248 2017-08-31 15:14:44Z alverson $

\newcommand{\ptref}{\ensuremath{p_\mathrm{T}^\text{Ref}}\xspace}
\newcommand{\vecpt}{\ensuremath{\vec{p}_\mathrm{T}}\xspace}
\newcommand{\ivecpt}[1]{\ensuremath{\vec{p}_{\mathrm{T},#1}}\xspace}
\newcommand{\ptmissref}{\ensuremath{p_\text{T,Ref}^\text{miss}}\xspace}
\newcommand{\ptmissof}[1]{\ensuremath{p_\text{T,#1}^\text{miss}}\xspace}
\newcommand{\vecptmiss}{\ensuremath{\vec{p}_\mathrm{T}^{\kern1pt\text{miss}}}\xspace}
\newcommand{\vecptmissof}[1]{\ensuremath{\vec{p}_\text{T,#1}^{\kern1pt\text{miss}}}\xspace}
\newcommand{\vecptmissref}{\ensuremath{\vec{p}_\text{T,Ref}^{\kern1pt\text{miss}}}\xspace}
\newcommand{\kT}{\ensuremath{k_\mathrm{T}}\xspace}
\providecommand{\Phm}{\ensuremath{\Ph^-}\xspace}
\newcommand{\Pnut}{\ensuremath{\PGn_{\Pgt}}}
\newcommand{\Pggx}{\ensuremath{\PGg^{*}}\xspace}

\cmsNoteHeader{PRF-14-001} \title{Particle-flow reconstruction and global event description with the CMS detector}

\date{\today}

\abstract{
The CMS apparatus was identified, a few years before the start of the LHC operation at CERN,
to feature properties well suited
to particle-flow (PF) reconstruction: a highly-segmented tracker,
a fine-grained electromagnetic calorimeter, a hermetic hadron calorimeter,
a strong magnetic field, and an excellent muon spectrometer.
A fully-fledged PF reconstruction algorithm tuned to the CMS detector was
therefore developed and has been consistently used in physics analyses for
the first time at a hadron collider. For each collision, the comprehensive
list of final-state particles identified and reconstructed by the algorithm
provides a global event description that leads to unprecedented CMS performance
for jet and hadronic $\tau$ decay reconstruction, missing transverse momentum
determination, and electron and muon identification. This approach also
allows particles from pileup interactions to be identified and
enables efficient pileup mitigation methods. The data collected by CMS at a
centre-of-mass energy of 8\TeV show excellent agreement with the simulation
and confirm the superior PF performance at least up to an average of 20 pileup
interactions.
}

\hypersetup{%
pdfauthor={CMS Collaboration},%
pdftitle={Particle-flow reconstruction and global event description with the CMS detector},%
pdfsubject={CMS Particle Flow},%
pdfkeywords={CMS, physics,  PF, reconstruction, performance}}

\maketitle
\section{Introduction}
\label{sec:introduction}
\begin{figure}[b]
\centering
\includegraphics[width=1.0\columnwidth]{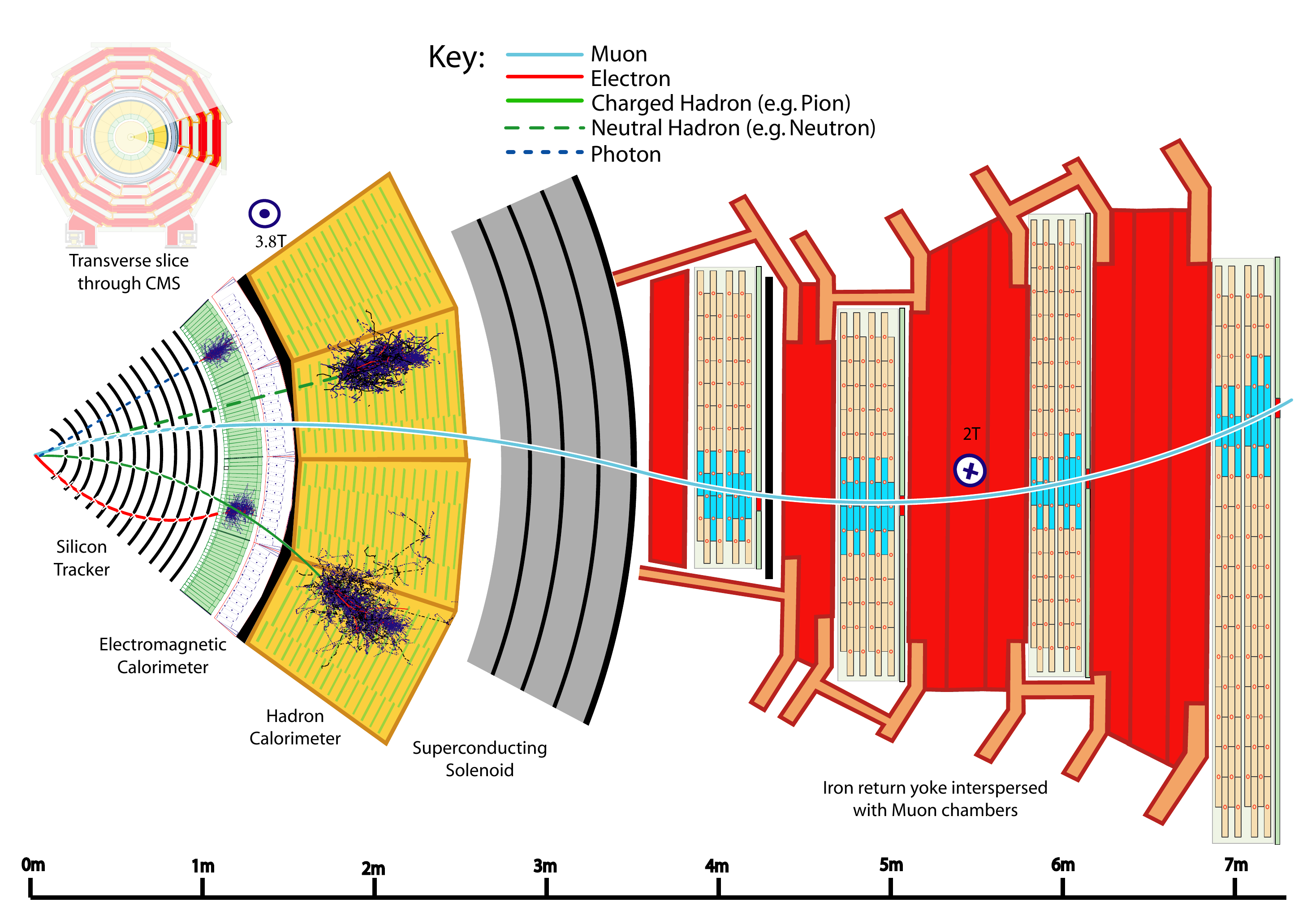}
\caption{\label{fig:CMSSlice} A sketch of the specific particle
interactions in a transverse slice of the CMS detector, from the beam
interaction region to the muon detector.  The muon and the charged pion
are positively charged, and the electron is negatively charged.}
\end{figure}

Modern general-purpose detectors at high-energy colliders are based on the
concept of cylindrical detection layers, nested around the beam axis.
Starting from the beam interaction region, particles first enter a
tracker, in which charged-particle trajectories (\textit{tracks}) and origins
(\textit{vertices}) are reconstructed from signals (\textit{hits}) in the sensitive
layers. The tracker is immersed in a magnetic field that bends the
trajectories and allows the electric charges and momenta of charged particles
to be measured. Electrons and photons are then absorbed in an electromagnetic
calorimeter (ECAL). The corresponding electromagnetic showers are detected
as \textit{clusters} of energy recorded in neighbouring cells, from which the
energy and direction of the particles can be determined. Charged
and neutral hadrons may initiate a hadronic shower in the ECAL as well,
which is subsequently fully absorbed in the hadron calorimeter (HCAL).
The corresponding clusters are used to estimate their energies and
directions. Muons and neutrinos traverse the calorimeters with little or
no interactions. While neutrinos escape undetected, muons produce hits
in additional tracking layers called muon detectors, located outside the calorimeters.
This simplified view is graphically
summarized in Fig.~\ref{fig:CMSSlice}, which displays a sketch of a
transverse slice of the CMS detector~\cite{cms_paper}.

\vfill\eject

This apparent simplicity has led to a tradition at hadron colliders
of reconstructing \textit{physics objects} based---at least to a large extent---on the signals collected by a given detector as follows:
\begin{itemize}
\item \textit{Jets} consist of hadrons and photons, the energy of which
can be inclusively measured by the calorimeters without any attempt to separate individual jet particles.
Jet reconstruction can therefore be performed without any contribution from the tracker and
the muon detectors. The same argument applies to the \textit{missing
transverse momentum}\footnote{The CMS coordinate
system is oriented such that the $x$ axis points to the centre of the LHC
ring, the $y$ axis points vertically upward, and the $z$ axis is in the
direction of the counterclockwise proton beam, when looking at the
LHC from above. The origin is centred at the nominal collision point inside
the experiment. The azimuthal angle $\varphi$ (expressed in radians in this
paper) is measured from the $x$ axis in the $(x,y)$ plane, and the  radial
coordinate in this plane is denoted $r$. The polar angle $\theta$ is defined
in the $(r,z)$ plane with respect to the $z$ axis and the pseudorapidity is
defined as $\eta = -\ln\tan\left(\theta/2\right)$. The component of the
momentum transverse to the $z$ axis is denoted $\pt$. The missing transverse
momentum \ptmiss is the vectorial sum of the undetectable particle transverse
momenta. The transverse energy is defined as $\ET=E\sin\theta$.} (\ptmiss)
reconstruction.
\item The reconstruction of \textit{isolated photons and electrons} primarily
concerns the ECAL.
\item The \textit{tagging} of jets originating from hadronic $\Pgt$ decays
and from b quark hadronization is based on the properties of the
pertaining charged particle tracks, and thus mostly involves
the tracker.
\item The identification of \textit{muons} is principally based on the
information from the muon detectors.
\end{itemize}

\begin{figure}[htbp]
  \begin{centering}
\includegraphics[width=0.6\textwidth] {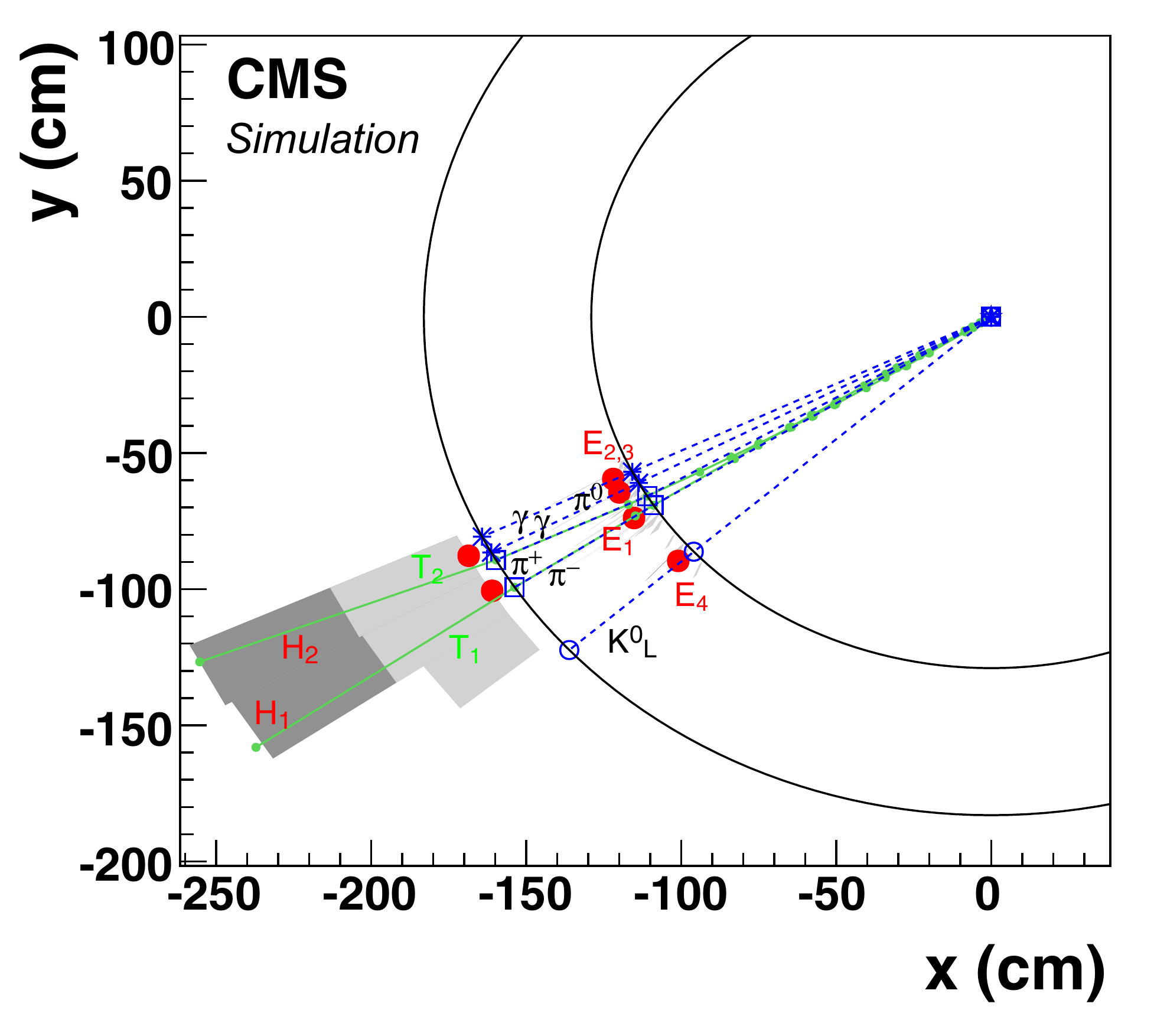}
 \includegraphics[width=0.49\textwidth]{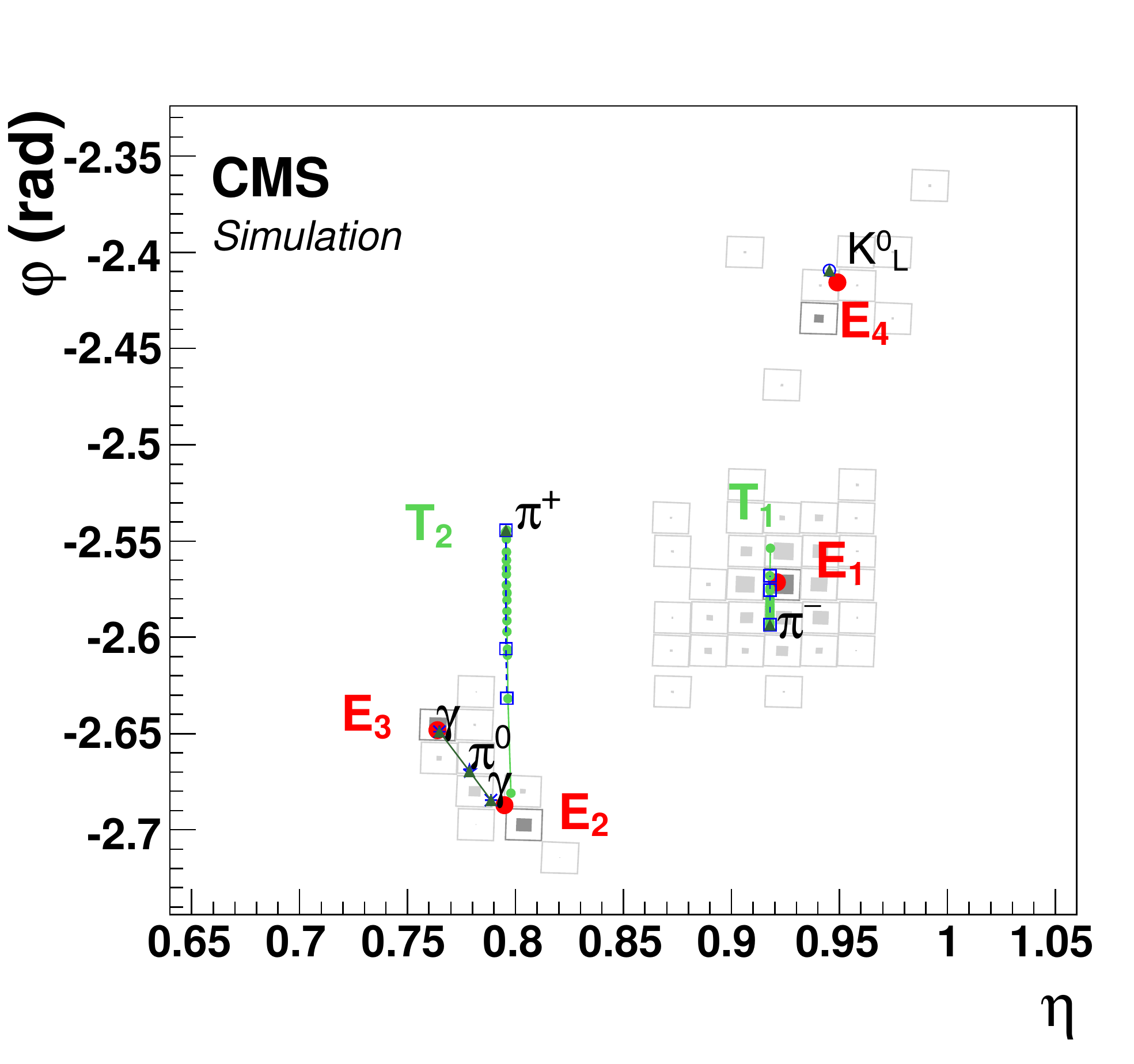}
 \includegraphics[width=0.49\textwidth]{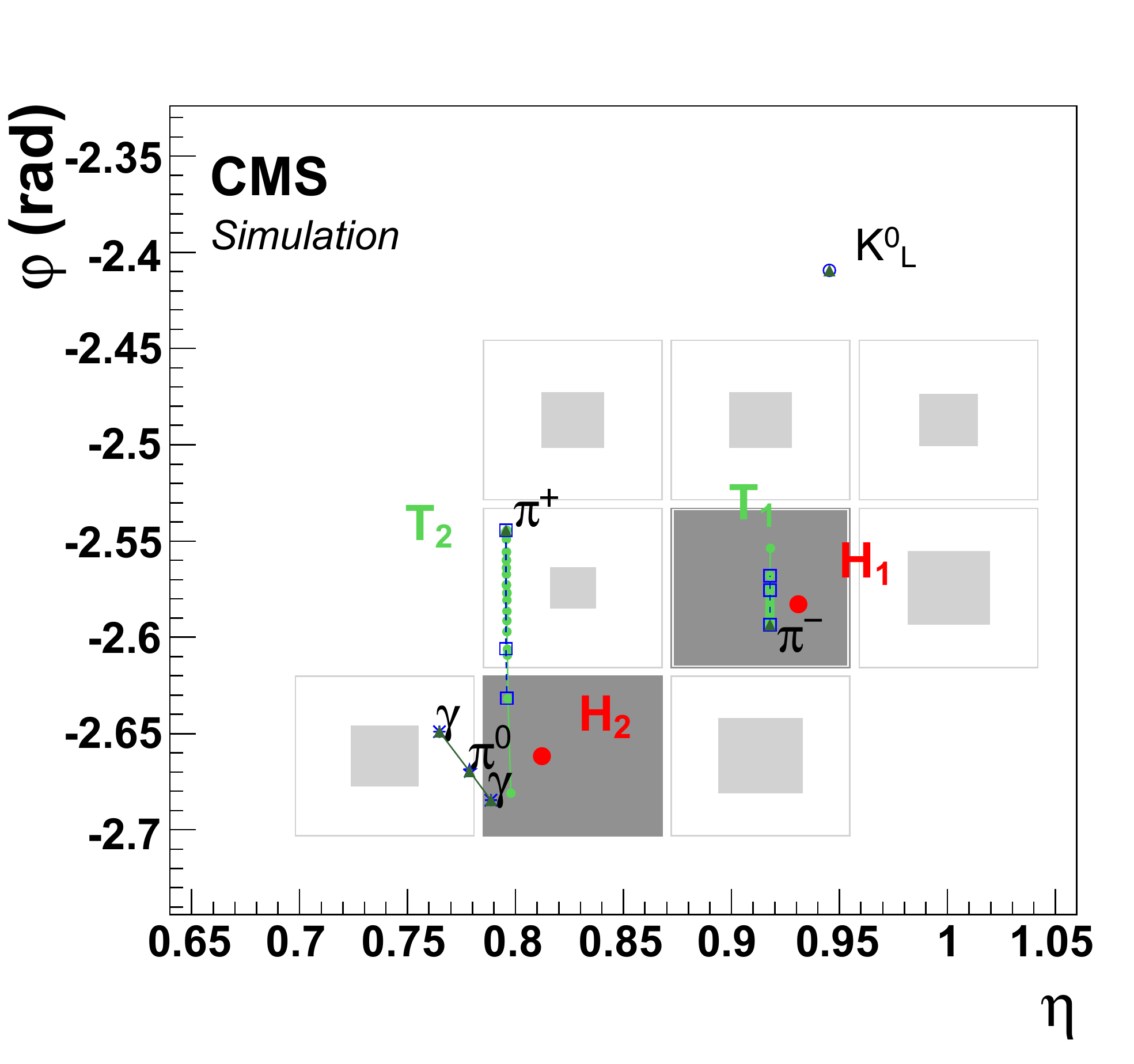}
    \caption{Event display of an illustrative jet made of five particles only in the $(x,y)$ view
(upper panel), and in the ($\eta,\varphi$) view on the ECAL surface (lower left) and the
HCAL surface (lower right). In the top view, these two surfaces are represented as
circles centred around the interaction point. The $\mathrm{K}^0_\mathrm{L}$,
the $\pi^-$, and the two photons from the $\pi^0$ decay are detected as
four well-separated ECAL clusters denoted $\mathrm{E}_{1,2,3,4}$. The
$\pi^+$ does not create a cluster in the ECAL. The two charged pions are reconstructed
as charged-particle tracks $\mathrm{T}_{1,2}$, appearing as vertical solid
lines in the ($\eta,\varphi$) views and circular arcs in the $(x,y)$ view.
These tracks point towards two HCAL clusters $\mathrm{H}_{1,2}$.
In the bottom views, the ECAL and HCAL cells are represented as squares,
with an inner area proportional to the logarithm of the cell energy.
Cells with an energy larger than those of the neighbouring cells are shown in dark grey.
In all three views, the cluster positions are represented by dots, the simulated
particles by dashed lines, and the positions of their impacts on the
calorimeter surfaces by various open markers.}
    \label{fig:cms_detector_display}
  \end{centering}
\end{figure}

A significantly improved event description can be achieved by correlating the
basic \textit{elements} from all detector layers (tracks and clusters) to identify
each final-state particle, and by combining the corresponding measurements to
reconstruct the particle properties on the basis of this identification.
This holistic approach is called \textit{particle-flow (PF) reconstruction}.
Figure~\ref{fig:cms_detector_display}
provides a foretaste of the benefits from this approach. This figure
shows a jet simulated in the CMS detector
with a transverse momentum of 65\GeV.
This jet is made of only five particles for illustrative purposes: two charged hadrons
(a $\pi^+$ and a $\pi^-$), two photons (from the decay of a $\pi^0$),
and one neutral hadron (a $\mathrm{K}^0_\mathrm{L}$). The charged hadrons are
identified by a geometrical connection (\textit{link}) in the $(\eta,\varphi)$
views between one track and one or more calorimeter clusters, and by the absence of signal
in the muon detectors. The combination of the measurements in the tracker and
in the calorimeters provides an improved determination of the energy and
direction of each charged hadron, dominated by the superior tracker
resolution in that particular event. The photons and neutral hadrons
are in general identified by ECAL and HCAL clusters with no track link.
This identification allows the cluster energies to be calibrated more
accurately under either the photon or the hadron hypothesis. No attempt is
made to distinguish the various
species of neutral and charged hadrons in the PF reconstruction. Electrons
and muons are not present in this jet. Electrons would be identified by a track
and an ECAL cluster, with a momentum-to-energy ratio compatible with unity,
and not connected to an HCAL cluster. Muons would be identified by a track
in the inner tracker connected to a track in the muon detectors.

The PF concept was developed and used
for the first time by the ALEPH experiment at LEP~\cite{Buskulic:1994wz}
and is now driving the design of detectors for possible future $\Pep\Pem$
colliders, ILC and CLIC~\cite{Thomson:2009rp,Ruan:2014paa},
FCC-ee~\cite{Gomez-Ceballos:2013zzn}, and CEPC~\cite{CEPC}. Attempts to repeat
the experience at hadron colliders had not met with success so far. A key
ingredient in this approach is the fine spatial granularity of the detector
layers. Coarse-grained detectors may cause the signals from different particles
to merge, especially within jets, thereby reducing the particle identification
and reconstruction capabilities. Even in that case, however, the tracker
resolution can be partially exploited by locally subtracting from the
calorimeter energy either the energy expected from charged hadrons or the
energy measured within a specific angle from the charged hadron trajectories.
Such \textit{energy-flow} algorithms~\cite{EFDELPHI, EFH1, EFZEUS, EFD0, EFCDF,
EFCMS, EFATLAUS, EFATLAS} are used in general to improve the determination of
selected hadronic jets or hadronic tau decays. If, on the other hand, the
subdetectors are sufficiently segmented to provide good separation between
individual particles, as shown for CMS in Fig.~\ref{fig:cms_detector_display},
a global event description becomes possible, in which all particles are
identified. From the list of identified particles, optimally reconstructed
from a combined fit of all pertaining measurements, the physics objects can be
determined with superior efficiencies and resolutions.

Prior to the LHC startup, however, it was commonly feared that the intricacy
of the final states arising from proton-proton or heavy ion collisions
would dramatically curb the advantages of the PF paradigm. The capacity to
individually identify the particles from the hard scatter was indeed expected
to be seriously downgraded by the proton or ion debris, the particles from
pileup interactions
(proton-proton interactions concurrent to the hard scatter in the same or different bunch crossings),
the particle proximity inside high-energy jets, the secondary interactions in the tracker material,
etc.
Detailed Monte Carlo (MC) simulations performed in 2009,
and the commissioning of the  algorithm in the first weeks of LHC data
taking at $\sqrt{s} = 0.9$ and $2.36$\TeV
 in December 2009, and at $7$\TeV
 in March 2010,  demonstrated the adequacy of the CMS
detector design for PF reconstruction of proton-proton collisions,
with benefits similar to those observed in $\Pep\Pem$ collisions.
The holistic approach also gave ways to quickly
cross-calibrate the various subdetectors, to validate their measurements,
and to identify and mask detector backgrounds. The PF reconstruction
was ready for use in physics analyses in June 2010, and was implemented
in the high level trigger and in heavy ion collision analyses
in 2011. Since then, practically all CMS physics results have been based
on PF reconstruction, and the future detector upgrade designs
are routinely assessed by reference to it.

This paper is organized as follows. In Section~\ref{sec:cms_detector},
the properties of the CMS detector are summarized in view of its
PF capabilities. The implementation of the PF concept
for CMS is the subject of the following two sections.
Section~\ref{sec:reconstruction_pf_elements} describes the basic elements
needed for a proper particle reconstruction through its specific signals in
the various subdetectors. The algorithm that links the basic elements together
and the subsequent particle identification are presented in
Section~\ref{sec:particle_id_reco}. The expected performance of the resulting
physics objects is compared to that of the traditional methods in
Section~\ref{sec:expected_performance}, in the absence of pileup interactions.
Finally, the physics object performance observed in data, and the mitigation
of the effects of pileup interactions---for which the final state particles,
also exclusively reconstructed by the PF approach, provide precious additional
handles---are underlined in Section~\ref{sec:commissioning_and_pileup}.

\section{The CMS detector }
\label{sec:cms_detector}

The CMS detector~\cite{cms_paper} turns out to be well-suited to PF, with:

\begin{itemize}
\item a large magnetic field, to separate the calorimeter energy
  deposits of charged and neutral particles in jets;
\item a fine-grained tracker, providing a pure and efficient
charged-particle trajectory reconstruction in jets with $\pt$ up to
around 1\TeV, and therefore an excellent measurement of $\sim$65\% of the jet energy;
\item a highly-segmented ECAL, allowing energy deposits from particles in jets
(charged ha\-drons, neutral hadrons, and photons) to be clearly separated
from each other up to a jet $\pt$ of the order of 1\TeV. The resulting
efficient photon identification, coupled to the high ECAL energy resolution,
allows for an excellent measurement of another $\sim$25\% of the jet energy;
\item a hermetic HCAL with a coarse segmentation, still sufficient to separate
charged and neutral hadron energy deposits in jets up to a jet $\pt$ of
200--300\GeV, allowing the remaining 10\% of the jet energy to be
reconstructed, although with a modest resolution;
\item an excellent muon tracking system, delivering an efficient and pure
  muon identification, irrespective of the surrounding particles.
\end{itemize}

The characteristics of the magnet and of the CMS subdetectors relevant to PF
are described in this section.

\subsection{The magnet}

The central feature of the CMS design is a large superconducting solenoid
magnet~\cite{CMS:1997fm}. It delivers an axial and uniform magnetic field of
3.8\unit{T} over a length of 12.5\unit{m} and a free-bore radius of 3.15\unit{m}.  This
radius is large enough to accommodate the tracker and both the ECAL and HCAL,
thereby minimizing the amount of material in front of the calorimeters. This
feature is an advantage for PF reconstruction, as it eliminates the energy
losses before the calorimeters caused by particles showering in the coil
material and facilitates the link between tracks and calorimeter clusters.
At normal incidence, the bending power of 4.9\unit{$\mathrm{T}\cdot\mathrm{m}$} to the inner surface of
the calorimeter system provides strong separation between charged- and
neutral-particle energy deposits. For example, a charged particle with $\pt = 20\GeV$
is deviated in the transverse plane by 5\unit{cm} at the ECAL surface, a
distance large enough to resolve its energy deposit from that of a photon
emitted in the same direction.

\subsection{The silicon inner tracker}
\label{sec:silicontracker}

The full-silicon inner tracking system~\cite{CMS:1998aa,CMS:2000aa} is a
cylinder-shaped detector with an outer radius of 1.20\unit{m} and a length of
5.6\unit{m}. The barrel (each of the two endcaps) comprises three (two) layers of
pixel detectors, surrounded by ten (twelve) layers of micro-strip detectors.
The 16\,588 silicon sensor modules are finely segmented into 66 million
$150{\times}100\mum$ pixels and 9.6 million $80$-to-$180\mum$-wide
strips. This fine granularity offers separation of closely-spaced particle
trajectories in jets.

\begin{figure}[tb]
\includegraphics[width=0.49\textwidth]{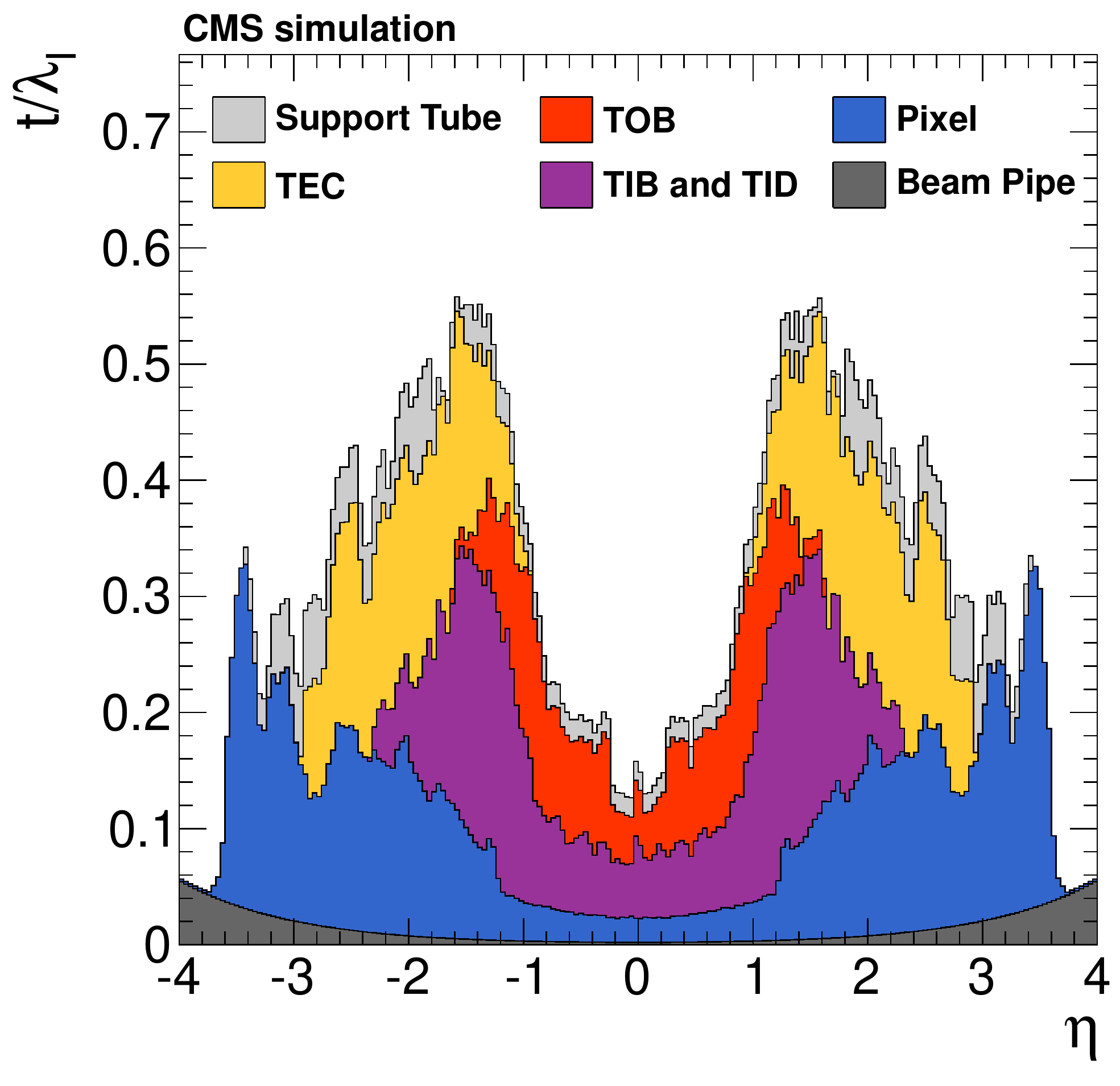}
\includegraphics[width=0.49\textwidth]{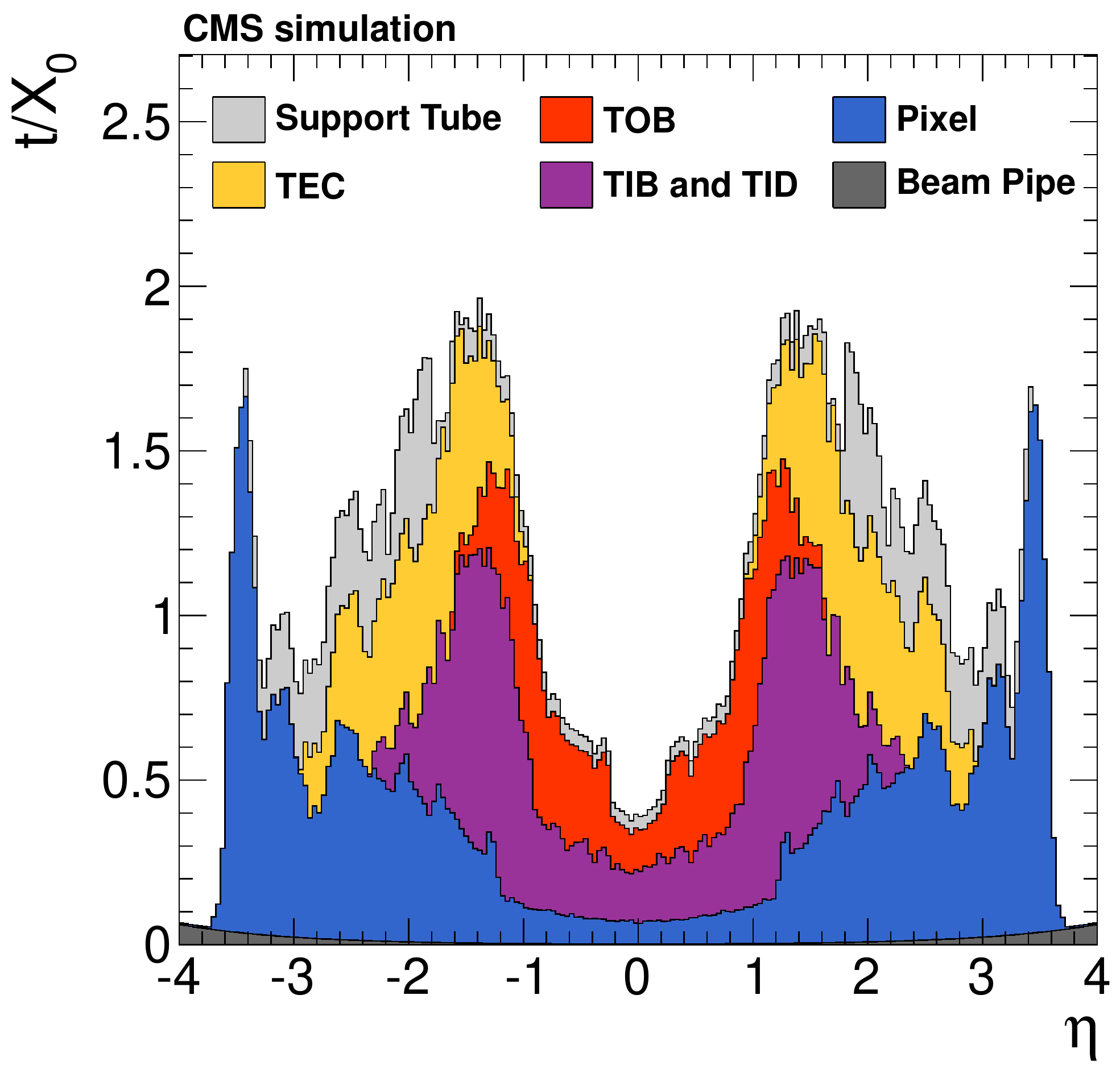}
\caption{Total thickness $t$ of the inner tracker material expressed in units
of interaction lengths $\lambda_l$ (left) and radiation lengths $X_0$ (right),
as a function of the pseudorapidity $\eta$. The acronyms TIB, TID, TOB, and
TEC stand for ``tracker inner barrel'', ``tracker inner disks'', ``tracker
outer barrel'', and ``tracker endcaps'', respectively. The two figures are
taken from Ref.~\cite{cms_tracking_paper}.
}
\label{fig:cms_detector:tracker_material}
\end{figure}

As displayed in Fig.~\ref{fig:cms_detector:tracker_material},
these layers and the pertaining services (cables, support, cooling)
represent a substantial amount of material in front of the calorimeters,
up to 0.5 interaction lengths or 1.8 radiation lengths. At $\abs{\eta} \approx 1.5$, the
probability for a photon to convert or for an electron to emit a bremsstrahlung
photon by interacting with this material is about 85\%. Similarly, a hadron
has a 20\% probability to experience a nuclear interaction before reaching
the ECAL surface. The large number of emerging secondary particles
turned out to be a major source of complication in the PF reconstruction
algorithm.
It required harnessing the full granularity and redundancy of the silicon
tracker measurements for this complication to be eventually overcome.

The tracker measures the \pt of charged hadrons at normal incidence with a resolution
of 1\% for $\pt<20\GeV$.
The relative resolution then degrades with increasing \pt to reach the calorimeter energy resolution for track momenta of several hundred \GeV.
Because the fragmentation of high-\pt partons typically produces many
charged hadrons at a lower \pt, the tracker is expected to contribute significantly to
the measurement of the momentum of jets with a \pt up to a few TeV.
\subsection{The electromagnetic calorimeter}

The ECAL~\cite{CMS:1997ema,CMS:2002xia} is a hermetic homogeneous calorimeter
made of lead tungstate (PbWO$_4$) crystals. The barrel covers $\abs{\eta} < 1.479$
and the two endcap disks $1.479 < \abs{\eta} < 3.0$ . The barrel (endcap) crystal
length of 23 (22)\unit{cm} corresponds to 25.8 (24.7) radiation lengths,
sufficient to contain more than 98\% of the energy of electrons and photons
up to 1\TeV. The crystal material also amounts to about one interaction
length, causing about two thirds of the hadrons to start showering in the ECAL
before entering the HCAL.

The crystal transverse size  matches the small Moli\`ere radius of PbWO$_4$,
{2.2\unit{cm}}. This fine transverse granularity makes it possible
to fully resolve hadron and photon energy deposits as close as 5\unit{cm} from one
another, for the benefit of exclusive particle identification in jets.
More specifically, the front face of the barrel
crystals has an area of $2.2{\times}2.2\unit{cm}^2$, equivalent to
$0.0174{\times}0.0174$ in the ($\eta,\varphi$) plane. In the endcaps,
the crystals are arranged instead in a rectangular $(x,y)$ grid, with a
front-face area of $2.9{\times}2.9\unit{cm}^2$. The intrinsic energy resolution
of the ECAL barrel was measured with an ECAL supermodule directly exposed to an
electron beam, without any attempt to reproduce the material of the tracker in
front of the ECAL~\cite{Adzic:2007mi}. The relative energy resolution
is parameterized as a function of the electron energy as
\begin{linenomath}
\begin{equation}
\frac{\sigma}{E}  = \frac{2.8\%}{\sqrt{E/\GeV}}  \oplus \frac{12\%}{E/\GeV}  \oplus 0.3\%.
\label{eq:ECALresolution}
\end{equation}
\end{linenomath}
Because of the very small stochastic term inherent to homogeneous calorimeters,
the photon energy resolution is excellent in the 1--50\GeV range typical of
photons in jets.

The ECAL electronics noise $\sigma_\text{noise}^\mathrm{ECAL}$ is measured to
be about $40\,(150)\MeV$  per crystal in the barrel (endcaps). Another important
source of spurious signals arises from particles directly ionizing the
avalanche photodiodes (APD), aimed at collecting the crystal scintillation
light~\cite{1748-0221-8-03-C03020}.
This effect gives rise to single-crystal spikes with a relative amplitude about
$10^5$ times larger than the scintillation light. Such spikes would be
misidentified by the PF algorithm as photons with an energy up to 1\TeV.
Since these spikes mostly affect a single crystal and more rarely two
neighbouring crystals, they are rejected by requiring the energy deposits
to be compatible with arising from a particle shower:
the ratios $E_4/E_1$ and $E_6/E_2$ should exceed 5\% and 10\% respectively,
where $E_1$ ($E_2$) is the energy collected in the considered crystal
(crystal pair) and $E_4$ ($E_6$) is the energy collected in the four (six)
adjacent crystals. The timing of the energy deposits in excess of 1\GeV
is also required to be compatible with the beam crossing time to better
than $\pm$2\unit{ns}.

A much finer-grained detector, known as preshower, is installed in front
of each endcap disk. It consists of two layers, each comprising a lead
radiator followed by a plane of silicon strip sensors. The two lead radiators
represent approximately two and one radiation lengths, respectively. The two
planes of silicon sensors have orthogonal strips with a pitch of 1.9\unit{mm}. When
either a photon or an electron passes through the lead, it initiates an
electromagnetic shower. The granularity of the detector and the small radius
of the initiating shower provide an accurate measurement of the shower
position. Originally, the aim of the superior granularity of the preshower was
twofold: \textit{(i)} resolve the photons from $\pi^0$ decays so as to
discriminate them from prompt photons; and \textit{(ii)} indicate the
presence of a photon or an electron in the ECAL by requiring an associated
signal in the preshower.
Parasitic signals, however, are generated by the large number of neutral pions
produced by hadron interactions in the tracker material, followed
by photon conversions and electron bremsstrahlung. These signals substantially
affect the preshower identification and separation capabilities. In the
PF algorithm, these capabilities can therefore not be fully exploited, and
the energy deposited in the preshower is simply added to that of the closest
associated ECAL cluster, if any, and discarded otherwise.

\subsection{The hadron calorimeter}

The HCAL~\cite{CMS:1997tfa} is a hermetic sampling calorimeter consisting of
several layers of brass absorber and plastic scintillator tiles. It surrounds
the ECAL, with a barrel ($\abs{\eta}<1.3$) and two endcap disks ($1.3<\abs{\eta}<3.0$).
In the barrel, the HCAL absorber thickness amounts to almost six interaction
lengths at normal incidence, and increases to over ten interaction lengths at
larger pseudorapidities. It is complemented by a tail catcher (HO), installed
outside the solenoid coil. The HO material (1.4 interaction lengths at
normal incidence) is used as an additional absorber. At small pseudorapidities
($\abs{\eta}<0.25$), this thickness is enhanced to a total of three interaction
lengths by a 20\unit{cm}-thick layer of steel. The total depth of the
calorimeter system (including ECAL) is thus extended to a minimum of twelve
interaction lengths in the barrel. In the endcaps, the thickness amounts to
about ten interaction lengths.

The HCAL is read out in individual towers with a cross section
$\Delta \eta{\times}\Delta \varphi = 0.087{\times}0.087$ for $\abs{\eta}<1.6$
and $0.17{\times}0.17$ at larger pseudorapidities. The combined (ECAL+HCAL) calorimeter
energy resolution was measured in a pion test beam~\cite{ehcal_test_beam} to be
\begin{linenomath}
\begin{equation}
\frac{\sigma}{E} = \frac{110\%}{\sqrt{E}} \oplus 9\%,
\label{eq:HCALresolution}
\end{equation}
\end{linenomath}
where $E$ is expressed in \GeV.

The typical HCAL electronics noise $\sigma^\mathrm{HCAL}_\text{noise}$  is measured
to be $\approx 200\,\MeV$ per tower. Additionally, rare occurrences of
high-amplitude, coherent noise were observed in the HCAL
barrel~\cite{hcal_noise}. This coherent noise was understood as follows.
The barrel is made of two half-barrels covering positive and negative $z$,
respectively. Each half barrel is made of 18 identical azimuthal wedges,
each of which contains four rows of 18 towers with the same $\varphi$ value.
All towers in a row are read out by a single pixelated hybrid photodiode
(HPD). The four HPDs serving a wedge are installed in a readout box (RBX).
Discharges in the HPD affect blocks of up to 18 cells at the same $\varphi$
value in a half-barrel, while a global pedestal drifting in an RBX may affect
all 72 towers in the wedge. Since this coherent HCAL noise would be
misinterpreted as high-energy neutral hadrons by the PF algorithm, the
affected events are identified by their characteristic topological features
and rejected at the analysis level.

The HCAL is complemented by hadron forward (HF) calorimeters situated at
$\pm 11$\unit{m} from the interaction point that extend the angular coverage
on both sides up to $\abs{\eta} \simeq 5$. The HF consists of a steel absorber
composed of grooved plates. Radiation-hard quartz fibres are inserted in the
grooves along the beam direction and are read out by photomultipliers.
The fibres alternate between long fibres running over the full thickness of
the absorber (about 165\unit{cm}, corresponding to typically ten interaction
lengths), and short fibres covering the back of the absorber and starting
at a depth of 22\,\cm from the front face. The signals from short and
long fibres are grouped so as to define calorimeter towers with a cross section
$\Delta \eta{\times}\Delta \varphi = 0.175{\times}0.175$ over most of the
pseudorapidity range.  In each calorimeter tower, the signals from the short
and long fibres are used to estimate the electromagnetic and hadronic
components of the shower. If $L$ ($S$) denotes the energy measured in the
long (short) fibres,  the energy of the electromagnetic component, concentrated
in the first part of the absorber, can be approximated by $L-S$, and the
energy of the hadronic component is the complement, \ie $2S$. Spurious
signals in the HF, caused for example by high-energy beam-halo muons directly
hitting the photomultiplier windows, are reduced by rejecting \textit{(i)}
high-energy $S$ deposits not backed up by a $L$ deposit in the same tower;
\textit{(ii)} out-of-time $S$ or $L$ deposits of more than 30\GeV,
\textit{(iii)} $L$ deposits larger than 120\GeV with $S < 0.01 L$ in the same
tower; \textit{(iv)} isolated $L$ deposits larger than 80\GeV,
with small $L$ and $S$ deposits in the four neighbouring towers.

\subsection{The muon detectors}

Outside the solenoid coil, the magnetic flux is returned through a yoke
consisting of three layers of steel interleaved with four muon detector
planes~\cite{CMS:1997dma,cms_muon_paper}. Drift tube (DT) chambers and
cathode strip chambers (CSC) detect muons in the regions $\abs{\eta}<1.2$ and
$0.9 <\abs{\eta}< 2.4$, respectively, and are complemented by a system of
resistive plate chambers (RPC) covering the range $\abs{\eta}<1.6$. The
reconstruction, described in Section~\ref{sec:reco_muons}, involves a global
trajectory fit across the muon detectors and the inner tracker. The
calorimeters and the solenoid coil represent a large amount of material
before the muon detectors and thus induce multiple scattering. For this
reason, the inner tracker dominates the momentum measurement up to a \pt
of about 200\GeV.

\section{Reconstruction of the particle-flow elements}
\label{sec:reconstruction_pf_elements}

This section describes the advanced algorithms specifically set up
for the reconstruction of the basic PF elements: the reconstruction of the
trajectories of charged particles in the inner tracker is discussed first;
the specificities of electron and muon track reconstruction are then introduced;
finally, the reconstruction and the calibration of calorimeter clusters in the
preshower, the ECAL, and the HCAL, are presented.

\subsection{Charged-particle tracks and vertices}
\label{sec:charged_particles_tracks_and_vertices}

Charged-particle track reconstruction was originally
aimed~\cite{Bayatian:922757} at measuring the momentum of energetic and
isolated muons, at identifying energetic and isolated hadronic $\Pgt$
decays, and at tagging b quark jets. Tracking was therefore primarily
targeting energetic particles and was limited to well-measured tracks.
A combinatorial track finder based on Kalman Filtering (KF)~\cite{Adam:934067}
was used to reconstruct these tracks in three stages: initial seed
generation with a few hits compatible with a charged-particle
trajectory; trajectory building (or pattern recognition) to gather hits
from all tracker layers along this charged-particle trajectory; and final
fitting to determine the charged-particle properties: origin, transverse
momentum, and direction. To be kept for further analysis, the tracks
had to be seeded with two hits in consecutive layers in the pixel detector, and
were required to be reconstructed with at least eight hits in total (each
contributing to less than 30\% of the overall track goodness-of-fit $\chi^2$) and
with at most one missing hit along the way. In addition, all tracks were
required to originate from within a cylinder of a few mm radius centred
around the beam axis and to have $\pt$ larger than $0.9\GeV$.

\begin{figure}[t]
\centering
\includegraphics[width=0.49\columnwidth]{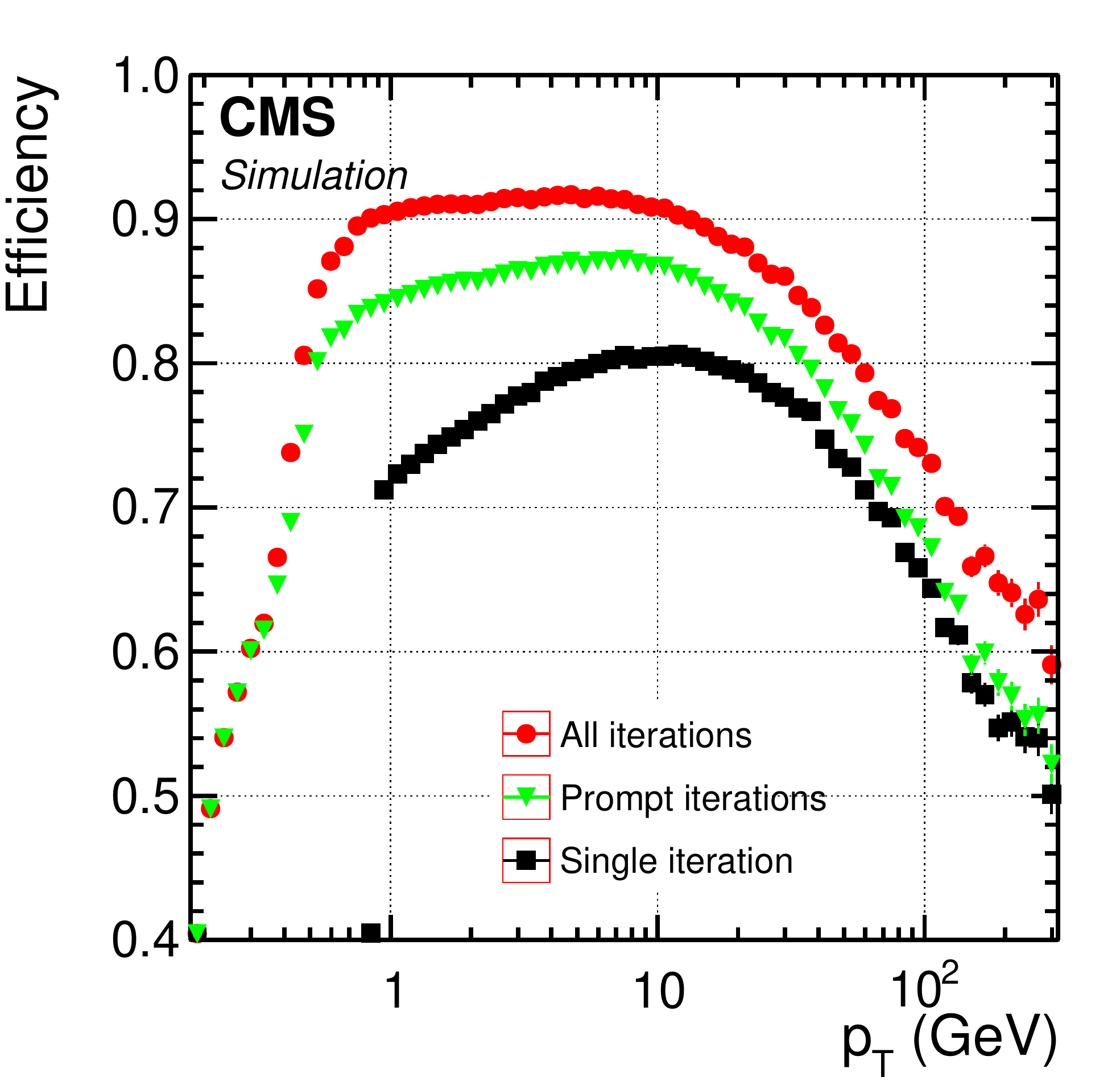}
\includegraphics[width=0.49\columnwidth]{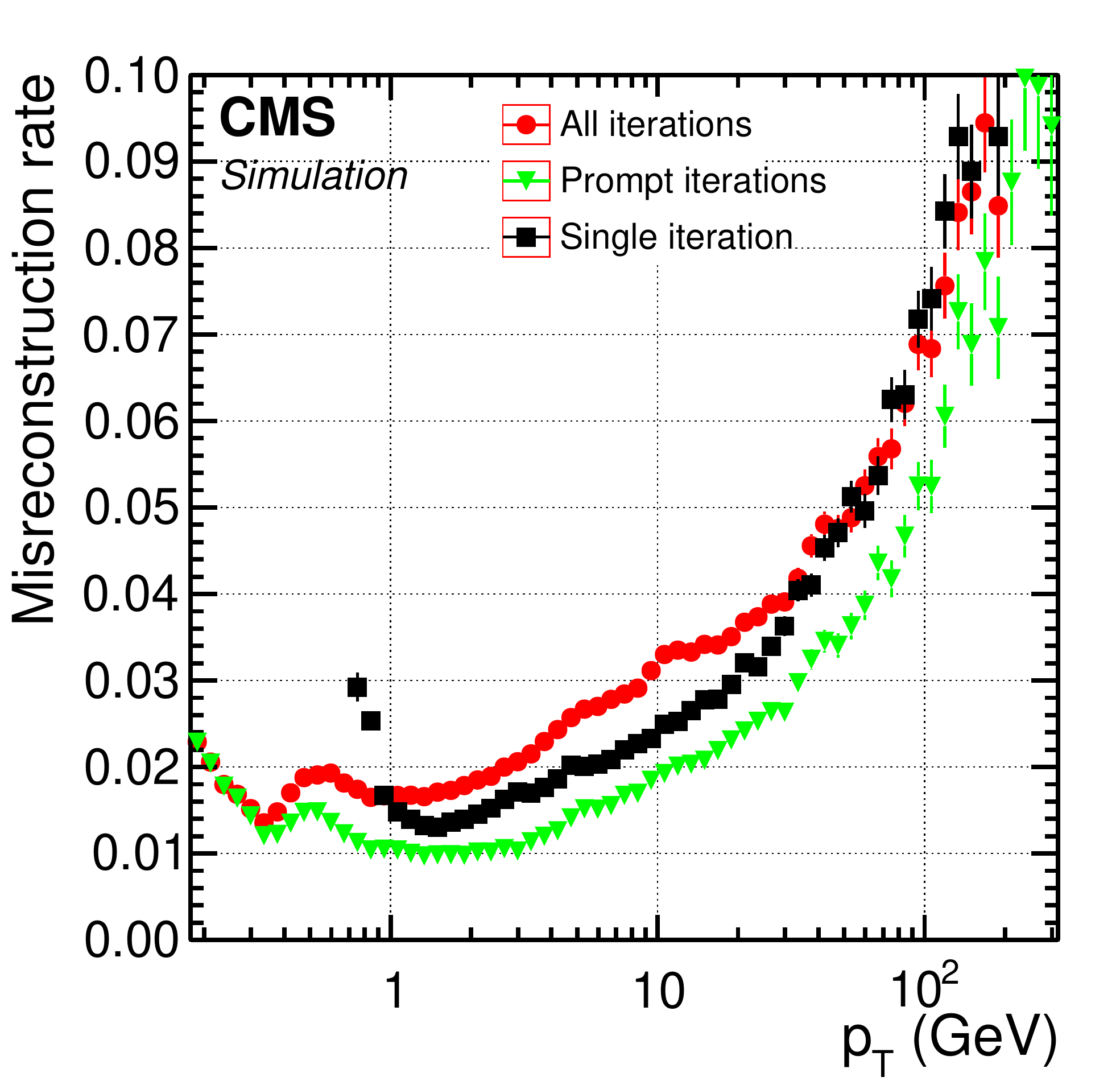}

\caption{\label{fig:IterativeTrackingEfficiency} Efficiency (left) and
misreconstruction rate (right) of the global combinatorial track finder
(black squares); and of the iterative tracking method (green triangles: prompt
iterations based on seeds with at least one hit in the pixel detector; red
circles: all iterations, including those with displaced seeds), as a
function of the track \pt, for charged hadrons in multijet events without pileup interactions.
Only tracks with $\abs{\eta} < 2.5$ are considered in the efficiency and
misreconstruction rate determination. The efficiency is displayed for
tracks originating from within 3.5\unit{cm} of the beam axis and $\pm 30$\unit{cm}
of the nominal centre of CMS along the beam axis.}
\end{figure}

The performance in terms of reconstruction efficiency and misreconstruction
rate of this global combinatorial track finder can be found in
Ref.~\cite{Bayatian:922757} for muons and charged pions within jets and is
shown in Fig.~\ref{fig:IterativeTrackingEfficiency} for charged hadrons in a sample of
simulated QCD multijet events as a function of the reconstructed track $\pt$.
The efficiency is defined as the fraction of simulated tracks reconstructed with at
least 50\% of the associated simulated hits, and with less than 50\% of unassociated
simulated hits. The misreconstruction rate is the fraction of reconstructed tracks
that cannot be associated with a simulated track.
The stringent track quality criteria are instrumental in keeping the
misreconstructed track rate at the level of a few per cent, but limit the
reconstruction efficiency to only 70--80\% for charged pions with $\pt$
above $1\GeV$, compared to 99\% for isolated muons. Below a few tens of \GeV,
the difference between pions and muons is almost entirely accounted
for by the possibility for pions to undergo a nuclear interaction within
the tracker material. For a charged particle to accumulate eight hits along
its trajectory, it must traverse the beam pipe, the pixel detector, the
inner tracker, and the first layers of the outer tracker before the
first significant nuclear interaction. The probability for a hadron to
interact within the tracker material, before reaching the eight-hits
threshold---causing the track to be missed---can be inferred from
Fig.~\ref{fig:cms_detector:tracker_material} (left) and ranges between 10 and
30\%. The tracking efficiency is further reduced for $\pt$ values above
10\GeV: these high-$\pt$ particles are found mostly in collimated
jets, in which the tracking efficiency is limited by the silicon detector
pitch, \ie by the capacity to disentangle hits from overlapping particles.

Each charged hadron missed by the tracking algorithm would be solely (if at
all) detected by the calorimeters as a neutral hadron, with reduced efficiency,
largely degraded energy resolution, and biased direction due to the bending of
its trajectory in the magnetic field. As two thirds of the energy in a jet are
on average carried by charged hadrons, a 20\% tracking inefficiency would
double the energy fraction of identified neutral hadrons in a jet from 10\% to
over 20\% and therefore would degrade the jet energy and angular resolutions---expected from PF reconstruction to be dominated by the modest
neutral-hadron energy resolution---by about 50\%. Increasing the track
reconstruction efficiency while keeping the misreconstructed rate unchanged
is therefore critical for PF event reconstruction.

The tracking inefficiency can be substantially reduced by accepting tracks with
a smaller $\pt$ (to recover charged particles with little probability to
deposit any measurable energy in the calorimeters) and with fewer hits (to
catch particles interacting with the material of the tracker inner layers).
This large improvement, however, comes at the expense of an
exponential increase of the combinatorial rate of misreconstructed
tracks~\cite{2009NuPhS.197..275A}: the misreconstruction rate is multiplied
by a factor of five when the $\pt$ threshold is loosened to 300\MeV and
increases by another order of magnitude when the total number of hits
required to make a track is reduced to five. It reaches a value of up to
80\% when the two criteria are loosened together. These misreconstructed
tracks, made of randomly associated hits, have randomly
distributed momenta and thus would cause large energy excesses in PF
reconstruction.

\subsubsection{Iterative tracking}
\label{sec:iterative}

To increase the tracking efficiency while keeping the misreconstructed track
rate at a similar level, the combinatorial track finder was applied in several
successive iterations~\cite{cms_tracking_paper}, each with moderate efficiency
but with as high a purity as possible. At each step, the reduction of the
misreconstruction rate is accomplished with quality criteria on the track
seeds, on the track fit $\chi^2$, and on the track compatibility with
originating from one of the reconstructed primary vertices, adapted to
the track $\pt$, $\abs{\eta}$, and number of hits $n_\text{hits}$. In
practice, no quality criteria are applied to tracks reconstructed with
at least eight hits, as the misreconstruction rate is already small enough
for these tracks. The hits associated with the selected tracks are masked in
order to reduce the probability of random hit-to-seed association in the
next iteration. The remaining hits may thus be used in the next iteration
to form new seeds and tracks with relaxed quality criteria, increasing in
turn the total tracking efficiency without degrading the purity. The
same operation is repeated several times with progressively more complex
and time-con\-su\-ming seeding, filtering, and tracking algorithms.

\begin{table}[b]
\topcaption{Seeding configuration and targeted tracks of the ten tracking
iterations. In the last column, $R$ is the targeted distance between the
track production position and the beam axis.}\label{tab:iterations}
  \centering
    \begin{tabular}{c l l l}
      \hline
      Iteration &      Name    &      Seeding          & Targeted Tracks \\
      \hline
      1     & InitialStep      & pixel triplets        & prompt, high \pt \\
      2     & DetachedTriplet  & pixel triplets        & from b hadron decays,
$R\lesssim 5\unit{cm}$ \\
      3     & LowPtTriplet     & pixel triplets        & prompt, low  \pt \\
      4     & PixelPair        & pixel pairs           & recover high \pt \\
      5     & MixedTriplet     & pixel+strip triplets  & displaced,
$R\lesssim 7\unit{cm}$ \\
      6     & PixelLess        & strip triplets/pairs  & very displaced,
$R\lesssim 25\unit{cm}$  \\
      7     & TobTec           & strip triplets/pairs  & very displaced,
$R\lesssim 60\unit{cm}$  \\
      8     & JetCoreRegional  & pixel+strip pairs     & inside high \pt~jets \\
      9     & MuonSeededInOut  & muon-tagged tracks    & muons  \\
     10     & MuonSeededOutIn  & muon detectors        & muons  \\
      \hline
    \end{tabular}

\end{table}

The seeding configuration and the targeted tracks of each of the ten iterations
are summarized in Table~\ref{tab:iterations}. The tracks from the first three
iterations are seeded with triplets of pixel hits, with additional criteria
on their distance of closest approach to the beam axis. The resulting
high purity allows the requirements on $n_\text{hits}$ and on the track $\pt$
to be loosened to typically three and 200\MeV, respectively. With an overall
efficiency of $\sim$80\%, the fractions of hits masked for the next iterations
amount to 40\% (20\%) in the pixel (strip) detector. The fourth and fifth
iterations aim at recovering tracks with one or two missing hits in the
pixel detector. They address mostly detector inefficiencies, but also
particle interactions and decays within the pixel detector volume. The next
two iterations are designed to reconstruct very displaced tracks. Without
pixel hits to seed the tracks, they can only be processed after the first five
iterations, which offer an adequate reduction of the number of leftover hits
in the strip detector. The eighth iteration addresses specifically the dense
core of high-\pt jets. In these jets, hits from nearby tracks may merge and
be associated with only one track---or even none because of their poorly
determined position---causing the tracking efficiency to severely decrease.
Merged pixel hit clusters, found in narrow regions compatible with the
direction of high-energy deposits in the calorimeters, are split into
several hits. Each of these hits is paired with one of the remaining hits in
the strip detector to form a seed for this iteration. The last two iterations
are specifically designed to increase the muon-tracking reconstruction
efficiency with the use of the muon detector information in the seeding step.

As shown in
Fig.~\ref{fig:IterativeTrackingEfficiency}, the \textit{prompt} iterations,
which address tracks seeded with at least one hit in the pixel detector
(iterations 1, 2, 3, 4, 5, and 7), recover about half of the tracks with
$\pt$ above 1\GeV missed by the global combinatorial track finder,
with slightly smaller misreconstruction rate levels. These iterations also
extend the acceptance to the numerous particles with $\pt$ as small
as 200\MeV, typically below the calorimeter thresholds. (Particles with a \pt
between 200 and 700\MeV never reach the calorimeter barrel, but follow a
helical trajectory to one of the calorimeter endcaps.) With such performance,
and also because track reconstruction was found to be twice as fast with
several iterations than in a single step (because of the much smaller
number of seeds identified at each step), iterative tracking quickly
became the default method for CMS. Despite the significant improvement,
the tracking efficiency at high \pt remains limited. The consequences
for jet energy and angular resolutions are minute, as the calorimeter
resolutions are already excellent at these energies. The significant
increase of the misreconstructed track rate at high \pt is dealt with
when the information from the calorimeters and the muon system becomes
available, as described in Section~\ref{sec:particle_id_reco}.

\subsubsection{Nuclear interactions in the tracker material}
\label{sec:reco_pf_elements_nuclint}

Nuclear interactions in the tracker material may lead to either a kink
in the original hadron trajectory, or to the production of a number of
secondary particles. On average, two thirds of these secondary particles
are charged. Their reconstruction efficiency is enhanced by the sixth
and seventh iterations of the iterative tracking. The tracking efficiency and
misreconstruction rate with all iterations included are displayed in
Fig.~\ref{fig:IterativeTrackingEfficiency}. While the displaced-track
iterations typically add 5\% to the tracking efficiency, they also increase
the total misreconstruction rate by 1\% for tracks with $\pt$ between
1 and 20\GeV. The relative misreconstruction rate of these iterations
is therefore at the level of 20\%.

\begin{figure}[htbp]
\centering
\includegraphics[width=0.48\textwidth]{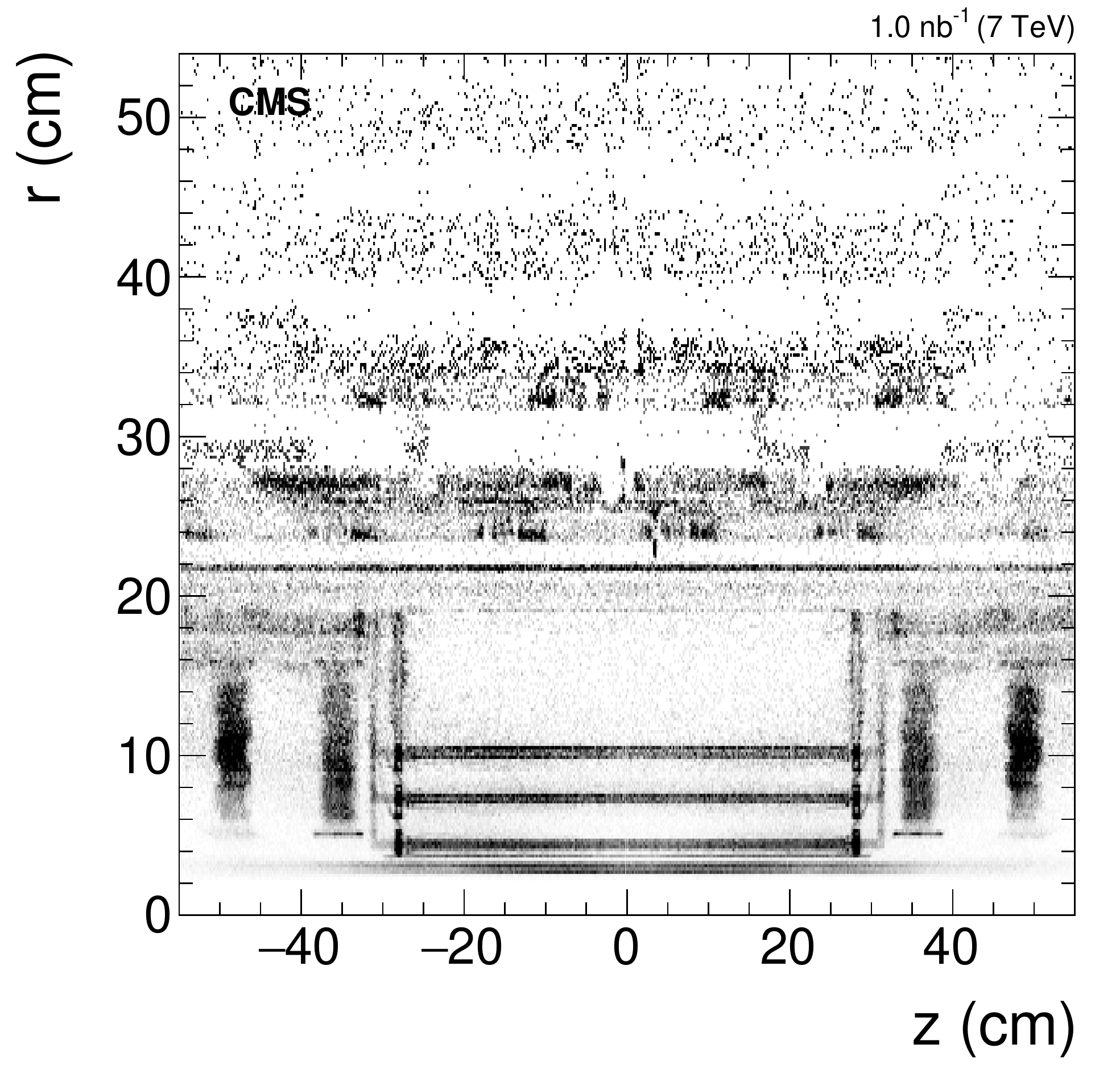}
\includegraphics[width=0.48\textwidth]{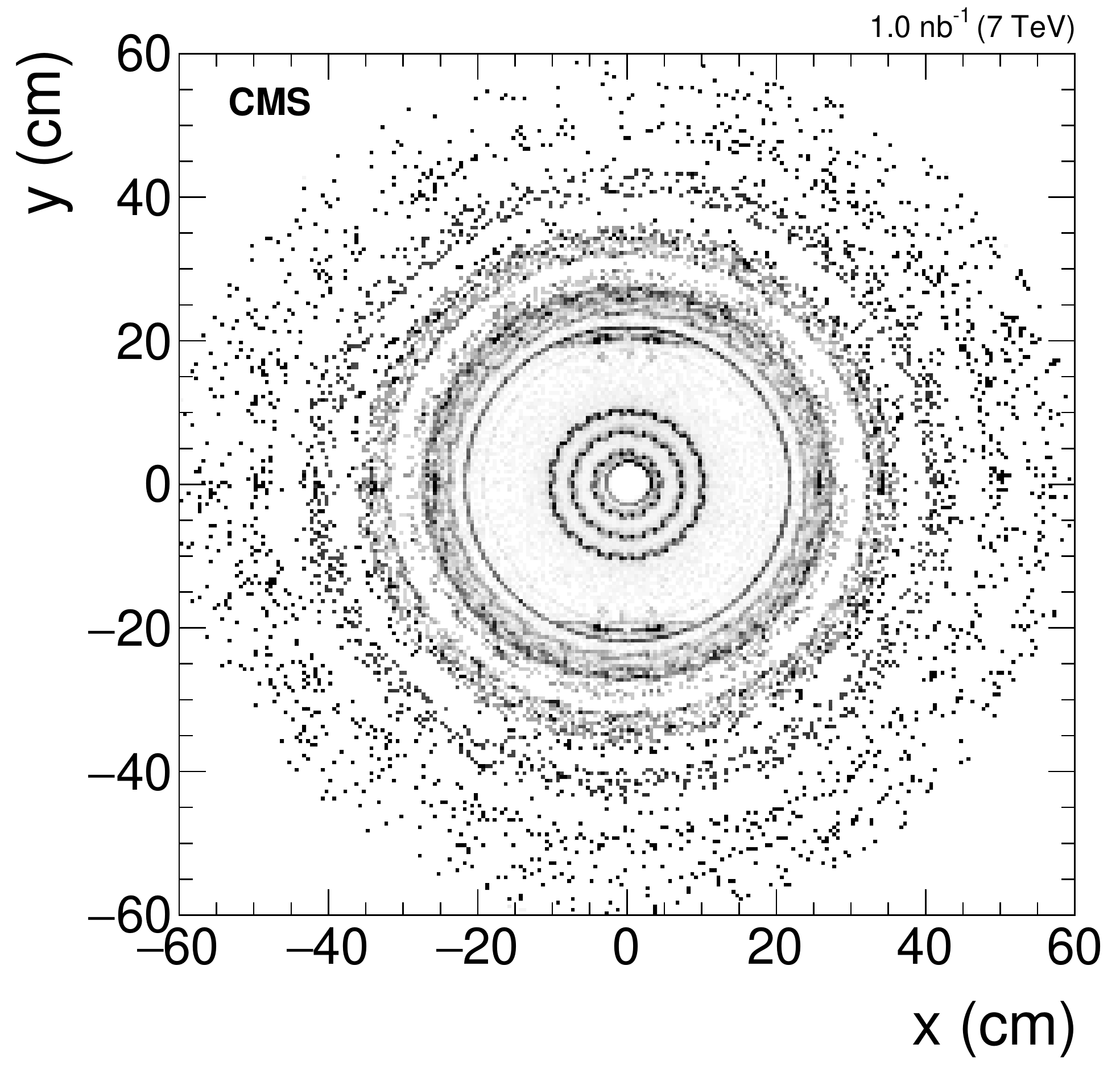}
\caption{\label{fig:NuclearInteractions} Maps of nuclear interaction
vertices for data collected by CMS in 2011 at $\sqrt{s} = 7$\TeV,
corresponding to an integrated luminosity of 1\nbinv,
in the longitudinal (left) and transverse (right) cross sections of the
inner part of the tracker, exhibiting its structure in concentric layers
around the beam axis.}
\end{figure}

A dedicated algorithm was thus developed to identify tracks linked
to a common secondary displaced vertex within the tracker
volume~\cite{CMS-PAS-TRK-10-003,CMSTrackerCommissionning}.
Figure~\ref{fig:NuclearInteractions} shows the positions of these reconstructed
nuclear interaction vertices in the inner part of the tracker.
The observed pattern matches well the tracker layer structure and material.
The misreconstruction rate is further reduced with a specific
treatment of these tracks in the PF algorithm, described in
Section~\ref{sec:particle_id_reco}.

\subsection{Tracking for electrons}
\label{sec:reco_electrons}

Electron reconstruction, originally aimed at characterizing energetic,
well-isolated electrons, was naturally based on the ECAL
measurements, without emphasis on the tracking capabilities. More
specifically, the traditional electron seeding strategy (hereafter called the
\textit{ECAL-based} approach)~\cite{Khachatryan:2015hwa} makes use of energetic
ECAL clusters ($\ET > 4$\GeV). The cluster energy and position
are used to infer the position of the hits expected in the innermost
tracker layers under the assumptions that the cluster is produced either
by an electron or by a positron. Because of the significant tracker thickness
(Fig.~\ref{fig:cms_detector:tracker_material} right), most of the electrons
emit a sizeable fraction of their energy in the form of bremsstrahlung
photons before reaching the ECAL.
The performance of the method therefore depends on the ability to gather
all the radiated energy, and only that energy.
The energy of the electron and of possible bremsstrahlung photons is
collected by grouping into a \textit{supercluster}
the ECAL clusters reconstructed in a small window in $\eta$ and
an extended window in $\varphi$ around the electron direction
(to account for the azimuthal bending of the electron in the magnetic field).

For electrons in jets, however, the energy and position of the associated
supercluster are often biased by the overlapping contributions from other
particle deposits, leading to large inefficiencies. In addition,
the backward propagation from the supercluster to the interaction region is
likely to be compatible with many hits from other charged particles in
the innermost tracker layers, causing a substantial misreconstruction rate.
To keep the latter under control, the ECAL-based
electron seeding efficiency has to be further limited, \eg by strict
isolation requirements, to values that are unacceptably small in jets
when a global event description is to be achieved. Similarly, for
electrons with small $\pt$, whose tracks are significantly bent
by the magnetic field, the radiated energy is spread over such an
extended region that the supercluster cannot include all deposits.
The missed deposits bias the position of the supercluster and prevent it
from being matched with the proper hits in the innermost tracker layers.

To reconstruct the electrons missed by the ECAL-based approach, a
\textit{tracker-based} electron seeding method was developed  in the context
of PF reconstruction. The iterative tracking (Section \ref{sec:iterative})
is designed to have a large efficiency for these electrons: nonradiating
electrons can be tracked as efficiently as muons and radiating electrons
produce either shorter or lower \pt tracks largely recovered by the
loose requirements on the number of hits and on the \pt to form a
track. All the tracks from the iterative tracking are therefore used as
potential seeds for electrons, if their $\pt$ exceeds 2\GeV.

The large probability for electrons to radiate in the tracker material is
exploited to disentangle electrons from charged hadrons. When the energy
radiated by the electron is small, the corresponding track can be
reconstructed  across the whole tracker with a well-behaved $\chi^2$ and be
safely propagated to the ECAL inner surface, where it can be matched with the
closest ECAL cluster. (Calorimeter clustering and track-cluster matching in PF
are described in Sections~\ref{sec:reconstruction_pf_elements_caloclusters}
and~\ref{sec:particle_id_reco_link}, respectively.) For these tracks to form
an electron seed, the ratio of the cluster energy to the track momentum is
required to be compatible with unity. In the case of soft photon emission,
the pattern recognition may still succeed in collecting most hits along
the electron trajectory, but the track fit generally leads to a large
$\chi^2$ value. When energetic photons are radiated, the pattern recognition
may be unable to accommodate the change in electron momentum,  causing the
track to be reconstructed with a small number of hits. A preselection
based on the number of hits and the fit $\chi^2$ is therefore applied and
the selected tracks are fit again with a Gaussian-sum filter (GSF)~\cite{GSF}.
The GSF fitting is more adapted to electrons than the KF used in the
iterative tracking, as it allows for sudden and substantial
energy losses along the trajectory. At this stage, a GSF with only
five components is used, in order to keep the computing time under
control. A final requirement is applied to the score of a
boosted-decision-tree (BDT) classifier that combines the discriminating
power of the number of hits, the $\chi^2$ of the GSF track fit and its
ratio to that of the KF track fit, the energy lost along the GSF track,
and the distance between the extrapolation of the track to the ECAL inner
surface and the closest ECAL cluster.

\begin{figure}[htbp]
\centering
\includegraphics[width=0.49\columnwidth]{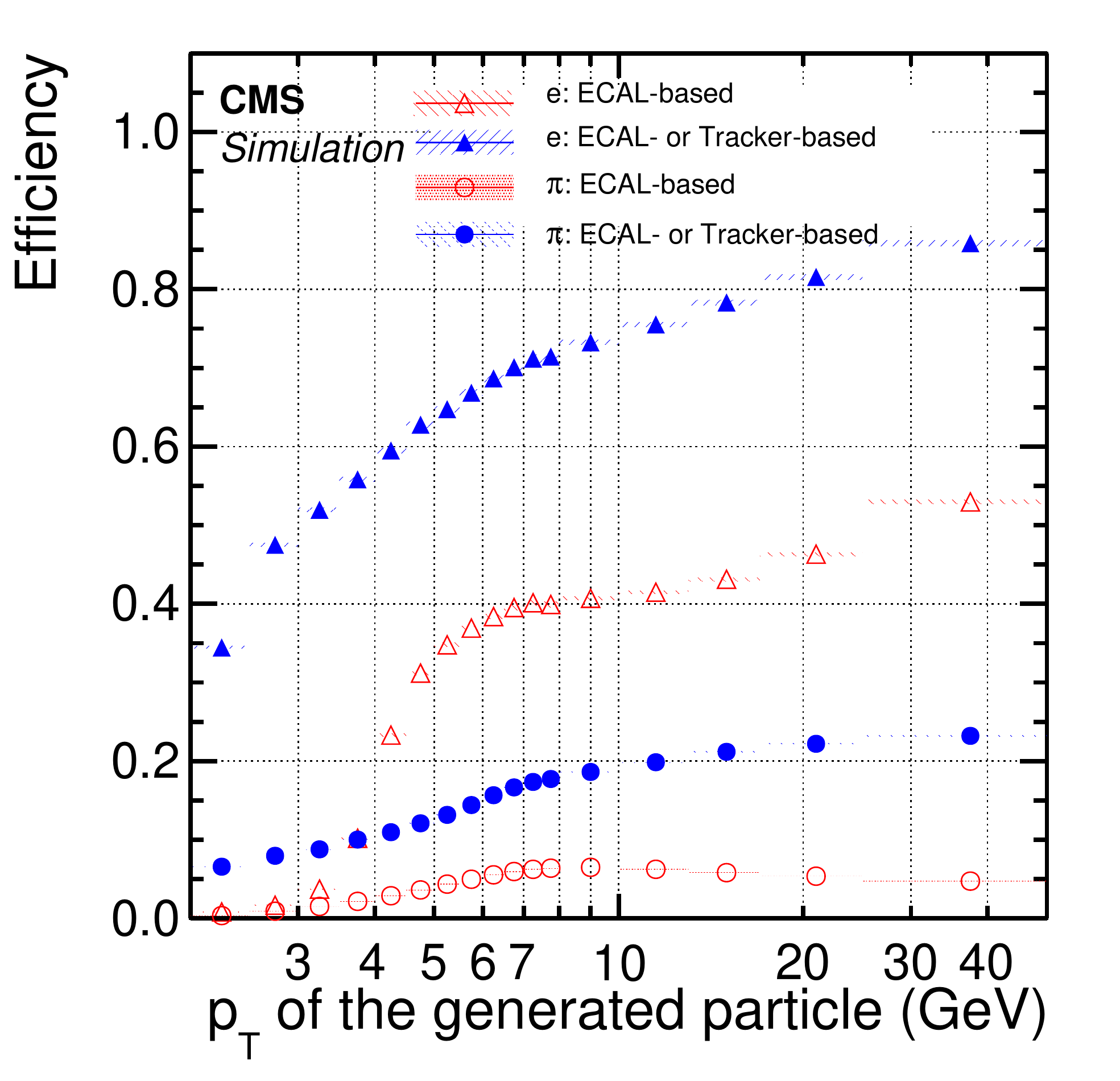}
\includegraphics[width=0.49\columnwidth]{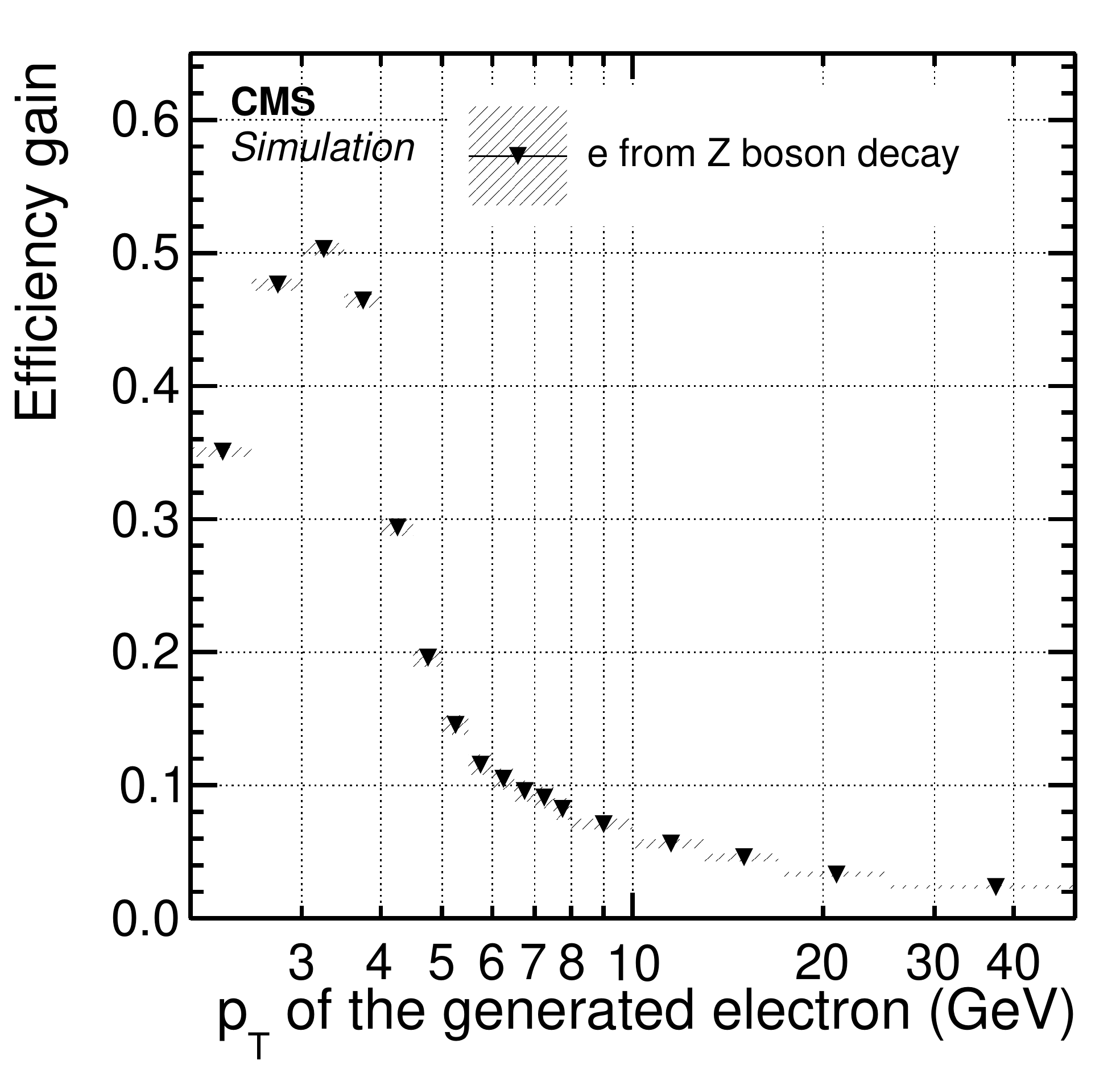}

\caption{\label{fig:ElectronSeedingEfficiency} Left: Electron seeding
efficiency for electrons (triangles) and pions (circles) as a function
of $\pt$, from a simulated event sample enriched in b quark jets
with \pt between 80 and 170\GeV, and with at least one semileptonic b
hadron decay. Both the efficiencies for ECAL-based seeding only
(hollow symbols) and with the tracker-based seeding added (solid symbols)
are displayed. Right: Absolute efficiency gain from the tracker-based seeding
for electrons from Z boson decays as a function of $\pt$.
The shaded bands indicate the \pt bin size and the statistical uncertainties
on the efficiency.
}
\end{figure}

The electron seeds obtained with the tracker- and ECAL-based procedures are
merged into a unique collection and are submitted to the full electron
tracking with twelve GSF components. The significant increase of seeding
efficiency brought by the tracker-based approach is shown in the left panel
of Fig.~\ref{fig:ElectronSeedingEfficiency} for electrons in b quark jets.
The probability for a charged hadron to give rise to an electron seed is
displayed in the same figure. At this preselection stage, the addition of
the tracker-based seeding almost doubles the electron efficiency and extends
the electron reconstruction down to a \pt of 2\GeV. These improvements
come with an increase of misidentification rate, dealt with at a later
stage of the PF reconstruction, when more information becomes available
(Section~\ref{sec:particle_id_reco_electrons}). Here, the misidentification
rate is only a concern for the electron track reconstruction computing time,
kept within reasonable limits by the preselection. For isolated
electrons, the ECAL-based seeding is already quite effective, but the
tracker-based seeding improves the overall efficiency by several per
cent, as shown in the right panel of Fig.~\ref{fig:ElectronSeedingEfficiency},
and makes it possible to reconstruct electrons with a  $\pt$ below
4\GeV.

The tracker-based seeding is also effective at selecting electrons and
positrons from conversions in the tracker material, for both prompt and
bremsstrahlung photons. The recovery of the converted photons of the latter
category and their association to their parent electrons is instrumental in
minimizing energy double counting in the course of the PF reconstruction.

\subsection{Tracking for muons}
\label{sec:reco_muons}

Muon tracking~\cite{Bayatian:922757,cms_muon_paper} is not specific to PF
reconstruction. The muon spectrometer allows muons to be identified with high
efficiency over the full detector acceptance. A high purity is granted by
the upstream calorimeters, meant to absorb other particles (except neutrinos).
The inner tracker provides a precise measurement of the momentum of these
muons. The high-level muon physics objects are reconstructed in a multifaceted
way, with the final collection being composed of three different muon types:

\begin{itemize}
 \item \textit{standalone muon}.  Hits within each DT or CSC detector
are clustered to form track segments, used as seeds for the pattern recognition
in the muon spectrometer, to gather all DT, CSC, and RPC hits along the muon
trajectory. The result of the final fitting is called a \textit{standalone-muon
track}.
 \item \textit{global muon}.  Each standalone-muon track is matched to
a track in the inner tracker (hereafter referred to as an \textit{inner track})
if the parameters of the two tracks propagated onto a common
surface are compatible. The hits from the inner track and from the
standalone-muon track are combined and fit to form a \textit{global-muon track}.
 At large transverse momenta, $\pt \gtrsim 200\GeV$, the global-muon fit
improves the momentum resolution with respect to the tracker-only fit.
 \item \textit{tracker muon}. Each inner track with $\pt$ larger than 0.5\GeV
and a total momentum $p$ in excess of 2.5\GeV is extra\-po\-lated to the
muon system. If at least one muon segment matches the extrapolated
track, the inner track qualifies as a \textit{tracker muon track}.
The track-to-segment matching is performed in a local $(x,y)$ coordinate
system defined in a plane transverse to the beam axis, where $x$ is the
better measured coordinate. The extrapolated track and the segment are
 matched either if the absolute value of the difference between their
positions in the $x$ coordinate is smaller than 3\unit{cm}, or if the ratio
of this distance to its uncertainty (\textit{pull}) is smaller than 4.
\end{itemize}

Global-muon reconstruction is designed to have high efficiency for muons
penetrating through more than one muon detector plane. It typically requires
segments to be associated in at least two muon detector planes. For
momenta below about 10\GeV, this requirement fails more often because
of the larger multiple scattering in the steel of the return yoke. For these
muons, the tracker muon reconstruction is therefore more efficient, as it
requires only one segment in the muon system~\cite{MUO-10-004}.

Owing to the high efficiency of the inner track and muon segment
reconstruction, about 99\% of the muons produced within the geometrical
acceptance of the muon system are reconstructed either as a global muon or
a tracker muon and very often as both. Global muons and tracker muons
that share the same inner track are merged into a single candidate. Muons
reconstructed only as standalone-muon tracks have worse momentum resolution
and a higher admixture of cosmic muons than global and tracker muons.

Charged hadrons may be misreconstructed as muons \eg if some of the
hadron shower remnants reach the muon system (\textit{punch-through}).
Different identification criteria can be
applied to the muon tracks in order to obtain the desired balance between
identification efficiency and purity. In the PF muon identification algorithm
(Section~\ref{sec:particle_id_reco_muons}), muon energy deposits in ECAL, HCAL,
and HO are associated with the muon track and this information is used to
improve the muon identification performance.

\subsection{Calorimeter clusters}
\label{sec:reconstruction_pf_elements_caloclusters}

The purpose of the clustering algorithm in the calorimeters is fourfold:
\textit{(i)} detect and measure the energy and direction of stable neutral
particles such as photons and neutral hadrons; \textit{(ii)} separate these
neutral particles from charged hadron energy deposits; \textit{(iii)} reconstruct
and identify electrons and all accompanying bremsstrahlung photons; and
\textit{(iv)} help the energy measurement of charged hadrons for which the
track parameters were not determined accurately, which is the case for
low-quality and high-$\pt$ tracks.

A specific clustering algorithm was developed for the PF event reconstruction,
with the aims of a high detection efficiency even for low-energy particles and
of separating close energy deposits, as illustrated in
Fig.~\ref{fig:cms_detector_display}. The clustering is performed separately
in each subdetector: ECAL barrel and endcaps, HCAL barrel and endcaps,
and the two preshower layers. In the HF, no clustering is performed:
the electromagnetic or hadronic components of each cell directly give rise
to an \textit{HF EM} cluster and an \textit{HF HAD} cluster. All parameters of the
clustering algorithm are described in turn below. Their values are
summarized in Table~\ref{tab:clustering_parameters}.

First, \textit{cluster seeds} are identified as cells with an energy larger than a
given seed threshold,  and larger than the energy of the neighbouring cells. The
cells considered as neighbours are either the four closest cells, which share a
side with the seed candidate, or the eight closest cells, including cells that only
share a corner with the seed candidate. Second, \textit{topological clusters} are
grown from the seeds by aggregating cells with at least a corner in
common with a cell already in the cluster and with an energy in excess
of a cell threshold set to twice the noise level.
In the ECAL endcaps, because the noise level
increases as a function of $\theta$, seeds are additionally required to
satisfy a threshold requirement on $\ET$.

An expectation-maximization algorithm based on a Gaussian-mixture model is
then used to reconstruct the clusters within a topological cluster. The
Gaussian-mixture model postulates that the energy deposits in the $M$
individual cells of the topological cluster arise from $N$ Gaussian energy
deposits where $N$ is the number of seeds. The parameters of the model are
the amplitude $A_i$ and the coordinates in the $(\eta, \varphi)$ plane of
the mean $\vec{\mu_i}$ of each Gaussian, while the width $\sigma$ is fixed to different values
depending on the considered calorimeter. The expectation-maximization
algorithm is an iterative algorithm with two steps at each iteration. During
the first step, the parameters of the model are kept constant and the
expected fraction $f_{ji}$ of the energy $E_j$ measured in the cell at
position $\vec{c_j}$ arising from the $i$th Gaussian energy deposit is
calculated as
\begin{equation}
{ f_{ji} = \frac{ A_i \re^{-(\vec{c_j}-\vec{\mu_i})^2/(2
      \sigma^2)}}{ \sum_{k=1}^{N} A_k \re^{-(\vec{c_j}-\vec{\mu_k})^2/(2
      \sigma^2)}}}.
\end{equation}

The parameters of the model are determined during the second step in an
analytical maximum-likelihood fit yielding
\begin{equation}
  A_i = \sum_{j=1}^{M} f_{ji} E_j, \\
  \vec{\mu_i} = \sum_{j=1}^{M}f_{ji} E_j \vec{c_j}.
\end{equation}
The energy and position of the seeds are used as initial values for the
parameters of the corresponding Gaussian functions and the expectation
maximization cycle is repeated until convergence. To stabilize the
algorithm, the seed energy is entirely attributed to the corresponding
Gaussian function at each iteration. After convergence, the positions and
energies of the Gaussian functions are taken as cluster parameters.

In the lower-right panel of Fig.~\ref{fig:cms_detector_display}, for example,
two cluster seeds
(dark grey) are identified in the HCAL within one topological cluster
formed of nine cells. The two seeds give rise to two HCAL clusters,
the final positions of which are indicated by two red dots. These
reconstructed positions match the two charged-pion track extrapolations
to the HCAL. Similarly, the bottom-left ECAL topological cluster in
the lower-left panel of Fig.~\ref{fig:cms_detector_display} arising
from the $\pi^0$ is
split in two clusters corresponding to the two photons from the $\pi^0$
decay.

\begin{table}
  \topcaption{Clustering parameters for the ECAL, the HCAL, and the preshower.
    All values result from optimizations based on the simulation of single
photons, $\pi^{0}$, $\mathrm{K}^{0}_\mathrm{L}$, and jets.
  }
  \label{tab:clustering_parameters}
  \centering
  \begin{tabular}{l|rr|rr|r}
    \hline
    & \multicolumn{2}{c|}{ECAL} & \multicolumn{2}{c|}{HCAL} & Preshower \\
    &  barrel  &  endcaps  &  barrel  &  endcaps  &    \\
    \hline
 Cell $E$ threshold (\MeVns{})       &  80    &  300   & 800  &  800    & 0.06  \\
    \hline
 Seed \# closest cells            &  8     &  8     & 4    &  4      & 8    \\
 Seed $E$ threshold (\MeVns{})       & 230    &  600   & 800  &  1100   & 0.12 \\
 Seed $\ET$ threshold (\MeVns{})&  0     &  150   & 0    &  0      & 0    \\
    \hline
 Gaussian width (cm)              &  1.5   &  1.5   & 10.0 &  10.0   & 0.2  \\
    \hline
  \end{tabular}
\end{table}

\subsection{Calorimeter cluster calibration}
\label{sec:reconstruction_pf_elements_calocalib}

In the PF reconstruction algorithm, photons and neutral hadrons are
reconstructed from calorimeter clusters. Calorimeter clusters separated
from the extrapolated position of any charged-particle track in the
calo\-ri\-meters constitute a clear signature of neutral particles. On the
other hand, neutral-particle energy deposits overlapping with charged-particle
clusters can only be detected as calorimeter energy excesses with respect
to the sum of the associated charged-particle momenta.
An accurate calibration of the calorimeter response to photons and hadrons
is instrumental in maximizing the probability to identify these neutral
particles while minimizing the rate of misreconstructed energy excesses, and
to get the right energy scale for all neutral particles. The calibration of
electromagnetic and hadron clusters is described in
Sections~\ref{sec:reconstruction_pf_elements_ecalib}
and~\ref{sec:reconstruction_pf_elements_hcalib}.

\subsubsection{Electromagnetic deposits}
\label{sec:reconstruction_pf_elements_ecalib}

A first estimate of the absolute calibration of the ECAL response to electrons
and photons, as well as of the cell-to-cell relative calibration,
has been determined with test beam data, radioactive
sources, and cosmic ray measurements, all of which were collected prior to the start of collision data taking.
The ECAL calibration was then refined with
collision data collected at $\sqrt{s} = 7$ and $8$\TeV~\cite{EcalCalibration}.

The clustering algorithm described in
Section~\ref{sec:reconstruction_pf_elements_caloclusters} applies several
thresholds to the ECAL cell energies. Consequently, the energy measured in
clusters of ECAL cells is expected to be somewhat smaller than that of the
incoming photons, especially at low energy, and than that of the superclusters
used for the absolute ECAL calibration. A residual energy calibration,
required to account for the effects of these thresholds, is determined from
simulated single photons. This generic calibration is applied to all ECAL clusters
prior to the hadron cluster calibration discussed in the next section, and to
the particle identification step described in Section~\ref{sec:particle_id_reco}.
Specific additional electron and photon energy corrections, on the other hand, are
applied after the electron and photon reconstruction described in
Section~\ref{sec:particle_id_reco_electrons}. Large samples of single
photons with energies varying from 0.25 to 100\GeV were processed
through a \GEANTfour simulation~\cite{geant4} of the CMS detector. Only
the photons that do not experience a conversion prior to their entrance
in the ECAL are considered in the analysis, in order to deal with the
calibration of single clusters.

In the ECAL barrel, an analytical function of the type $f(E,\eta) = g(E)
h(\eta)$, where $E$ is the energy and $\eta$ the pseudorapidity of
the cluster, is fitted to the two-dimensional distribution of the average
ratio $\langle E^\text{true} / E \rangle$ in the $(E,\eta)$ plane, where
$E^\text{true}$ is the true photon energy. This function is, by construction,
the residual correction to be applied to the measured cluster
energy. It is close to unity at high energy, where threshold effects
progressively vanish. The correction can be as large as $+$20\% at low energy.

In the ECAL endcaps, the crystals are partly shadowed by the preshower. The
calibrated cluster energy is therefore expressed as a function of the energies
measured in the ECAL ($E_\mathrm{ECAL}$) and in the two preshower layers
($E_\mathrm{PS1}$ and $E_\mathrm{PS2}$) as
\begin{linenomath}
\begin{equation}
E^\text{calib} = \alpha(E^\text{true},\eta^\text{true}) E_\mathrm{ECAL}
+ \beta(E^\text{true},\eta^\text{true}) \left[ E_\mathrm{PS1}
+  \gamma(E^\text{true},\eta^\text{true}) E_\mathrm{PS2} \right].
\end{equation}
\end{linenomath}

The calibration parameters $\alpha$, $\beta$, and $\gamma$ depend on the
energy $E^\text{true}$ and the pseudorapidity $\eta^\text{true}$ of the generated photon
and are chosen in each $(E^\text{true},\eta^\text{true})$ bin to minimize the following
$\chi^2$,
\begin{linenomath}
\begin{equation}
\chi^2 = \sum_{i=1}^{N_\text{events}}
\frac{\left( E^\text{calib}_i - E_i^\text{true} \right)^2}{\sigma_i^2}.
\end{equation}
\end{linenomath}

In this expression, $\sigma_i$ is an estimate of the energy measurement
uncertainty for the $i$th photon, with a dependence on $E_i^\text{true}$ similar to
that displayed in Eq.~(\ref{eq:ECALresolution}), but with stochastic and noise
terms typically four times larger than in the barrel. Analytical functions
of the type $g^\prime(E^\text{true}) h^\prime(\eta^\text{true})$ are used to fit
the equivalent three
calibration parameters for the endcaps. A similar $\chi^2$ minimization,
with only two parameters, is performed for the photons that leave energy
only in one of the two preshower layers. The case where no energy is measured
in the preshower, which includes the endcap region outside the preshower
acceptance, is handled with the same method as that used for the ECAL barrel.

When it comes to evaluating the calibration parameters for actual clusters in
the preshower fiducial region, $\eta^\text{true}$ is estimated from the ECAL cluster
pseudorapidity, and $E^\text{true}$ is approximated by a linear combination of
$E_\mathrm{ECAL}$, $E_\mathrm{PS1}$, and $E_\mathrm{PS2}$, with fixed coefficients.
These calibration parameters
correct the ECAL energy by $+5\%$ at the largest photon energies---meaning that an
energetic photon loses on average 5\% of its energy in the preshower
material---and up to $+40\%$ for the smallest photon energies.
In all ECAL regions and for all energies, the calibrated energy agrees on
average with the true photon energy to within $\pm 1\%$.

\begin{figure}[hbtp]
  \centering
  \includegraphics[width=.48\textwidth]{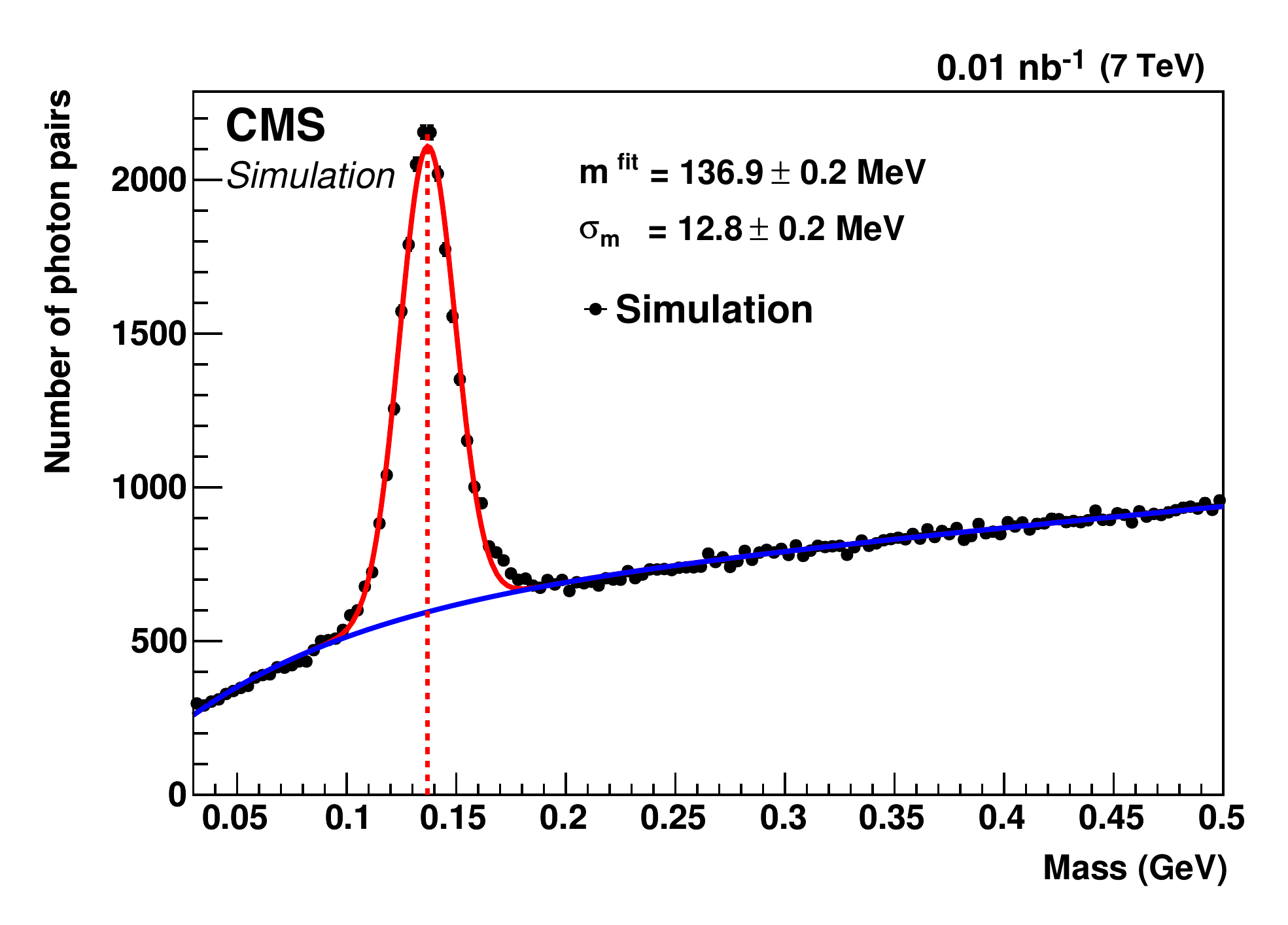}
  \includegraphics[width=.48\textwidth]{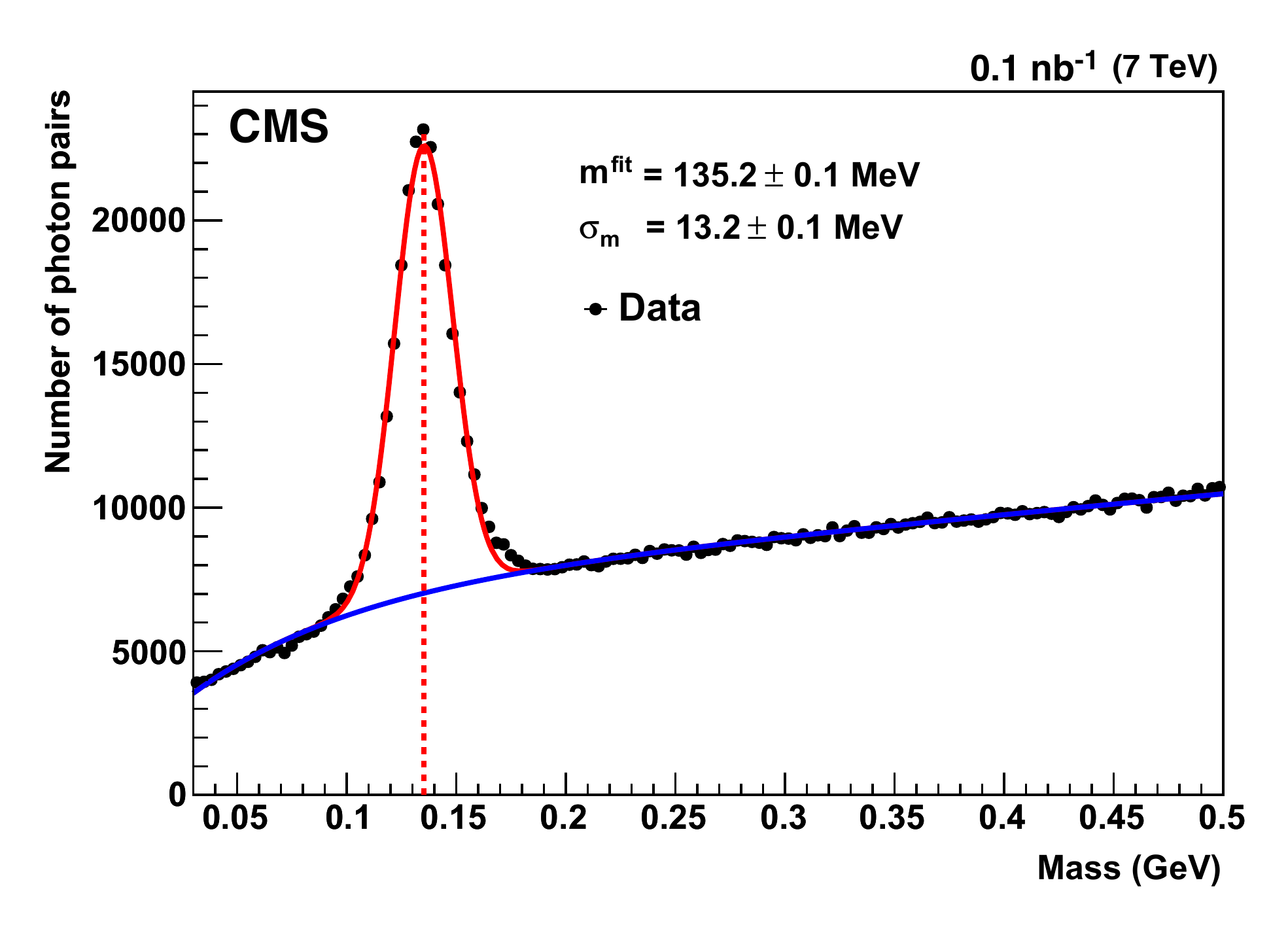}
   \caption{Photon pair invariant mass distribution in the barrel
($\abs{\eta} <$ 1.0) for the simulation (left) and the data (right).
The $\pi^0$ signal is modelled by a Gaussian (red curve) and the background
by an exponential function (blue curve). The Gaussian mean value (vertical
dashed line) and its standard deviation are denoted $m^\text{fit}$
and $\sigma_\mathrm{m}$, respectively.}
  \label{fig:pi0}
\end{figure}

Both the absolute photon energy calibration and the uniformity of the
response can be checked with the abundant $\pi^0$ samples produced in pp
collisions. To reconstruct these neutral pions, all ECAL clusters with a
calibrated energy in excess of 400\MeV and identified as photons as described
in Section~\ref{sec:particle_id_reco_hadphot} are paired. The total energy of
the photon pair is required to be larger than 1.5\GeV. The resulting photon
pair invariant mass distribution is displayed in Fig.~\ref{fig:pi0}, for
simulated events and for the  first LHC data recorded in 2010 at
$\sqrt{s}=7$\TeV. The per-cent level agreement of the fitted mass resolutions
in data and simulation, and that of the fitted mass values with the nominal $\pi^0$
mass, demonstrate the adequacy of the simulation-based ECAL cluster
calibration for low-energy photons.

\subsubsection{Hadron deposits}
\label{sec:reconstruction_pf_elements_hcalib}

Hadrons generally deposit energy in both ECAL and HCAL. The ECAL is already
calibrated for photons as described in the previous section, but has a
substantially different response to hadrons. The initial calibration of the HCAL
was realized with
test beam data for 50\GeV charged pions not interacting in the ECAL, but
the calorimeter response depends on the fraction of the shower energy
deposited in the ECAL, and is not linear with energy. The ECAL and HCAL
cluster energies therefore need to be substantially recalibrated to get an
estimate of the true hadron energy.

The calibrated calorimetric energy associated with a hadron is expressed as
\begin{linenomath}
\begin{equation}
E_\text{calib} = a + b(E) f(\eta)  E_\mathrm{ECAL} + c(E) g(\eta) E_\mathrm{HCAL},
\end{equation}
\end{linenomath}
where $E_\mathrm{ECAL}$ and $E_\mathrm{HCAL}$ are the energies measured in the ECAL
(calibrated as described in Section~\ref{sec:reconstruction_pf_elements_ecalib})
and the HCAL, and where $E$ and $\eta$ are the true energy and pseudorapidity of
the hadron. The coefficient $a$ (in \GeV) accounts for the energy lost
because of the energy thresholds of the clustering algorithm and is taken
to be independent of $E$. Similarly to what is done in
Section~\ref{sec:reconstruction_pf_elements_ecalib}, a large sample of
simulated single neutral hadrons (specifically, $\mathrm{K}^0_\mathrm{L}$) is used to
determine the calibration coefficients $a$, $b$, and $c$, as well as the
functions $f$ and $g$. Hadrons that  interact with the tracker material
are rejected.  In a first pass, the functions $f(\eta)$ and $g(\eta)$ are
fixed to unity.  For a given value of $a$ and in each bin of $E$,
the $\chi^2$ defined as
\begin{linenomath}
\begin{equation}
\chi^2 = \sum_{i=1}^N \frac{\left(E_i^\text{calib} -E_i\right)^2}
{\sigma_i^2},
\end{equation}
\end{linenomath}
where $E_i$ and $\sigma_i$ are the true energy and the expected calorimetric
energy resolution of the $i$th single hadron, is minimized with respect
to the coefficients $b$ and $c$. The energy dependence of the energy
resolution $\sigma_i$, as displayed in Eq.~(\ref{eq:HCALresolution}), is
determined iteratively. Prior to the first iteration of the $\chi^2$
minimization, a Gaussian is fitted to the distribution of $E_\mathrm{ECAL}+E_\mathrm{HCAL}-E$ in each bin of true energy.
The coefficients of Eq.~(\ref{eq:HCALresolution})
are then fitted to the evolution of the Gaussian standard deviation as a
function of $E$. These two operations are repeated in the subsequent
iterations, for which the calibrated energy, $E_\text{calib}$, is substituted
for the raw energy, $E_\mathrm{ECAL}+E_\mathrm{HCAL}$. The procedure converges
at the second iteration.

The barrel and endcap regions are treated separately to account for
different thresholds and cell sizes. In each region, the determination
of $b$ and $c$ is performed separately for hadrons leaving energy
solely in the HCAL (in which case only $c$ is determined) and those
depositing energy in both ECAL and HCAL. No attempt is made to calibrate
the hadrons leaving energy only in the ECAL, as such clusters are
identified as photon or electron clusters by the PF algorithm. For each
of the four samples, the relatively small residual dependence of the
calibrated energy on the particle pseudorapidity is corrected for in a
third iteration of the $\chi^2$ minimization with second-order polynomials
for $f(\eta)$ and $g(\eta)$, and with $b(E)$ and $c(E)$ taken from the
result of the second iteration.

\begin{figure}[tbh]
  \begin{centering}
  \includegraphics[width=0.48\textwidth]{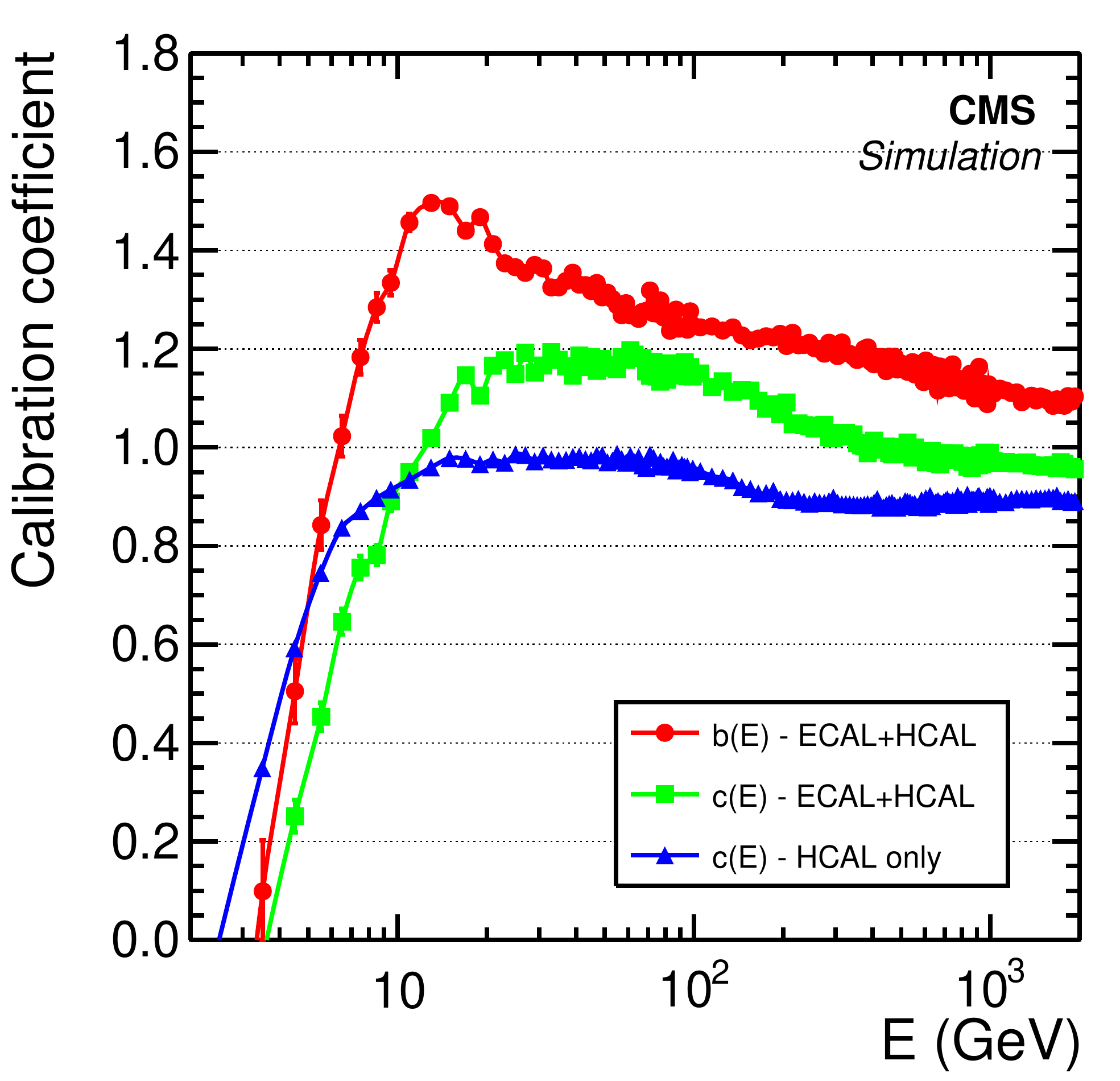}
  \includegraphics[width=0.48\textwidth]{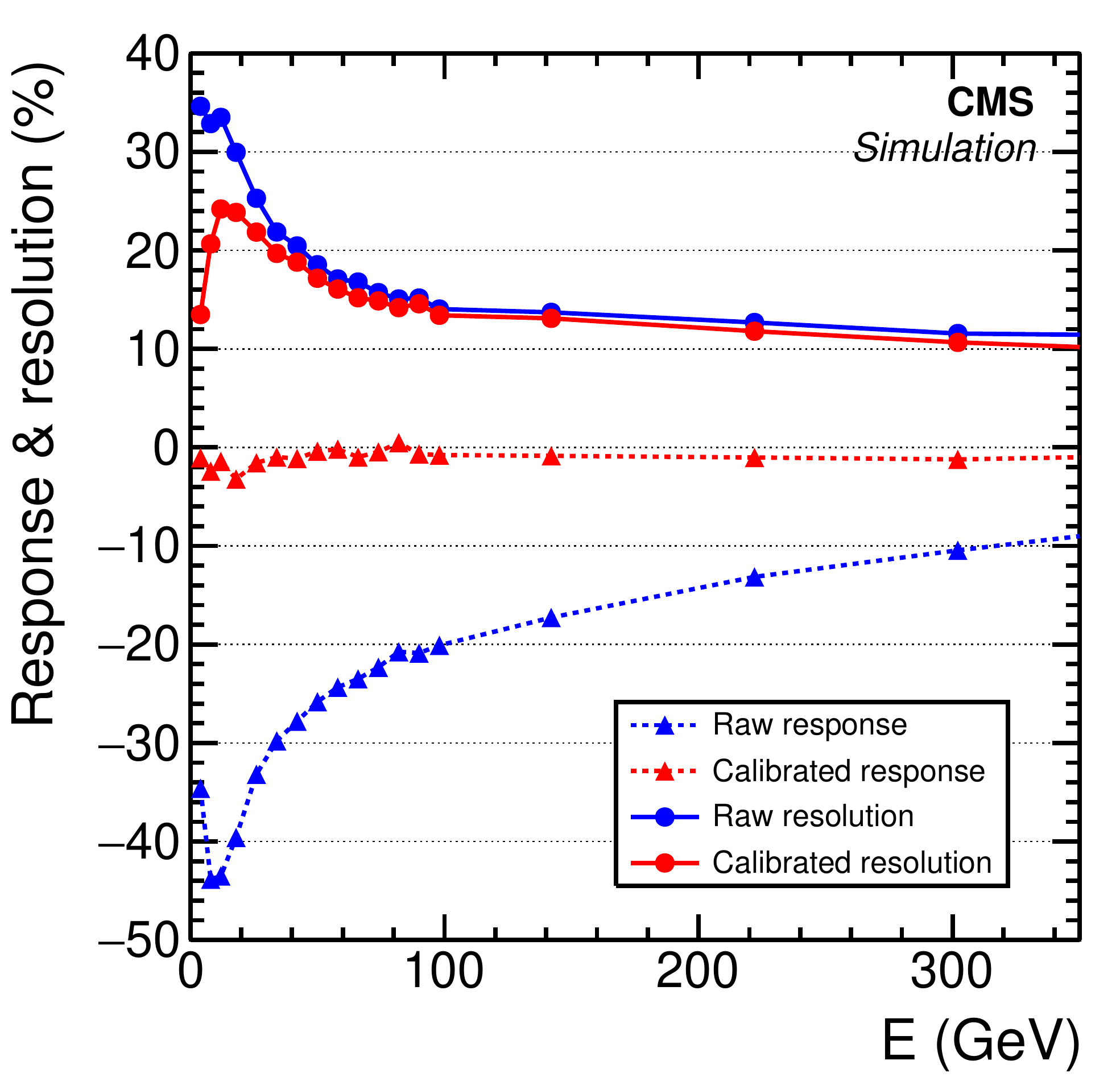}
  \caption{Left: Calibration coefficients obtained from single hadrons in the
barrel as a function of their true energy $E$, for hadrons depositing
energy only in the HCAL (blue triangles), and for hadrons depositing energy
in both the ECAL and HCAL, for the ECAL (red circles) and for the HCAL
(green squares) clusters. Right: Relative raw (blue) and calibrated (red)
energy response (dashed curves and triangles) and resolution (full curves
and circles) for single hadrons in the barrel, as a function of their true
energy $E$. Here the raw (calibrated) response and resolution are
obtained by a Gaussian fit to the distribution of the relative difference
between the raw (calibrated) calorimetric energy and the true hadron energy.}
  \label{fig:calibration}
\end{centering}
\end{figure}

To avoid the need for an accurate estimate of the true hadron energy $E$
(which might not be available in real data), the constant $a$ is chosen to
minimize the dependence on $E$ of the coefficients $b$ and $c$, for $E$ in
excess of 10\GeV. It is estimated to amount to 2.5\GeV for hadrons showering
in the HCAL only, and 3.5\GeV for hadrons interacting in both ECAL and HCAL.
The left panel of Fig.~\ref{fig:calibration} shows the coefficients
$b$ and $c$, determined for each energy bin in the barrel region, as a
function of the true hadron energy. The residual dependence of these
coefficients on $E$ is finally fitted to adequate continuous functional
forms $b(E)$ and $c(E)$, for later use in the course of the PF reconstruction.
As expected, the coefficient $c$ is close to unity for 50\GeV hadrons
leaving energy only in the HCAL. The larger values of the coefficient
$c$ for the hadrons that leave energy also in ECAL make up for the
energy lost in the dead material between ECAL and HCAL, which amounts
to about half an interaction length. The fact that the coefficients
$b$ and $c$ depend on the true energy up to very large values is a
consequence of the nonlinear calorimeter response to hadrons.

The right panel of Fig.~\ref{fig:calibration} shows that the \textit{calibrated response},
defined as the mean relative difference between the calibrated energy and the true energy,
is much closer to zero than the raw response, which underestimates hadron energies by
up to 40\% at low energy.
The calibration procedure therefore restores the linearity of the calorimeter
response. The relative calibrated energy resolution, displayed in the same
figure, also exhibits a sizeable improvement with respect to the raw
resolution at all energies. For hadrons with an energy below 10\GeV, the
resolution rapidly improves when the energy decreases.
This remarkable behaviour is an effect of the convergence of the $b$ and $c$
coefficients to zero in this energy range, which itself is an artefact of
the presence of the $a$ constant in the calibration procedure. The explanation
is as follows. Hadrons with energy below 10\GeV often leave too little
energy in the calorimeters to exceed the thresholds of the clustering
algorithm. As a consequence, those that leave energy do so because of
an upward fluctuation in the showering process. Such fluctuations are
calibrated away by the small $b$ and $c$ values. The procedure effectively
replaces the energy of soft hadrons, measured with large fluctuations,
with a constant $a$, de facto closer to the actual hadron energy.

Isolated charged hadrons selected from early data recorded at
$\sqrt{s}= 0.9$, $2.2$, and $7$\TeV have been used to check that the
calibration coefficients determined from the simulation are adequate for
real data. Section~\ref{sec:particle_id_reco_hadphot} describes
how the calibration is applied for the identification and reconstruction of
nonisolated particles. Finally, it is worth stressing at this point that
this calibration affects only 10\% of the measured event energy. The
latter is therefore expected to be modified, on average, by only a few per
cent by the calibration procedure.

\section{Particle identification and reconstruction}
\label{sec:particle_id_reco}

\subsection{Link algorithm}
\label{sec:particle_id_reco_link}

A given particle is, in general, expected to give rise to several PF elements
in the various CMS subdetectors. The reconstruction of a particle therefore
first proceeds with a \textit{link algorithm} that connects the PF elements from
different subdetectors. The event display of
Fig.~\ref{fig:cms_detector_display} illustrates most of the possible
configurations for charged hadrons, neutral hadrons, and photons. The
probability for the algorithm to link elements from one particle only is
limited by the granularity of the various subdetectors and by the number of
particles to resolve per unit of solid angle. The probability to link all
elements of a given particle is mostly limited by the amount of material
encountered upstream of the calorimeters and the muon detector, which may lead
to trajectory kinks and to the creation of secondary particles.

The link algorithm can test any pair of elements in the event.
In order to prevent the computing time of the link algorithm from growing
quadratically with the number of particles,
the pairs of elements considered by the link procedure are restricted to the
nearest neighbours in the ($\eta,\varphi$) plane, as obtained
with a $k$-dimensional tree~\cite{Bentley:kdtree}.
The specific conditions required to link two elements depend on their nature,
and are listed in the next paragraphs. If two elements are found to be
linked, the algorithm defines a distance between these two elements, aimed
at quantifying the quality of the link.  The link algorithm then produces
\textit{PF blocks} of elements associated either by a direct link or by an
indirect link through common elements.

More specifically, a link between a track in the central tracker and a
calorimeter cluster is established as follows. The track is first extrapolated
from its last measured hit in the tracker to---within the corresponding
angular acceptance---the two layers of the preshower, the ECAL
at a depth corresponding to the expected maximum of a typical longitudinal
electron shower profile, and the HCAL at a depth corresponding
to one interaction length. The track is linked to a cluster if its extrapolated
position is within the cluster area, defined by the union of the areas
of all its cells in the $(\eta, \varphi)$ plane for the HCAL and the
ECAL barrel, or in the $(x,y)$ plane for the ECAL endcaps and the preshower.
This area is enlarged by up to the size of a cell in each direction, to
account for the presence of gaps between calorimeter cells or cracks between
calorimeter modules, for the uncertainty in the position of the shower
maximum, and for the effect of multiple scattering on low-momentum charged
particles. The link distance is defined as the distance between the
extrapolated track position and the cluster position in the $(\eta, \varphi)$
plane. In case several HCAL clusters are linked to the same track,
or if several tracks are linked to the same ECAL cluster, only the
link with the smallest distance is kept.

To collect the energy of photons emitted by electron bremsstrahlung, tangents
to the GSF tracks are extrapolated to the ECAL from the intersection points
between the track and each of the tracker layers. A cluster is linked to the
track as a potential bremsstrahlung photon if the extrapolated tangent position
is within the boundaries of the cluster, as defined above, provided that
the distance between the cluster and the GSF track extrapolation in $\eta$
is smaller than 0.05. These bremsstrahlung photons, as well as prompt photons,
have a significant probability to convert to an $\Pep\Pem$ pair in the
tracker material. A dedicated conversion finder~\cite{Khachatryan:2015iwa}
was therefore developed to create links between any two tracks compatible
with originating from a photon conversion. If the converted photon direction,
obtained from the sum of the two track momenta, is found to be compatible
with one of the aforementioned track tangents, a link is created between
each of these two tracks and the original track.

Calorimeter cluster-to-cluster links are sought between HCAL clusters
and ECAL clusters, and between
ECAL clusters and preshower clusters in the preshower acceptance. A link is
established when the cluster position in the more granular calorimeter
(preshower or ECAL) is within the cluster envelope in the less granular
calorimeter (ECAL or HCAL). The link distance is also defined as the distance
between the two cluster positions, in the $(\eta, \varphi)$ plane for an
HCAL-ECAL link, or in the $(x,y)$ plane for an ECAL-preshower link. When
multiple HCAL clusters are linked to the same ECAL cluster, or when multiple
ECAL clusters are linked to the same preshower clusters, only the link with
the smallest distance is kept. A trivial link between an ECAL cluster and
an ECAL supercluster is established when they share at least one ECAL cell.

Charged-particle tracks may also be linked together through a common secondary
vertex, for nuclear-interaction reconstruction
(Section~\ref{sec:reco_pf_elements_nuclint}). The relevant displaced vertices
are retained if they feature at
least three tracks, of which at most one is an incoming track, reconstructed with
tracker hits between the primary vertex and the displaced vertex. The
invariant mass formed by the outgoing tracks must exceed 0.2\GeV. All the
tracks sharing a selected nuclear-interaction vertex are linked together.

Finally, a link between a track in the central tracker and information in
the muon detector is established as explained in Section~\ref{sec:reco_muons}
to form global and tracker muons.

In the event shown in Fig.~\ref{fig:cms_detector_display}, the track T$_1$ is
linked to the ECAL cluster E$_1$ and to the HCAL clusters H$_1$ (with a smaller
link distance) and H$_2$ (with a larger link distance), while the track T$_2$
is linked only to the HCAL clusters H$_2$ and H$_1$. These two tracks form a
first PF block with five PF elements: T$_1$, E$_1$, and H$_1$ (corresponding
to the generated $\pi^-$); and T$_2$ and H$_2$ (corresponding to the generated
$\pi^+$). The other three ECAL clusters are not linked to any track or cluster
and thus form three PF blocks on their own, corresponding to the generated
pair of photons from the $\pi^0$ decay, and to the neutral kaon. Owing to
the granularity of the CMS subdetectors, the majority of the PF blocks
typically contain a handful of elements originating from one or few
particle(s): the logic of the subsequent PF algorithm is therefore not affected
by the particle multiplicity in the event and the computing time increases
only linearly with multiplicity.

In each PF block, the identification and reconstruction sequence proceeds
in the following order. First, muon candidates are identified and
reconstructed as described in Section~\ref{sec:particle_id_reco_muons},
and the corresponding PF elements (tracks  and clusters) are removed
from the PF block. The electron identification and reconstruction follows,
as explained in Section~\ref{sec:particle_id_reco_electrons}, with the aim
of collecting the energy of all bremsstrahlung photons.  Energetic and
isolated photons, converted or unconverted, are identified in the same step.
The corresponding tracks and ECAL or preshower clusters are excluded from
further consideration.

At this level, tracks with a \pt uncertainty in excess of the calorimetric
energy resolution expected for charged hadrons (Fig.~\ref{fig:calibration})
are masked, which allows the rate of misreconstructed tracks at large \pt
(Fig.~\ref{fig:IterativeTrackingEfficiency}) to be adequately reduced.
In multijet events, 0.2\% of the tracks are rejected by this requirement,
on average. About 10\% of these rejected tracks originate from genuine
high-\pt charged hadrons, with a \pt estimate incompatible with the true \pt value.
Their energies are
measured in that case more accurately in the calorimeters than in the
tracker. The remaining elements in  the block are then subject to a
cross-identification of charged hadrons, neutral hadrons, and photons,
arising from parton fragmentation, hadronization, and decays in jets. This step
is described in Section~\ref{sec:particle_id_reco_hadphot}.

Hadrons experiencing a nuclear interaction in the tracker material create
secondary particles. These hadrons are identified and reconstructed as
summarized in Section~\ref{sec:particle_id_reco_secondaries}. When an incoming
track is identified, it is used to refine the reconstruction outcome, but is
otherwise ignored in the track-cluster link algorithm as well as in
the particle reconstruction algorithms described in
Sections~\ref{sec:particle_id_reco_muons} to~\ref{sec:particle_id_reco_hadphot}.

Finally, when the global event description becomes available, \ie when all
blocks have been processed and all particles have been identified, the
reconstructed event is revisited by a post-processing step described in
Section~\ref{sec:particle_id_reco_postprocessing}.

\subsection{Muons}
\label{sec:particle_id_reco_muons}

In the PF algorithm, muon identification proceeds by a set of selections
based on the global and tracker muon properties.
Isolated global muons are first selected by considering additional inner tracks
and calorimeter energy deposits with a distance $\Delta R$ to the muon direction
in the $(\eta,\varphi)$ plane smaller than 0.3. The sum of the \pt of the
tracks and of the $\ET$ of the deposits is required not to exceed 10\% of
the muon \pt. This isolation criterion alone is sufficient to adequately
reject hadrons that would be misidentified as muons, hence no further
selection is applied to these muon candidates.

Muons inside jets, for example those from semileptonic heavy-flavour decays
or from charged-hadron decays in flight, require more stringent
identification criteria.  Indeed, for charged hadrons misidentified as muons
\eg because of punch-through, the PF algorithm will tend to
create additional spurious neutral particles from the calorimeter deposits.
Unidentified muons, on the other hand, will be considered to be charged
hadrons, and will tend to absorb the energy deposits of nearby neutral
particles.

For nonisolated global muons, the tight-muon selection~\cite{MUO-10-004}
is applied. In addition, it is required either that at least three matching
track segments be found in the muon detectors, or that the calorimeter
deposits associated with the track be compatible with the muon
hypothesis. This selection removes the majority of high-\pt hadrons
misidentified as muons because of punch-through, as well as accidental
associations of tracker and standalone muon tracks.

Muons that fail the tight-muon selection due to a poorly reconstructed inner
track, for example because of hit confusion with other nearby tracks, are
salvaged if the standalone muon track fit is of high quality and is
associated with a large number of hits in the muon detectors (at least 23
DT or 15 CSC hits, out of 32 and 24, respectively). Alternatively, muons may
also fail the tight-muon selection due to a poor global fit. In this case, if
a high-quality fit is obtained with at least 13 hits in the tracker, the muon
is selected, provided that the associated calorimeter clusters be compatible
with the muon hypothesis.

The muon momentum is chosen to be that of the inner track if its $\pt$
is smaller than 200\GeV. Above this value, the momentum is chosen according
to the smallest $\chi^2$ probability from the different track fits: tracker
only, tracker and first muon detector plane, global, and global without the
muon detector planes featuring a high occupancy~\cite{MUO-10-004}.

The PF elements that make up these identified muons are masked against further
processing in the corresponding PF block, \ie are not used as building
elements for other particles. As discussed in
Sections~\ref{sec:particle_id_reco_hadphot}
and~\ref{sec:particle_id_reco_postprocessing},
muon identification and reconstruction is not complete at this point.
For example, charged-hadron candidates
are checked for the compatibility of the measurements of their momenta in the
tracker and their energies in the calorimeters. If the track momentum is found
to be significantly larger than the calibrated sum of the linked calorimeter
clusters, the muon identification criteria are revisited, with somewhat looser
selections on the fit quality and on the hit or segment associations.

\subsection{Electrons and isolated photons}
\label{sec:particle_id_reco_electrons}

Electron reconstruction is based on combined information from the inner
tracker and the calo\-ri\-meters. Due to the large amount of material
in the tracker, electrons often emit brems\-strahlung photons and photons
often convert to $\Pep\Pem$ pairs, which in turn emit brems\-strahlung
photons, etc.  For this reason, the basic properties and the technical
issues to be solved for the tracking and the energy deposition patterns
of electrons and photons are similar. Isolated photon reconstruction is
therefore conducted together with electron reconstruction. In a given
PF block, an electron candidate is seeded from a GSF track, as described in
Section~\ref{sec:reco_electrons}, provided that the corresponding ECAL
cluster is not linked to three or more additional tracks. A photon
candidate is seeded from an ECAL supercluster with $\ET$ larger
than 10\GeV, with no link to a GSF track.

For ECAL-based electron candidates and for photon candidates, the sum of the
energies measured in the HCAL cells with a distance to the supercluster
position smaller than 0.15 in the ($\eta,\varphi$) plane must not exceed
10\% of the supercluster energy. To ensure an optimal energy containment,
all ECAL clusters in the PF block linked either to the supercluster or to
one of the GSF track tangents are associated with the candidate.
Tracks linked to these ECAL clusters are associated in turn if the track
momentum and the energy of the HCAL cluster linked to the track are compatible
with the electron hypothesis. The tracks and ECAL clusters belonging to
identified photon conversions linked to the GSF track tangents are associated
as well.

The total energy of the collected ECAL clusters is corrected for the energy
missed in the association process, with analytical functions of $E$ and $\eta$.
These corrections can be as large as 25\% at $\abs{\eta} \approx 1.5$ where the
tracker thickness is largest, and at low $\pt$. This corrected energy is
assigned to the photons, and the photon direction is taken to be that of
the supercluster. The final energy assignment for electrons is obtained
from a combination of the corrected ECAL energy with the momentum of the
GSF track and the electron direction is chosen to be that of the GSF
track~\cite{Baffioni2007}.

Electron candidates must satisfy additional identification criteria.
Specifically, up to fourteen variables---including the amount of energy
radiated off the GSF track, the distance between the GSF track extrapolation
to the ECAL entrance and the position of the ECAL seeding cluster, the ratio
between the energies gathered in HCAL and ECAL by the track-cluster association
process, and the KF and GSF track $\chi^2$ and numbers of hits---are combined
in BDTs trained separately in the ECAL barrel and endcaps acceptance, and for
isolated and nonisolated electrons.

Photon candidates are retained if
they are isolated from other tracks and calorimeter clusters in the
event, and if the ECAL cell energy distribution and the ratio between the HCAL
and ECAL energies are compatible with those expected from a photon shower.
The PF selection is looser than the requirements typically  applied at
analysis level to select isolated photons. The reconstruction of less
energetic or nonisolated photons is discussed in
Section~\ref{sec:particle_id_reco_hadphot}.

All tracks and clusters in the PF block used to reconstruct electrons and
photons are masked against further processing. Tracks identified
as originating from a photon conversion but not used in the process are
masked as well, as they are typically poorly measured and likely
to be misreconstructed tracks. The distinction
between electrons and photons in the PF global event description can be
different from a selection optimized for a specialized analysis. To deal
with this complication, the complete history of the electron and photon
reconstruction is tracked and saved, to allow a different event
interpretation to be made without running the complete PF algorithm
again.

\subsection{Hadrons and nonisolated photons}
\label{sec:particle_id_reco_hadphot}

Once muons, electrons, and isolated photons are identified and removed from
the PF blocks, the remaining particles to be identified are hadrons from
jet fragmentation and hadronization. These particles may be detected as
charged hadrons ($\pi^\pm$, $\mathrm{K}^\pm$, or protons), neutral hadrons
(\eg $\mathrm{K}^0_\mathrm{L}$ or neutrons), nonisolated photons
(\eg from $\pi^0$ decays),
and more rarely additional muons (\eg from early decays of charged hadrons).

The ECAL and HCAL clusters not linked to any track give rise to photons and
neutral hadrons. Within the tracker acceptance ($\abs{\eta}< 2.5$),
all these ECAL clusters are turned into photons and all
these HCAL clusters are turned into neutral hadrons. The precedence given
in the ECAL to photons over neutral hadrons is justified by the observation
that, in hadronic jets, 25\% of the jet energy is carried by photons, while
neutral hadrons leave only 3\% of the jet energy in the ECAL. (This fraction
is reduced by one order of magnitude for taus, for which decays to final
states with neutral hadrons are Cabibbo-suppressed to a branching ratio of
about 1\%.) Beyond the tracker acceptance, however, charged and neutral hadrons
cannot be distinguished and they leave in total 25\% of the jet energy in
the ECAL. The systematic precedence given to photons for the ECAL energy is
therefore no longer justified. For this reason, ECAL clusters linked to a
given HCAL cluster are assumed to arise from the same (charged- or neutral-)
hadron shower, while ECAL clusters without such a link are classified as
photons. These identified photons and hadrons are calibrated as described in
Sections~\ref{sec:reconstruction_pf_elements_ecalib}
and~\ref{sec:reconstruction_pf_elements_hcalib}. The estimated true energy
of each identified particle, needed for the determination of the calibration
coefficients, is taken to be the raw calorimetric energy, \ie
$E_\mathrm{ECAL}$ for photons, $E_\mathrm{HCAL}$ for hadrons inside the tracker acceptance,
 and $E_\mathrm{ECAL}+E_\mathrm{HCAL}$ for hadrons outside the tracker acceptance.
The HF EM and HF HAD clusters are added to the particle list as
\textit{HF photons} and \textit{HF hadrons} without any further calibration.

Each of the remaining HCAL clusters of the PF block is linked to one or
several tracks (not linked to any other HCAL cluster) and these
tracks may in turn be linked to some of the remaining ECAL clusters
(each linked to only one of the tracks). The calibrated calorimetric energy
is determined with the procedure described in
Section~\ref{sec:reconstruction_pf_elements_hcalib} from the energy of the
HCAL cluster and the total energy of the ECAL clusters, under the single
charged-hadron hypothesis. The true energy, needed
to determine the calibration coefficients $b$ and $c$, is estimated
to be either the sum of the momenta of the tracks, or the sum of the raw
ECAL and HCAL energies, whichever is larger. The sum of the track momenta
is then compared to the calibrated calorimetric energy in order to determine
the particle content, as described below.

If the calibrated calorimetric energy is in excess of the sum of the track
momenta by an amount larger than the expected calorimetric energy resolution
for hadrons, the excess may be interpreted as the presence of photons
and neutral hadrons. Specifically, if the excess is smaller than the total
ECAL energy and larger than 500\MeV, it is identified as a photon with an
energy corresponding to this excess after recalibration under the photon
hypothesis, as described in Section~\ref{sec:reconstruction_pf_elements_ecalib}.
Otherwise, the recalibrated ECAL energy still gives rise to a photon,
and the remaining part of the excess, if larger than 1\GeV, is identified as
a neutral hadron. Each track  gives rise to a charged hadron,
the momentum and energy of which are directly taken from the corresponding
track momentum, under the charged-pion mass hypothesis.

If the calibrated calorimetric energy is compatible with the sum of the
track momenta, no neutral particle is identified. The charged-hadron momenta
are redefined by a $\chi^2$ fit of
the measurements in the tracker and the calorimeters, which reduces to a
weighted average if only one track is linked to the HCAL cluster. This
combination is particularly relevant when the track parameters are measured
with degraded resolutions, \eg at very high energies or at large
pseudorapidities. It ensures a smooth transition between the low-energy
regime, dominated by the tracker measurements, and the high-energy regime,
dominated by the calorimetric measurements. The resulting energy resolution
is always better than that of the calorimetric energy measurement,
even at the highest energies.

In rare cases, the calibrated calorimetric energy is significantly smaller
than the sum of the track momenta. When the difference is larger than three
standard deviations, a relaxed search for muons, which deposit little energy
in the calorimeters, is performed. All global muons remaining after the
selection described in Section~\ref{sec:particle_id_reco_muons}, and for
which an estimate of the momentum exists with a relative precision better
than 25\%, are identified as PF muons and the corresponding tracks are masked.
The redundancy of the measurements in the tracker and the calorimeters thus
allows a few more muons to be found without increasing the misidentified
muon rate. If the track momentum sum is still significantly larger than
the calorimetric energy, the excess in momentum is often found to arise
from residual misreconstructed tracks with a $\pt$ uncertainty
in excess of 1\GeV. These tracks are sorted in decreasing order of their
$\pt$ uncertainty and are sequentially masked either until no such
tracks remain in the PF block or until the momentum excess disappears,
whichever comes first. Less
than 0.3 per mil of the tracks in multijet events are affected by this
procedure. In general, after these two steps, either the compatibility
of total calibrated calorimetric energy with the reduced sum of the track
momenta is restored, or a calorimetric energy excess appears. These cases
are treated as described above.

The event of Fig.~\ref{fig:cms_detector_display} is interpreted by the PF
algorithm as follows. The three ECAL clusters E$_2$, E$_3$, and E$_4$, are
within the tracker acceptance, and thus no link with any HCAL cluster is
created. As they are not linked to any track either, the three corresponding
PF blocks give rise to one photon each. The first two correspond to the photons
from the generated $\pi^0$ decay, and the third one to the energy deposited in the
ECAL by the generated $\mathrm{K}^0_\mathrm{L}$, which is therefore misidentified by the
algorithm and calibrated as a photon. The fourth PF block consists of the two
tracks T$_1$ and T$_2$, the ECAL cluster E$_1$, and the two HCAL clusters
H$_1$ and H$_2$. The track T$_1$ is initially linked to E$_1$, as well as to
the two HCAL clusters. Only the link to the closest HCAL cluster, H$_1$, is
kept. Similarly, only the link of T$_2$ to H$_2$ is kept. The clusters H$_1$
and E$_1$, and the track T$_1$ give rise to a charged hadron, corresponding
to the generated $\pi^-$, the direction of which is that of T$_1$. The
calibrated calorimetric energy is obtained under the charged-hadron
hypothesis, from the E$_1$ and H$_1$ raw energies, with an estimate of
the true hadron energy given by the momentum of T$_1$. As the calibrated
energy is found to be compatible with the momentum of T$_1$, no neutral
particle is identified and the charged hadron energy is obtained from
the weighted average of the track momentum and the calibrated calorimetric
energy. Similarly, the cluster H$_2$ and the track T$_2$ give rise to a
second charged hadron, corresponding to the generated $\pi^+$.

\subsection{Nuclear interactions in the tracker material}
\label{sec:particle_id_reco_secondaries}

A hadron interaction in the tracker material often results in the creation
of a number of charged and neutral secondary particles originating from a
secondary interaction vertex. One such secondary vertex is
reconstructed (Section~\ref{sec:reco_pf_elements_nuclint}) and identified
(Section~\ref{sec:particle_id_reco_link}) on average in a typical
top-quark pair event.
The secondary particles,
whether or not the secondary vertex is identified,
are reconstructed as charged particles (mostly charged hadrons, but also
muons and electrons), photons, and neutral hadrons by the PF algorithm,
as explained in Sections~\ref{sec:particle_id_reco_muons}
to~\ref{sec:particle_id_reco_hadphot}.

When the secondary charged-particle tracks are linked together by an
identified nuclear-inter\-action vertex, the secondary charged particles are
replaced in the reconstructed particle list by a single primary charged
hadron. Its direction is obtained from the vectorial sum of the momenta
of the secondary charged particles, its energy is given by the sum of their
energies (denoted $E_\text{sec}$), and its mass is set to the charged-pion mass.
The nuclear-inter\-action vertex may also include an incoming track, not used
so far in the PF reconstruction. The direction of the primary charged hadron
is taken in that case to be that of the incoming track. If, in
addition, the momentum of the incoming track $p_\text{prim}$ is well measured,
it is used to estimate the energy of undetected
secondary particles, reconstructed neither as secondary charged particles
nor as neutral particles. The energy of the primary charged hadron is
then estimated as
\begin{linenomath}
\begin{equation}
E = E_\text{sec} + f(\eta, p_\text{prim}) p_\text{prim}.
\end{equation}
\end{linenomath}
The small fraction of undetected energy $f(\eta, p_\text{prim})$ in this expression
is obtained from the simulation of single charged-hadron events.

\subsection{Event post-processing}
\label{sec:particle_id_reco_postprocessing}

Although the particles reconstructed and identified by the algorithms
presented in Sections~\ref{sec:particle_id_reco_link}
to~\ref{sec:particle_id_reco_secondaries} are the result of an optimized
combination of the information from all subdetectors, a small, but nonzero,
probability of particle misidentification and misreconstruction cannot
be avoided. In general, these individual particle mishaps tend to average
out and are hardly noticeable when global event quantities are evaluated.
In some rare cases, however, an artificially large missing transverse momentum,
\ptmiss, is reconstructed in the event.
This large \ptmiss, most often caused by a misidentified or
misreconstructed high-\pt muon, may lead the event to be wrongly selected
by a large set of new physics searches, and therefore needs to be understood and corrected.
The strategy for the post-processing algorithm consists of three
steps: the high-\pt particles that may lead to a large artificial \ptmiss are
selected; the correlation of the particle transverse momentum and direction
with the \ptmiss amplitude and direction is quantified; the identification
and the reconstruction of these particles are a posteriori modified, if this
change is found to reduce the \ptmiss by at least one half.

The first cause of muon-related artificial \ptmiss  is the presence of genuine
muons from cosmic rays traversing CMS in coincidence with an LHC beam
crossing. These cosmic muons are identified when their trajectories are more
than 1\unit{cm} away from the beam axis, and are removed from the particle
list if the measured \ptmiss is consequently reduced. Muons from semileptonic
decays of b hadrons also can, albeit rarely, be reconstructed more than
1\unit{cm} away from the beam axis and therefore be considered by this rejection
algorithm. In these semileptonic decays, however, the direction of
the missing momentum caused by the accompanying neutrino is strongly
correlated with the muon direction, and the removal of the muon would
further increase this missing momentum instead of reducing it. As the
direction of the rest of the \ptmiss in these rare events, if any, is
uncorrelated with that of the b hadron, such muons are in practice always
kept in the particle list.

The second cause of muon-related artificial \ptmiss, still from genuine muons,
is a severe misreconstruction of the muon momentum. Such a misreconstruction
is identified by significant differences between the available estimates of
the muon momentum (Section~\ref{sec:particle_id_reco_muons}). Large
differences may be caused by a wrong inner track association, an interaction
in the steel yoke, a decay in flight, or substantial synchrotron radiation.
In this case, the choice of the momentum done by the PF algorithm is reviewed
for muons with $\pt > 20$\GeV. If the \ptmiss is reduced by at least half,
the momentum estimate that leads to the smallest \ptmiss value is taken.

The third cause of muon-related artificial \ptmiss is particle
misidentification. For example, a punch-through charged hadron
can be misidentified as a muon. In that case, an energetic neutral hadron,
resulting from the energy deposited by the charged hadron in the
calorimeters, is wrongly added to the particle list and leads to
significant \ptmiss in the opposite direction. If both the muon momentum
and the neutral hadron energy are larger than 100\GeV, the neutral
hadron is removed from the particle list, the muon is changed to a charged
hadron, and the charged-hadron momentum is taken to be that of the inner
track, provided that it allows the \ptmiss to be reduced by at least one half.

An energetic tracker or global muon ($\pt > 20$\GeV) can also fail the strict
identification criteria of Section~\ref{sec:particle_id_reco_muons} and still
be missed by the recovery algorithm of
Section~\ref{sec:particle_id_reco_hadphot}, because it overlaps with an
energetic neutral hadron with similar energy. In that case, the  muon
candidate is misidentified as a charged hadron in the course of
the PF reconstruction, and the neutral hadron disappears in the process,
leading to significant \ptmiss in the same direction. These charged hadrons
are turned into muons and a neutral hadron is added to the particle list
with the associated calorimetric energy, if the \ptmiss is reduced by at
least half in the operation.

These criteria were originally designed to reduce the fraction of events
with large \ptmiss in standard model multijet events from data and simulated
samples, in the context of a search for new physics in hadronic events with
large \ptmiss at $\sqrt{s} = 7$\TeV. A systematic visual inspection of the
events observed with unexpectedly large \ptmiss values in these early data
proved to be particularly instrumental in identifying undesired features,
either in the software producing inputs to the PF algorithm, or in the PF
algorithm itself, or even in the detector hardware. These shortcomings were
taken care of immediately with software fixes or workarounds (either in the
PF algorithm itself or in the post-processing step described above), which
consequently improved the core response and resolution of the physics objects
described in Section~\ref{sec:expected_performance}.
Physics events with genuine \ptmiss, such as semileptonic \ttbar
events in data, or simulated processes predicted by new physics theories
(supersymmetry, heavy gauge bosons, etc.),
were checked to be essentially unaffected by the post-processing algorithm.
The reason is twofold: On the one hand, the fraction of misreconstructed or
misidentified muons is minute (typically smaller than 0.1 per mil) and
on the other, the presence of genuine \ptmiss, uncorrelated with these
reconstruction shortcomings, causes the already rare reassignments proposed
by the post-processing algorithm not to reduce, in general, the observed
\ptmiss value.

\section{Performance in simulation}
\label{sec:expected_performance}

The particles identified and reconstructed by the PF algorithm,
described in Section~\ref{sec:particle_id_reco},
can be used straightforwardly in physics analyses.
In the absence of pileup interactions---the case studied in this section---these particles are meant to match the stable particles of the final state of the collision.

\begin{figure}[htbp]
\centering
\includegraphics[width=\textwidth]{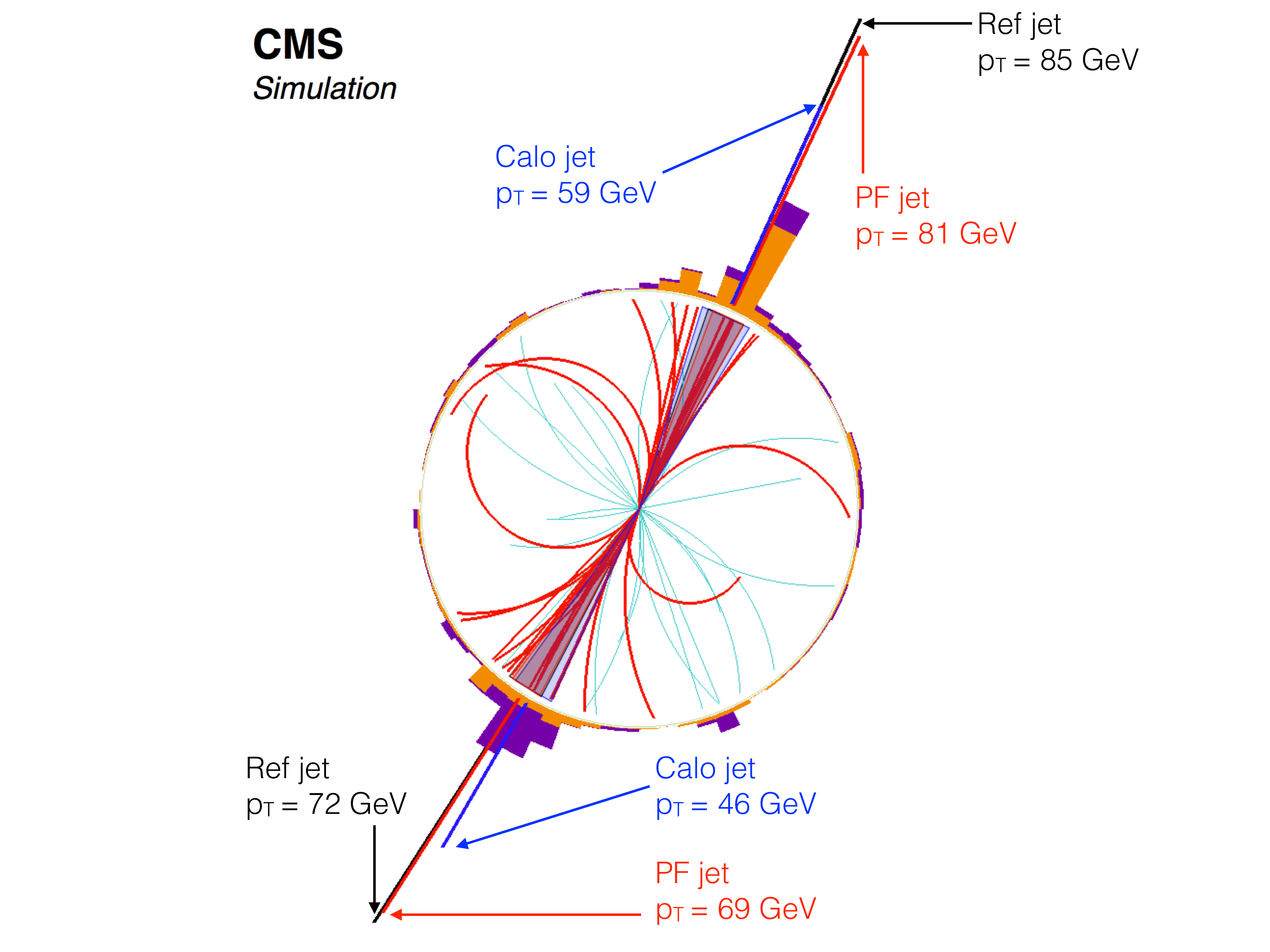}\\
\caption{Jet reconstruction in a simulated dijet event.
The particles clustered in the two PF jets are displayed with a thicker line.
For clarity, particles with $\pt<1\GeV$ are not shown.
The PF jet \vecpt, indicated as a radial line, is
compared to the \vecpt of the corresponding generated (Ref) and
calorimeter (Calo) jets.
In all cases, the four-momentum of the jet is obtained by summing the
four-momenta of its constituents, and no jet energy correction
is applied.
}
\label{fig:dijet_display}
\end{figure}

In this section, the performance of the PF reconstruction is assessed
with pp collision events generated with \PYTHIA 8.205~\cite{pythia6, pythia8p2} at a
centre-of-mass energy of 13\TeV. All events are processed by the CMS \GEANTfour
simulation without any pileup effects, and by the CMS reconstruction algorithms.
The reconstructed particles are used to build the \textit{physics objects},
namely jets, the missing transverse momentum \ptmiss, muons, electrons, photons, and taus.
They are also used to compute other quantities related to these physics objects, such as particle isolation.
These physics objects and observables are compared to the ones obtained from the stable particles produced by the event generator so as to evaluate the response,
the resolution, the efficiency, and the purity of the PF
reconstruction.
To quantify the improvements from PF,
these quantities are also evaluated for the physics objects
reconstructed with the techniques used prior to the PF development.
An example of such a comparison is given in Fig.~\ref{fig:dijet_display},
which displays a simulated dijet event.
In this event, the jets of reconstructed particles are closer in energy and direction to
the jets of generated particles than the calorimeter jets.

The comparison with the data recorded by CMS at a centre-of-mass energy of 8\TeV and the influence of pileup interactions on the PF reconstruction performance are presented in Section~\ref{sec:commissioning_and_pileup}.

\subsection{Jets }
\label{sec:expected_performance_jets}

The jet performance is quantified with a sample of QCD multijet events.
Jets are reconstructed with the anti-\kt algorithm (radius parameter $R=0.4$)~\cite{Cacciari:2008gp,Cacciari:2011ma}.
The algorithm clusters either all particles reconstructed by the PF algorithm (PF jets),
or the sum of the ECAL and HCAL energies deposited in the calorimeter towers~\footnote{A
calorimeter tower is composed of an HCAL tower and the 25 underlying ECAL crystals.} (Calo jets),
or all stable particles produced by the event generator excluding neutrinos (Ref jets).
Particle-flow jets are studied down to a \pt of $15\GeV$, while Calo jets with a \pt lower than 20\GeV are deemed unreliable and are rejected.

Each PF (Calo) jet is matched to the closest Ref jet in the $(\eta, \varphi)$ plane,
with $\Delta R < 0.1 (0.2)$.
The $\Delta R$ limit of 0.1 for PF jets is justified by the jet direction resolution being twice as good for PF jets as it is for Calo jets, as can be seen in Figure 10.
This choice results in a similar matching efficiency for both PF and Calo jets.
The improved angular resolution for PF jets is mainly due to the precise determination of the charged-hadron directions and momenta.
In calorimeter jets, the energy deposits of charged hadrons are spread along the $\varphi$ direction by the magnetic field, leading to an additional degradation of the azimuthal angular resolution.

\begin{figure}[htb]
  \centering
  \includegraphics[width=0.49\textwidth]{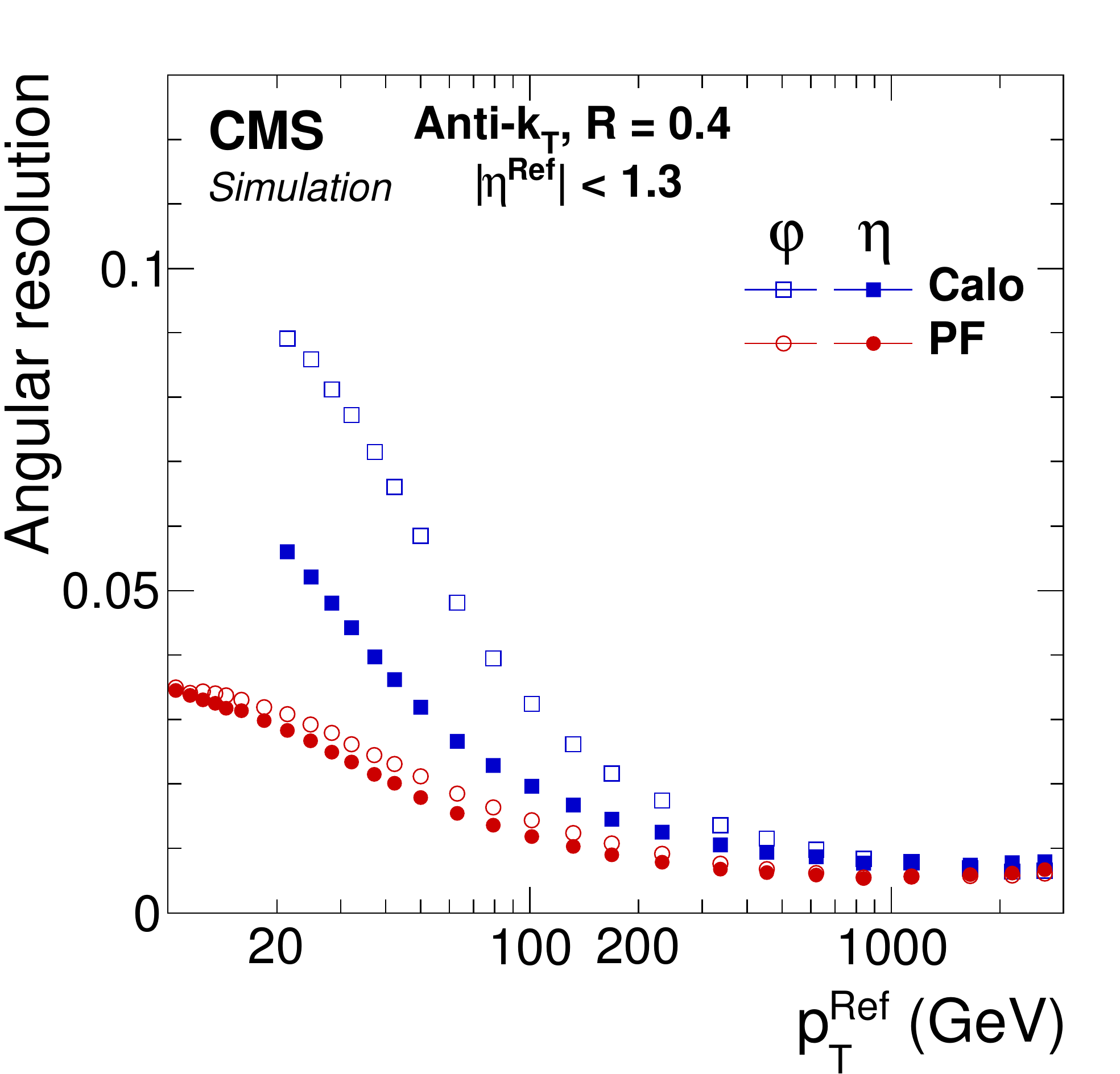}
  \includegraphics[width=0.49\textwidth]{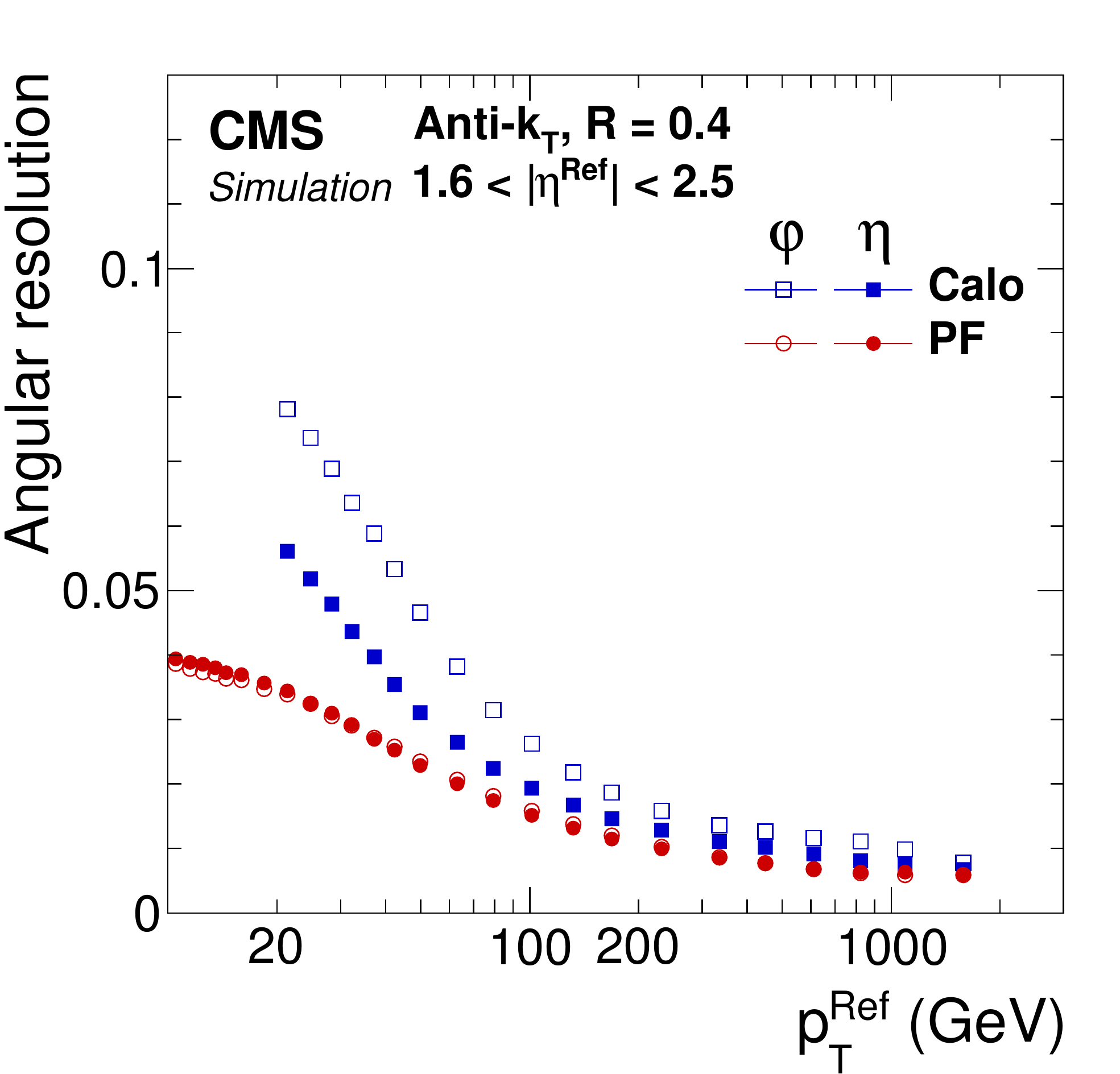}
  \caption{Jet angular resolution in the barrel (left) and endcap (right) regions, as a function of the \pt of the reference jet.
    The $\varphi$ resolution is expressed in radians. }
  \label{fig:expected_performance_jets_angular}
\end{figure}

\begin{figure}
  \centering
  \includegraphics[width=0.49\textwidth]{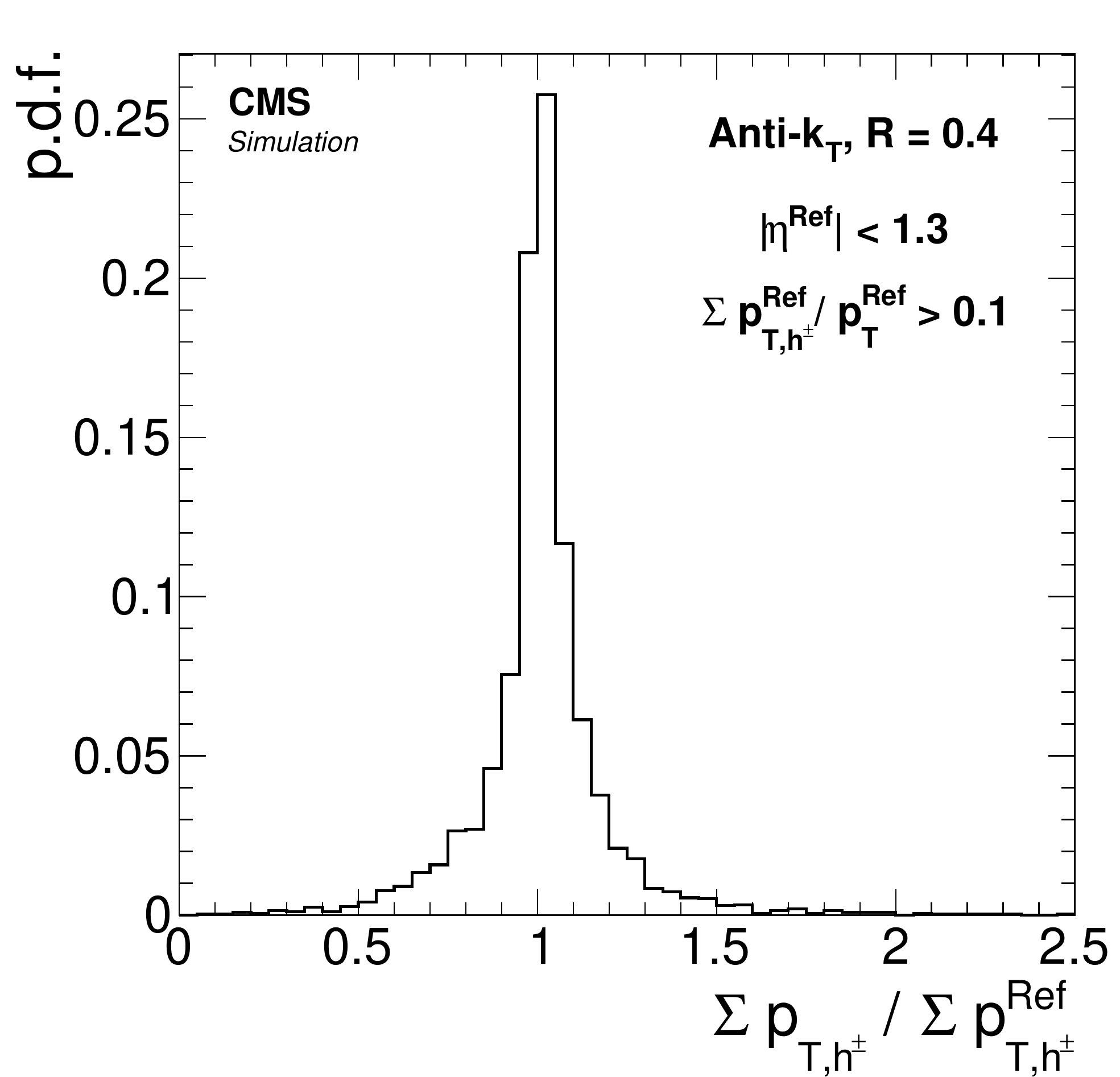}
  \includegraphics[width=0.49\textwidth]{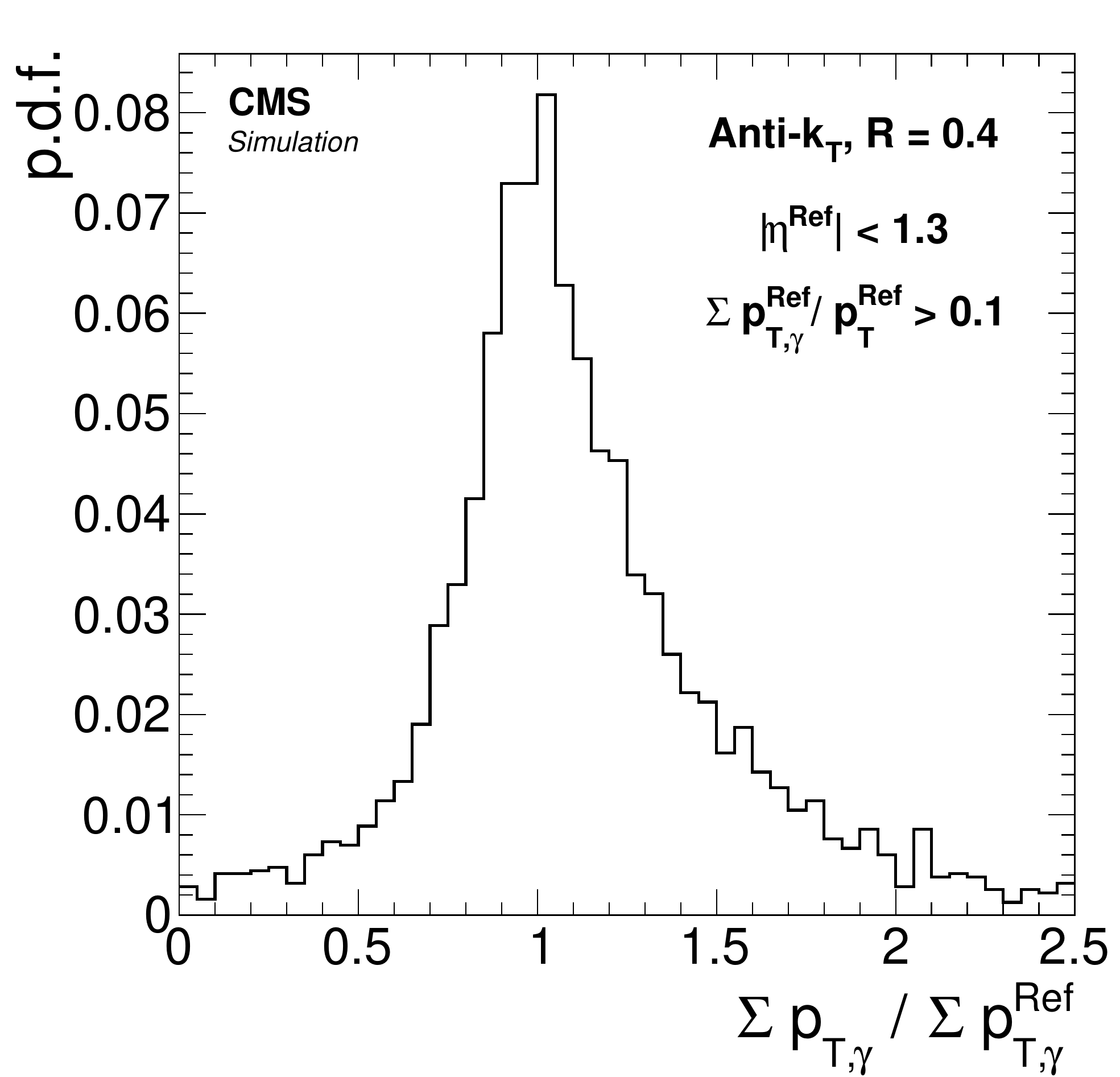}
  \includegraphics[width=0.49\textwidth]{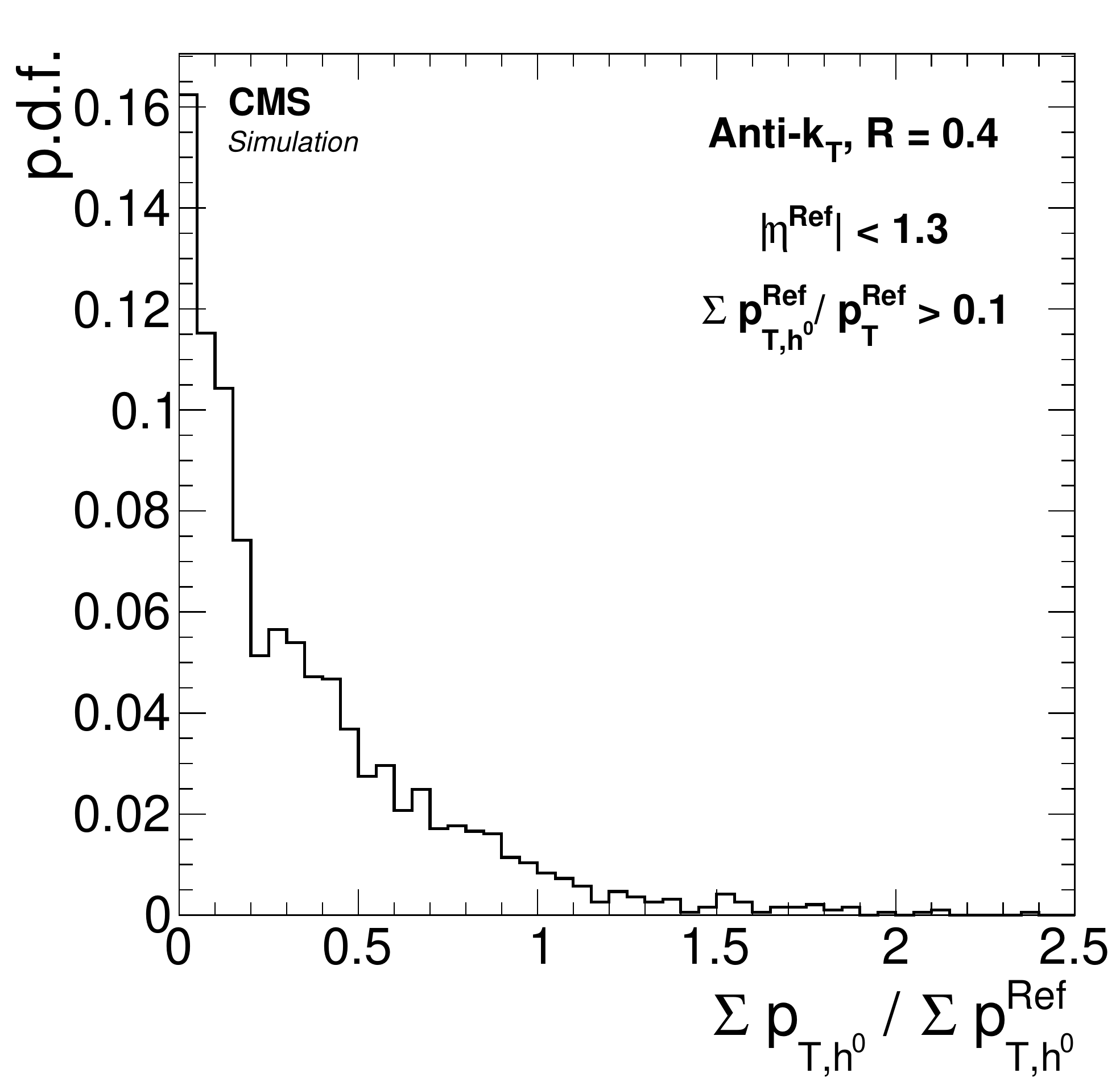}
  \includegraphics[width=0.49\textwidth]{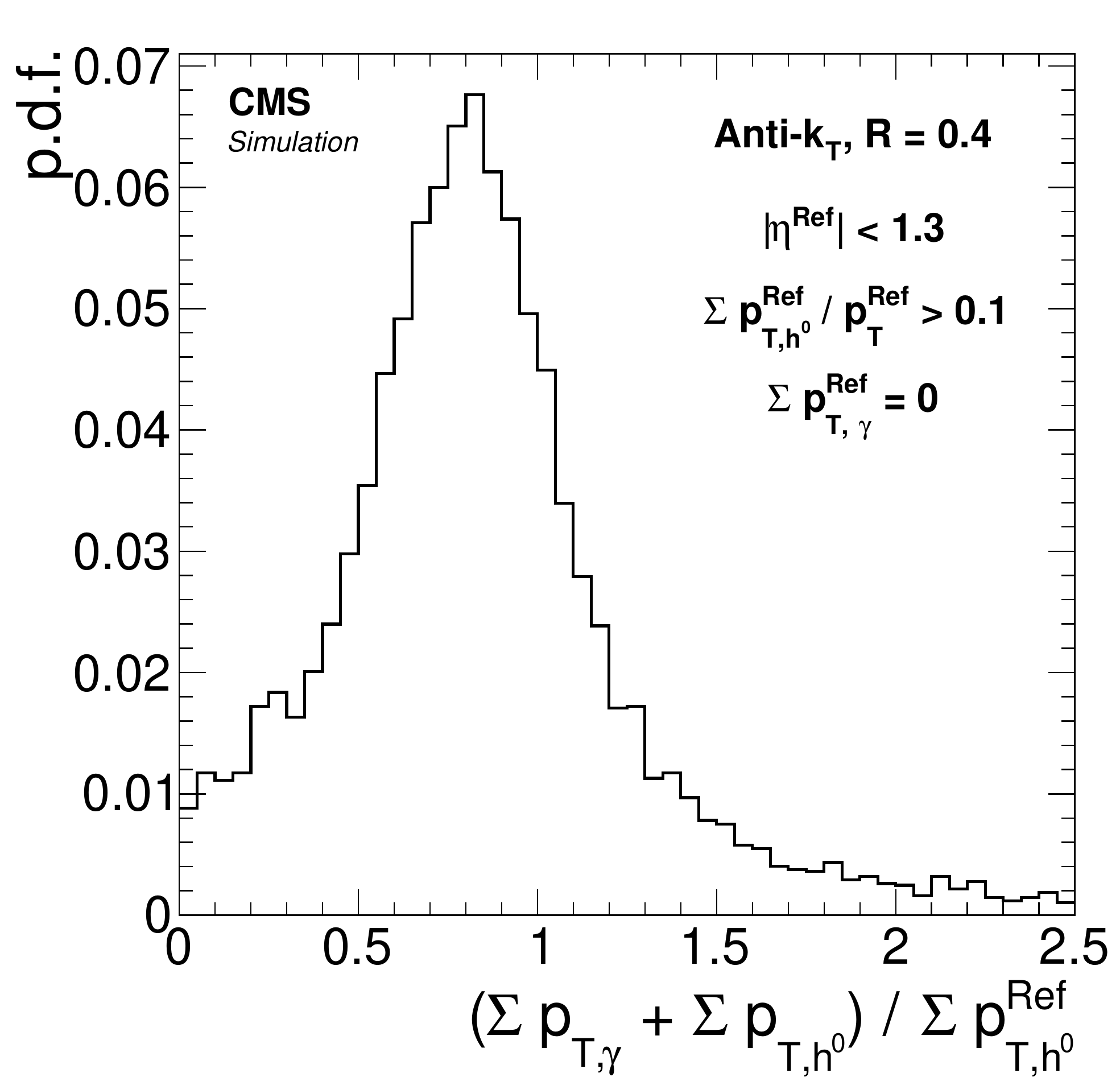}
  \caption{Distribution of the ratio between the reconstructed and reference transverse momenta, $\Sigma p_{\mathrm{T},i} / \Sigma p_{\mathrm{T},i}^\text{Ref}$, for charged hadrons (top left), photons (top right), neutral hadrons (bottom left), and for all neutral particles in Ref jets with no photon (bottom right).
The Ref jet is required to have at least 10\% of its \pt carried by particles of type $i$,
and to be located in the barrel. }
  \label{fig:expected_performance_jets_particleresponse}
\end{figure}

On average, 65\% of the jet energy is carried by charged hadrons, 25\% by photons, and 10\% by neutral hadrons.
The ability of the PF algorithm to identify these particles within jets is studied by comparing the jet energy fractions measured in PF jets to those of the corresponding Ref jet.
The distribution of the ratio between the reconstructed and reference energy fraction is shown in Fig.~\ref{fig:expected_performance_jets_particleresponse} for charged hadrons, photons, and neutral hadrons in barrel jets.
An important part of the \pt carried by neutral hadrons is reconstructed as coming from photons because the energy deposits of neutral hadrons in the ECAL are systematically identified as photons for the reasons given in Section~\ref{sec:particle_id_reco_hadphot}.
However, around 80\% of the neutral hadron energy is recovered, which is demonstrated by summing up the energy of reconstructed photons and neutral hadrons for Ref jets without photons.
The remaining 20\% of the energy is lost because the energy deposited by neutral hadrons in the ECAL
is identified as originating from photons.
It is therefore calibrated under the electromagnetic hypothesis to a scale that is underestimated by 20 to 40\%,
as indicated by the value of the calibration coefficient $b(E)$ in Fig.~\ref{fig:calibration},
which would have been used under the hadron hypothesis.

\begin{figure}[p]
  \centering
\includegraphics[width=0.5\textwidth]{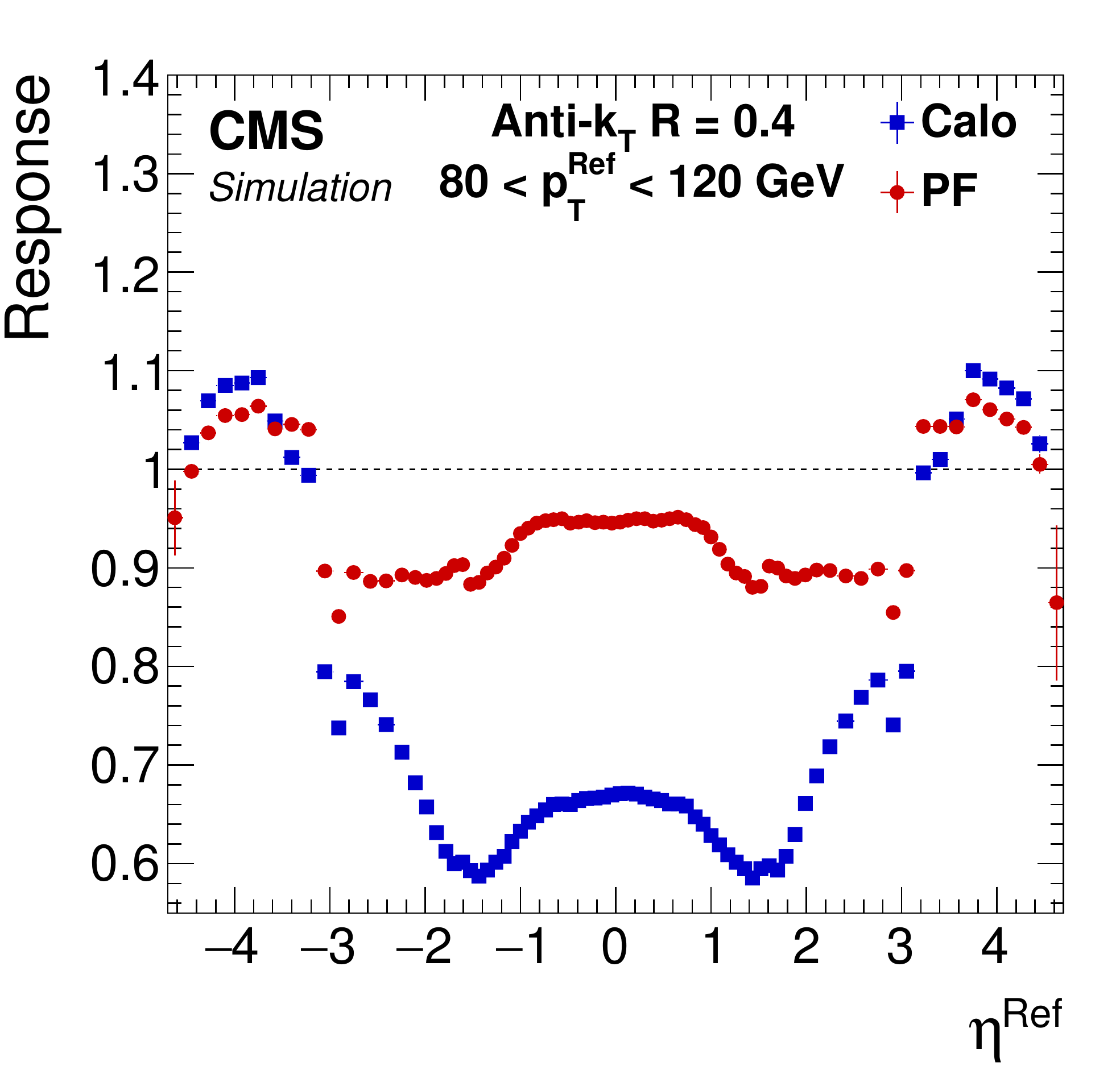}\\
  \includegraphics[width=0.49\textwidth]{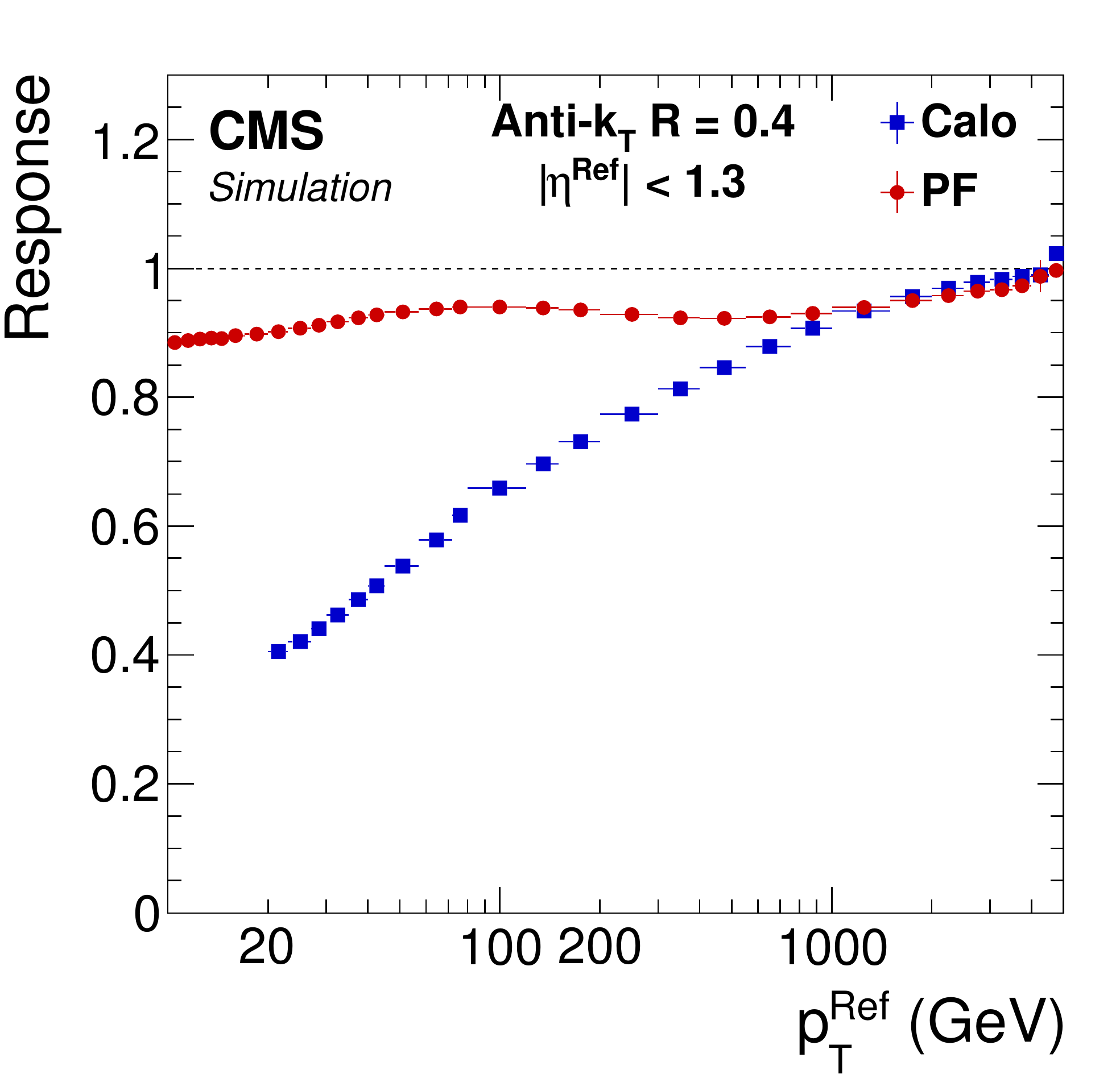}
  \includegraphics[width=0.49\textwidth]{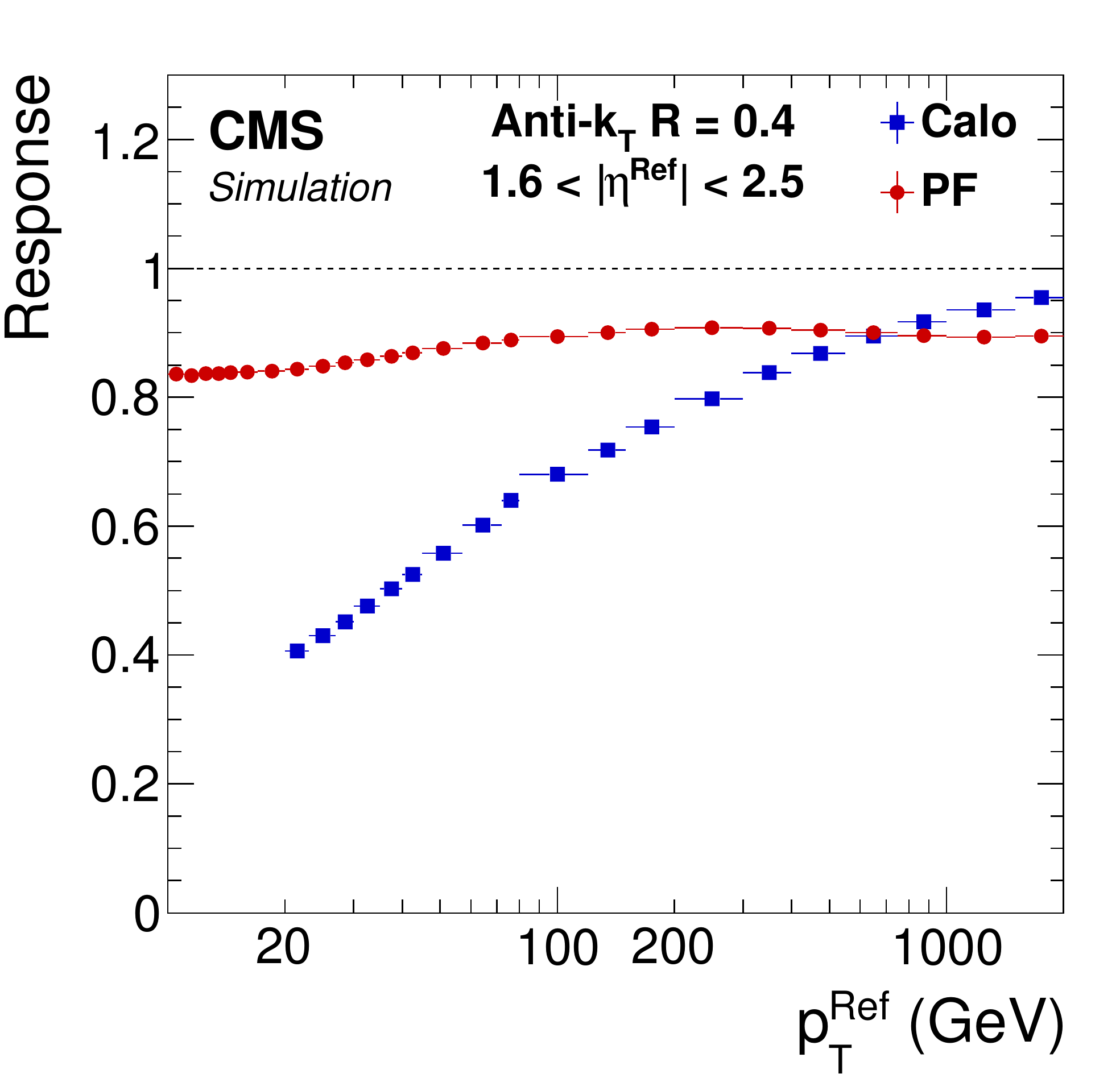}
  \caption{
    Jet response as a function of $\eta^\text{Ref}$ for the range $80 < \ptref < 120\GeV$ (top) and as a function of \ptref in the barrel (left) and in the endcap (right) regions.
  }
  \label{fig:expected_performance_jets_response}
\end{figure}

\begin{figure}[p]
  \centering
  \includegraphics[width=0.49\textwidth]{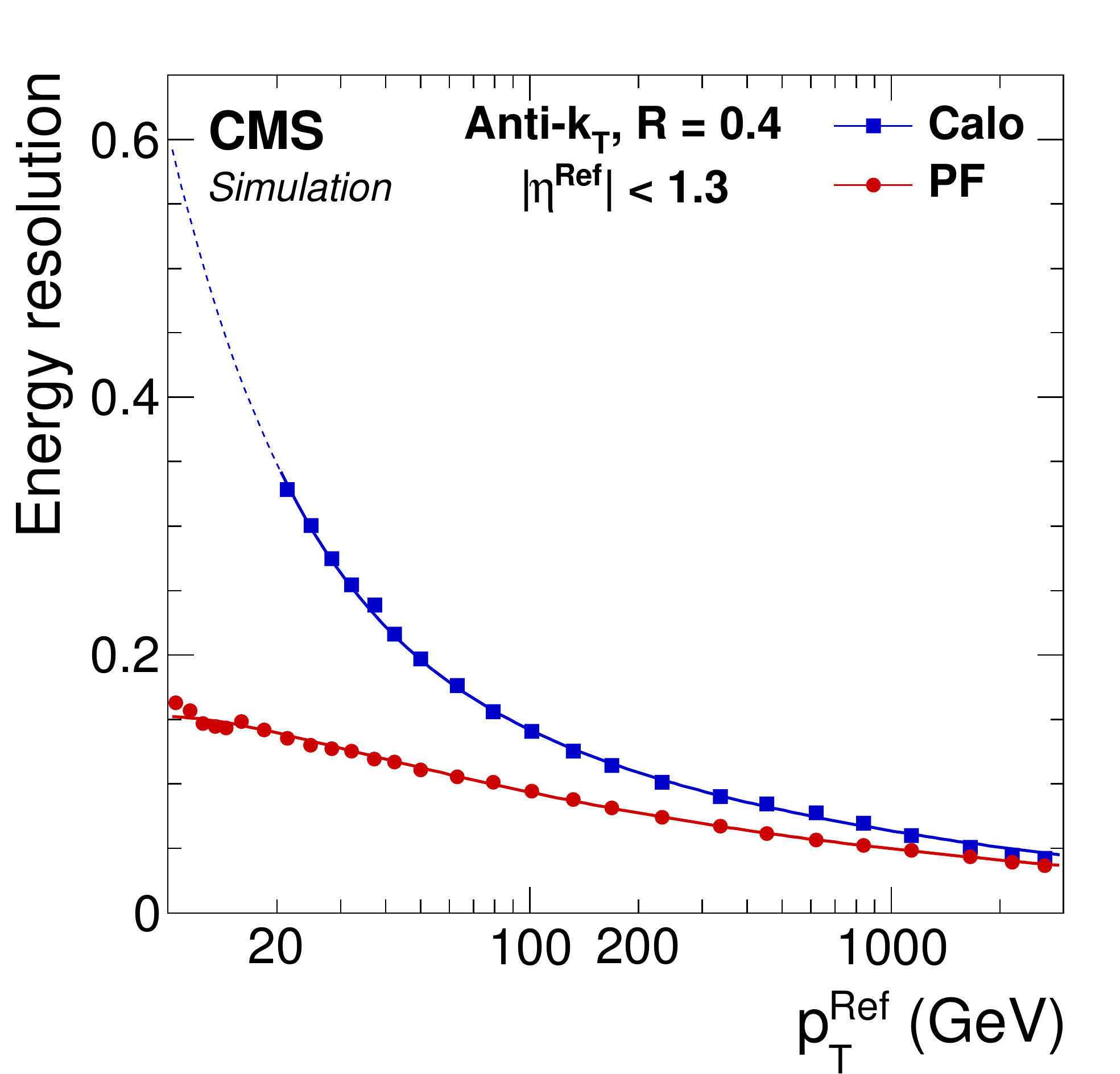}
  \includegraphics[width=0.49\textwidth]{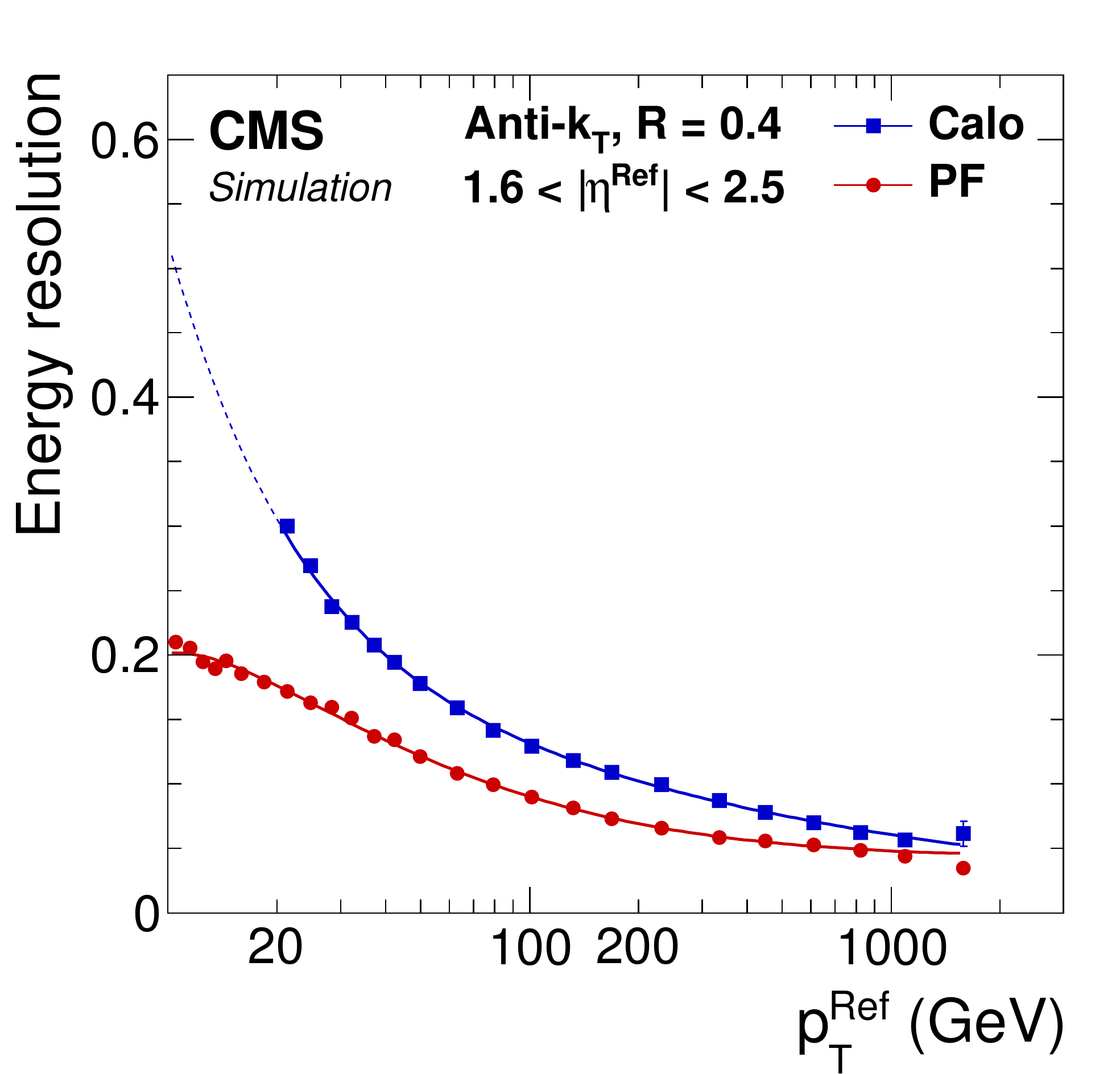}
  \caption{
    Jet energy resolution as a function of \ptref in the barrel (left) and in the endcap (right) regions.
The lines, added to guide the eye, correspond to fitted functions with ad hoc parametrizations.
  }
  \label{fig:expected_performance_jets_resolution}
\end{figure}

The raw jet energy response, defined as the mean ratio of the reconstructed jet energy to the reference jet energy, is shown in Fig.~\ref{fig:expected_performance_jets_response}.
The PF jet response is almost constant as a function of the jet \pt and is close to unity across the whole detector acceptance.
A jet energy correction procedure is used to bring the jet energy response to unity, which removes any dependence on \pt and $\eta$~\cite{JME-13-004}.
After this correction, the jet energy resolution, defined as the Gaussian width of the ratio between the corrected and reference jet energies, is shown in Fig.~\ref{fig:expected_performance_jets_resolution}.

The improvements in angular resolution, energy response, and energy resolution result mostly from a more precise and accurate measurement of the jet charged-hadron momentum in the PF algorithm.
In Calo jets, the charged-hadron energy is measured by the ECAL and HCAL with a resolution of $110\%/\sqrt{E/\GeV} \oplus 9\%$ and is underestimated for three reasons.
First, since low-\pt charged hadrons are swept away by the magnetic field,
their energy deposits typically remain unclustered or end up in a different jet.
Second, hadrons with an energy lower than 10\GeV have a low probability to be detected in the HCAL because of shower fluctuations and early showers in the ECAL.
Third, because the deposits of charged and neutral hadrons in the ECAL cannot be separated from the electromagnetic deposits without the PF algorithm,
they remain calibrated at the electromagnetic scale for the reasons given above.
With the PF algorithm, on the other hand, charged hadrons are reconstructed with the right direction, the correct energy scale, and with a much superior resolution in angle and momentum.

\begin{figure}[htbp]
  \centering
  \includegraphics[width=0.49\textwidth]{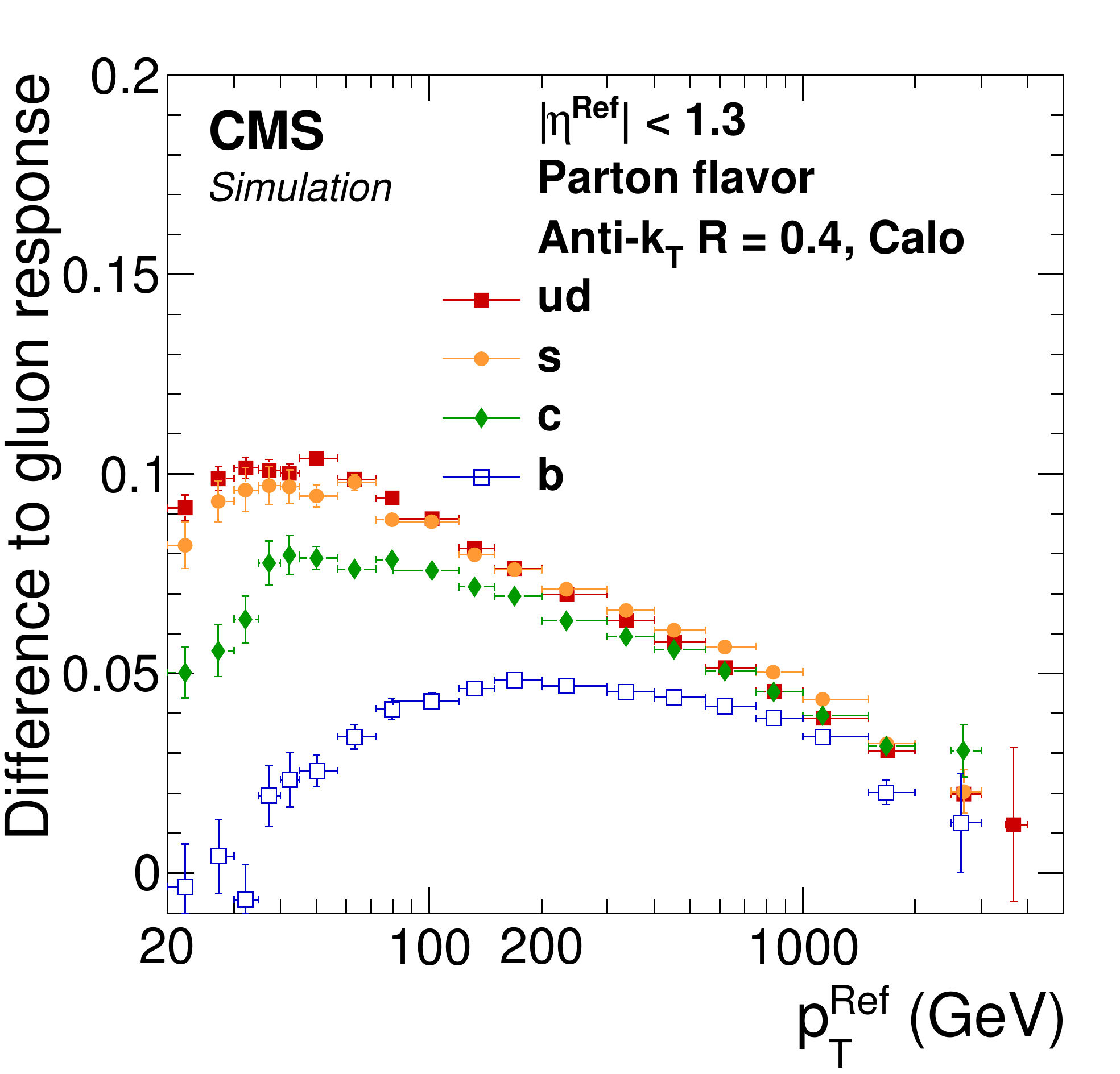}
  \includegraphics[width=0.49\textwidth]{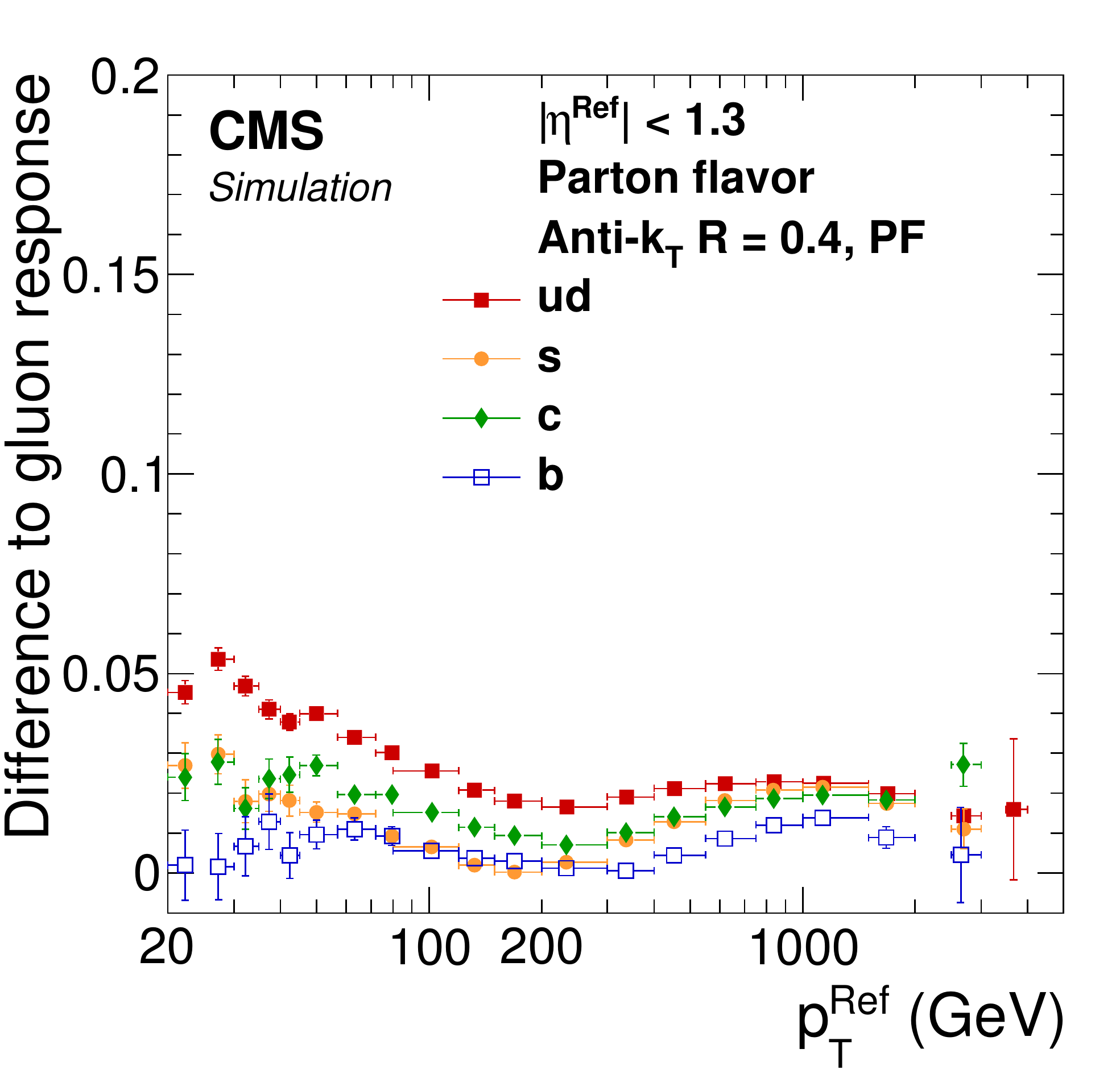}
  \caption{Absolute difference in jet energy response between quark and gluon jets as a function of \ptref for Calo jets (left) and PF jets (right). }
  \label{fig:expected_performance_jets_flavour}
\end{figure}

The particle content of jets in terms of particle type and energy distribution is described by the fragmentation functions and depends on the flavour of the parton that initiated the jet.
Gluon jets, especially, feature on average more low-energy particles than quark jets~\cite{Abreu:1999af},
which results in a lower jet energy response.
Because the flavour of the parton that initiated the jet cannot be determined with sufficient confidence in most physics analyses, the same jet energy correction is applied to all jets,
and the difference in response between quark and gluon jets is considered as a source of systematic uncertainty.
The relative difference in response is shown in Fig.~\ref{fig:expected_performance_jets_flavour} for Calo and PF jets.
For the reasons detailed above, the low-energy particles in gluon jets are more likely to be captured in PF jets,
and the difference between quark and gluon jet energy response is therefore smaller than for Calo jets.

\subsection{Missing transverse momentum}
\label{sec:expected_performance_met}

The presence of particles that do not interact with the detector
material, \eg neutrinos,
is indirectly revealed by missing transverse momentum, often referred to as missing transverse energy~\cite{JME-13-003-JINST}.
The raw missing transverse momentum vector is defined in such a way as
to balance the vectorial sum of the transverse momenta of all particles,
\begin{equation}
  \vecptmissof{PF} (\text{raw}) = - \sum_{i=1}^{N_\text{particles}} \ivecpt{i}.
\end{equation}
The jet-energy-corrected missing transverse momentum,
\begin{equation}
  \vecptmissof{PF} = - \sum_{i=1}^{N_\text{particles}} \ivecpt{i} - \sum_{j=1}^{N_\text{PF jets}} ({\vec p^\text{corr}_{\mathrm{T},j}} - \ivecpt{j}),
\end{equation}
includes a term that replaces the raw momentum $\ivecpt{j}$  of each PF jet with $\ivecpt{j} >10\GeV$ by its corrected value $\vec p^\text{corr}_{\mathrm{T},j}$.
As can be seen from Fig.~\ref{fig:expected_performance_jets_response},
the PF response to jets is close to unity,
which makes this correction term small.

Prior to the deployment of PF reconstruction, the missing
transverse momentum was evaluated as
\begin{equation}
 \vecptmissof{Calo}  = - \sum_{i=1}^{N_\text{cells}} \ivecpt{i} - \sum_{j=1}^{N_\text{Calo jets}} ({\vec p^\text{corr}_{\mathrm{T},j}} - \ivecpt{j}) - \sum_{k=1}^{N_\text{muons}} \ivecpt{k}.
\end{equation}
The first term, which corresponds to the raw calorimeter missing transverse momentum, balances the total transverse momentum vector measured by the calorimeters.
In this term, the transverse momentum $\ivecpt{i}$ of a given cell is
calculated under the assumption that the energy measured by the cell is
deposited by a massless particle coming from the origin of the CMS coordinate system.
The jet momentum correction term,
computed with all Calo jets with $\pt>20\GeV$,
is substantial given the relatively low response of Calo jets.
The second correction term accounts for the presence of identified muons with $\pt>10\GeV$;
it is necessary because muons do not leave significant energy in the calorimeters.

\begin{figure}[htbp]
\centering
\includegraphics[width=0.49\textwidth]{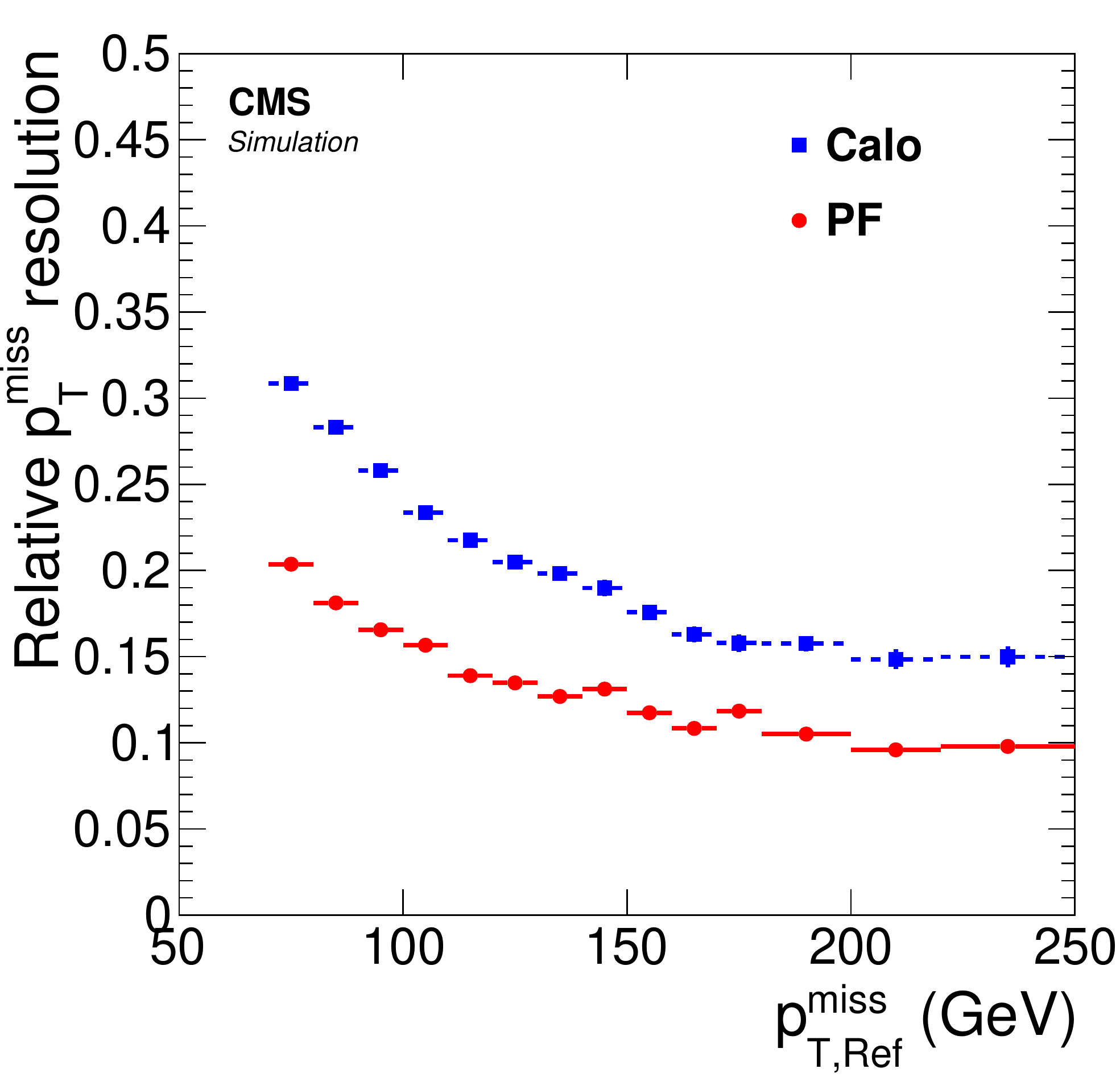}
\includegraphics[width=0.49\textwidth]{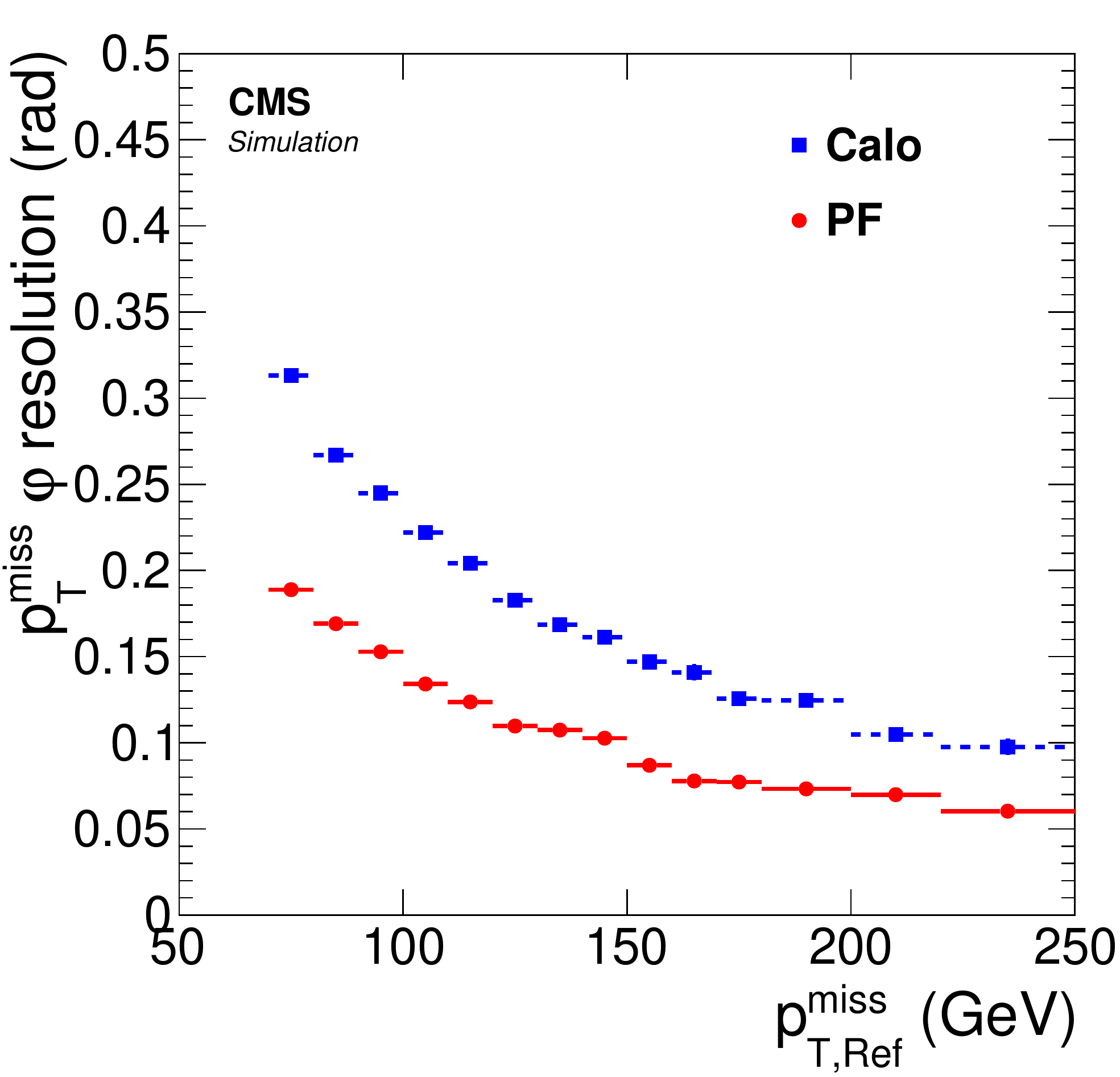}
\caption{
Relative \ptmiss resolution and resolution on the \vecptmiss direction as a function of \ptmissof{Ref} for a simulated \ttbar sample.
}
\label{fig:expected_performance_met_phi_resolution}
\end{figure}

The performance improvement brought by PF reconstruction is
quantified with a sample of \ttbar events by comparing \vecptmissof{PF} and
\vecptmissof{Calo} to the reference \vecptmissof{Ref},
calculated with all stable particles from the event generator,
excluding neutrinos. The \ptmiss resolution must be studied for events in
which the \ptmiss response has been calibrated to unity. The $\ptmissref$ is
therefore required to be larger than $70\GeV$, a value above which  the
jet-energy corrections are found to be sufficient to adequately calibrate
the PF and Calo \ptmiss response.
Figure~\ref{fig:expected_performance_met_phi_resolution} shows
the relative $\ptmiss$ resolution and the $\vecptmiss$ angular resolution,
obtained with a Gaussian fit in each bin of $\vecptmissref$.

\subsection{Electrons}
\label{sec:expected_performance_electrons}

The electron seeding and the subsequent reconstruction steps are described in Sections~\ref{sec:reco_electrons} and~\ref{sec:particle_id_reco_electrons}.
In the reconstruction, electron candidates are only required to satisfy loose identification criteria so as to ensure high identification efficiency for genuine electrons,
with the potential drawback of a large misidentification probability for charged hadrons interacting mostly in the ECAL.
In this section, as is typically done in physics analyses, the electron identification is tightened with a threshold on the classifier score of a BDT trained for electrons selected without any trigger requirement~\cite{Khachatryan:2015hwa}.

\begin{figure}[tbh]
\centering
\includegraphics[width=0.45\columnwidth]{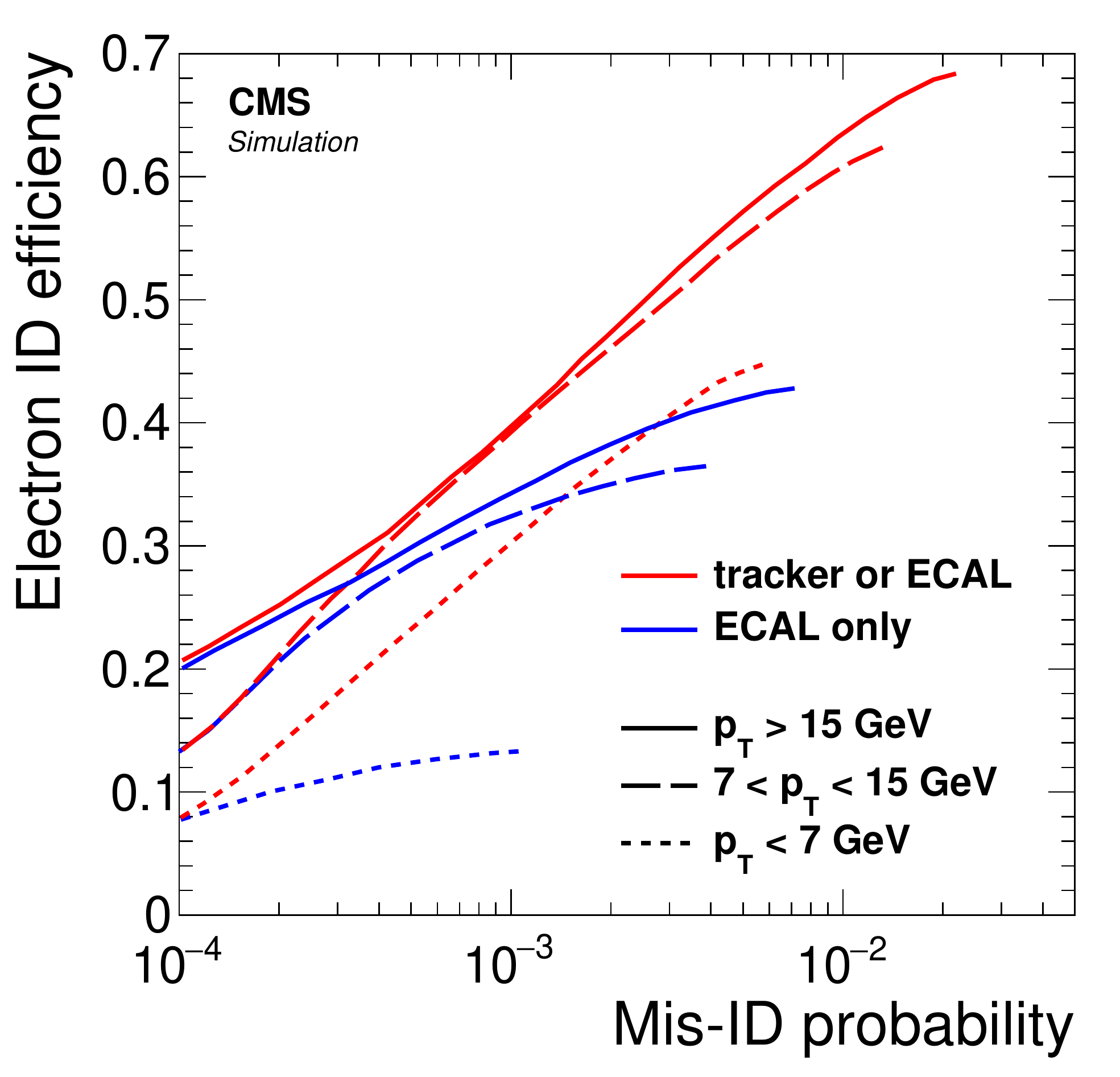}
\includegraphics[width=0.45\columnwidth]{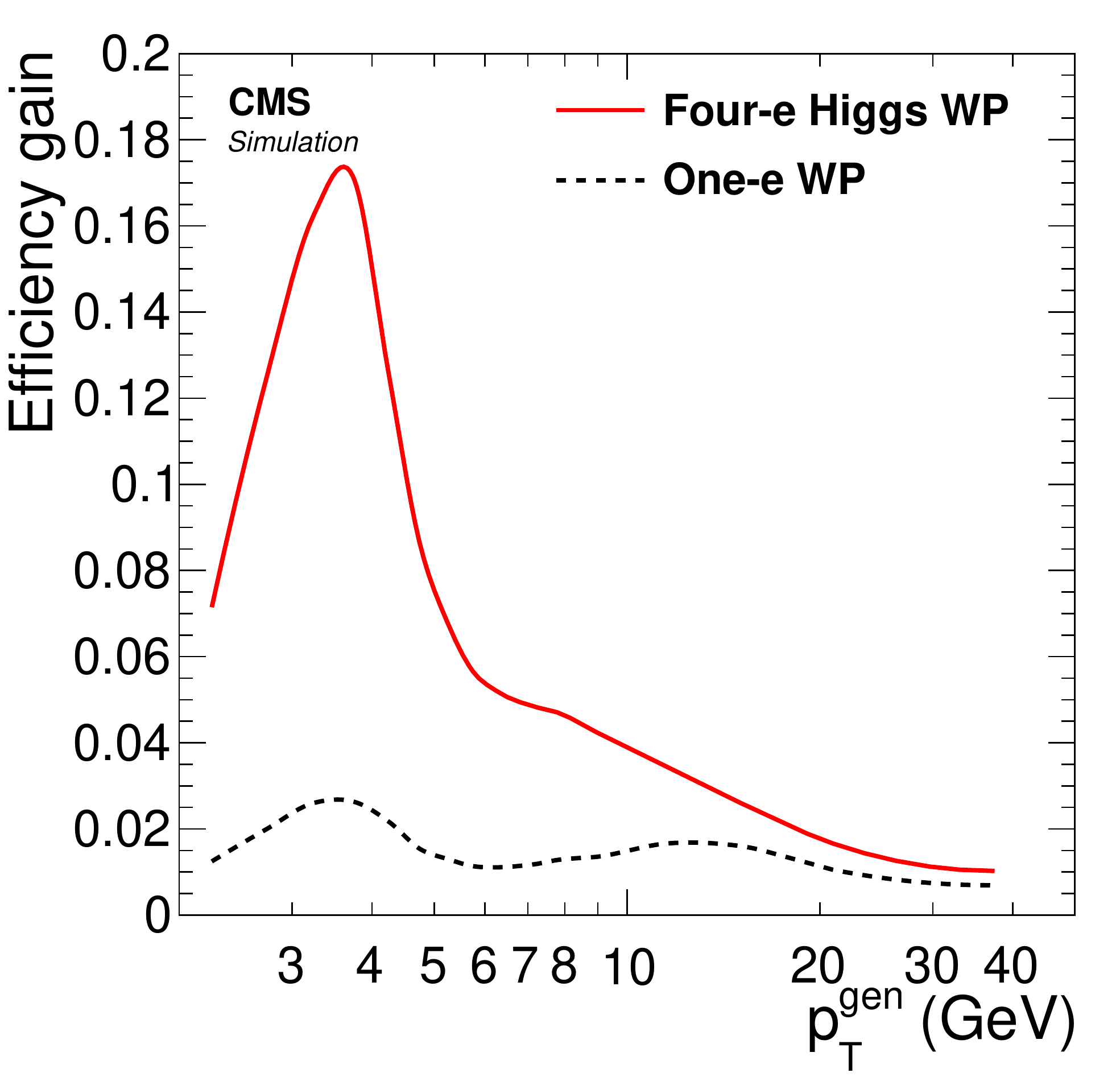}
\caption{\label{fig:ElectronIdEfficiency}
Left: Efficiency to reconstruct electrons from b hadron decays (signal) versus the probability to misidentify a hadron as an electron (background). The solid, long-dashed, and short-dashed lines refer to electrons and hadrons with \pt larger than 15, within $[7,15]$,  and lower than $7\GeV$, respectively. The curves correspond to a threshold scan on the BDT classifier score for ECAL-based seeded electrons and for tracker- or ECAL-based seeded electrons.
Right: Absolute gain in reconstruction and identification efficiency provided by the tracker-based seeding procedure for two working points (WP) corresponding to different values of the threshold on the BDT classifier score. The solid line corresponds to the value used in the $\PH \to \Z\Z \to 4\,\Pe$ analyses and the dashed line to the value typically used in analyses of single-electron final states.
In all cases, the classifier score of the BDT trained for electrons selected without any trigger requirement is used.
}
\end{figure}

The gain brought by the use of the tracker-based seeding in addition to the ECAL-based seeding is quantified in Fig.~\ref{fig:ElectronIdEfficiency}, for electrons in jets and for isolated electrons produced in the decay of heavy resonances.
The left plot shows the reconstruction and identification efficiency for electrons in jets as a function of the hadron misidentification probability.
Electrons and hadrons are selected from the same simulated sample of multijet events,
with $\pt > 2\GeV$ and $\abs{\eta}<2.4$.
Electrons are additionally required to come from the decay of b hadrons.
The electron efficiency is significantly improved, paving the way for b quark jet identification algorithms based on the presence of electrons in jets.

The absolute gain in efficiency for isolated electrons is quantified in the right plot for electrons from Z boson decays in a simulated Drell--Yan sample, and for two different working points.
The first working point, used in the search for $\PH\to ZZ\to 4\,\Pe$~\cite{higgs-cms, PhysRevD.89.092007},
provides very high electron efficiency in order to maximize the selection efficiency for events with four electrons.
At this working point, the addition of the tracker-based seeding adds almost 20\% to the identification efficiency of low-\pt electrons.
In the context of the $\PH\to ZZ\to 4\,\Pe$ analysis,
in which all four electrons are required to have $\pt>7\GeV$,
the tracker-based seeding adds 7\% to the selection efficiency of signal events.
The second working point, typical of single-electron analyses, aims at reducing the large multijet background.
In these analyses that only consider electrons with $\pt > 20\GeV$ due to triggering requirements, the gain in signal efficiency is about 1\%.
For both working points, the addition of the tracker-based seeding increases the hadron misidentification probability by less than a factor of 1.2 for \pt larger than $10\GeV$, and by less than a factor of 2 for \pt between 5 and $10\GeV$.

\subsection{Muons}
\label{sec:expected_performance_muons}

The PF muon identification, described in
Section~\ref{sec:particle_id_reco_muons}, is designed to retain \textit{prompt
muons} (from \eg decays of W and Z bosons or quarkonia states), muons from
\textit{heavy hadrons} (from decays of beauty or charm hadrons), and muons
from \textit{light hadrons} (from decays in flight of $\pi$ or K mesons),
with the highest possible efficiency. On the other hand, it has to
minimize the probability to misidentify a charged hadron as a muon,
\eg because of punch-through.

\begin{figure}[htb]
\centering
\includegraphics[width=0.48\textwidth]{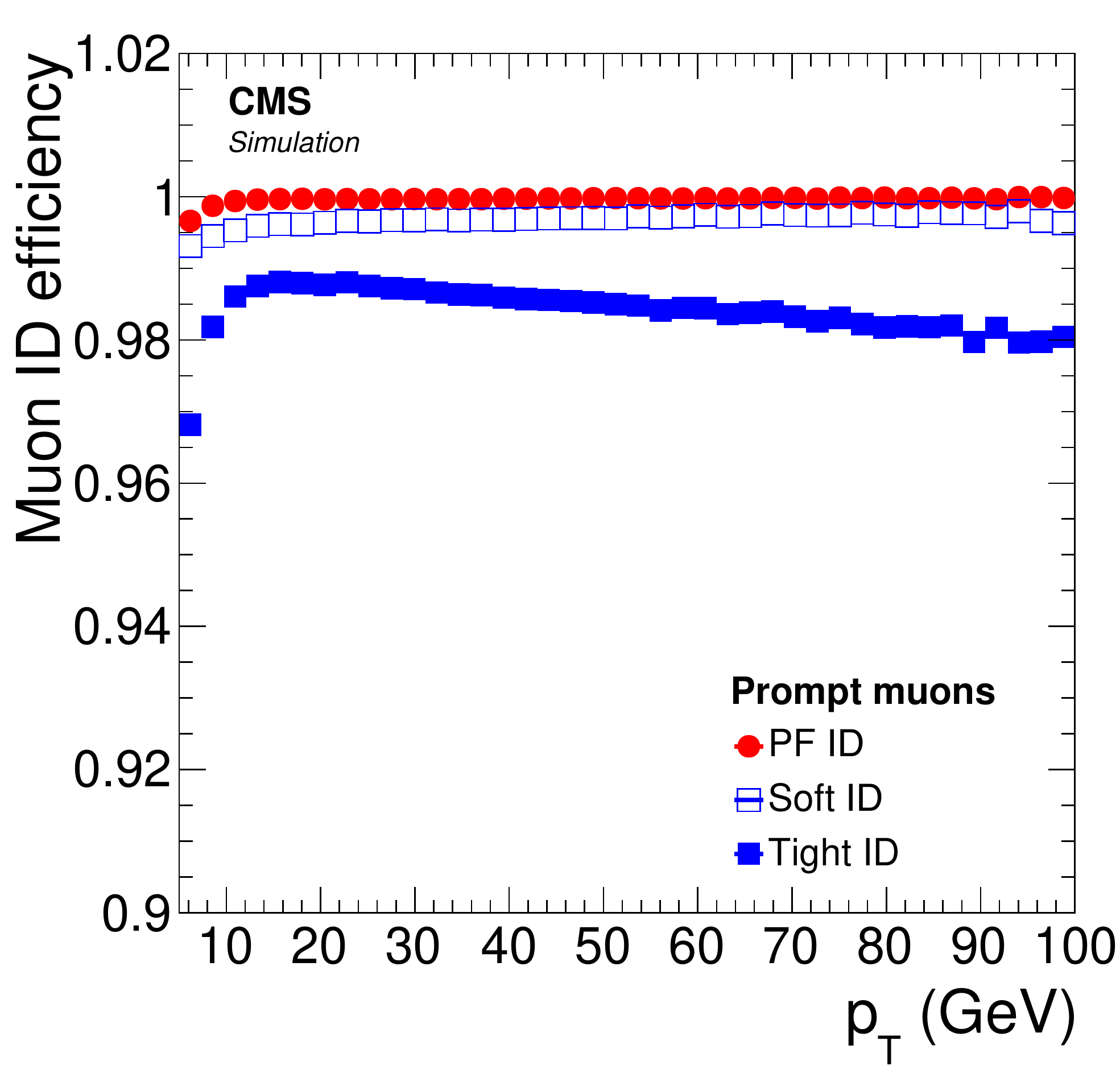}
\includegraphics[width=0.48\textwidth]{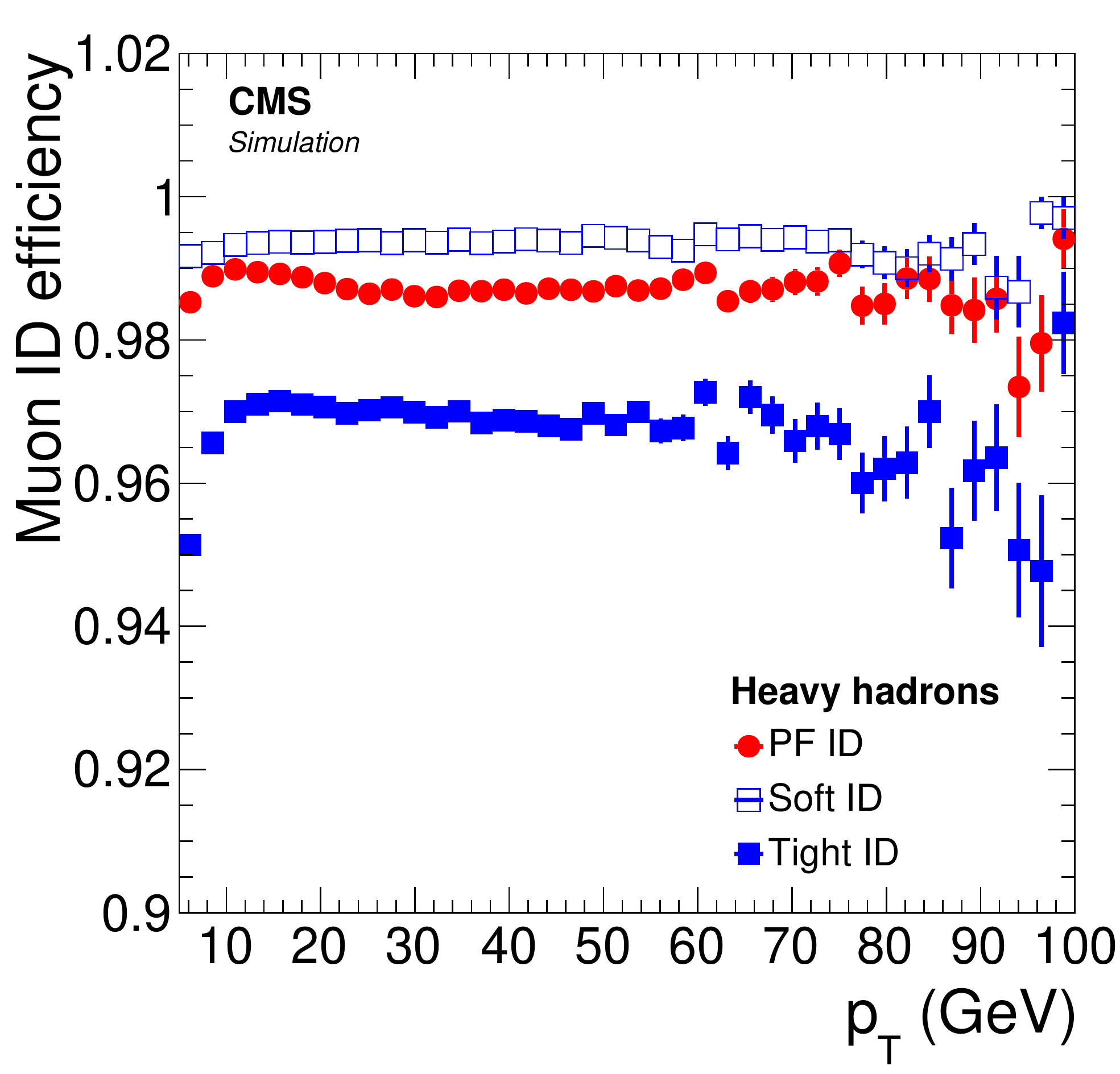}\\
\includegraphics[width=0.48\textwidth]{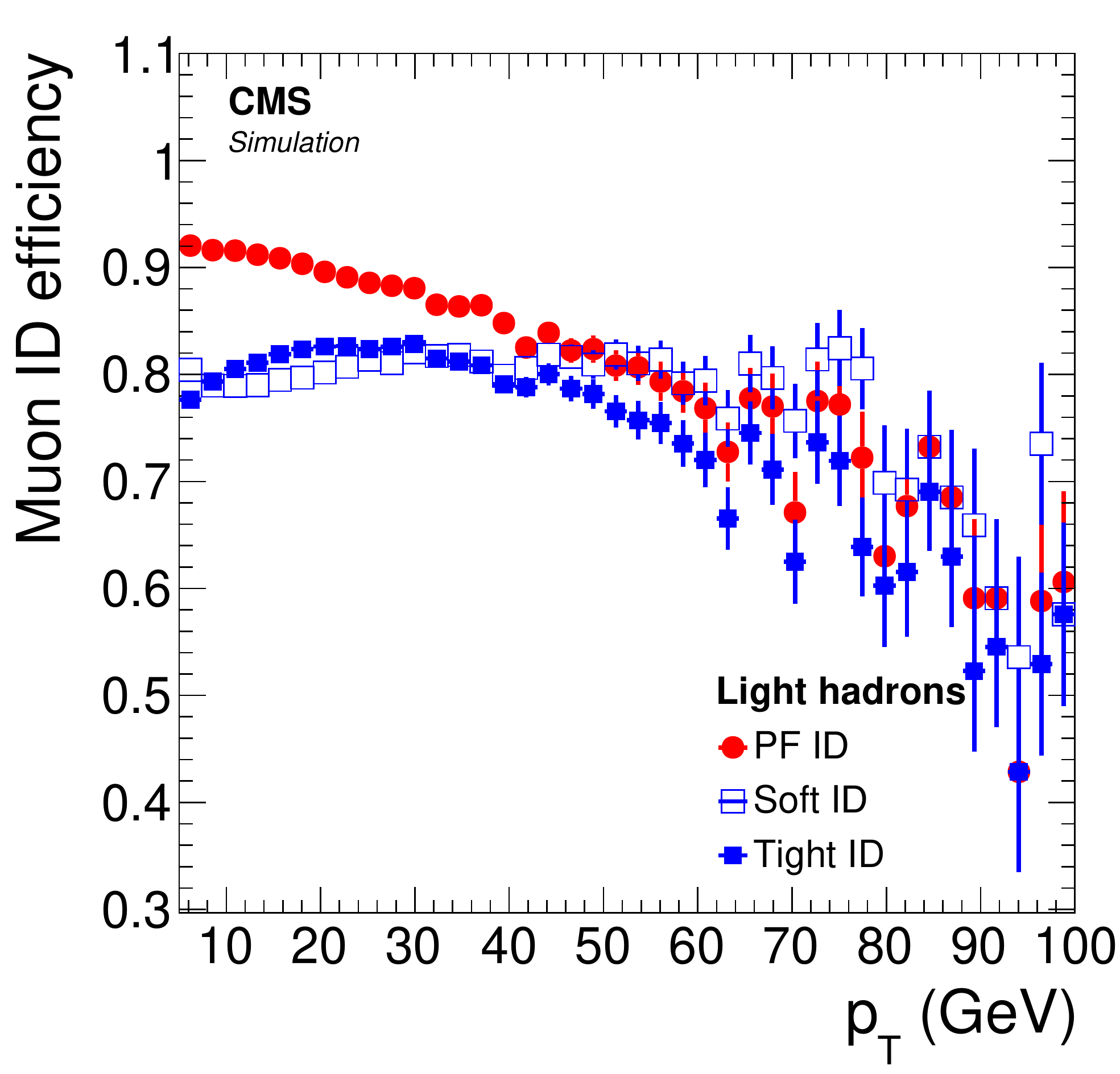}
\includegraphics[width=0.48\textwidth]{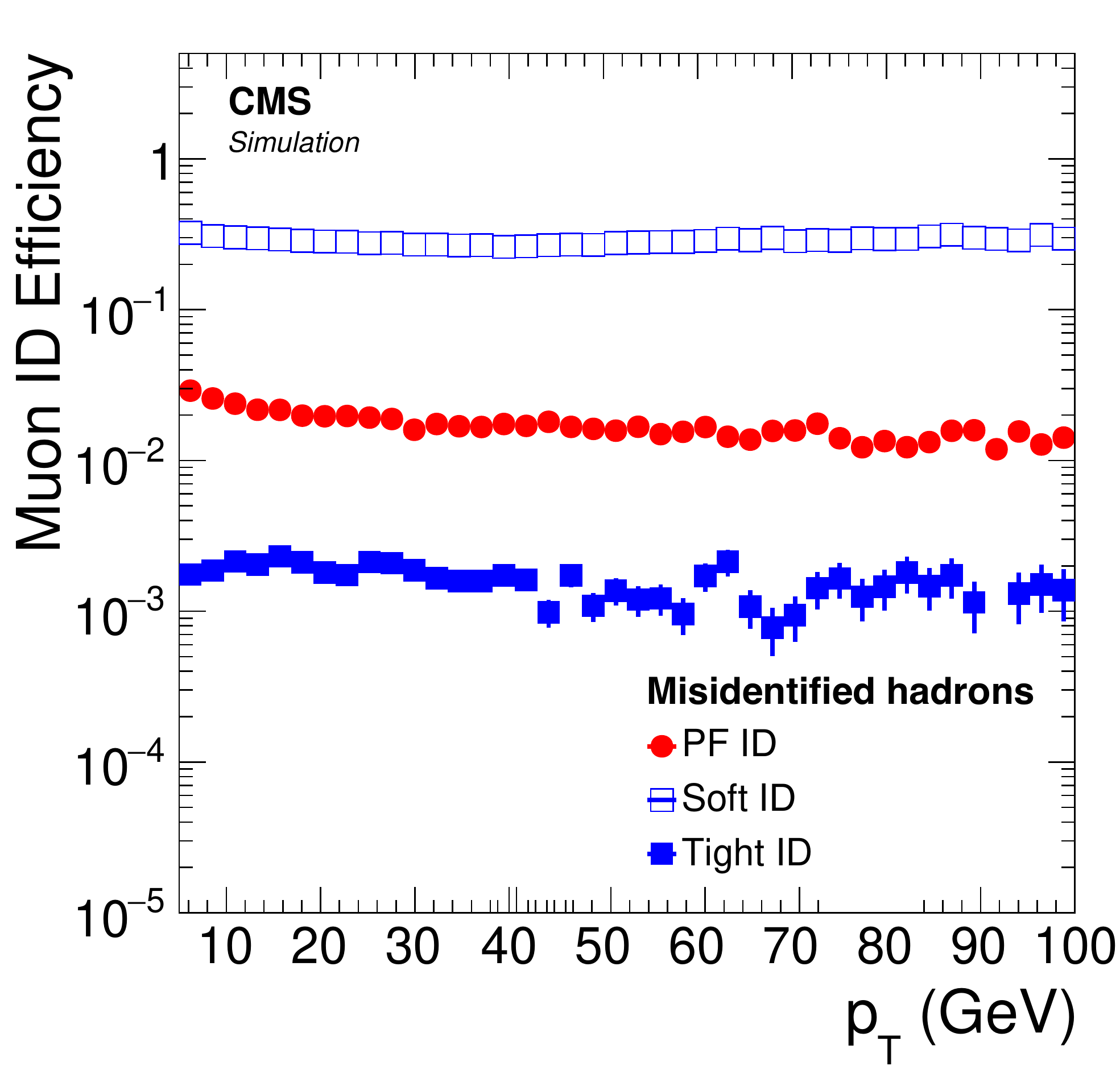}\\
\caption{Efficiency for different
    algorithms (PF, soft, and tight) to identify a
    simulated muon track that has been reconstructed as a
    tracker muon, as a function of the \pt of the
    reconstructed track. From top left to bottom right the
    efficiency of the three identification algorithms is shown for prompt muons,
    for muons from heavy-flavour decays, for muons from light-flavour decays,
    and for misidentified hadrons.}
\label{fig:muonsexpected}
\end{figure}

A Drell--Yan $\mu^+\mu^-$ event sample is used to evaluate the prompt muon
identification efficiency, while a muon-enriched multijet QCD sample is used
for the other three types of muon candidates. Figure~\ref{fig:muonsexpected} compares
the muon identification efficiency obtained with the PF algorithm
to the efficiency of other algorithms available prior to the developments
carried out for PF identification:
\begin{itemize}
  \item \textit{The soft muon identification} aims to achieve efficient identification
of muons from decays of quarkonia states. This selection requires a tracker muon with
a tighter matching to the muon segment, with a pull below 3 in the $x$ and $y$
directions instead of a pull below 4 in the $x$ direction only as in the tracker muon selection.
Additionally, the inner track must be reconstructed from at least five
inner-tracker layers, including one pixel detector layer.
  \item \textit{The tight muon identification} specifically targets muons
from $\PZ$ and $\PW$ decays. This selection requires a global-muon
track with a $\chi^2$ per degree-of-freedom lower than 10 and at least
one hit in the muon detectors. In addition, the candidate should be a
tracker muon with at least two matched muon segments in different muon
stations and an inner track reconstructed from at least five inner-tracking
layers, including one pixel detector layer.
\end{itemize}
The regular soft and tight ID criteria also feature an upper threshold
on the muon-track impact parameter, aimed at rejecting muons
from charged-hadron decays in flight. This requirement would defeat the
purpose of PF identification, which aims at being as inclusive
as possible for a truly global description of the event. As it also
reduces the efficiency of the soft and tight ID criteria, it is not applied
here for a fairer comparison. Because these two algorithms require the
selected tracks to be tracker muons, the muon identification efficiency is
displayed in Fig.~\ref{fig:muonsexpected} for tracker muons only. Muons
reconstructed as global muons but not tracker muons are considered only by
the PF muon identification, increasing the number of identified muons
by about 2\% over the whole $\pt$ spectrum ($+$1\% in the heavy-flavour
category, $+$5\% in the light-hadron category, and $+$5\% in the
misidentified-hadron category).

The PF identification is the most efficient one for prompt muons.
The soft identification is 0.5\% more efficient on muons from semileptonic
decays of heavy hadrons,
but its much higher hadron misidentification rate (30\% instead of 2\%) makes
this selection unusable for PF. The calorimeter deposits from a charged hadron
misidentified as a muon are automatically identified as (spurious) neutral
particles in the PF algorithm, leading to a potentially large overestimation of
the corresponding jet energy. The PF muon identification, in this respect,
strikes a balance between efficiency and misidentification rate for PF
reconstruction and global event description.

\subsection{Lepton isolation }
\label{sec:expected_performance_isolation}

Lepton isolation is the main handle for selecting prompt muons and electrons produced in the electroweak decay of massive particles such as $\PZ$ or $\PW$ bosons
and for rejecting the large number of leptons produced in jets through the decay of heavy-flavour hadrons or the decay in flight of charged pions and kaons.
The isolation is quantified by estimating the total \pt of the particles emitted around the direction of the lepton.
The particle-based isolation relative to the lepton \pt is defined as
\begin{equation}
 I_\mathrm{PF} = \frac{1}{\pt}\left( \sum_{\Ph^\pm}  \pt^{\Ph^\pm} + \sum_{\gamma} \pt^{\gamma} + \sum_{\Ph^0} \pt^{\Ph^0} \right),
\label{eq:isolation}
\end{equation}
where the sums run over the charged hadrons (${\Ph^\pm}$), photons ($\gamma$),
and neutral hadrons (${\Ph^0}$) with a distance $\Delta R$ to the lepton smaller
than either 0.3 or 0.5 in the $(\eta, \varphi)$ plane.

\begin{figure}[htb!]
  \centering
  \includegraphics[height=0.6\textwidth,angle=0]{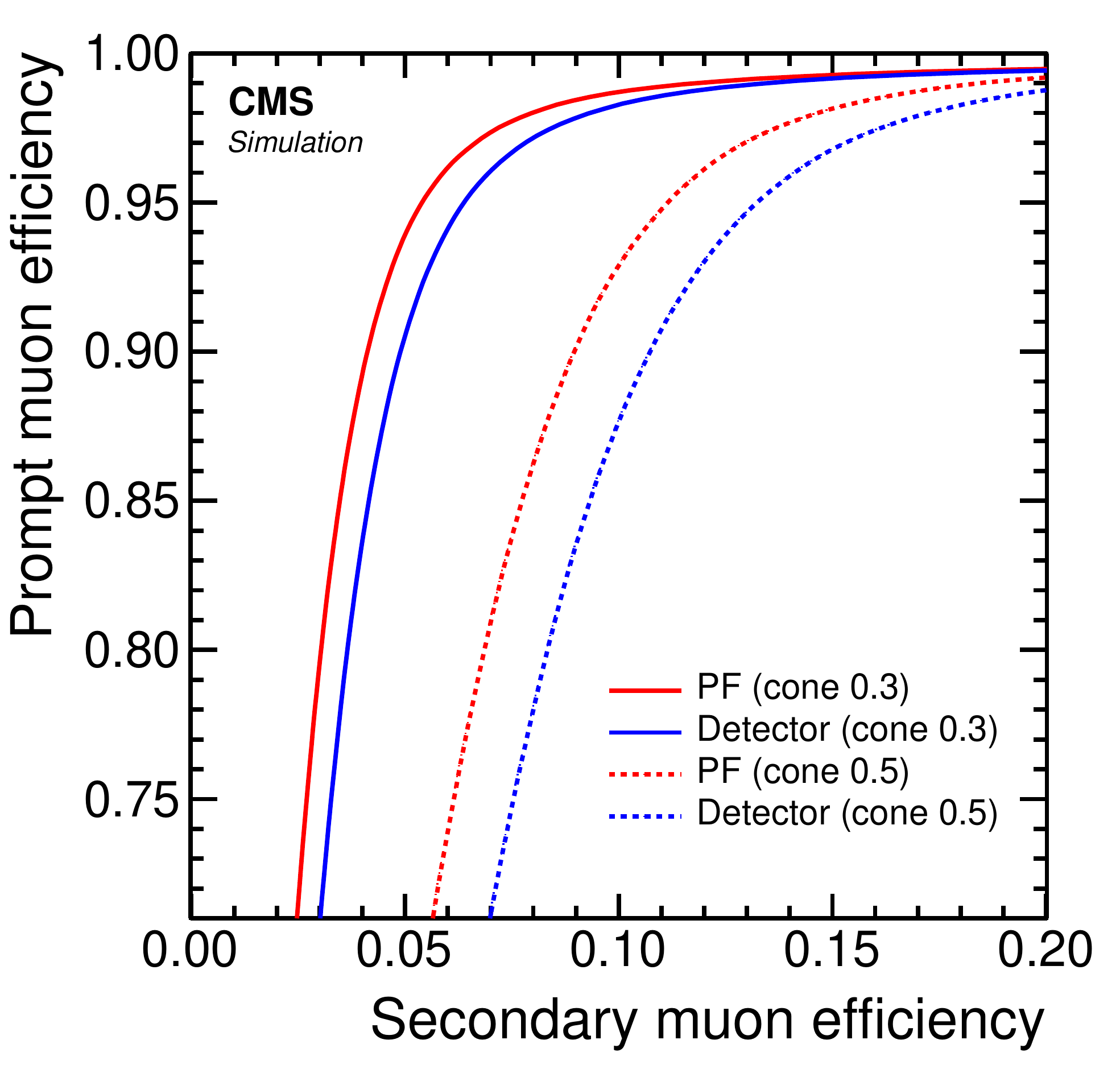}
  \caption{
Isolation efficiency for muons from $\PW$ boson decays versus isolation efficiency for
muons from secondary decays, as a function of the threshold on the isolation for the
detector- and particle-based methods. All muons come from simulated $\ttbar$ events and
are required to have a \pt larger than 15\GeV. The efficiencies are shown for two
choices of the maximum $\Delta R$ (isolation cone size): 0.3 and 0.5.
}
  \label{fig:IsoROCexp}
\end{figure}

The performance of the particle-based isolation is studied for muons identified in simulated $\ttbar$ events.
Figure~\ref{fig:IsoROCexp} shows the efficiency to select signal prompt muons as a function of the probability to select background secondary muons.
The performance of the particle-based isolation is compared to the performance of the detector-based isolation, computed from the \pt and energy of the neighbouring inner tracks and calorimeter deposits, respectively, as
\begin{linenomath}
\begin{equation}
 I_\text{det} = \frac{1}{\pt}\left( \sum_\text{tracks}  \pt^\text{track} + \sum_\mathrm{ECAL} \ET^\mathrm{ECAL} + \sum_\mathrm{HCAL} \ET^\mathrm{HCAL} \right).
\end{equation}
\end{linenomath}

The performance of the detector-based isolation is worse mainly
because the \pt carried by charged hadrons is counted
twice, through the tracks and through the calorimeter deposits.

\subsection[Hadronic decays of taus]{Hadronic $\Pgt$ decays}
\label{sec:expected_performance_taus}

The $\Pgt$  decay produces either a charged lepton ($\Pe$ or $\mu$) and two neutrinos,
or a few hadrons and one neutrino, with the branching fractions given in Table~\ref{tab:tauDecayModes}.
Hadronic $\Pgt$ decays, denoted as $\tauh$, can be differentiated from
quark and gluon jets by the multiplicity, the collimation, and the isolation of the decay products.

\begin{table}[htbp]
\centering
\topcaption{
  Branching fraction $\mathcal{B}$ of the main (negative) $\Pgt$  decay modes~\cite{PDG}.
  The generic symbol $\Phm$ represents a charged hadron, pion or kaon.
  In some cases, the decay products arise from an intermediate mesonic resonance.
}
\label{tab:tauDecayModes}
\begin{tabular}{lcc}
\hline
Decay mode & Meson resonance    & $\mathcal{B}$ $[\%]$    \\
\hline
$\Pgt^{-} \to \Pe^{-} \, \Pagne \, \Pnut$     &                         & 17.8 \\
$\Pgt^{-} \to \Pgm^{-} \, \Pagngm \, \Pnut$     &                         & 17.4 \\
\\[0.2ex]
$\Pgt^{-} \to \Phm \, \Pnut$                             &             & 11.5 \\
$\Pgt^{-} \to \Phm \, \PGpz \, \Pnut$                    & $\Pgr$ & 26.0 \\
$\Pgt^{-} \to \Phm \, \PGpz \, \PGpz \, \Pnut$            & $\textrm{a}_1(1260)$            & 10.8 \\
$\Pgt^{-} \to \Phm \, \Ph^+ \, \Phm \, \Pnut$            & $\textrm{a}_1(1260)$            & 9.8 \\
$\Pgt^{-} \to \Phm \, \Ph^+ \, \Phm \, \PGpz \, \Pnut$ &                         & 4.8 \\
\\[0.2ex]
Other modes with hadrons &                         & 1.8 \\
All modes containing hadrons &                         & 64.8 \\
\hline
\end{tabular}

\end{table}

The PF algorithm is able to resolve the particles arising
from the $\Pgt$ decay and to reconstruct the surrounding particles to
determine its isolation,
thereby providing valuable information for $\tauh$ identification.
The particles are used as input to the hadrons-plus-strips (HPS) algorithm~\cite{TAU-11-001}  to reconstruct and identify PF $\tauh$ candidates.
This algorithm, presented in detail in Ref.~\cite{TAU-14-001},
is seeded by jets of $\pt > 14$\GeV and $\abs{\eta} < 2.5$
reconstructed with the anti-$\kT$ algorithm ($R=0.4$).
The jet constituent particles are combined into $\tauh$ candidates compatible with one of the main $\Pgt$ decay modes,
$\Pgt^{-} \to \Phm \, \Pnut$,
$\Pgt^{-} \to \Phm \, \PGpz \, \Pnut$, $\Pgt^{-} \to \Phm \, \PGpz \, \PGpz \, \Pnut$,
and $\Pgt^{-} \to \Phm \, \Ph^+ \, \Phm \, \Pnut$.
The decay mode $\Pgt^{-} \to \Phm \, \Ph^+ \, \Phm \, \PGpz \, \Pnut$ is not considered owing to its relatively small branching fraction and high contamination from quark and gluon jets.
Because of the large amount of material in the inner tracker (Fig.~\ref{fig:cms_detector:tracker_material}),
photons from $\PGpz$ decays often convert before reaching the ECAL.
The resulting electrons and positrons can be identified as such by the PF algorithm or,
in case their track is not reconstructed, as photons displaced along the $\varphi$ direction because of the bending in the 3.8\unit{T} magnetic field. Neutral pions are therefore obtained by gathering reconstructed photons and electrons located in a small window of size $0.05{\times}0.20$ in the $(\eta,\varphi)$ plane.
Each $\tauh$ candidate is then required to have a mass compatible with its decay mode and to have unit charge.
Collimated $\tauh$ candidates are selected by requiring all charged hadrons and neutral pions to be within a circle of radius $\Delta R = (3.0\GeV)/\pt$ in the $(\eta, \varphi)$ plane called the signal cone.
The size of the signal cone is, however, not allowed to increase above 0.1 at low \pt, nor to decrease below 0.05 at high \pt.
It decreases with $\pt$ to account for the boost of the $\Pgt$ decay products.
Finally, the highest $\pt$ selected $\tauh$ candidate in the jet is retained.
The four-momentum of the $\tauh$ candidate is determined by summing the four-momenta of its constituent particles.
Its absolute isolation is quantified as explained in Section~\ref{sec:expected_performance_isolation} with all particles at a distance $\Delta R$ from the $\tauh$ smaller than 0.5
apart from the ones used in the reconstruction of the $\tauh$ itself,
and without normalizing by the \tauh \pt.
The loose, medium, and tight isolation working points are defined by requiring the absolute isolation to be smaller than 2.0, 1.0, and $0.8\GeV$, respectively.

Before the advent of PF reconstruction,
$\tauh$ candidates were reconstructed as collimated and isolated calorimetric jets, called Calo $\tauh$~\cite{PTDR2}.
Their reconstruction is seeded by Calo jets reconstructed with the anti-$\kT$ algorithm ($R=0.5$) and matched with at least one track with $\pt > 5$\GeV.
The region $\Delta R < 0.07$ around the jet is chosen as the signal cone,
and is expected to contain the charged hadrons and neutral pions from the $\Pgt$ decay.
The signal cone must contain either one or three tracks,
with a total electric charge equal to $\pm 1$.
Isolated $\tauh$ candidates are selected with the requirements that no track with $\pt > 1\GeV$ be found within an annulus of size $0.07 < \Delta R < 0.5$ centred on the highest \pt track, and that less than $5\GeV$ of energy be measured in the ECAL within the annulus $0.15 < \Delta R < 0.5$.

The performance of the HPS (PF) and Calo $\tauh$ algorithms
are compared in terms of identification efficiency, jet
misidentification rate, and momentum reconstruction.
Genuine $\tauh$ with a \pt between 20\GeV and 2\TeV are obtained
in the simulation from the Drell--Yan process and from the decay of a hypothetical heavy particle of mass 3.2\TeV.
For the jet misidentification rate,
a simulated QCD multijet sample covering the same \pt range is used.

The probability for the HPS (PF) algorithm to assign the correct decay mode to the reconstructed and identified $\tauh$ is shown in Table~\ref{tab:tau_matrix}.
The generated decay mode is typically found for about $90\%$ of the $\tauh$.
The largest decay-mode migrations, of the order of 10--15\%,
affect $\tauh$ candidates with a single charged hadron
and are due to the reconstruction of an incorrect number of $\PGpz$.

\begin{table}[htbp!]
\centering
\topcaption{
   Correlation between the reconstructed and generated decay modes,
   for $\tauh$ produced in simulated $\cPZ/\Pggx \to \Pgt\Pgt$ events.
   Reconstructed $\tauh$ candidates are required to be matched to a generated $\tauh$,
   to be reconstructed with $\pt > 20$\GeV and $\abs{\eta} < 2.3$
   under one of the HPS decay modes, and to
   satisfy the loose isolation working point.
 }
\label{tab:tau_matrix}
\begin{tabular}{l|c|c|c}
  \hline
  & \multicolumn{3}{c}{Generated} \\
  \cline{2-4}
  Reconstructed & $\Pgt^{-} \to \Phm  \Pnut$ & $\Pgt^{-} \to \Phm  \geq 1 \PGpz\,  \Pnut$ & $\Pgt^{-} \to \Phm  \Ph^+  \Phm  \Pnut$  \\
  \hline
  $\Pgt^{-} \to \Phm \Pnut$                             &     {0.89}   &  0.16   &  0.01 \\
  $\Pgt^{-} \to \Phm \geq 1 \PGpz\,\Pnut$      &    0.11    &  {0.83} &   0.02  \\
  $\Pgt^{-} \to \Phm \Ph^+ \Phm \Pnut$       &    0.00    &  0.01 & {0.97} \\
  \hline
\end{tabular}
\end{table}

\begin{figure}
\setlength{\unitlength}{1mm}
\centering
  \includegraphics*[height=0.28\textheight]{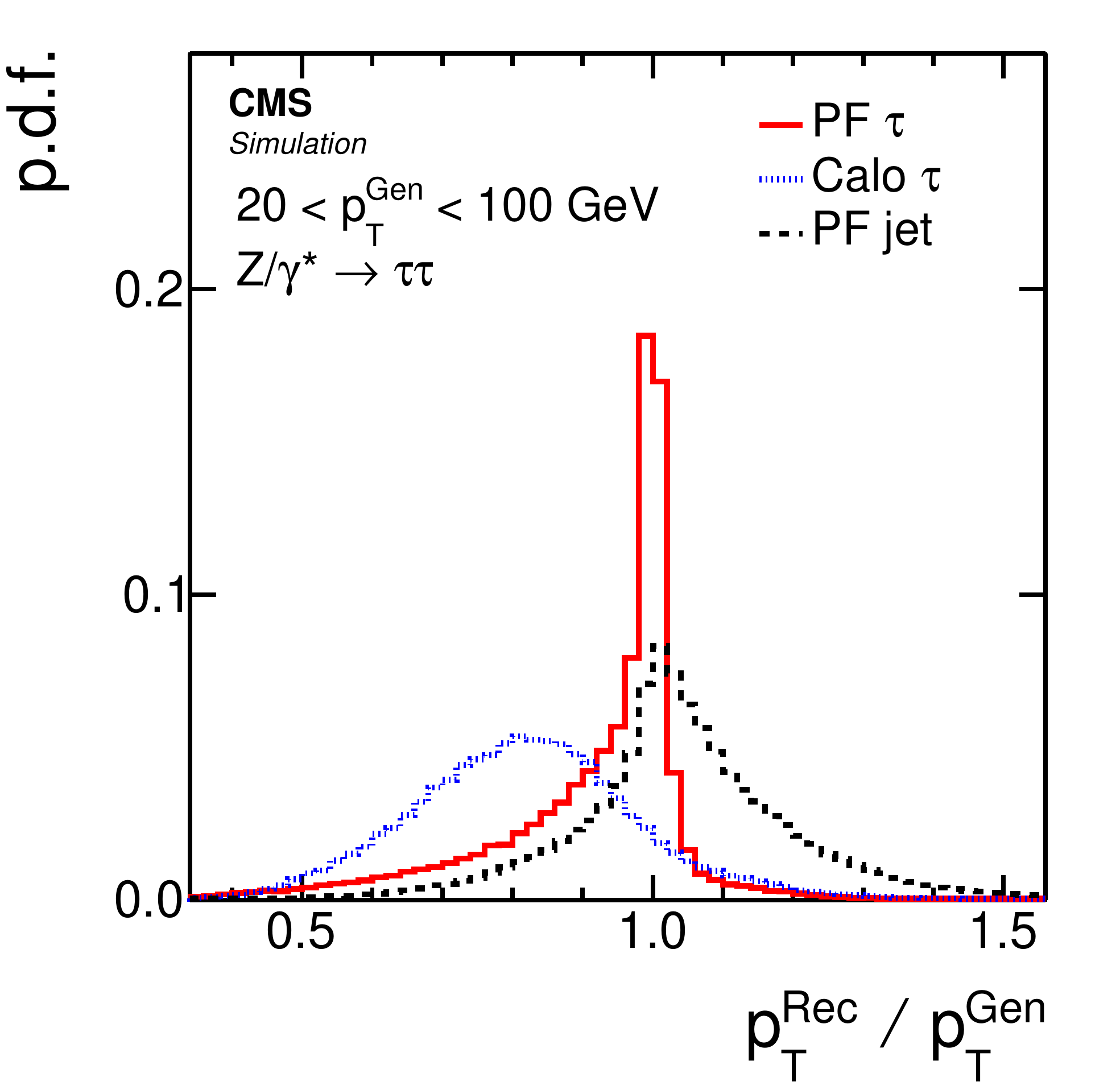}
  \includegraphics*[height=0.28\textheight]{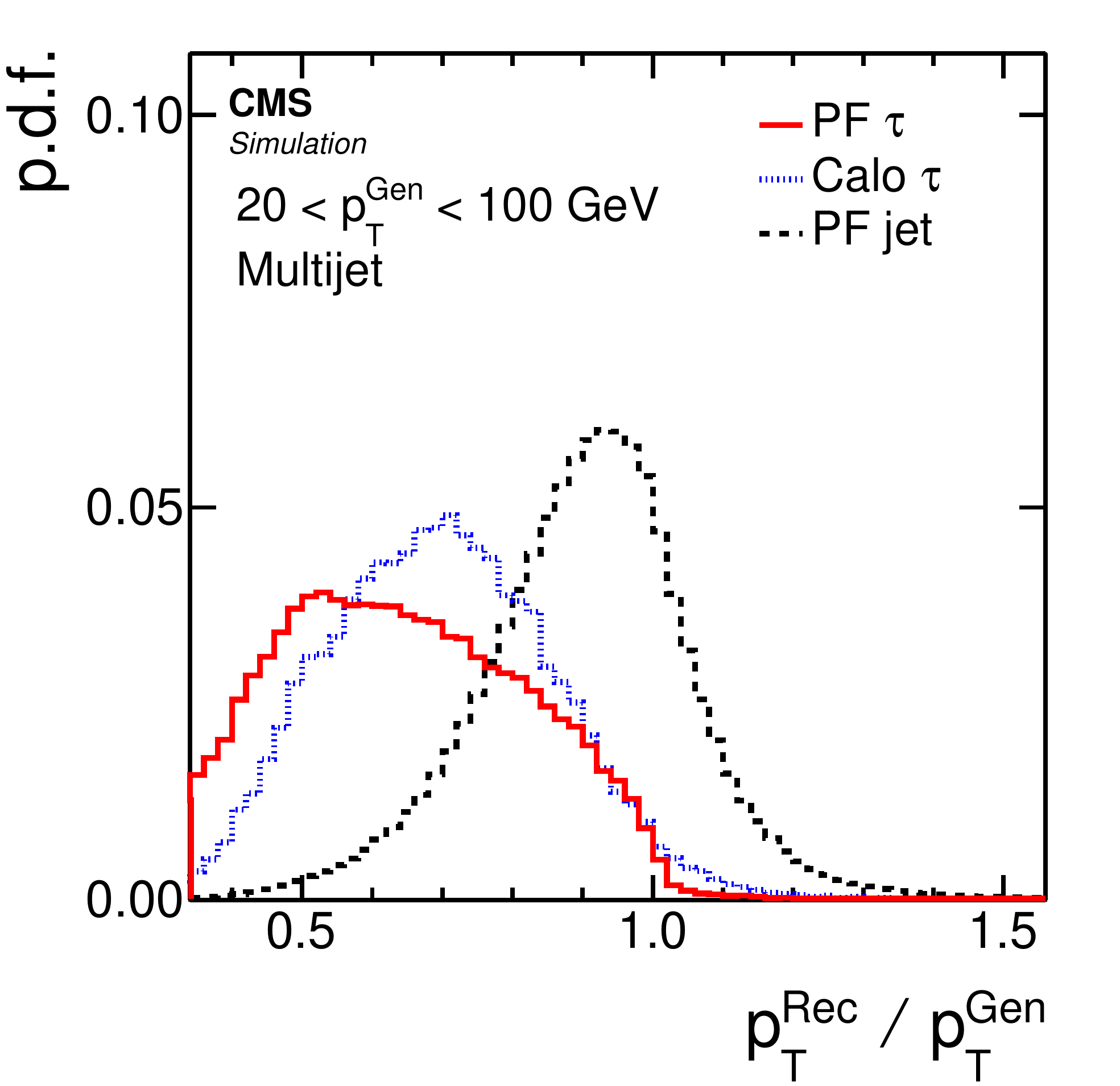}\\
  \includegraphics*[height=0.28\textheight]{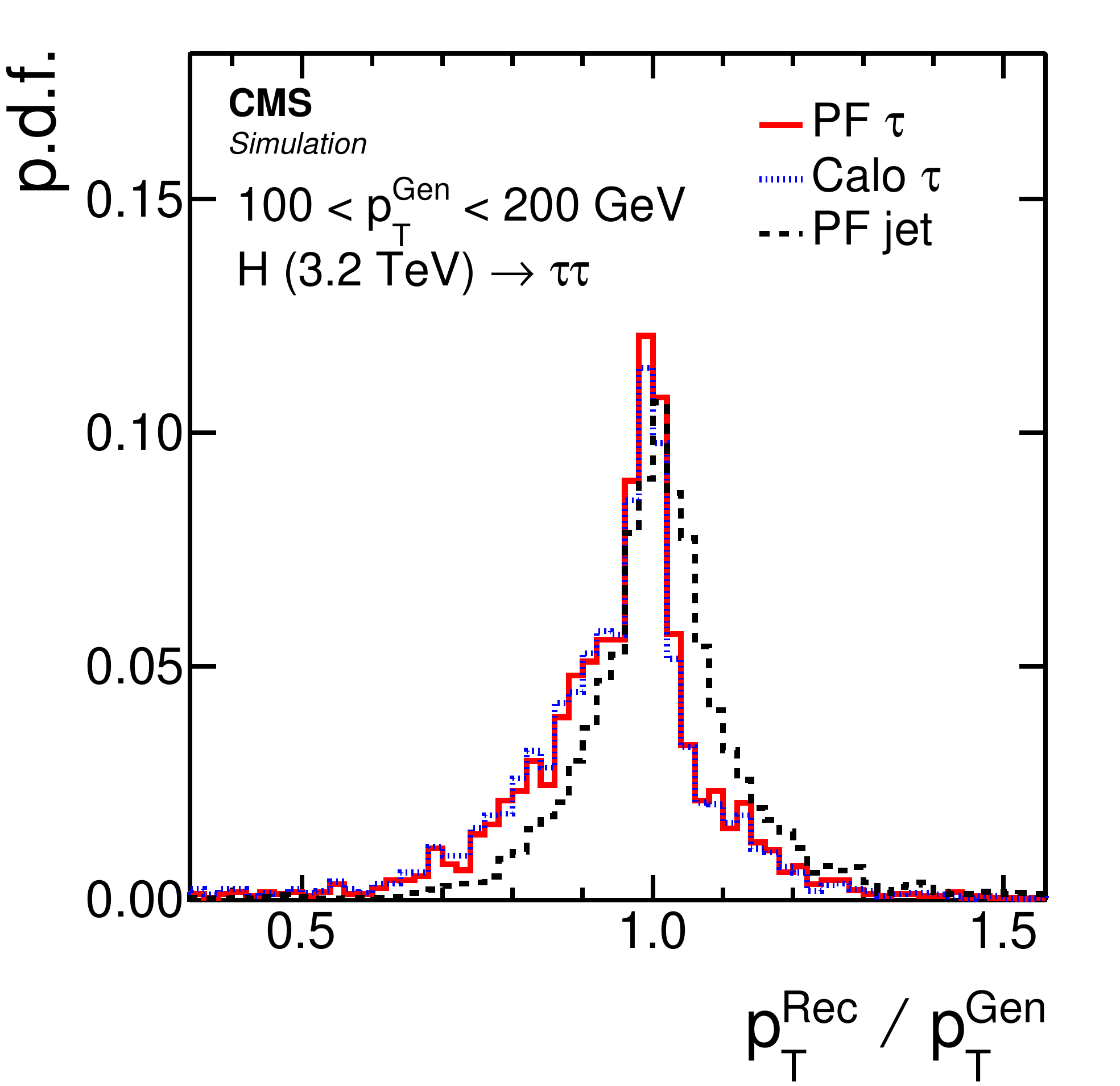}
  \includegraphics*[height=0.28\textheight]{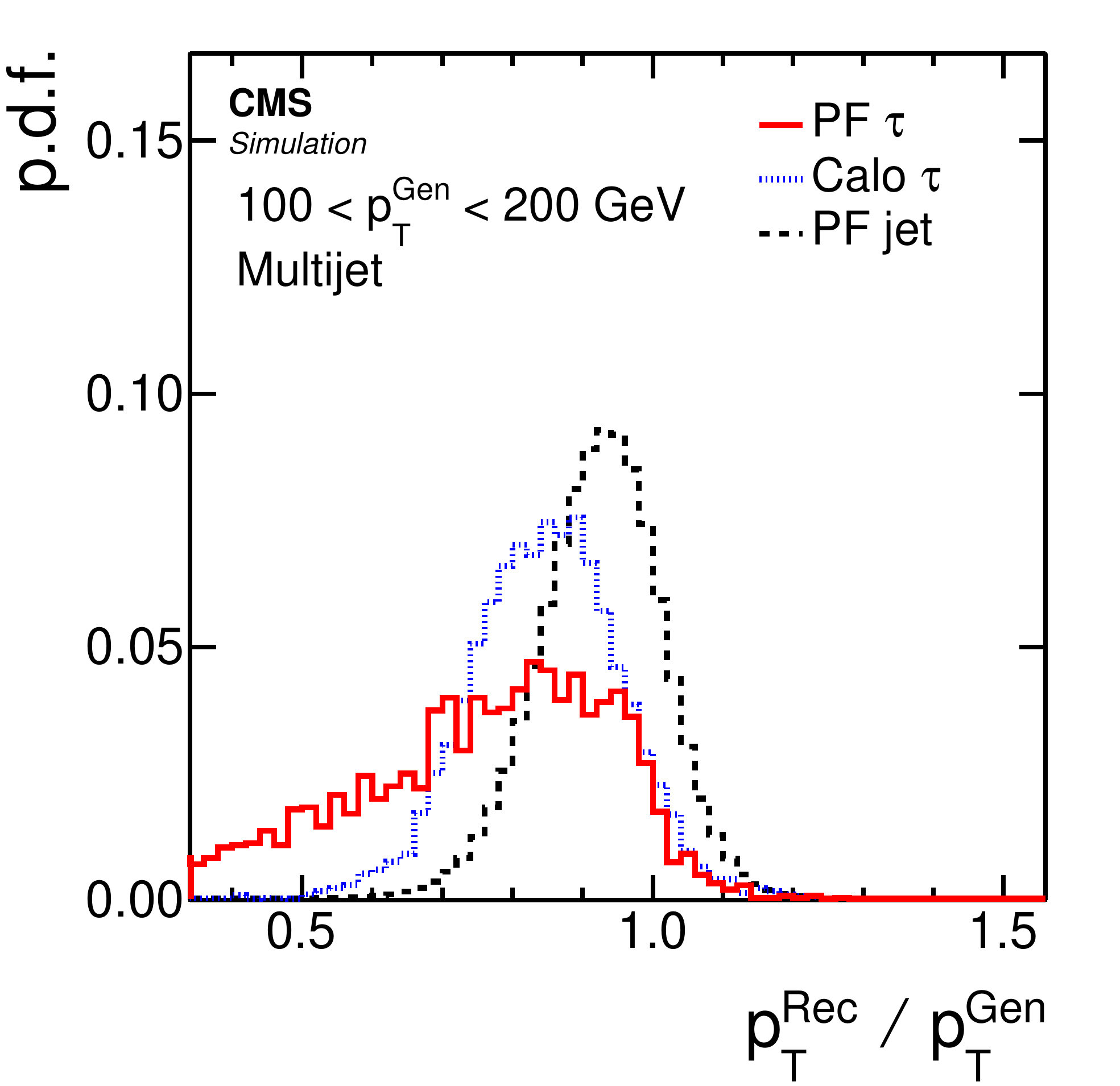}\\
  \includegraphics*[height=0.28\textheight]{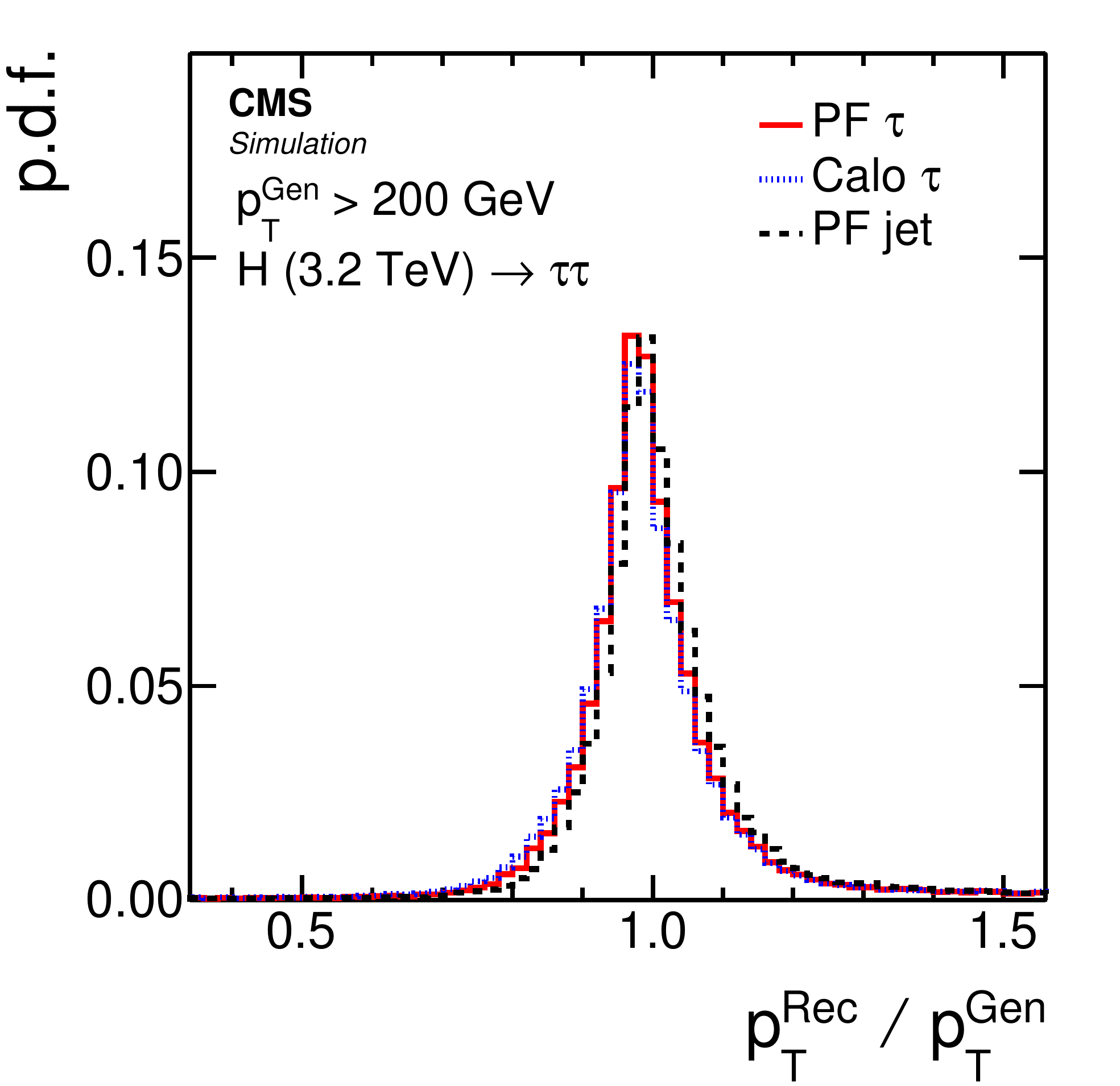}
  \includegraphics*[height=0.28\textheight]{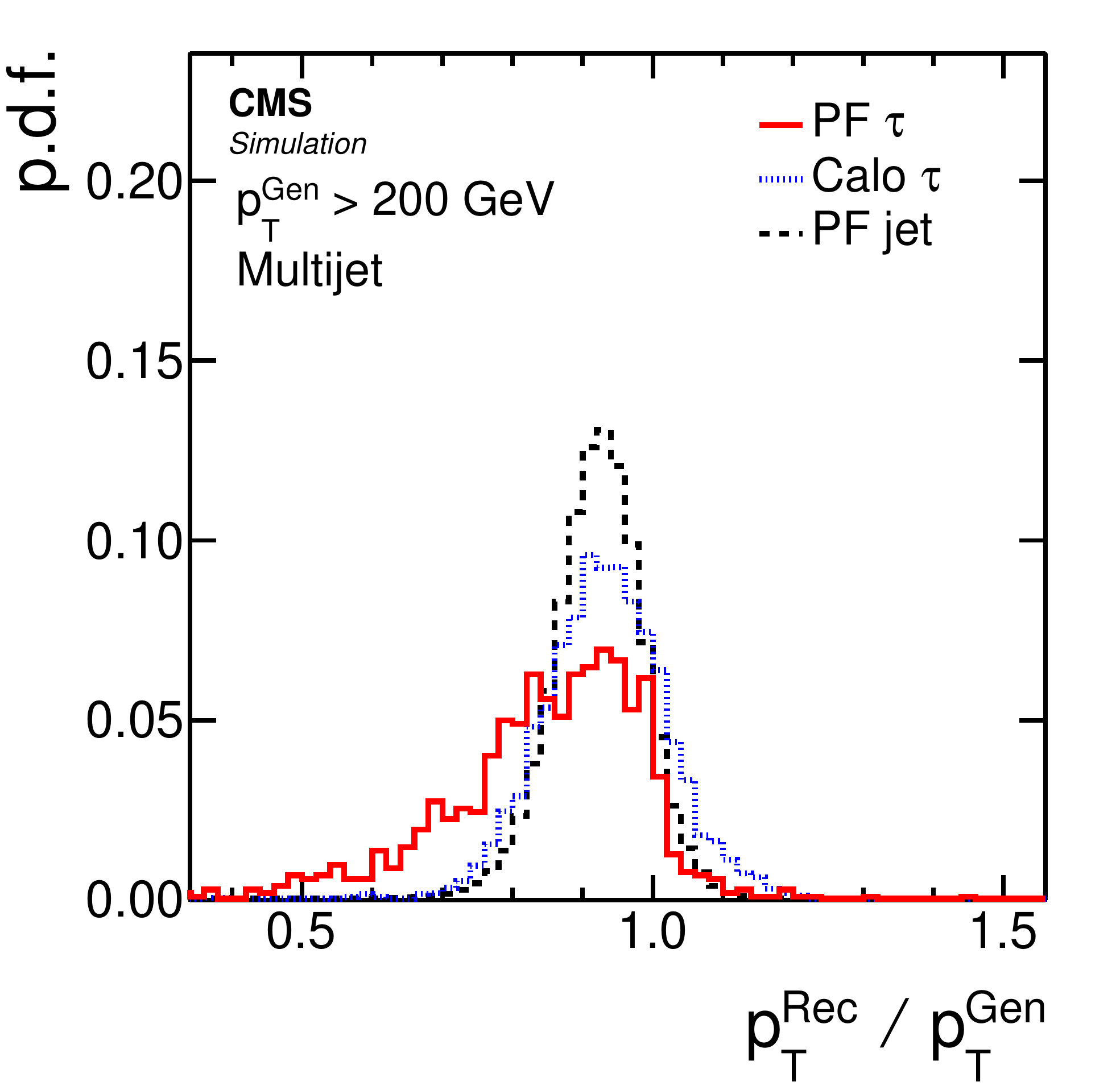}\\

\caption{
  Ratio of reconstructed-to-generator level $\pt$ for genuine $\tauh$ (left),
  and for quark and gluon jets that pass the $\tauh$ identification criteria (right),
  for different intervals in generator level $\pt$.
  In the PF $\Pgt$ case, the $\tauh$ candidates are reconstructed by the HPS algorithm and required to pass the loose isolation working point. In the Calo $\Pgt$ case, they are reconstructed solely with the calorimeters and required to pass the $\tauh$ identification criteria.
  The generator level \pt is taken to be either that of the $\tauh$ or that of the jet.
  For comparison, the ratio is also shown for the closest PF jet in the $(\eta, \varphi)$ plane.
}
\label{fig:expTauEnergyResponse_and_Resolution}
\end{figure}

The performance of the $\tauh$ momentum reconstruction from both the HPS (PF) and Calo algorithms is illustrated in Fig.~\ref{fig:expTauEnergyResponse_and_Resolution}.
The left side of the figure shows the distribution of the ratio between the reconstructed and generated $\tauh$ \pt.
Up to a generated $\pt$ of $100\GeV$, the HPS (PF) algorithm reconstructs the $\tauh$ momentum with a much better accuracy and precision than the calorimeters.
The asymmetry of the distribution is due to the cases in which some of the particles produced in the decay are left out because they would lead the $\tauh$ to fail the collimation or mass requirements.

The $\tauh$ is then reconstructed in a different decay mode and with a reduced momentum.
When all reconstructed particles in the jet matching the $\tauh$ are considered,
the distribution is more symmetric but the resolution degrades,
as some of the jet particles do not come from the $\Pgt$ decay.
In these events, simulated without pileup interactions, the additional particles
come from the underlying event and contribute less than 1\GeV on average to the jet
energy. As a consequence, the mean response is slightly shifted above unity for
a generated $\tauh$ \pt below 100\GeV. For larger \pt, the absolute contribution
from the underlying event becomes negligible and no shift can be observed.
As the generated \pt increases, the energy resolution of the HPS (PF) algorithm
converges to that of the Calo algorithm because the calorimeters start to dominate
the measurement of the momentum of charged hadrons. This effect occurs at a lower \pt
for $\tauh$ than for jets because, for typical $\tauh$ and jets at a given \pt, the jet
\pt is shared among many more charged hadrons at a lower \pt than in the $\tauh$ case.

The right side of Fig.~\ref{fig:expTauEnergyResponse_and_Resolution} shows the distributions obtained for quark or gluon jets misidentified as $\tauh$.
In this case, the $\tauh$ candidate is reconstructed with a fraction of the jet \pt as only a few jet particles can be selected by the HPS (PF) algorithm.
For this reason, while genuine $\tauh$ are reconstructed at the right momentum scale, misidentified $\tauh$ candidates tend to be pushed to lower \pt.
Therefore, the HPS (PF) algorithm reduces the probability for jets to pass
the \pt thresholds applied at analysis level, which leads to a lower
multijet background level than with the calorimeter-based $\tauh$
reconstruction.

\begin{figure}
\setlength{\unitlength}{1mm}
\centering
\includegraphics[width=0.45\textwidth]{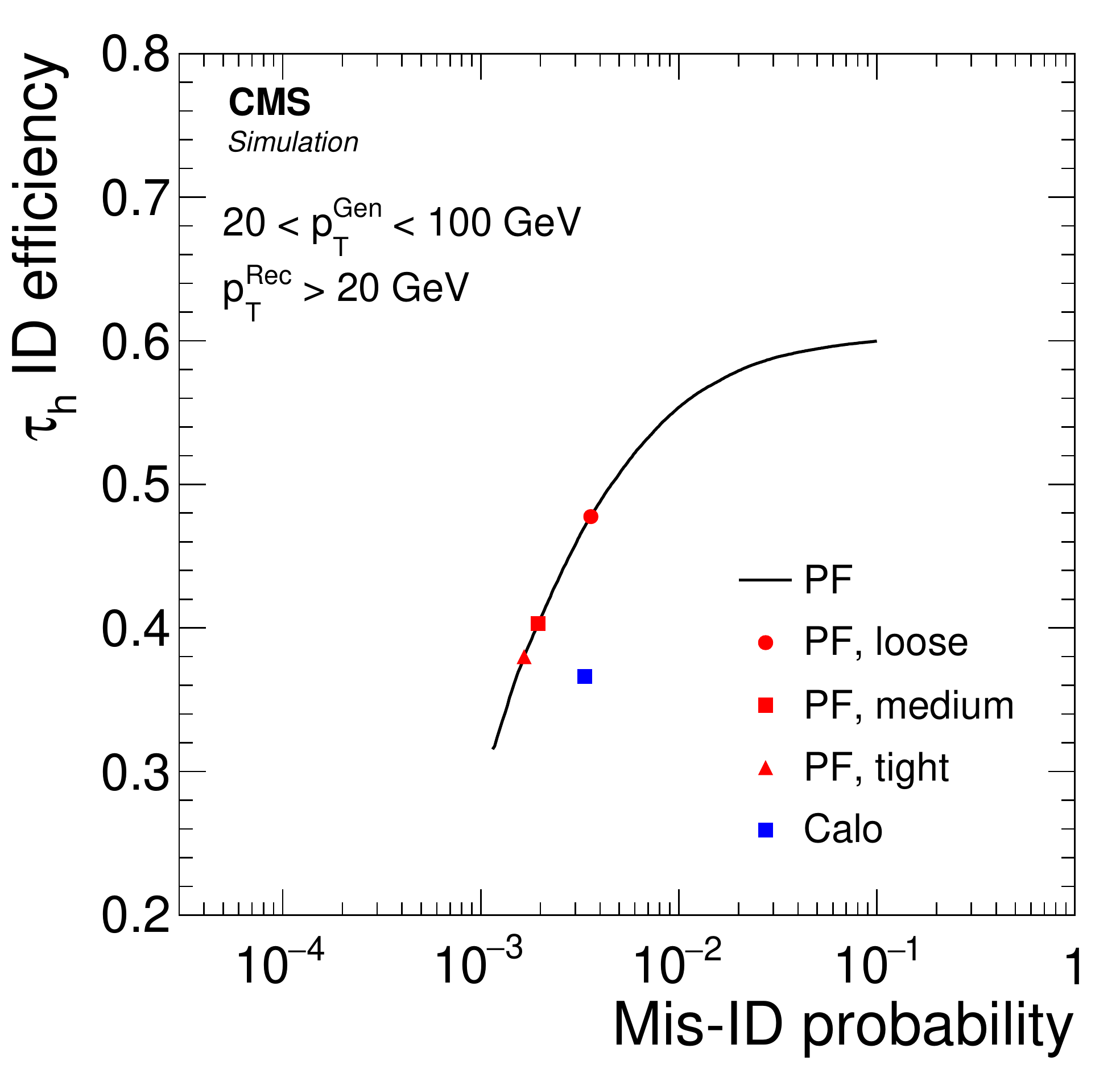}
\includegraphics[width=0.45\textwidth]{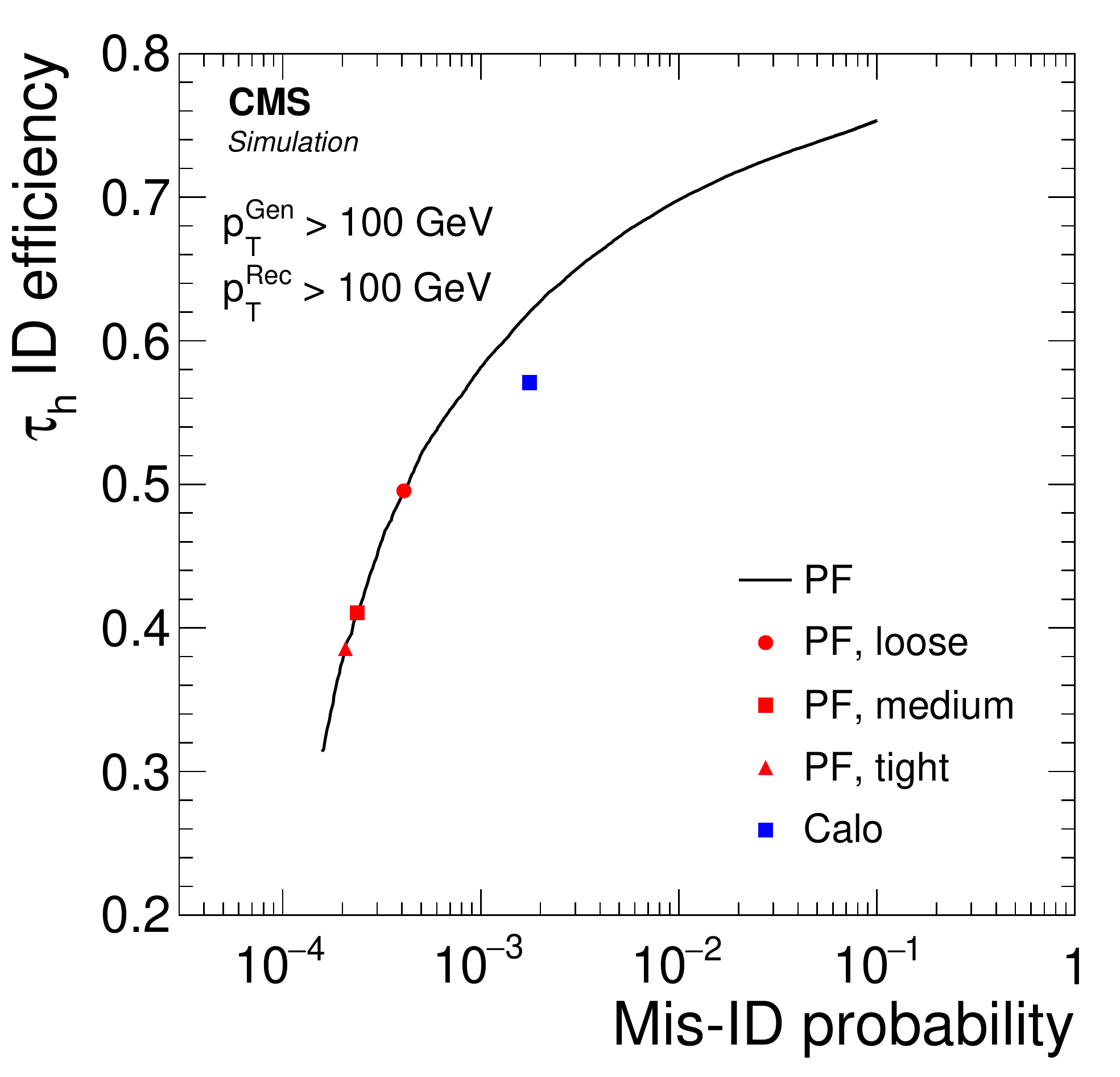}
\caption{
  Efficiency of the $\tauh$ identification versus misidentification probability for quark and gluon jets.
  The efficiency is measured for $\tauh$ produced at low \pt in simulated $\cPZ/\Pggx \to
  \Pgt\Pgt$ events (left), and at high \pt in the decay of a heavy
  particle $\PH ({3.2\TeV}) \to \Pgt\Pgt$ events (right).
  The misidentification probability is measured for quark and gluon jets in simulated multijet events.
  The line is obtained by varying the threshold on the absolute isolation for PF $\tauh$ identified with the HPS algorithm.
  On this curve, the three points indicate the
  loose, medium and tight isolation working points.
  The performance of the calorimeter-based $\tauh$ identification is depicted by a square away from the line.
}
\label{fig:tauROCcurves}
\end{figure}

\begin{figure}
\centering
\includegraphics[width=0.45\textwidth]{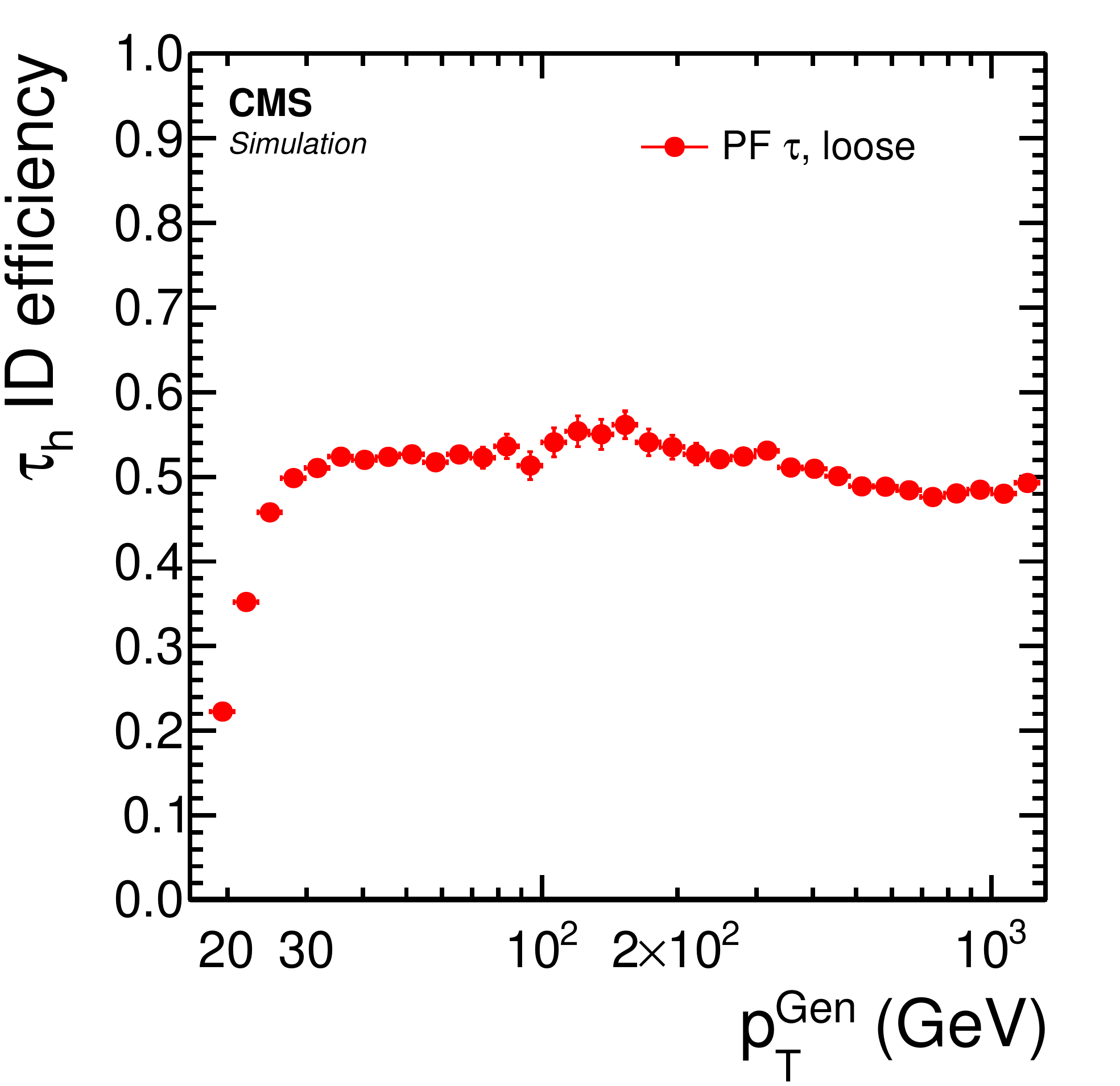}
\includegraphics[width=0.45\textwidth]{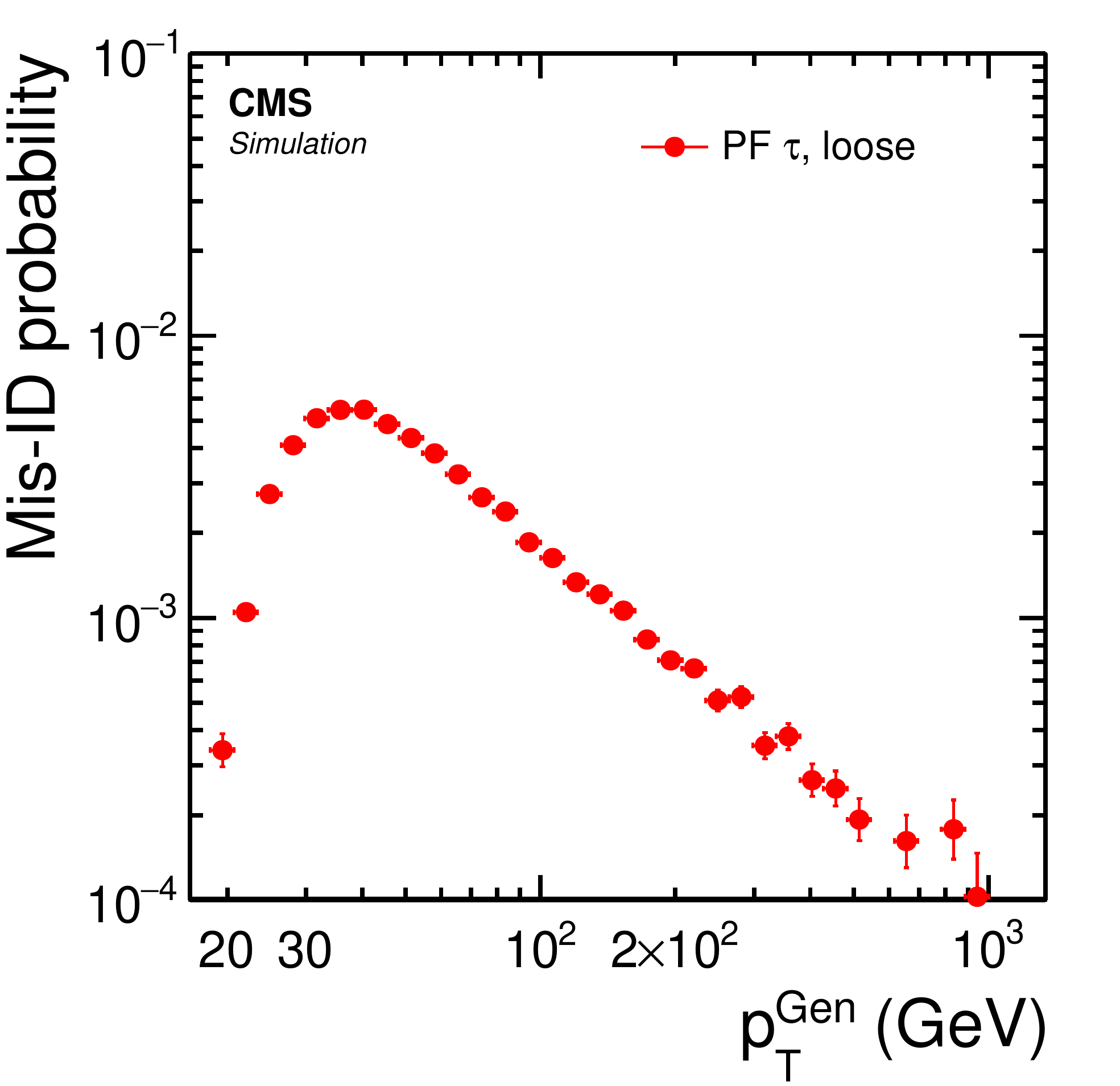}

\caption{
  Identification efficiency for genuine $\tauh$ (left), and $\tauh$
  misidentification probability for quark and gluon jets (right).
  Low-\pt $\tauh$ are obtained from simulated $\cPZ/\Pggx \to
  \Pgt\Pgt$ events
  and high-\pt $\tauh$ from simulated $\PH ({3.2\TeV}) \to \Pgt\Pgt$ events.
  Quark and gluon jets are obtained from simulated QCD multijet events.
  The $\tauh$ are required to be reconstructed by the HPS (PF) algorithm,
  to have $\pt > 20$\GeV and $\abs{\eta} < 2.3$,
  and to satisfy the loose $\tauh$ identification criteria.
}
\label{fig:expTauIdEfficiency_and_JetToTauFakeRate}
\end{figure}

The $\tauh$ identification efficiency is defined as the probability to reconstruct and identify a $\tauh$
matching a generated $\tauh$ within $\Delta R=0.3$.
As a baseline, both the reconstructed and generated $\tauh$ are required to have $\pt>20\GeV$ and $\abs{\eta}<2.3$.
With the same selection, the jet misidentification rate is defined as
the probability to reconstruct and identify a quark or gluon jet from
the multijet sample as a $\tauh$.
Figure~\ref{fig:tauROCcurves} shows the $\tauh$ efficiency as a function of the jet misidentification probability, for a varying threshold on the absolute isolation.
With respect to Calo $\tauh$ identification, the HPS (PF) algorithm
achieves a reduction of the jet misidentification probability by a
factor of $2$--$3$ for a given $\tauh$ identification efficiency.
For a given jet misidentification probability, the gain in efficiency ranges from 4 to 10\%.
The improvement in identification performance is due to three reasons.
First, the decay-mode selection reduces the momentum of jets misidentified as $\tauh$.
Second, with the PF reconstruction of the $\Pgt$ decay products,
mass and collimation criteria can be used in addition to isolation criteria.
Third, all the particles remaining after $\tauh$
reconstruction are used to evaluate the particle-based isolation, while the
detector-based isolation is computed without the tracks
and the calorimeter energy deposits in the signal cone.
Finally, the $\pt$ dependence of the $\tauh$ identification efficiency and jet misidentification probability is shown in Fig.~\ref{fig:expTauIdEfficiency_and_JetToTauFakeRate}.
As \pt rises above $30\GeV$, the HPS (PF) algorithm ensures a constant efficiency together with a sharp decrease of the jet misidentification probability.

In summary, the PF reconstruction of the $\Pgt$ decay products and of the neighbouring particles has led to a sizeable improvement of the $\tauh$ reconstruction and identification performance.
This performance has been further refined for the data-taking period
that started in 2015,
for example with identification techniques based on machine learning
that make use of additional information such as the impact parameter
of charged hadrons and the neutral-pion energy
profile with the strip~\cite{CMS-PAS-TAU-16-002}.

\subsection{Particle flow in the high-level trigger}
\label{sec:expected_performance_hlt}

The first level of the CMS trigger system~\cite{Khachatryan:2016bia}, composed of custom hardware processors, uses information from the calorimeters and muon detectors to select the most interesting events in a fixed time interval of less than 4\mus.
The high-level trigger (HLT) computer farm further decreases the event rate from around 100\unit{kHz} to about 1\unit{kHz}, before data storage for later offline reconstruction.
The HLT event selection imposes requirements on the number of physics objects with \pt over a given threshold.
The reconstruction of these objects at the HLT must be kept as close as possible to the offline reconstruction to limit the triggering inefficiency and the false trigger rate.
As exemplified in Sections~\ref{sec:expected_performance_jets} to~\ref{sec:expected_performance_taus}, the PF reconstruction provides physics objects with better resolution, efficiency, and purity than traditional reconstruction methods.
For this reason, PF reconstruction is used in the vast majority of physics analyses in CMS,
and also has been used at the HLT for optimal performance.

However, to cope with the incoming event rate, the online reconstruction of a single event at the HLT has to be done one hundred times faster than offline, within 140\unit{ms} on average.
Therefore, the reconstruction has to be simplified at the HLT.
Offline, most of the processing time is spent reconstructing the inner tracks for the PF algorithm
as explained in Section~\ref{sec:charged_particles_tracks_and_vertices}.
At the HLT, the tracking is reduced to three iterations,
dropping the time-consuming reconstruction of tracks with low \pt or arising from nuclear interactions in the tracker material.
These modifications preserve the reconstruction efficiency for tracks with $\pt>0.8\GeV$ originating from the primary vertex or from the decay of a heavy-flavour hadron.
After track reconstruction, a specific instance of the particle identification and reconstruction algorithm runs online,
with only two minor differences with respect to the offline algorithm described in Section~\ref{sec:particle_id_reco}:
the electron identification and reconstruction is not integrated in the PF algorithm,
and the reconstruction of nuclear interactions in the tracker is not performed.
These modifications lead to a slightly higher jet energy scale for jets featuring an electron or a nuclear interaction.
For QCD multijet events enriched with high-\pt jets and simulated without pileup,
the average time needed to perform the tracking is 0.6\unit{s} (52\%) offline
and 0.06\unit{s} (44\%) at the HLT, where the percentages are given with respect to
the total time spent in offline reconstruction and in HLT reconstruction, respectively,
under the assumption that the HLT PF reconstruction is performed for every event.
The average time needed for PF reconstruction is 0.07\unit{s} (6\%) offline,
and 0.03\unit{s} (24\%) at the HLT, in the same conditions.  Up to an average of
45 pileup interactions, the time spent for tracking and PF at the HLT is kept
below 20\% and 10\% of the total HLT computing time, respectively.

The ability of the HLT PF reconstruction to reproduce the offline results is tested with jets and $\tauh$ built from the reconstructed HLT particles, from a QCD multijet and a Drell--Yan sample, respectively.
While HLT jets are reconstructed in the same way as offline, the $\tauh$ reconstruction and identification proceeds differently, without decay mode reconstruction.
The $\tauh$ reconstruction is seeded by an HLT jet containing at least one charged hadron.
The direction of the highest-\pt charged hadron in the jet is used as the axis of a signal cone in which all neutral pions and up to two additional charged hadrons are collected to build the $\tauh$ four-momentum.
The charged particles in an annulus around the signal cone are used to quantify the isolation of the $\tauh$ candidate.
The $\tauh$ selection at the HLT is looser than the one usually applied offline in order to preserve the overall selection efficiency in the analysis.
For typical analyses based on a $\mu\tauh$ final state,
requiring a loosely isolated $\tauh$ at HLT in addition to an isolated muon reduces the background rate by a factor of about 20.

\begin{figure}[htbp]
  \centering
  \includegraphics[width=0.49\textwidth]{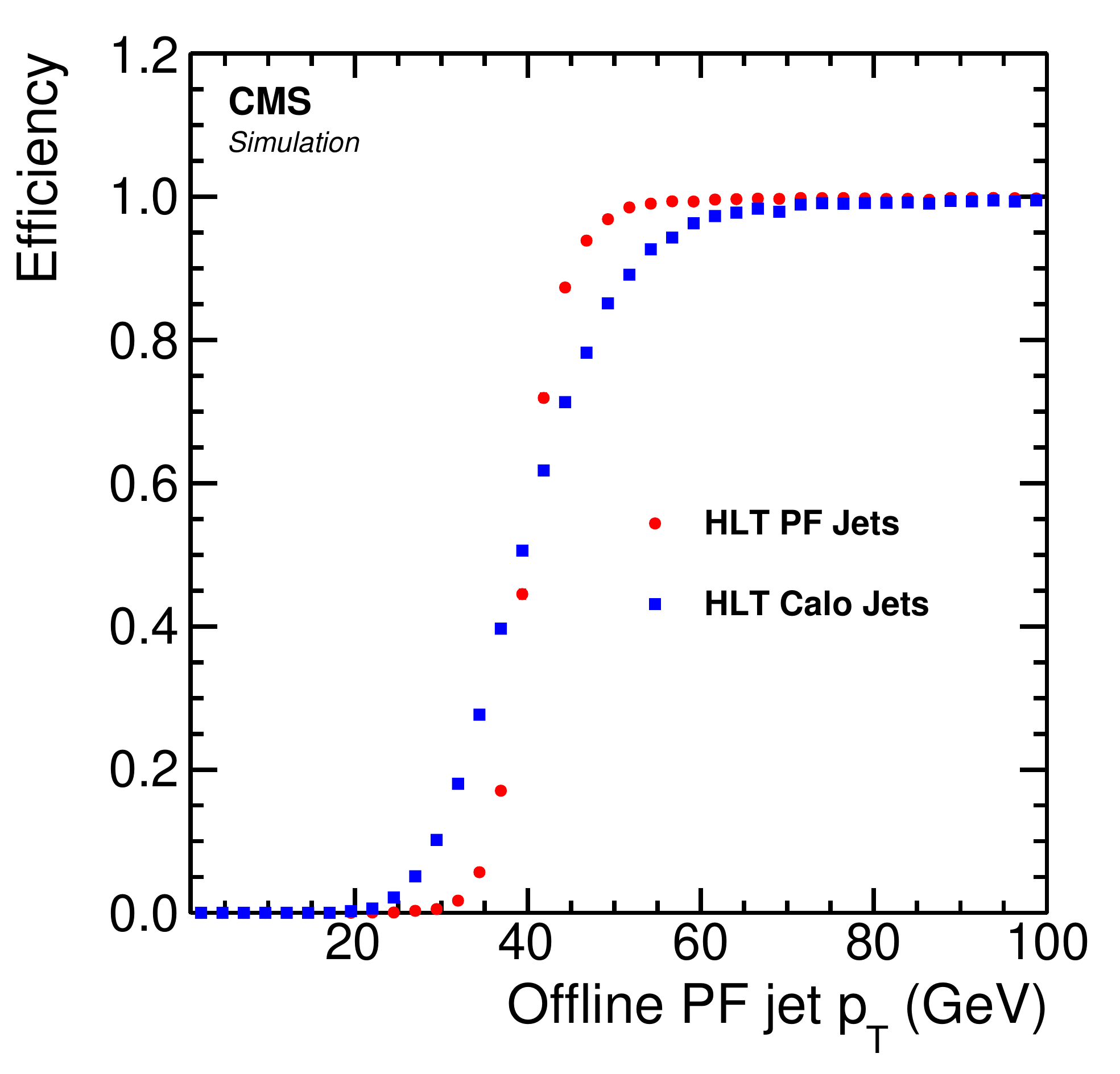}
  \includegraphics[width=0.49\linewidth]{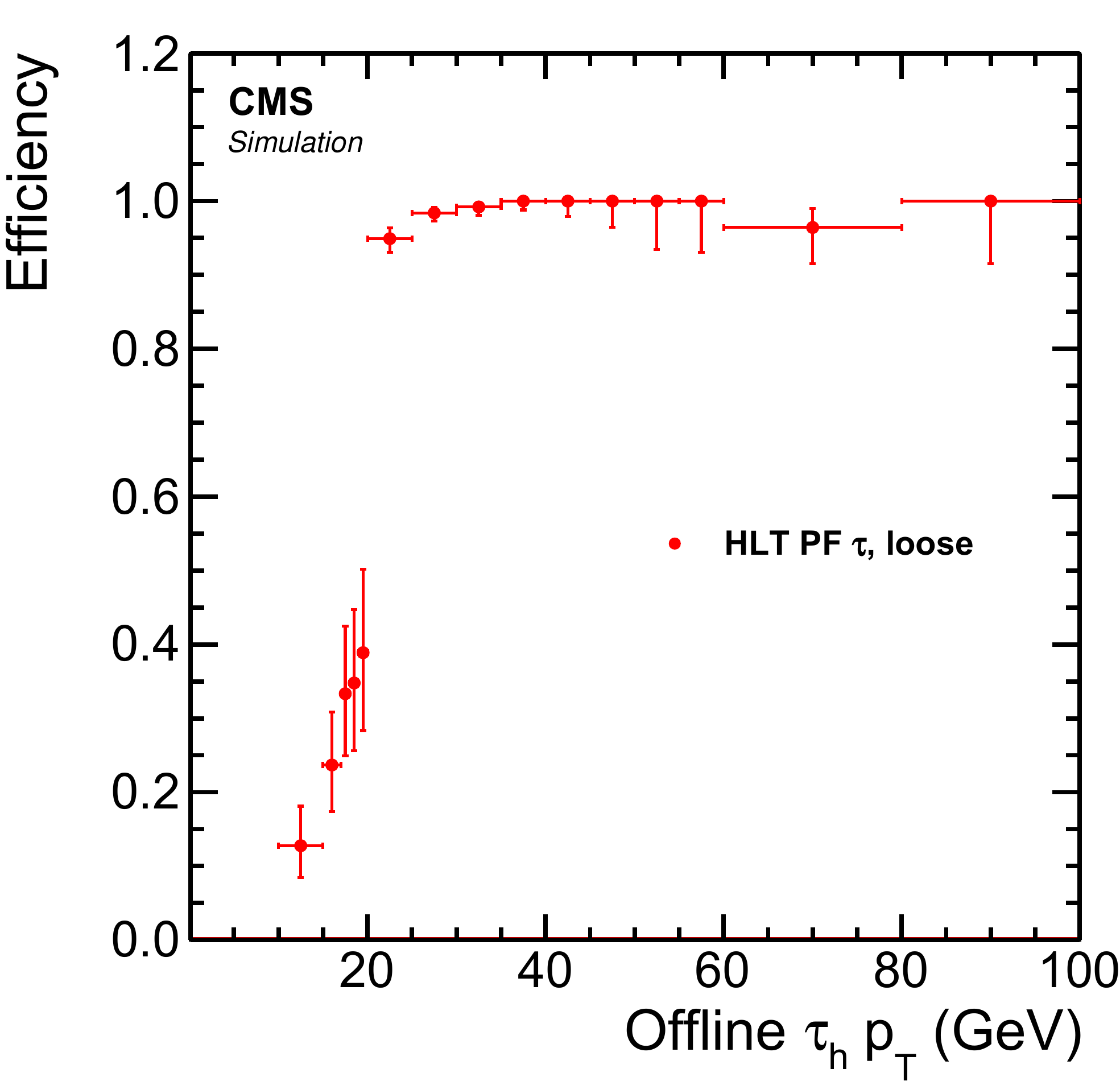}
  \caption{
    Left: Probability to find at HLT a jet with $\pt>40\GeV$ matching the jet reconstructed offline,
as a function of the offline jet \pt.  At the threshold, the curve is steeper for HLT PF jets (circles) than for HLT calorimeter jets (squares).
Right: Probability to find a $\tauh$ with $\pt>20\GeV$ at HLT matching the $\tauh$ reconstructed and identified offline with the loose isolation working point.
  }
  \label{fig:expected_perf_hlt}
\end{figure}

For offline jets and $\tauh$ of various \pt,
Fig.~\ref{fig:expected_perf_hlt} shows the probability to detect a matching physics object at the HLT within $\Delta R=0.3$, and with a \pt larger than typical HLT thresholds, 40\GeV for jets and 20\GeV for $\tauh$.
In the case of jets, this probability is compared to the one obtained for HLT calorimeter jets.
The consistent use of PF jets at the HLT allows for a sharper jet triggering efficiency curve than with calorimeter jets.
The $\tauh$ reconstructed offline is required to satisfy the criteria of the loose isolation working point.
At the HLT, the absence of decay mode identification
and the use of a loose isolation working point ensure a high triggering efficiency.
The sharp rise of the triggering efficiency curve at the threshold demonstrates
the excellent agreement between the $\tauh$ \pt reconstructed online and offline.

\section{Validation with data and pileup mitigation}
\label{sec:commissioning_and_pileup}

The previous section describes how PF improves the performance of physics object reconstruction in simulated events.
In this section, it is shown that the PF algorithm performs
as well with events recorded during Run~1, the first data-taking
period of the LHC.
The performance of reconstruction, identification, and isolation algorithms is compared for events simulated and recorded under Run~1 pileup conditions.
The PF algorithm was designed without taking pileup into account.
This section describes how the performance of object reconstruction and identification is affected by pileup, and how the collection of reconstructed particles can be used to mitigate the effects of pileup.

The results in this section are based on LHC Run~1 data recorded in 2012 at a centre-of-mass energy of 8\TeV and corresponding to an integrated luminosity of $19.7$\fbinv.
During this data-taking period, about 20 pileup interactions occurred on average per bunch crossing.
These interactions are spread along the beam axis around the centre of the CMS coordinate system, following a normal distribution with a standard deviation of about 5\unit{cm}.
The number of pileup interactions $\mu$ can be estimated either from the number of interaction vertices $N_\text{vtx}$ reconstructed with charged-particle tracks as input, with a vertex reconstruction efficiency of about 70\% for pileup interactions~\cite{JME-13-004},
or from a determination of the instantaneous luminosity of the given bunch crossing with dedicated detectors and, as additional input, the inelastic proton-proton cross section~\cite{Chatrchyan:2012nj}.

In the PF reconstruction, the particles produced in pileup interactions give rise to additional charged hadrons, photons, and neutral hadrons.
These result in an average additional \pt of about 1\GeV per pileup interaction and per unit area in the $(\eta,\varphi)$ plane.
As a consequence, reconstructed particles from pileup affect jets, \ETmiss, the isolation of leptons, and the identification of hadronic $\Pgt$ decays.
The measured energy deposits in the calorimeters used as input for particle reconstruction may also be directly affected by pileup interactions, including interactions from different bunch crossings.
The impact of these contributions is small under the pileup conditions considered.

The primary vertices, which are separated spatially along the beam axis, are ordered by the quadratic sum of the \pt of their tracks, $\sum \pt^2$.
The primary vertex with the highest $\sum \pt^2$ is identified as the hard-scatter vertex,
whereas the other vertices are considered as pileup vertices.
Charged hadrons reconstructed within the tracker acceptance can be identified as coming from pileup by associating their track with a pileup vertex.
If identified as coming from pileup,
these charged hadrons are removed from the list of reconstructed particles used to form physics objects.
This widely used algorithm is called \textit{pileup charged-hadron subtraction} and denoted as \textit{CHS}.

Photons and neutral hadrons as well as all reconstructed particles outside the tracker acceptance, however, cannot be associated with one of the reconstructed primary vertices with this technique.
To mitigate the impact of these particles on jets, lepton isolation, and $\tauh$ identification,
the uniformity of the \pt density of pileup interactions in the $(\eta,\varphi)$ plane allows the average \pt contributions expected from pileup to be subtracted.
The \pt density from pileup interactions $\rho$ can be calculated with jet clustering techniques~\cite{Cacciari:2007fd,Cacciari:2008gn,JME-13-004}, with the list of all reconstructed particles as input.
As an alternative, this contribution can be estimated locally, \eg around a given lepton, from the expected ratio of the neutral to the charged energy from pileup, typically 0.5.
After the end of Run~1, advanced pileup mitigation techniques have been explored~\cite{Bertolini:2014bba,CMS-PAS-JME-14-001}.
While not used extensively for analyses based on Run~1 data,
these techniques become increasingly important with the larger number of pileup interactions observed during the LHC Run~2.

Since the results in this section are based on data taken in 2012 and corresponding simulated events,
a few details of the physics object reconstruction are different from the choices discussed in the previous section, \eg the value of the radius parameter for jet clustering.
Like in Section~\ref{sec:expected_performance}, these results are derived for the objects and algorithms used in most CMS analyses, \ie jets, \MET, muons, lepton isolation, and reconstructed hadronic $\Pgt$ decays.
Results on electron reconstruction and identification can be found in Ref.~\cite{Khachatryan:2015hwa}.

\subsection{Jets}
\label{sec:jet_results}

Jets are reconstructed either from all reconstructed particles (PF jets) or from all reconstructed particles except charged hadrons associated with pileup vertices (PF+CHS jets).
Unless noted otherwise, jets are reconstructed with the anti-$k_t$ algorithm with a radius parameter $R = 0.5$.
The corrections for the difference in response between reconstructed and generated particle jets (Ref jets) are determined separately for PF jets and PF+CHS jets.
The expected average contribution from pileup is estimated with $\rho$ and the jet area~\cite{Cacciari:2008gn} as inputs, and is subtracted from the reconstructed jet.
This correction is about three times smaller for PF+CHS jets since CHS removes most of the charged hadrons from pileup,
which account roughly for two thirds of the pileup contribution.
Additional corrections are applied to the observed events to account for residual differences between data and simulation~\cite{JME-13-004}.

\begin{figure}[p!]
\centering
\includegraphics[width=0.45\textwidth]{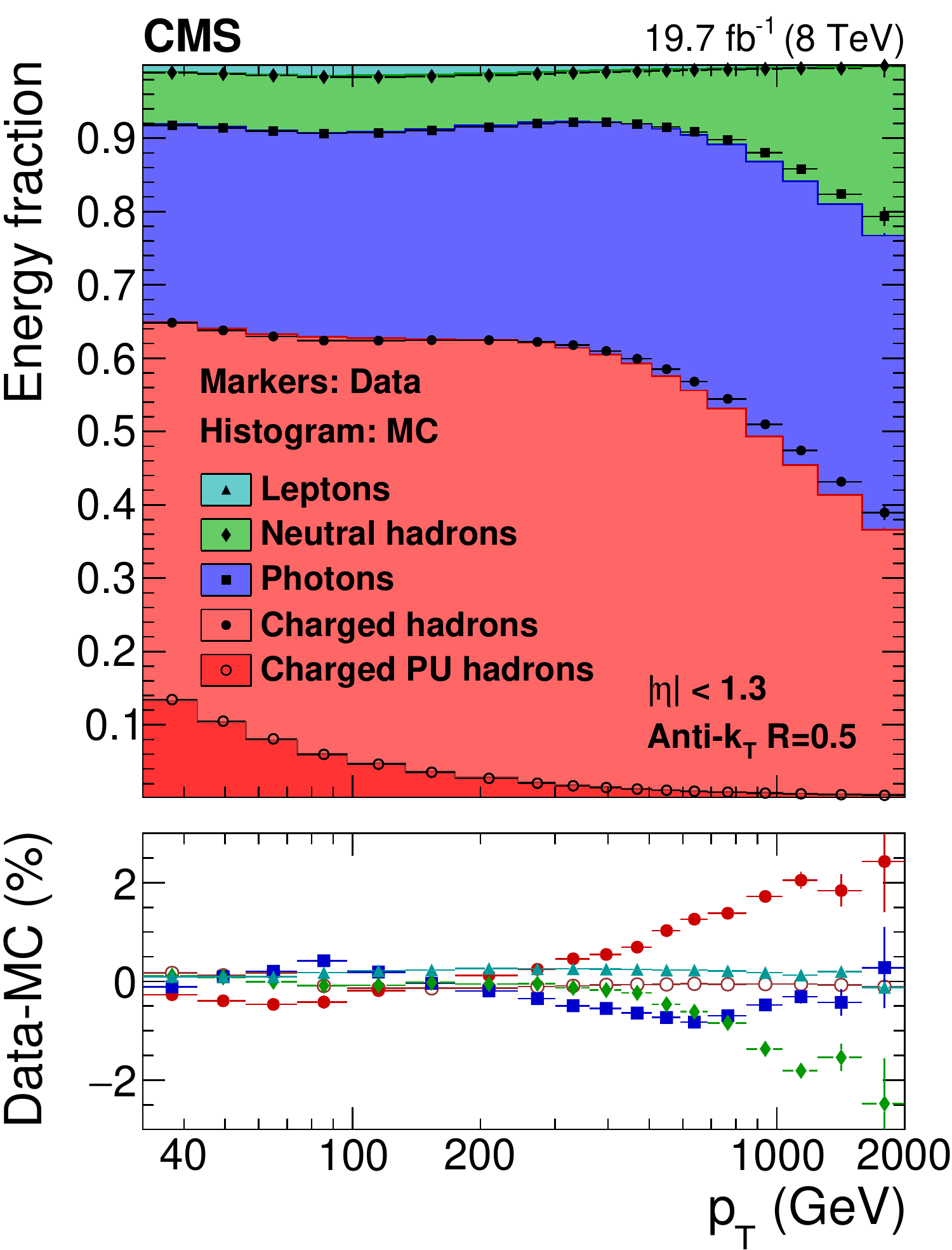}\hfill
\includegraphics[width=0.45\textwidth]{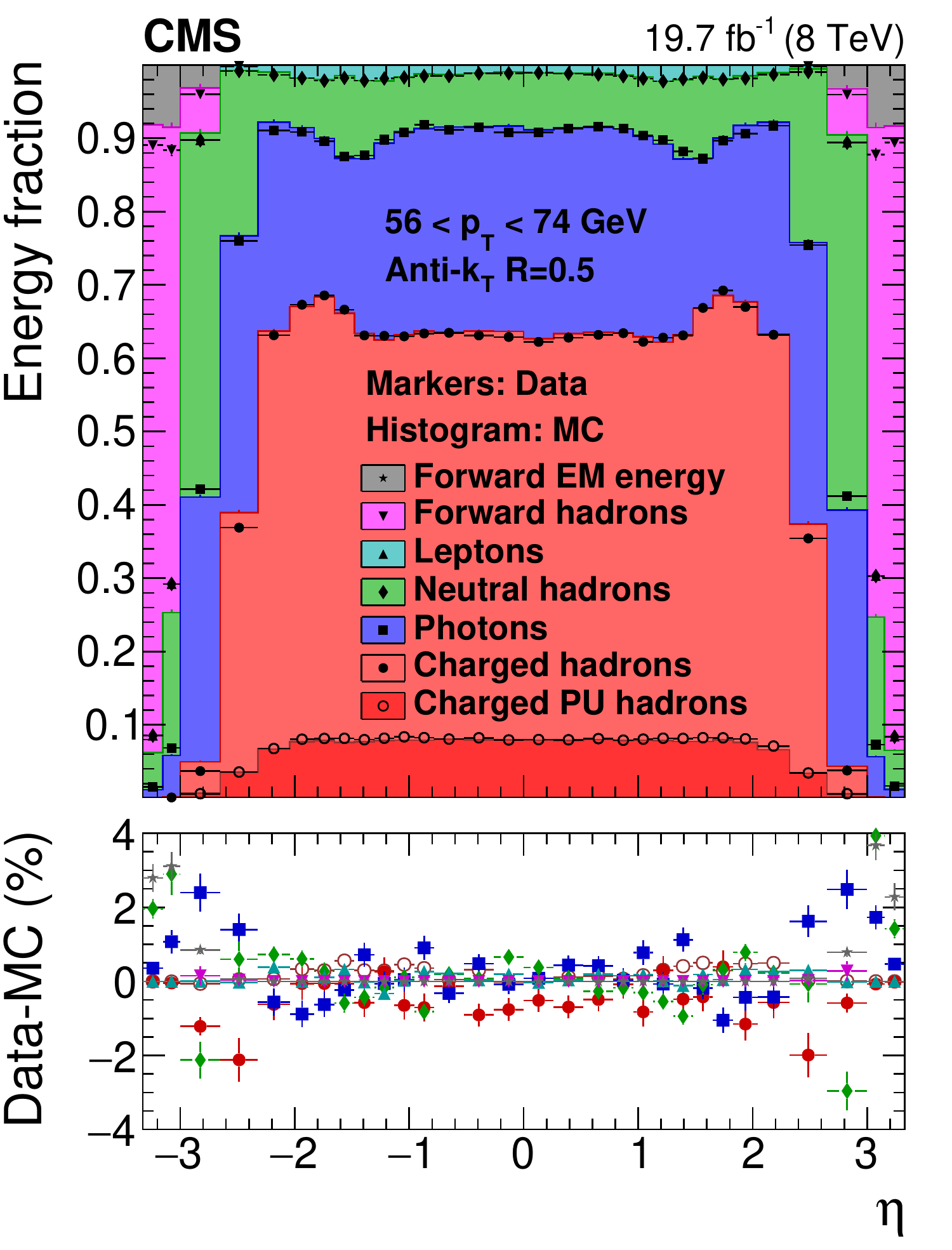}\\
\includegraphics[width=0.45\textwidth]{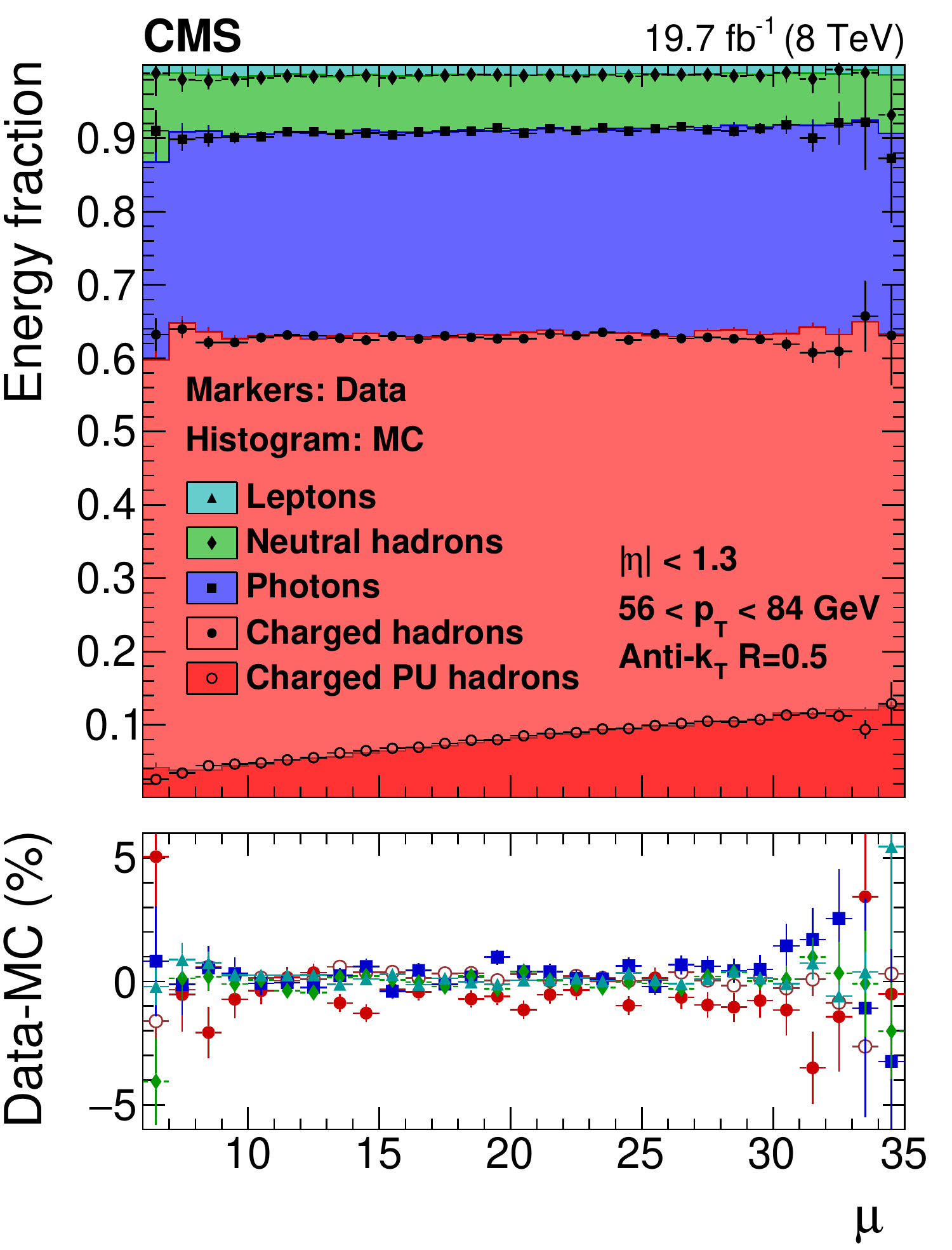}
\caption{Jet energy composition in observed and simulated events as a function of \pt (top left), $\eta$ (top right), and number of pileup interactions (bottom).
The top panels show the measured and simulated energy fractions stacked, whereas the bottom panels show the difference between observed and simulated events.
Charged hadrons associated with pileup vertices are denoted as \textit{charged PU hadrons}.
\label{fig:comp_combo_vs_pt_eta_pu}}
\end{figure}

The jet energy contributions from different types of particles are measured with the tag-and-probe technique~\cite{Khachatryan:2010xn} in back-to-back dijet events recorded by requiring at least one jet at the HLT.
The two jets with highest \pt in a given event must be separated by an angle $\Delta\varphi$ larger than 2.8\unit{rad} in the plane transverse to the beam axis.
Events with additional jets with $\pt^{\textrm{j}_3} > 5$\GeV and $\pt^{\textrm{j}_3} > 0.15(\pt^{\textrm{j}_1} + \pt^{\textrm{j}_2})$ are rejected to avoid biases from large parton radiation.
The tag jet is required to be in the barrel region and to correspond to the jet that triggered the data acquisition.
The energy contributions are measured from the probe jet,
whereas the value of the jet \pt is taken from the tag jet.
This procedure ensures that correlations of the jet energy fractions, \eg with upward fluctuations of the observed jet \pt, do not bias the measurement of these fractions.
Figure~\ref{fig:comp_combo_vs_pt_eta_pu} shows a comparison of the dependence of the PF jet composition on jet \pt, jet $\eta$, and the estimated number of pileup interactions between events observed in data and events simulated with \PYTHIA~\textsc{6.4}~\cite{pythia6}.
The number of pileup interactions is estimated from the number of clusters reconstructed
in the silicon pixel detectors~\cite{CMS-PAS-LUM-13-001}.
The composition as a function of jet \pt is given for central jets ($\abs{\eta} < 1.3$).
As opposed to the simulation results without pileup presented in Section~\ref{sec:expected_performance}, the measured jets have a significant energy contribution emerging from pileup.
As described in Section~\ref{sec:charged_particles_tracks_and_vertices}, the tracking efficiency drops within the densely populated jet core for high-\pt jets, leading to a reduction of the fraction of charged hadrons at high \pt.
The observed and simulated energy fractions agree within 1\% for $\pt < 500$\GeV, and within 2\% above.
The relative contribution from charged hadrons associated with pileup vertices is largest for low-\pt jets and becomes negligible in the \TeV range, as the contribution from pileup is expected to be fully uncorrelated with the hard scatter.
The composition with respect to $\eta$ is shown for jets with \pt between 56 and 74\GeV.
The simulated and observed fractions agree at the level of 1\% in the tracker acceptance and at the level of 2\% for $2.5 < \abs{\eta} < 3.0$.

The energy fractions as a function of the number of pileup interactions for central jets ($\abs{\eta} < 1.3$) with \pt between 56 and 84\GeV show a stable growth in the contribution of charged hadrons from pileup vertices.
The relative contributions from photons, neutral hadrons, and the sum of charged hadrons and charged hadrons from pileup vertices remain constant with increasing pileup.
This behaviour is due to the similar composition of QCD jets in the given \pt range and pileup in terms of the energy fractions from charged hadrons, neutral hadrons, and photons, which constitute about 99\% of the jet energy on average.
More details on the measurements of the jet composition are given in Ref.~\cite{JME-13-004}.

\begin{figure}[p!]
\centering
\includegraphics[width=0.49\textwidth]{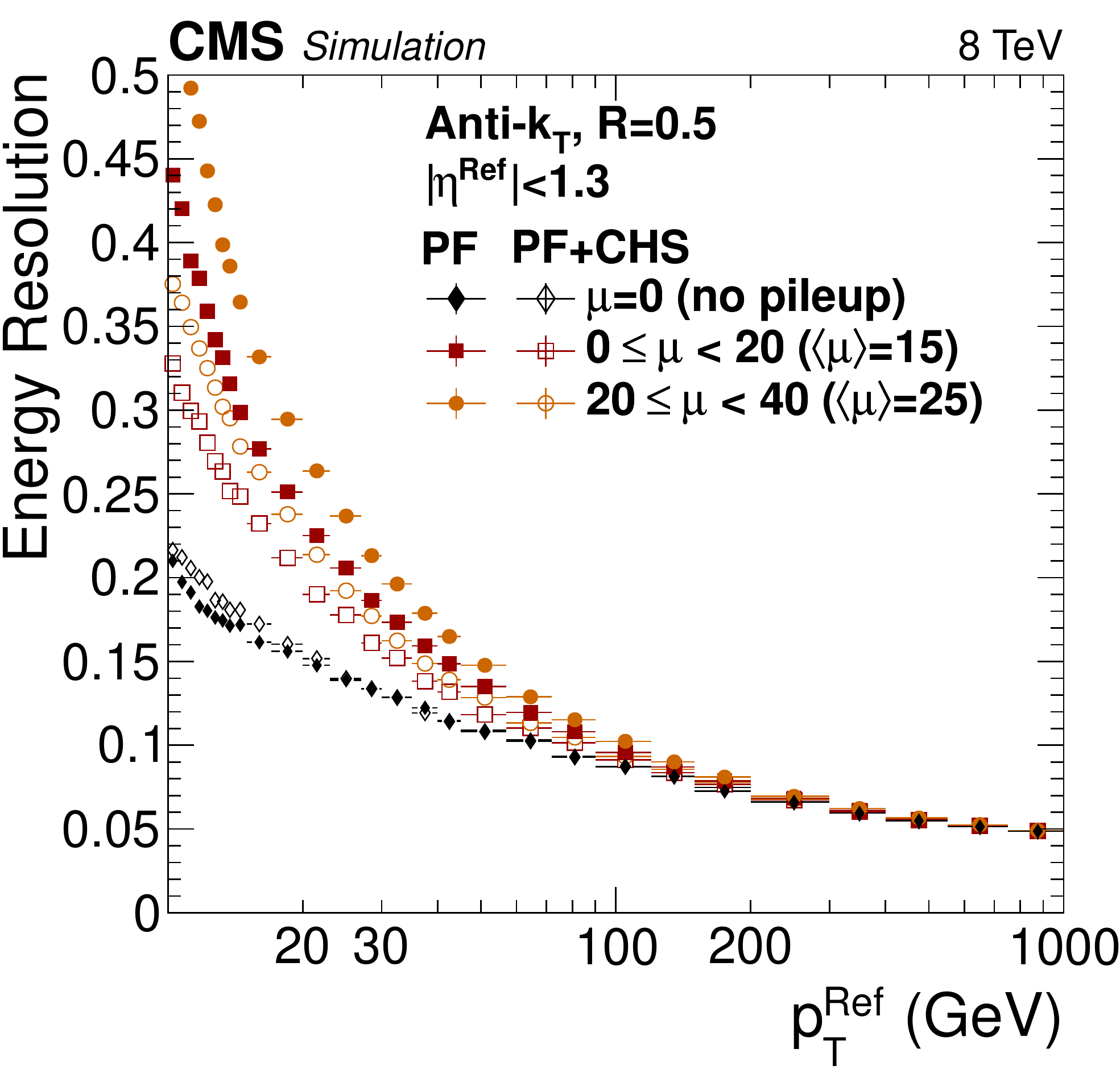}
\includegraphics[width=0.49\textwidth]{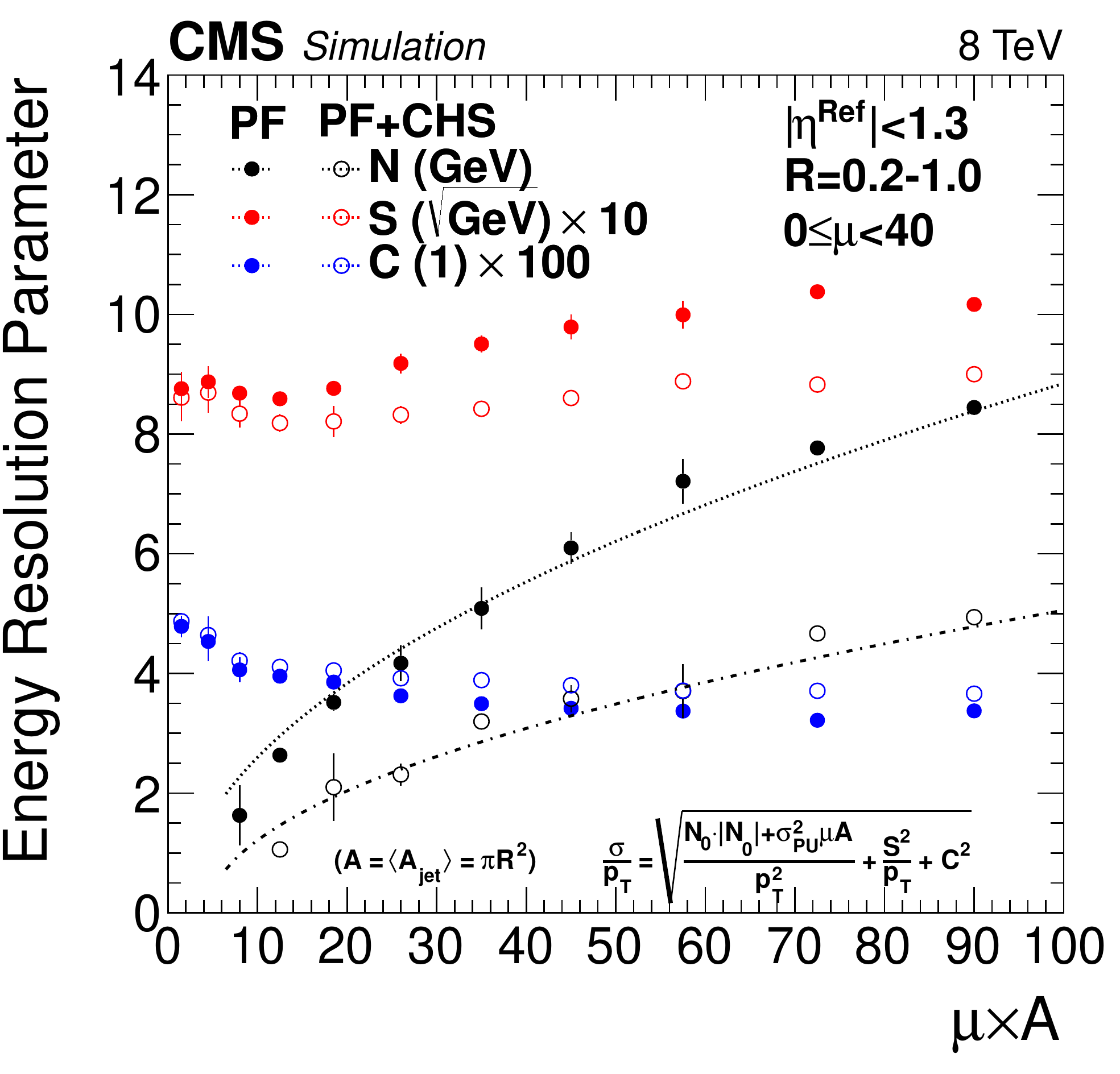}
\caption{Jet \pt resolution for PF+CHS jets (open markers) and PF jets (full markers) under three different pileup conditions (left), and jet energy resolution parameters (right).
The jet \pt resolution is shown as a function of $\pt^\text{Ref}$.
The jet energy resolution parameters (Eq.~(\ref{eq:jer_parameters})) are shown as a function of the number of pileup interactions $\mu$ times the jet area $A$ for PF jets and PF+CHS jets.
The three resolution parameters are determined in bins of $\mu$ for various radius parameters $R$, and then averaged in bins of $\mu A$.
\label{fig:jer_pu_chs_res_fit_param}}
\end{figure}

To investigate the impact of pileup on the jet energy resolution,
the resolution for central jets is displayed in the left panel of Fig.~\ref{fig:jer_pu_chs_res_fit_param} as a function of $\pt^\text{Ref}$ for simulated events under three different pileup conditions.
The resolution is defined as the width of a normal distribution obtained from a fit to the ratio of reconstructed and Ref jet \pt.
While the impact of pileup on the resolution for jets with $\pt$ larger than $100$\GeV is small,
the relative \pt resolution degrades significantly for lower \pt.
The application of CHS improves the jet energy resolution for these lower-\pt jets.
The improvement becomes larger for a higher number of pileup interactions.
As expected, the jet energy resolution is nearly identical for PF and PF+CHS jets if no pileup is present.
The small difference ($\sim$1\% at low \pt) can be attributed to the jet energy corrections that were obtained under the assumption that some amount of pileup is present.
Within this difference, this observation confirms that CHS does not remove charged hadrons from the hard interaction,
which would lead to a degradation of the jet energy resolution in the absence of pileup.

To understand the jet energy resolution in more detail,
the relative jet energy resolution is parameterized as the quadratic sum of
a pileup and noise term, a stochastic term, and a constant term,
\begin{equation}
\label{eq:jer_parameters}
    \frac{\sigma(\pt)}{\pt} = \frac{N}{\pt}  \oplus  \frac{S}{\sqrt{\pt}}  \oplus  C.
\end{equation}

The absolute contribution from pileup does not depend on the jet \pt and is hence only expected to affect the pileup and noise term $N$ of the relative energy resolution.
Because of the uniform distribution of pileup particles in the $(\eta,\varphi)$ plane,
the pileup contribution to the jet energy is proportional to the product of the number of pileup interactions and the jet area, $\mu A$,
which implies that the contribution to the jet energy resolution scales with $\sqrt{\mu A}$ in the limit of a large number of particles from pileup.
The resolution parameters are fitted in bins of $\mu$ for jets clustered with various radius parameters $R$, covering different areas in the $(\eta,\varphi)$ plane, and then averaged over bins of $\mu A$.
The resulting parameters are shown in the right panel of Fig.~\ref{fig:jer_pu_chs_res_fit_param} as a function of $\mu A$.
Both the constant and stochastic terms remain roughly constant as a function of $\mu A$
and are, as expected in the case that CHS only removes charged hadrons from pileup, of similar magnitude for PF and PF+CHS jets.
The combined pileup and noise term is parameterized as $N(\mu A) = \sqrt{N_0 |N_0| + \sigma_\text{pileup}^2 (\mu A})$, where $N_0$ is an additional empirical noise term. Allowing
$N_0$ to become negative improves the description of the resolution for small numbers
of pileup interactions.
The application of CHS reduces the pileup and noise term by almost a factor of two, consistent with the removal of two thirds of particles from pileup in the tracker volume.
More details on measurements of the jet energy resolution including a detailed discussion of the jet energy resolution parameters and a validation with observed data are given in Ref.~\cite{JME-13-004}.

\begin{figure}[p!]
\centering
\includegraphics[width=0.59\textwidth]{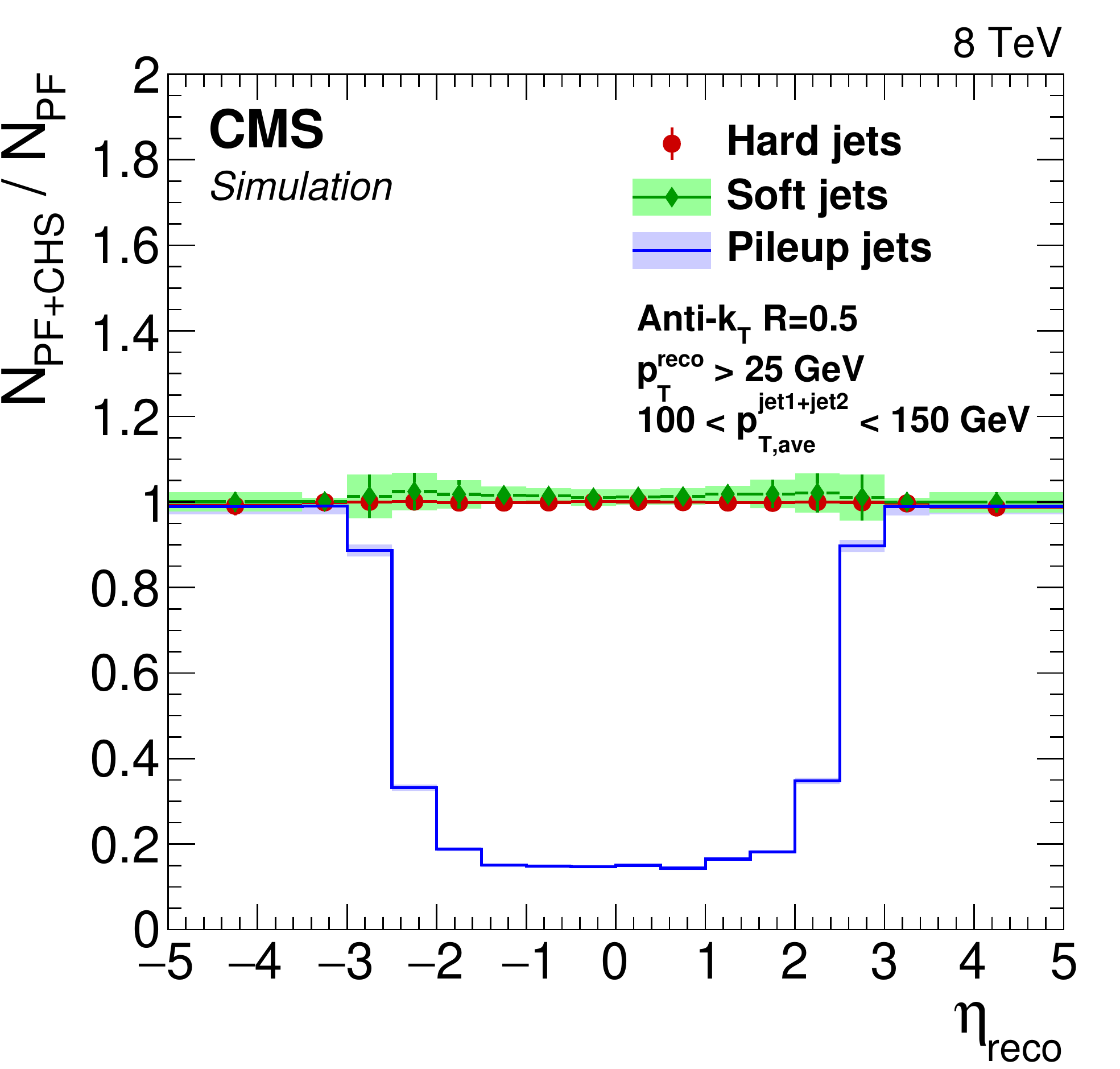}
\caption{Ratio of PF jet multiplicity with and without application of CHS,
for hard jets, pileup jets, and soft jets, as a function of the reconstructed jet pseudorapidity.
The uncertainty bands include both statistical uncertainties and uncertainties in the jet energy corrections.
\label{fig:comp_combo_vs_pu_chs_pu_reduction}}
\end{figure}

Pileup not only degrades the jet energy resolution, but can also lead to the emergence of additional jets with a \pt of a few tens of \GeV, in the following denoted as pileup jets.
These jets result from the overlap of two or more low-\pt jets from different pileup interactions, hence their \pt spectrum falls more steeply than the one of regular QCD jets~\cite{CMS-PAS-JME-13-005}.
The effect of CHS on the rate of pileup jets is studied in simulated QCD multijet events for reconstructed jets with $\pt > 25$\GeV.
Only events in which the \pt sum of the two highest-\pt jets j$_1$ and j$_2$ is between 200 and 300\GeV are considered.
All reconstructed jets are tentatively matched to a Ref jet built from the generated particles from the hard scatter, with $\pt^\text{Ref}>10$\GeV and a distance in the $(\eta, \varphi)$ plane smaller than 0.25.
Jets that cannot be matched to a Ref jet are classified as \textit{pileup jets}.
If j$_1$ and j$_2$ are matched, they are classified as \textit{hard jets}.
All other jets are classified as \textit{soft jets}.
The ratio of the numbers of PF+CHS and PF jets with $\pt > 25$\GeV is shown in Fig.~\ref{fig:comp_combo_vs_pu_chs_pu_reduction} as a function of jet $\eta$ for these three classes of jets.
In the tracker acceptance, CHS reduces the number of pileup jets by $\sim$85\% without affecting the multiplicity of either hard or soft jets.
Advanced information on the use of PF reconstruction for pileup mitigation can be found in Ref.~\cite{CMS-PAS-JME-14-001}.

\subsection{Missing transverse momentum}

The performance of \vecptmiss reconstruction is assessed with a sample of observed events selected in the dimuon final state, dominated by events with a $\cPZ$~boson decaying to two muons~\cite{JME-13-003-JINST}.
The data set is collected with a trigger requiring the presence of two muons passing \pt thresholds of 17 and 8\GeV, respectively.
The two reconstructed muons must fulfil $\pt > 20 $\GeV and $\abs{\eta} < 2.1$, satisfy isolation requirements, and have opposite charge.
Events where the invariant mass of the dimuon system is outside the $60<M_{\mu\mu}<120$\GeV window are rejected.

\begin{figure}[htb]
\centering
\includegraphics[width=.7\linewidth]{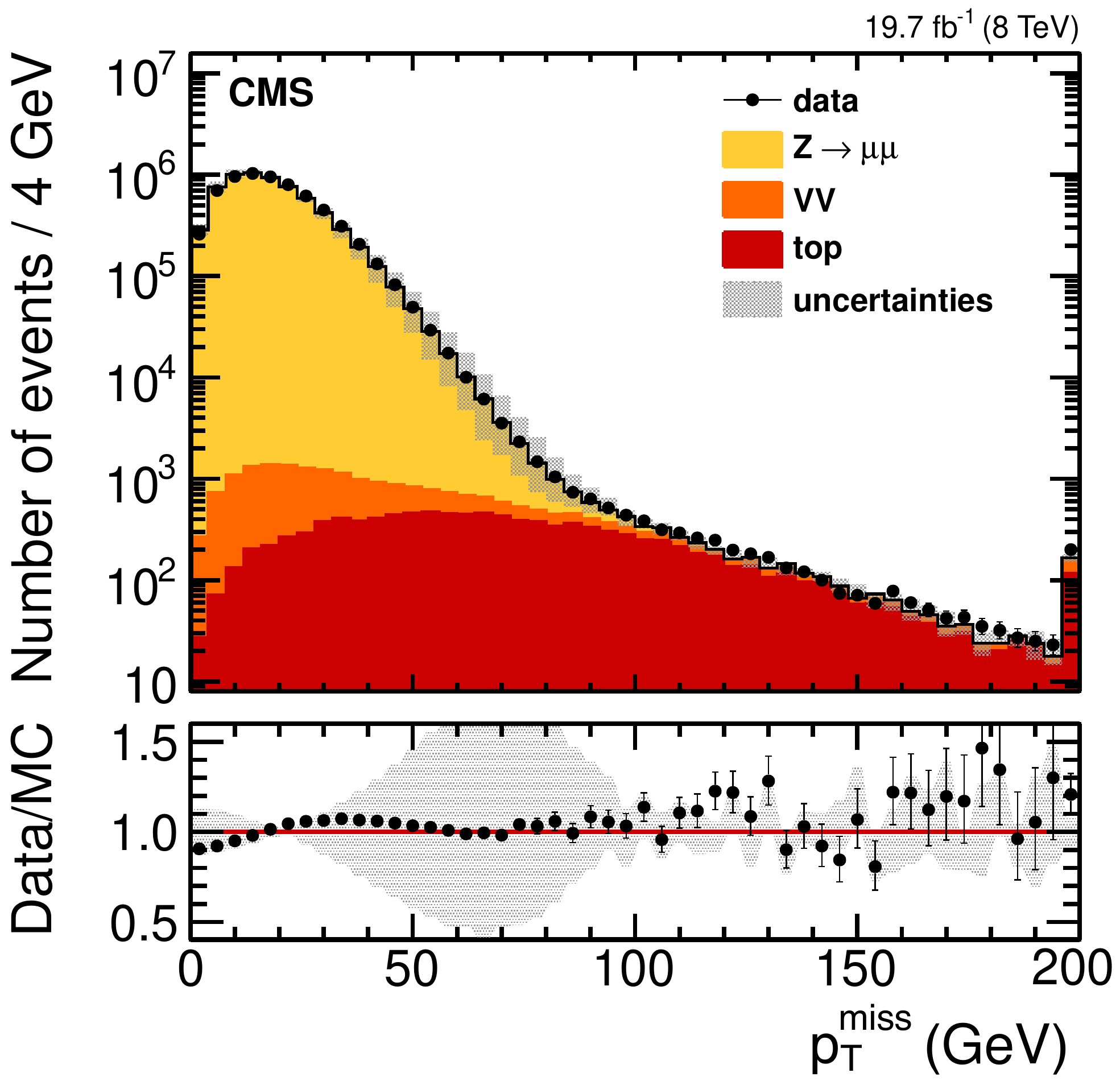}
\caption{Spectrum of PF $\ptmiss$ in a $\cPZ\to\mu\mu$ data set~\cite{JME-13-003-JINST}.
The observed data are compared to simulated $\cPZ\to\mu\mu$, diboson (VV), and $\ttbar$ plus single top quark events.
The lower panel shows the ratio of data to simulation, with the uncertainty bars of the points including the statistical uncertainties of both observed and simulated events and the grey uncertainty band displaying the systematic uncertainty in the simulation.
The last bin contains the overflow.
}
\label{fig:met_distribution}
\end{figure}

The expression of PF~$\vecptmiss$, defined in Section~\ref{sec:expected_performance_met},
includes a correction term that accounts for the response of the jets in the final state,
which also takes into account the expected contributions from pileup discussed in the previous section.
Here, two additional terms are introduced:
The first one corrects for the presence of many low-energy particles from pileup interactions,
and the second one for an observed asymmetry in the reconstructed PF~$\vecptmiss$ $\varphi$ distribution due to a shift in PF~\vecptmiss along the detector $x$ and $y$ axes.
This asymmetry is caused, amongst other reasons, by a shift between the centre of the CMS coordinate system and the beam axis.
Figure~\ref{fig:met_distribution} shows the spectrum of PF~\ptmiss in the $\Z\to\mu\mu$ event sample.
The simulation describes the observed distribution over more than four orders of magnitude.
The systematic uncertainty in the prediction includes contributions from uncertainties in the muon energy scale, the jet energy scale, the jet energy resolution, and the energy scale of low-energy particles.
A more detailed discussion of the uncertainties is given in Ref.~\cite{JME-13-003-JINST}.

The hadronic recoil $\vec u_\mathrm{T}$, defined as the vector sum of the transverse momenta of all reconstructed particles excluding the two muons from the $\cPZ$ boson decay, is used as a probe for the \ptmiss determination.
With the $\cPZ$ boson transverse momentum denoted as $\vec q_\mathrm{T}$,
momentum conservation in the transverse plane implies $\vec q_\mathrm{T} + \vec u_\mathrm{T} + \vecptmiss = 0$.
Muons are reconstructed with considerably higher precision than the hadronic recoil.
The precision of the \ptmiss reconstruction is therefore dominated by the precision
with which the hadronic recoil is reconstructed.
This precision is also representative of the resolution with which $\vecptmiss$ is reconstructed in events with prompt neutrinos, \eg in $\PZ\to\nu\nu$ decays.
The precision of the hadronic recoil reconstruction can be measured directly in $\PZ\to\mu\mu$ events under the assumption that there is no true source of missing transverse momentum.
The parallel ($u_\parallel$) and perpendicular ($u_\perp$) components of the hadronic recoil are defined with respect to $\vec q_\mathrm{T}$ in the transverse plane.
At high $q_\mathrm{T}$, the resolution of $u_\parallel$ is dominated by that of the jets recoiling against the direction of the $\PZ$ boson momentum,
whereas $u_\perp$ is more affected by random detector noise and by fluctuations of the underlying event.

Several algorithms were developed to mitigate the
deterioration of the resolution with increasing pileup~\cite{JME-13-003-JINST}.
Among those, the so-called \textit{No-PU PF \vecptmiss}
algorithm calculates \vecptmiss as a weighted sum of the different contributions to the event:
charged particles and neutral particles within jets identified as originating from the primary interaction vertex,
charged particles and neutral particles within jets identified as originating from pileup vertices,
other charged particles associated with the primary interaction vertex,
other charged particles not associated with the primary interaction vertex,
and other neutral particles.
The weights optimizing the \ptmiss resolution are found to be 1.0 except for a weight of 0.6 in the case of isolated neutral particles.
The MVA PF \vecptmiss algorithm combines the same inputs using a multivariate (MVA) regression technique to correct both the direction and the magnitude of the hadronic recoil.

\begin{figure}[htp]
\centering
\includegraphics[width=.7\linewidth]{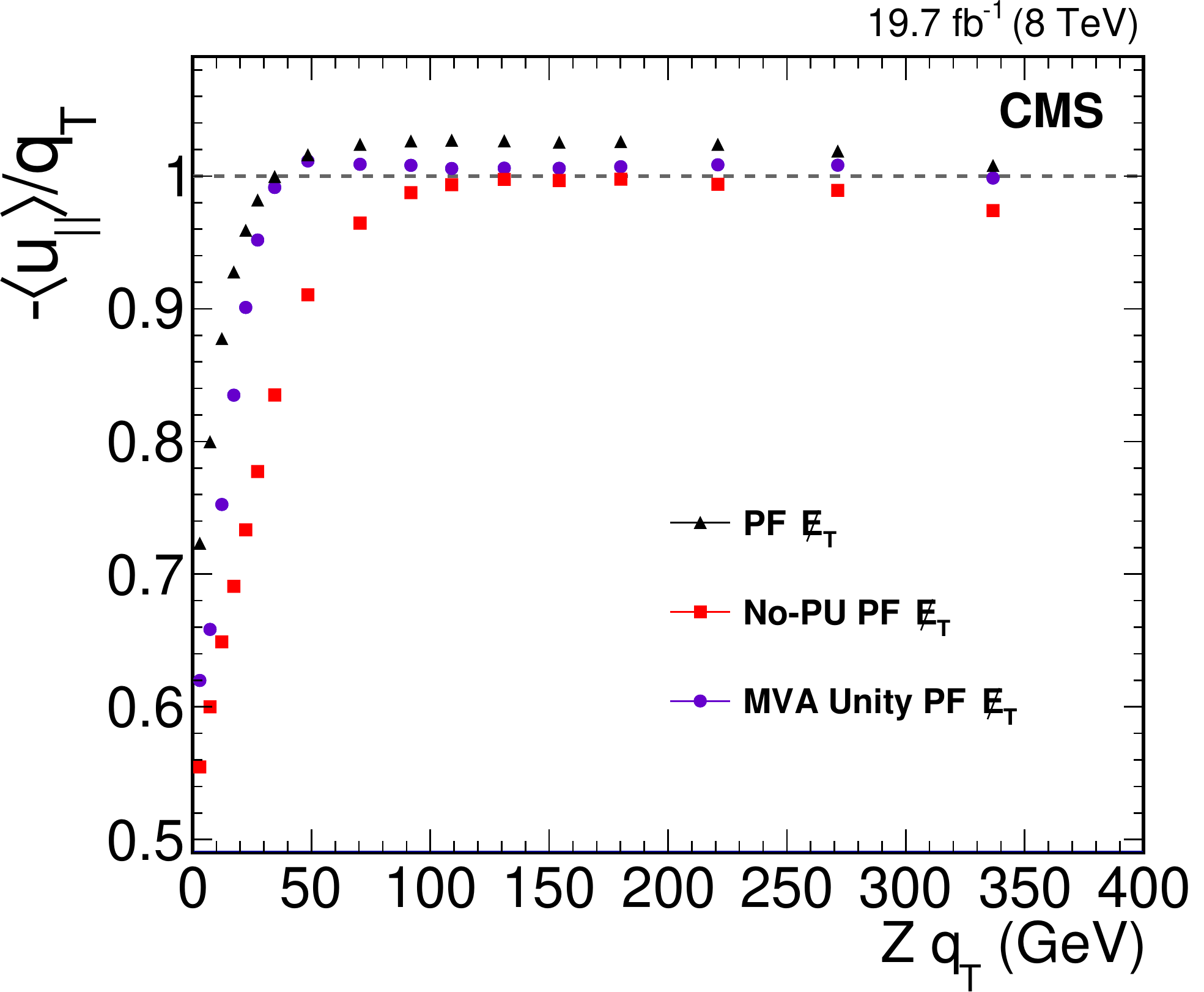}
\caption{Comparison of the average response of the parallel recoil component, $-\langle u_\parallel\rangle/q_\mathrm{T}$, for the PF \vecptmiss, No-PU PF \vecptmiss, and MVA PF \vecptmiss (denoted as \textit{MVA Unity PF \ETslash}) algorithms as a function of $q_\mathrm{T}$, as determined in $\cPZ\to\mu\mu$ events.
}
\label{fig:met_response}
\end{figure}

\begin{figure}[htp]
\centering
\includegraphics[width=.49\textwidth]{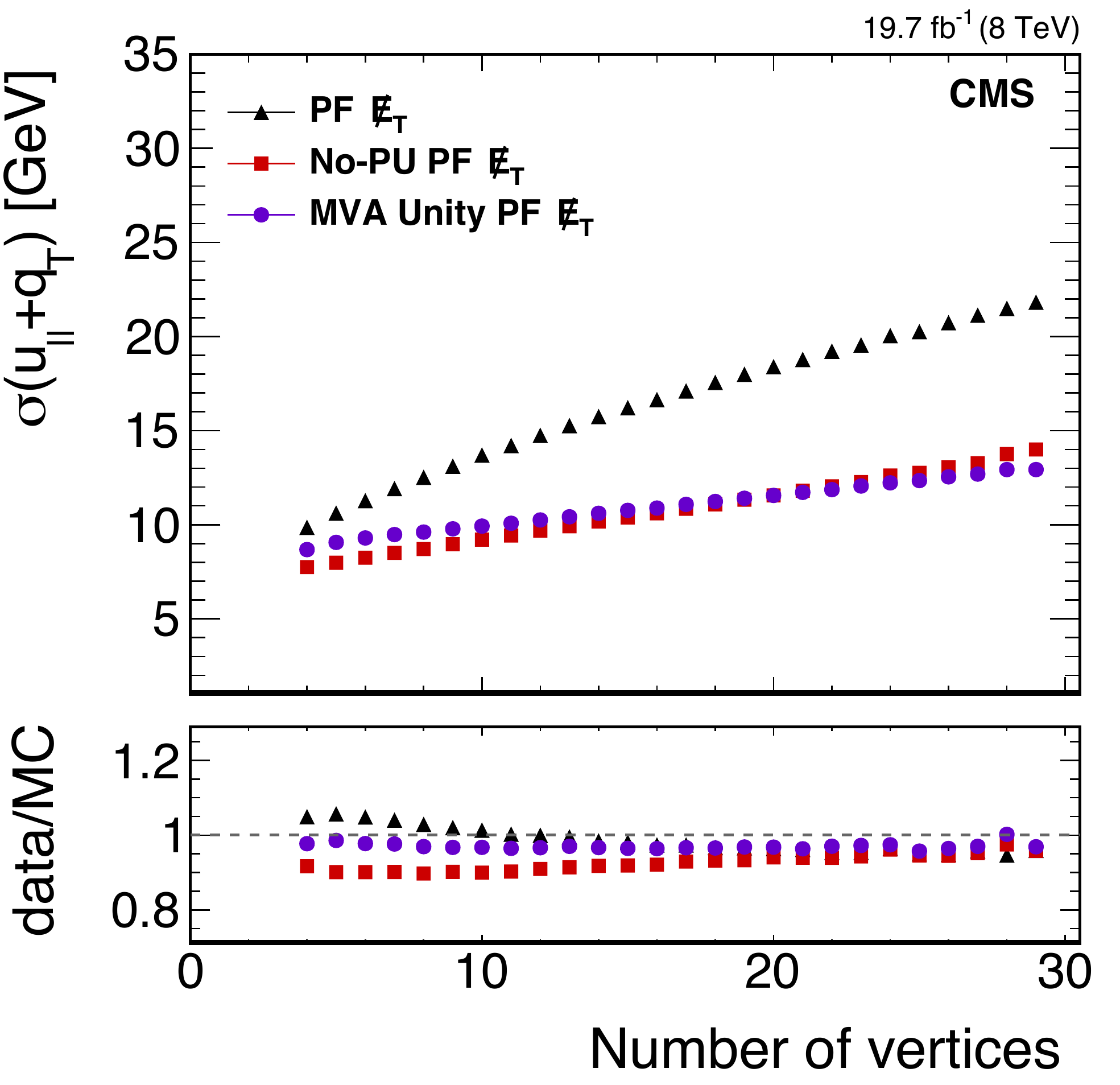}
\includegraphics[width=.49\textwidth]{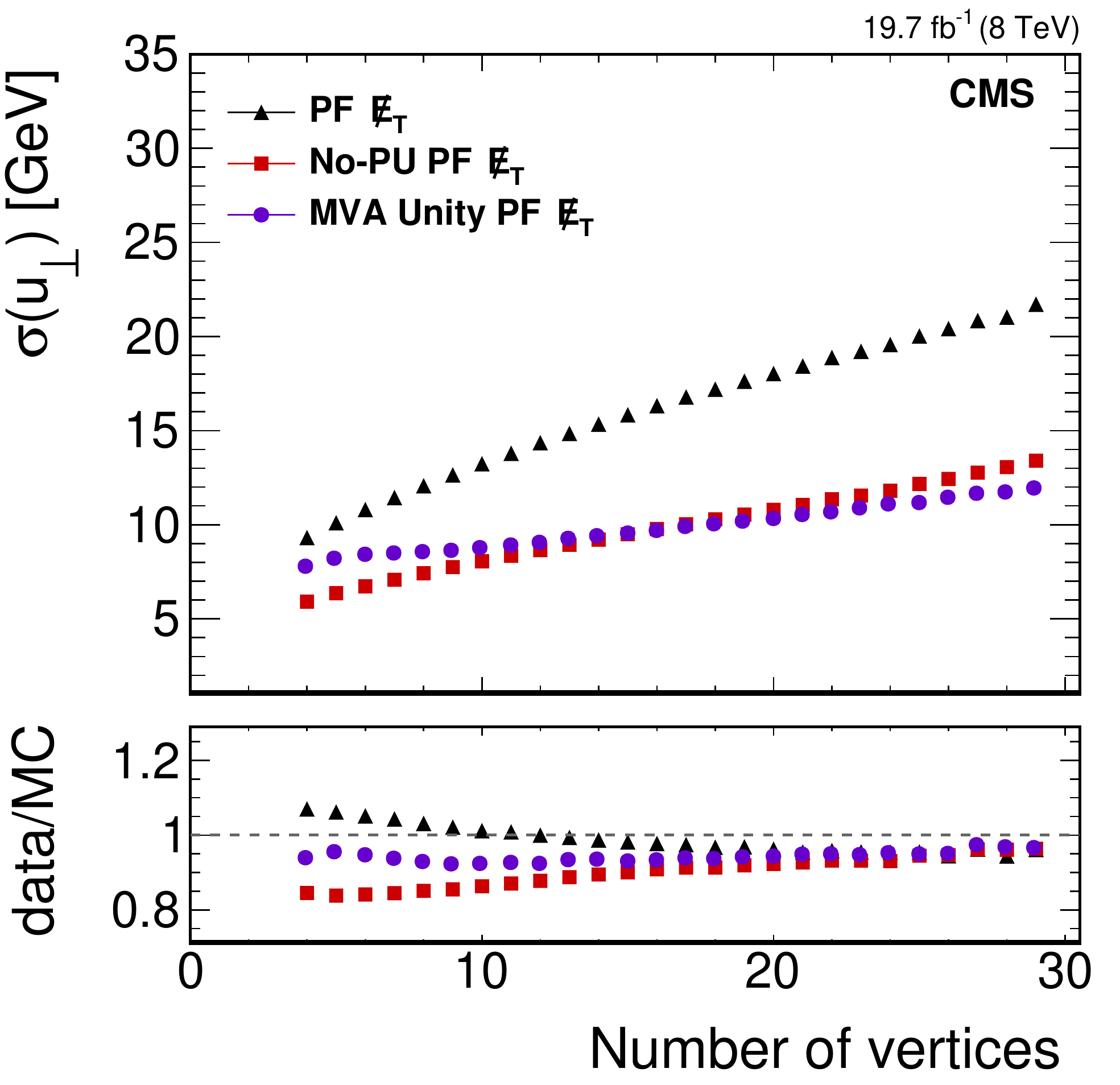}
\caption{Comparison of the resolutions of the parallel (left) and perpendicular (right) recoil components for the PF \vecptmiss, No-PU PF \vecptmiss, and MVA PF \vecptmiss algorithms as a function of the number of reconstructed vertices in $\cPZ\to\mu\mu$ events~\cite{JME-13-003-JINST}.
The upper frame of each figure shows the resolution in observed events;
the lower frame shows the ratio of data to simulation.
}
\label{fig:PFMetValidation}
\end{figure}

The response of the \vecptmiss algorithms is defined as the ratio of the average magnitude of the parallel recoil component and the magnitude of the $\PZ$~boson transverse momentum, $-\langle u_\parallel \rangle/q_\mathrm{T}$,
displayed in Fig.~\ref{fig:met_response} as a function of $q_\mathrm{T}$.
For $q_\mathrm{T} > 30$\GeV, the response agrees with unity within 5\% for the PF \vecptmiss and MVA PF \vecptmiss algorithms,
whereas a response near unity is only reached at $q_\mathrm{T} > 70$\GeV for the No-PU PF \vecptmiss algorithm.
The resolution of the hadronic recoil is assessed with a parametrization of the $u_\parallel + |\vec q_\mathrm{T}|$ or $u_\perp$ distributions by a Voigtian function,
defined by the convolution of a Breit--Wigner and a Gaussian function.
The resolution of each recoil component is obtained from the full width at half maximum of the Voigtian function divided by 2.35.
The event sample is divided according to vertex multiplicity,
and a fit to a Voigtian function is performed in each bin.
The resulting resolution curves of $u_\parallel$ and $u_\perp$ are shown in Fig.~\ref{fig:PFMetValidation} as a function of the number of reconstructed vertices in the event.
The resolutions for both No-PU PF \vecptmiss and MVA PF \vecptmiss reveal a considerably reduced dependence on the number of reconstructed vertices with respect to PF~\vecptmiss,
with an improvement of the resolution of each recoil component of almost a factor of two for 20 reconstructed vertices.

\subsection{Muons}
\label{subsec:commissioning_and_pilep_muons}

The performance of the PF muon identification is probed in samples of prompt muons from $\Z$~boson decays with a tag-and-probe technique.
Events are recorded with triggers requiring a single muon with \pt thresholds depending on the instantaneous luminosity.
The tag muons are well-identified muons matched to the muons identified at trigger level,
whereas the probes are muon candidates reconstructed with only the inner tracker to
avoid any potential bias of the measurement from the muon subdetectors~\cite{MUO-10-004}.
This procedure measures the efficiency to reconstruct a muon track in the muon
detectors, to link it with the inner track, and for this muon
to be identified by the PF algorithm.

\begin{figure}[htb!]
  \centering
\includegraphics[height=0.45\textwidth,angle=0]{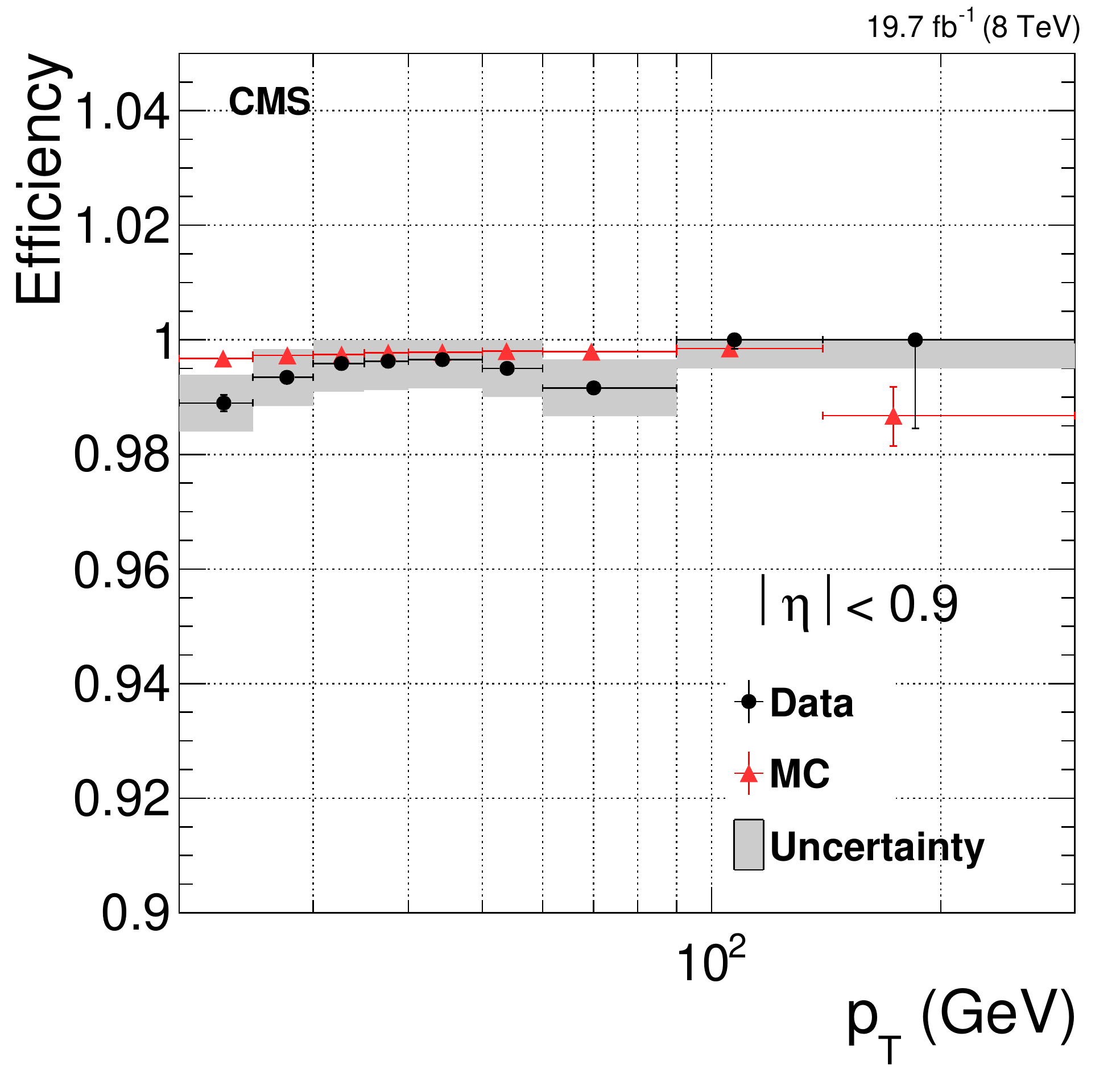}
\includegraphics[height=0.45\textwidth,angle=0]{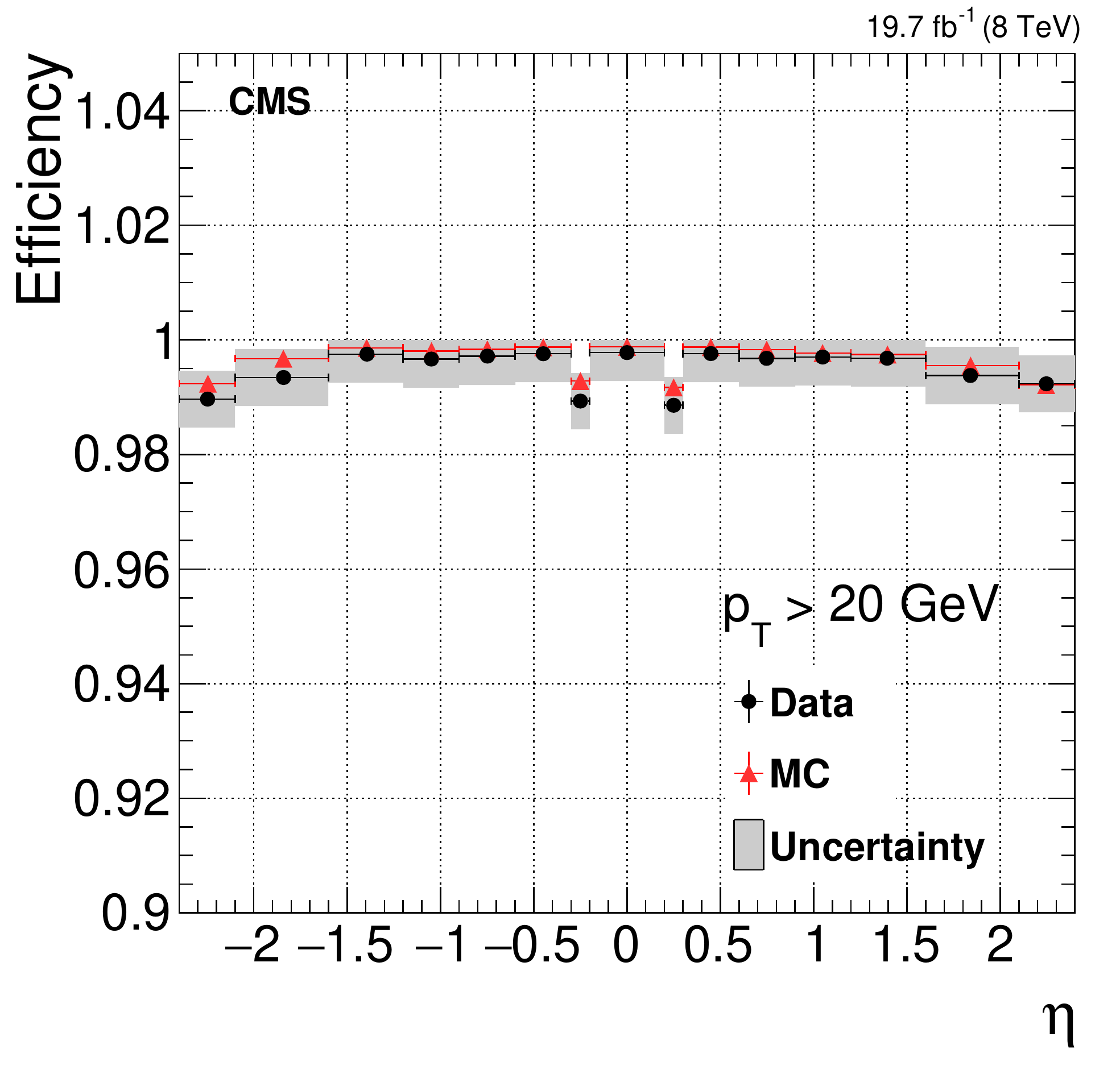} \\
\includegraphics[height=0.45\textwidth,angle=0]{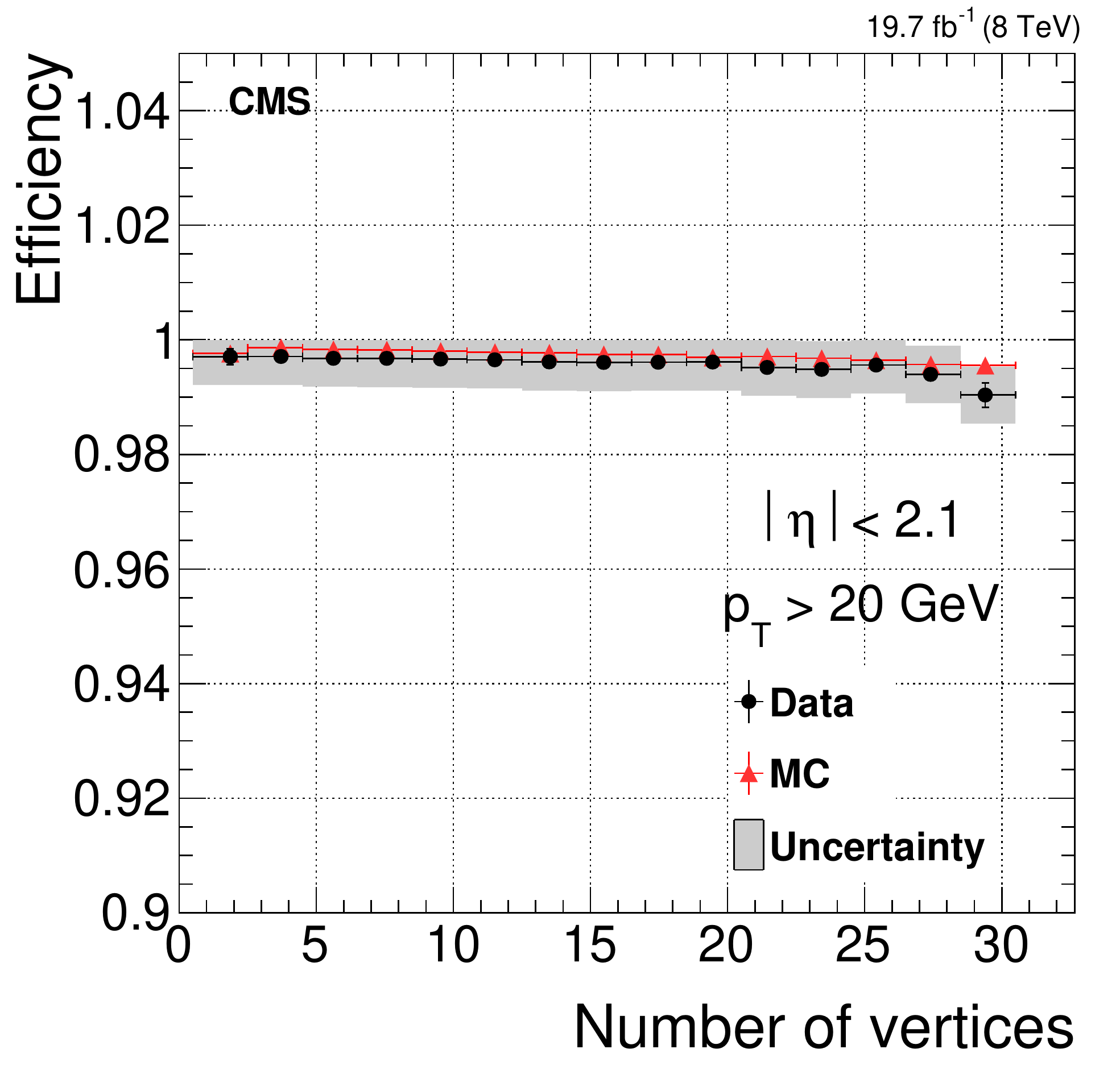}
  \caption{
Efficiency of the PF muon identification for muons from $\Z$~boson decays as a function
of \pt (top left), $\eta$ (top right), and $N_\text{vtx}$ (bottom). The efficiency
is measured for data and simulation with a tag-and-probe technique.
The uncertainty band includes the dominant source of systematic uncertainty, which
comes from imperfections in the parametrization of the signal and background dimuon
mass distributions.
}
  \label{fig:IdEff}
\end{figure}

Figure~\ref{fig:IdEff} (top left) compares the identification efficiencies measured in data and simulation as a function of muon \pt for muons with $20 < \pt < 250$\GeV from $\Z$~boson decays.
Only muons in the central barrel region with $\abs{\eta} < 0.9$ are considered.
Overall, there is an excellent agreement of observed and simulated efficiencies, and the data confirm that prompt muons are identified by the PF algorithm with an efficiency close to 100\%.
The efficiencies in data and simulation agree well within $1$\% for $\pt > 20$\GeV.
A similar agreement is displayed in Fig.~\ref{fig:IdEff} (top right) as a function of $\eta$.
The muon identification efficiency is only marginally affected by pileup, as shown in Fig.~\ref{fig:IdEff} (bottom), which displays the efficiency as a function of $N_\text{vtx}$.
Hence, no dedicated pileup mitigation strategies are deployed for muon identification.

\subsection{Lepton isolation}
\label{subsec:lepton_isolation}

Since the calculation of lepton isolation involves summing the \pt values of charged hadrons, photons, and neutral had\-rons,
lepton isolation is sensitive to pileup interactions,
which give rise to additional reconstructed particles inside the isolation cone.
For simplicity, the focus in this section is on muon isolation.
Electron isolation is calculated and verified with similar techniques.

To mitigate the deterioration of the isolation efficiency due to pileup,
the isolation as defined in Eq.~(\ref{eq:isolation}) is complemented in two ways.
First, only charged hadrons associated with the hard-scatter vertex (HS) are considered.
Second, the expected contributions from pileup are subtracted from the \pt sums of neutral hadrons and photons.
The pileup-mitigated absolute isolation for muons is defined as
\begin{linenomath}
\begin{equation}
I_\mathrm{PF}^\text{abs} \equiv \sum_{\Ph^\pm,\mathrm{HS}} \pt^{\Ph^\pm} + \max\left( 0, \sum_{\Ph^0}  \pt^{\Ph^0}
                                        +  \sum_{\gamma} {\pt^{\gamma}} - \Delta\beta \sum_{\Ph^\pm,\text{pileup}} \pt^{\Ph^\pm}  \right).
\label{eq:pf_reconstruction_isolation}
\end{equation}
\end{linenomath}

The expected contribution of photons and neutral hadrons from pileup is estimated from
the scalar sum of the transverse momenta of charged hadrons in the cone that are identified as coming from pileup vertices, $\sum_{\Ph^\pm,\,\text{pileup} }\pt$.
This sum is multiplied by the factor $\Delta\beta = 0.5$, which corresponds approximately to the ratio of neutral particle to charged hadron production
in inelastic proton-proton collisions, as estimated from simulation.
The relative lepton isolation is defined as $I_\mathrm{PF} = I_\mathrm{PF}^\text{abs}/\pt^{\ell}$.

\begin{figure}[htb!]
  \centering
\includegraphics[height=0.45\textwidth,angle=0]{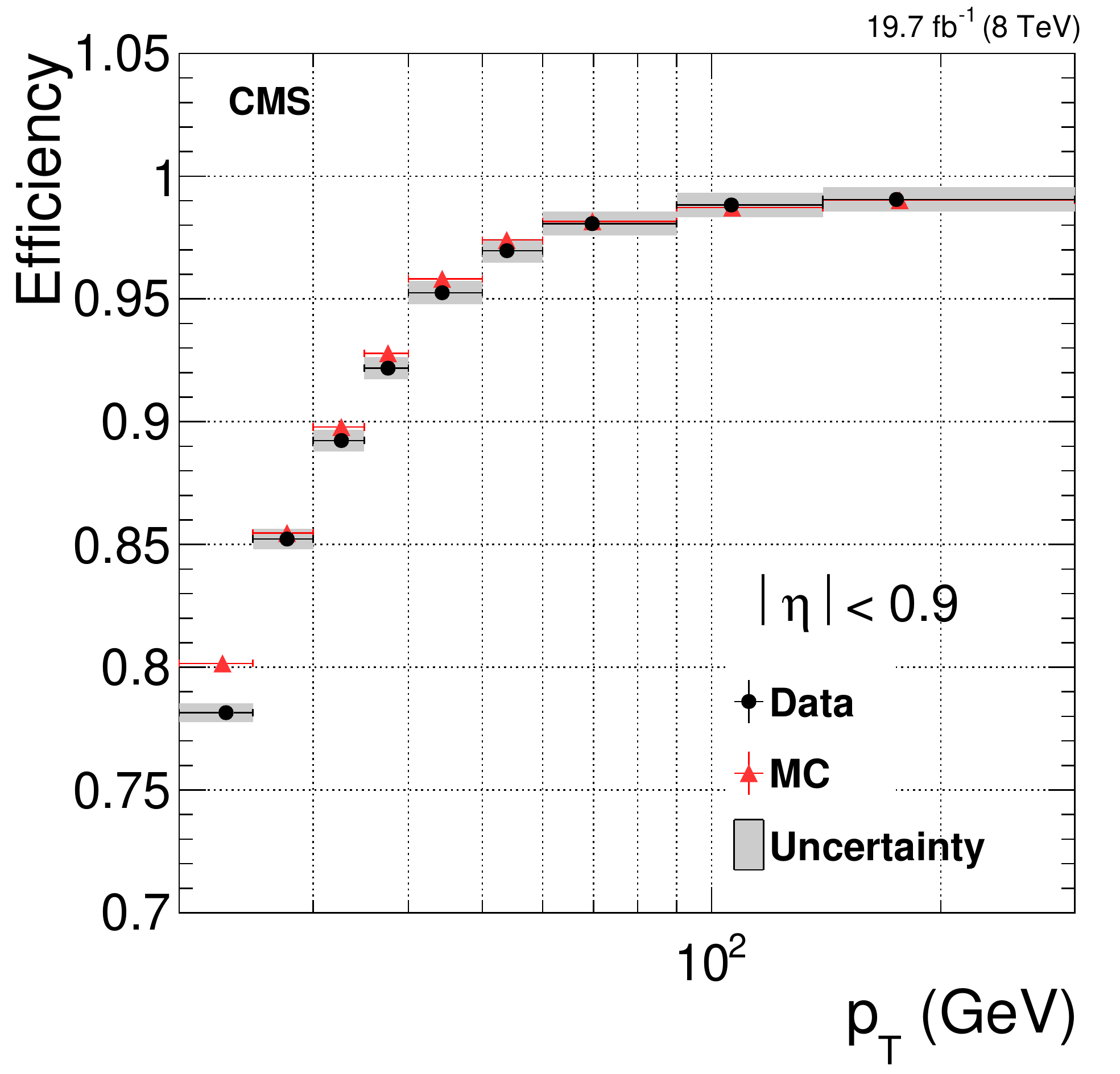}
\includegraphics[height=0.45\textwidth,angle=0]{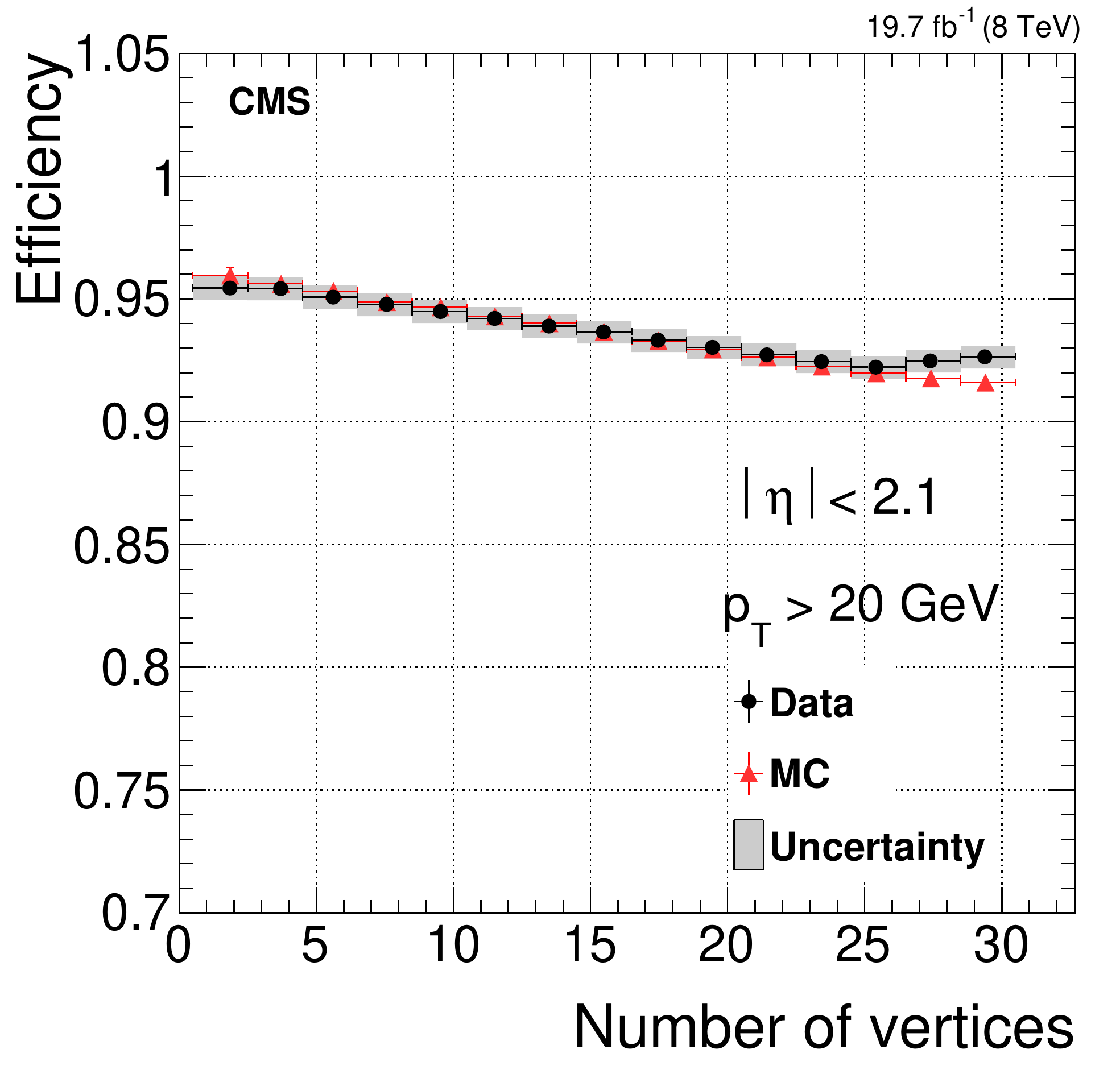}
  \caption{
Isolation efficiency for muons from $\Z$ boson decays as a function of \pt (left)
and $N_\text{vtx}$ (right).
}
  \label{fig:IsoEff}
\end{figure}

The efficiency of the muon isolation is measured in a sample of muons from $\cPZ$~boson decays with a tag-and-probe technique.
Events are selected according to the same criteria as for the measurement of the muon identification efficiency discussed in Section~\ref{subsec:commissioning_and_pilep_muons}.
In addition, since the goal of lepton isolation is to identify prompt muons, the \textit{tight} muon identification criteria described in Section~\ref{sec:expected_performance_muons} are applied to the probe muons.
The efficiencies to pass the muon isolation criterion $I_\mathrm{PF} < 0.12$ are
presented in Fig.~\ref{fig:IsoEff} as a function of muon $\pt$ for muons with $\abs{\eta} < 0.9$ and as a function of $N_\text{vtx}$ for muons with $\pt > 20$\GeV and $\abs{\eta} < 2.1$.
The simulated and observed efficiencies agree over the full spectra within uncertainties except for muons with $20 < \pt < 25$\GeV, where the observed efficiencies are 2\% below the expectation from simulation.
The muon isolation efficiency slightly decreases with $N_\text{vtx}$.
This decrease can be understood from the definition of the isolation:
while the expected average contribution from pileup is subtracted, an increasing amount of pileup makes it more likely for the remaining pileup contribution to fluctuate up, leading to a relative isolation larger than the cutoff value.

\subsection[Hadronic tau decays]{Hadronic $\Pgt$ decays}

Hadronic $\Pgt$ decays provide an ideal probe for commissioning several aspects of the PF reconstruction.
The reconstruction of $\tauh$ candidates in the different decay modes and the isolation discriminators test the reconstruction and identification efficiencies for charged hadrons and photons,
whereas observables that are sensitive to the $\tauh$ energy scale probe the energy scales of charged hadrons and photons.
To mitigate the impact from pileup, the expected contribution from pileup photons in the computation of the $\tauh$ isolation is subtracted with the same strategy as for the lepton isolation (Eq.~(\ref{eq:pf_reconstruction_isolation})). As opposed to the definition
of muon isolation, neutral hadrons are disregarded in the isolation sum.
Charged hadrons associated with pileup vertices, which are used for the pileup
mitigation only, are included if their distance to the $\tauh$ is smaller
than 0.8 in the $(\eta,\varphi)$ plane. The larger cone size makes it easier
to collect pileup charged hadrons for a more precise estimation of the pileup
contribution.
For the $\tauh$ isolation, an empiric $\Delta\beta$ factor of $0.46$ is used.
More details on $\tauh$ reconstruction and identification as well as on the validation with collision data discussed in the following are given in Ref.~\cite{TAU-14-001}.

The efficiency with which hadronic $\Pgt$ decays are reconstructed and identified by the HPS algorithm
is measured with $\cPZ/\Pggx \to \Pgt\Pgt$ events.
The events are selected in the channel in which one $\Pgt$ decays into a muon and the other decays hadronically.
These events are recorded with single-muon triggers.
The muon is required to satisfy $\pt > 25$\GeV and $\abs{\eta} < 2.1$ and
to pass tight identification and isolation criteria.
The $\tauh$ candidate is not required to pass any specific $\tauh$ reconstruction and identification criteria.
Instead, a loose $\tauh$ selection is applied to the collection of jets that seed the $\tauh$ reconstruction:
the jets are required to satisfy $\pt^\text{jet} > 20$\GeV and $\abs{\eta_\text{jet}} < 2.3$,
to be separated from the muon by $\Delta R(\mu, \text{jet}) > 0.5$, and to contain at least one track with $\pt > 5$\GeV and an electric charge opposite to that of the muon.
Furthermore, tight kinematic selection criteria are applied to reduce the contributions from background processes~\cite{TAU-14-001}.
Events containing additional prompt muons or electrons are rejected.

In this sample of selected $\cPZ/\Pggx \to \Pgt\Pgt$ events,
the $\tauh$ identification efficiency is obtained with a tag-and-probe technique.
The contribution of the $\cPZ/\Pggx \to \Pgt\Pgt$ signal to the events where the probe $\tauh$ candidate either passes or fails the $\tauh$ identification discriminator under study
is determined by fitting the distribution of the visible $\mu\tauh$ mass
with binned shape templates for the different signal and background processes.
Systematic uncertainties are represented by nuisance parameters in the fit.

\begin{figure}[htb]
\centering
\includegraphics[width=0.49\textwidth]{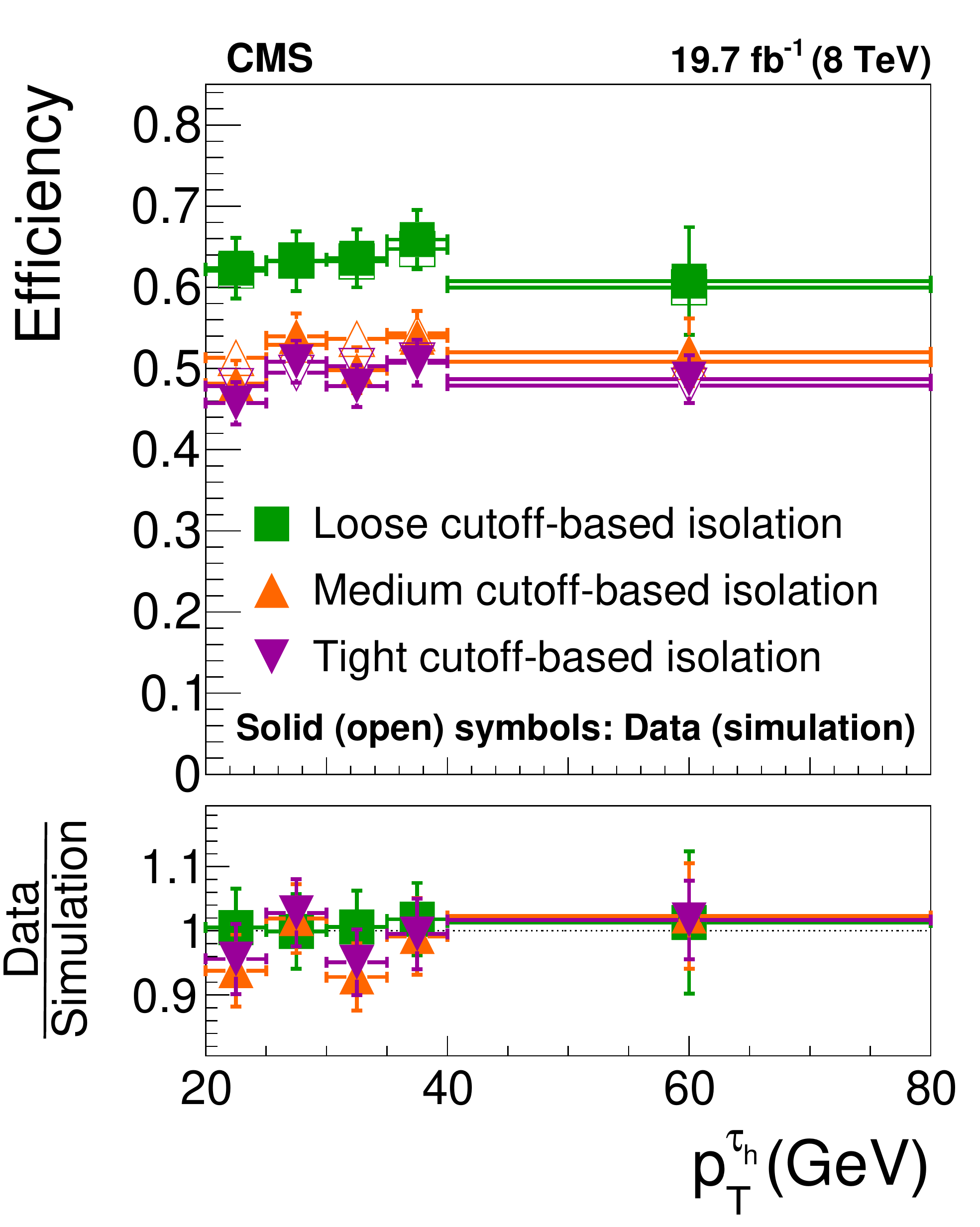}
\includegraphics[width=0.49\textwidth]{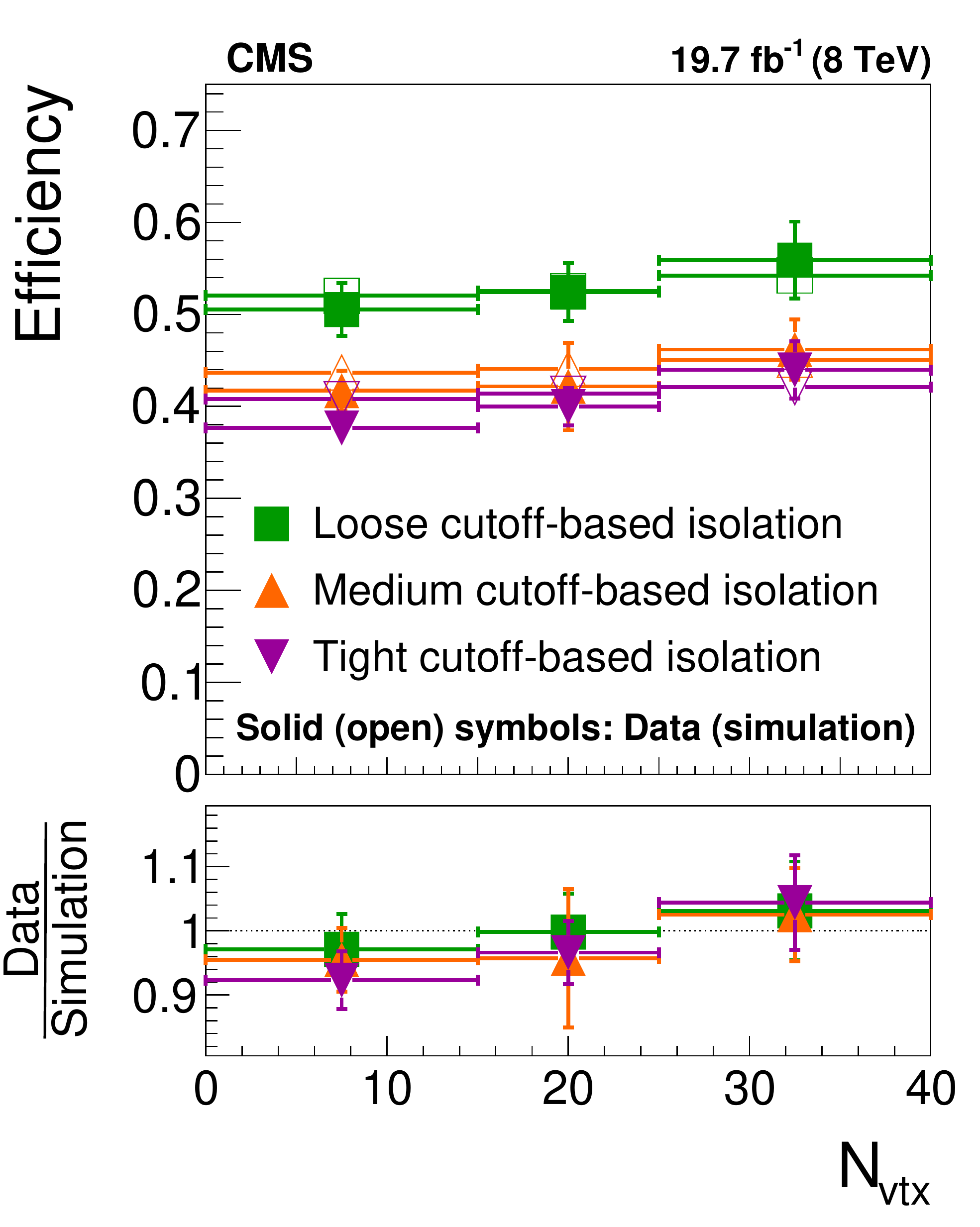}
\caption{
  Efficiency for hadronic $\Pgt$ decays to pass the loose, medium and
  tight working points of the HPS $\tauh$ identification algorithm,
  as measured with the tag-and-probe technique in recorded and simulated $\cPZ/\Pggx \to \Pgt\Pgt$ events~\cite{TAU-14-001}.
  The efficiency is presented as a function of $\tauh$ $\pt$ (left), and as function of the reconstructed vertex multiplicity (right).
}
\label{fig:commissioning_and_pileup:tau_id_eff}

\end{figure}

The $\tauh$ identification efficiencies measured in data are in agreement with the predictions of the simulation.
The efficiencies measured as a function of the reconstructed $\tauh$ $\pt$ and as a function of $N_\text{vtx}$,
the number of reconstructed vertices in the event, are shown in Fig.~\ref{fig:commissioning_and_pileup:tau_id_eff}.
The slight increase of the identification efficiency for higher numbers of reconstructed vertices
is caused by a small overcorrection of the pileup subtraction in the calculation of the $\tauh$ isolation.

The rate with which quark and gluon jets are misidentified as hadronic $\Pgt$ decays has been measured with a sample of QCD multijet events.
The events were recorded with a single-jet trigger with a $\pt$ threshold of $320$\GeV.
At least one further jet of $\pt > 20$\GeV and $\abs{\eta} < 2.3$
is required.
The misidentification rate is given by the fraction of jets with $\pt > 20\GeV$ and $\abs{\eta} < 2.3$ that result in a $\tauh$ with $\pt > 20\GeV$ and $\abs{\eta} < 2.3$ passing the $\tauh$ decay mode reconstruction and $\tauh$ isolation criteria.
The jet that passes the trigger is excluded from the computation of the misidentification rate
in case only one jet in the event satisfies the trigger requirement.
If two or more jets in the event pass the trigger requirement,
all jets fulfilling $\pt > 20$\GeV and $\abs{\eta} < 2.3$ in the event are included in the computation.
This procedure ensures that the measured $\text{jet}\to \tauh$ misidentification rates are not biased by trigger requirements.

\begin{figure}[htb]
\centering
\includegraphics[width=0.49\textwidth]{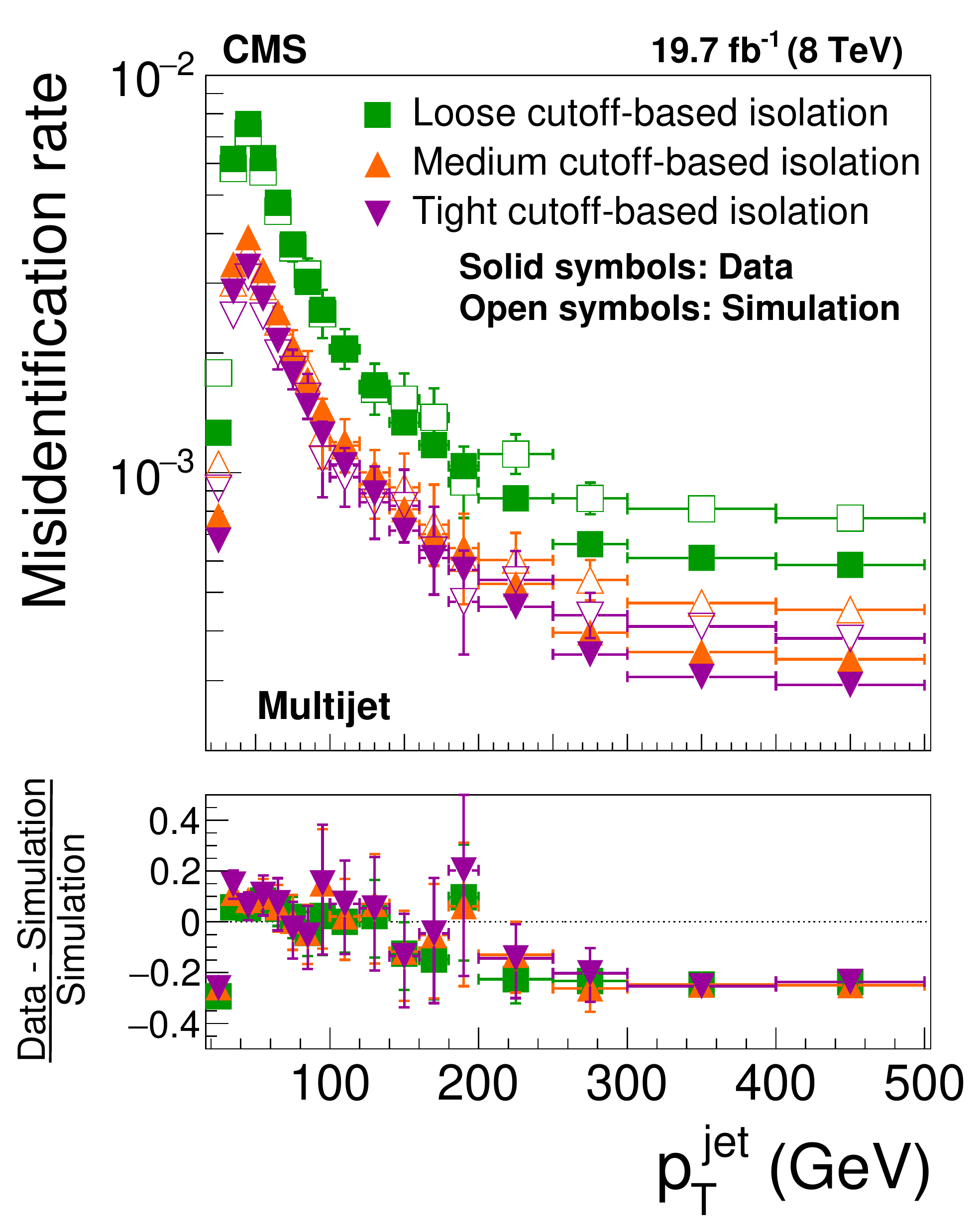}
\includegraphics[width=0.49\textwidth]{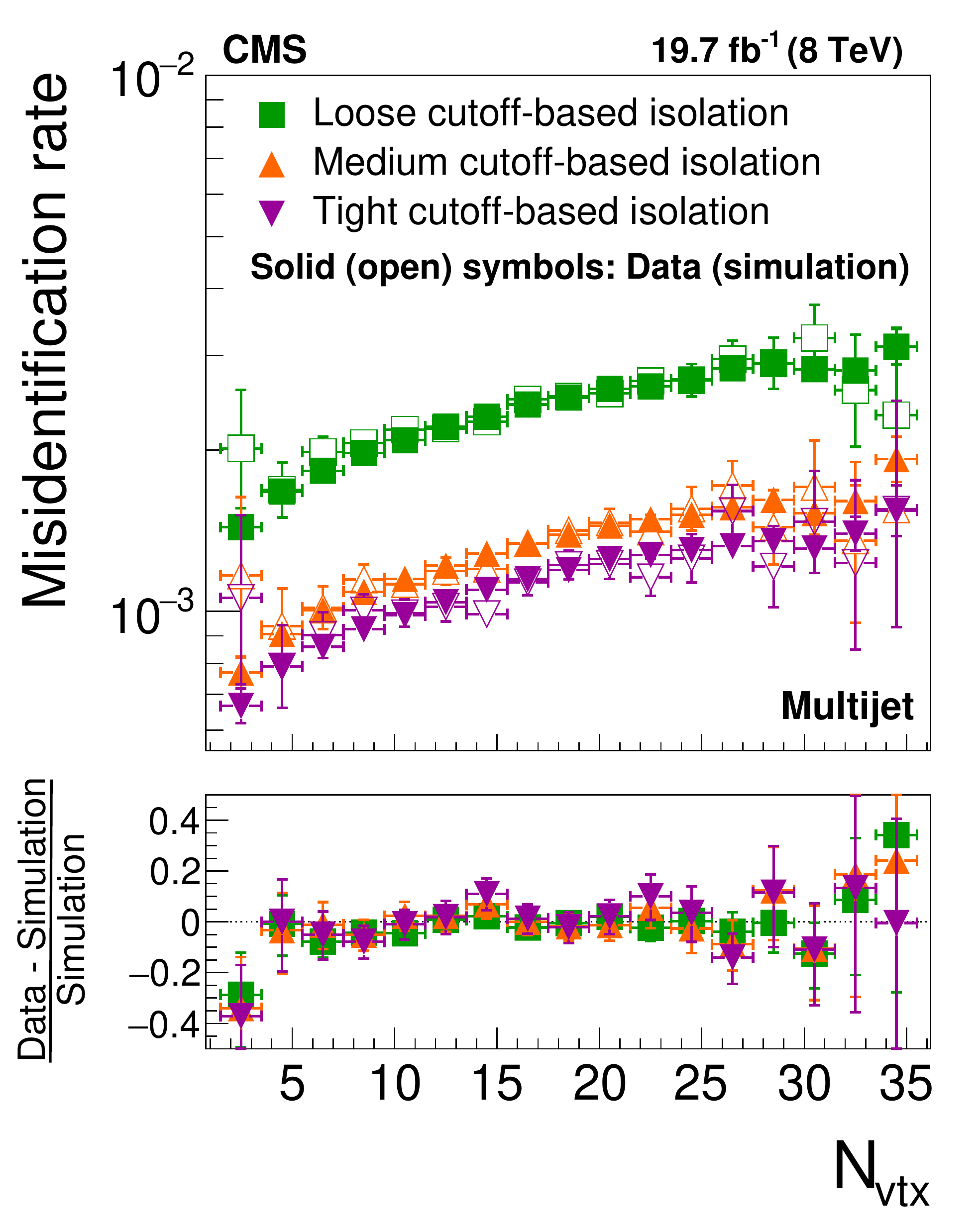}
\caption{
  Probability for quark and gluon jets to pass the $\tauh$ reconstruction and $\tauh$ isolation criteria
  as a function of jet $\pt$ (left) and number of reconstructed vertices (right)~\cite{TAU-14-001}.
  The misidentification rates measured in QCD multijet data are compared to the simulation.
}
\label{fig:commissioning_and_pileup:jet_tau_fake}

\end{figure}

The misidentification rates measured as function of jet $\pt$ and as function of vertex multiplicity are shown in Fig.~\ref{fig:commissioning_and_pileup:jet_tau_fake}.
The contributions from background processes, predominantly arising from $\ttbar$ production,
are accounted for in the simulation.
The probability for jets to pass the $\tauh$ identification criteria strongly depends on $\pt$ and moderately increases as function of pileup.
This increase is due to the $\Delta\beta$ pileup corrections introduced above.
The jet $\to \tauh$ misidentification rates measured in data agree with the simulation within $20\%$.
A trend versus $\pt$ is observed in the data-to-simulation ratio:
while the jet $\to \tauh$ misidentification rates measured in data exceed the expectation at low $\pt$,
the misidentification rates measured at high $\pt$ are below the prediction.
This trend is likely due to the modelling of hadronization processes by the event generator,
in this case \PYTHIA~\textsc{6.4} with tune~\textsc{Z2*}~\cite{Khachatryan:2015pea}.
The observed differences between data and simulation in the probability for jets to get misidentified as hadronic $\Pgt$ decays are applied as corrections to simulated events in physics analyses.

\begin{figure}[htb]
\centering
\includegraphics[width=0.49\textwidth]{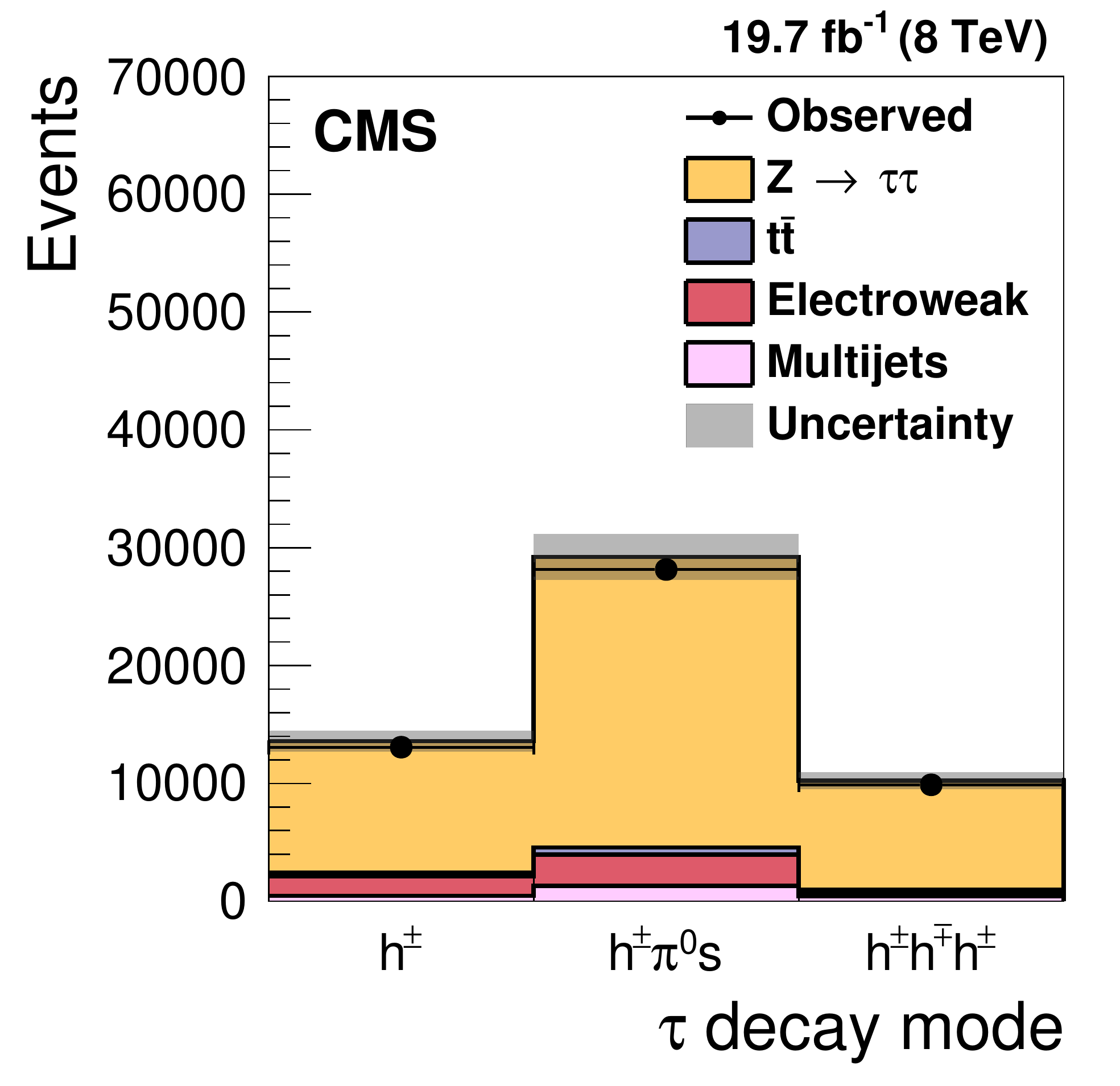}
\includegraphics[width=0.49\textwidth]{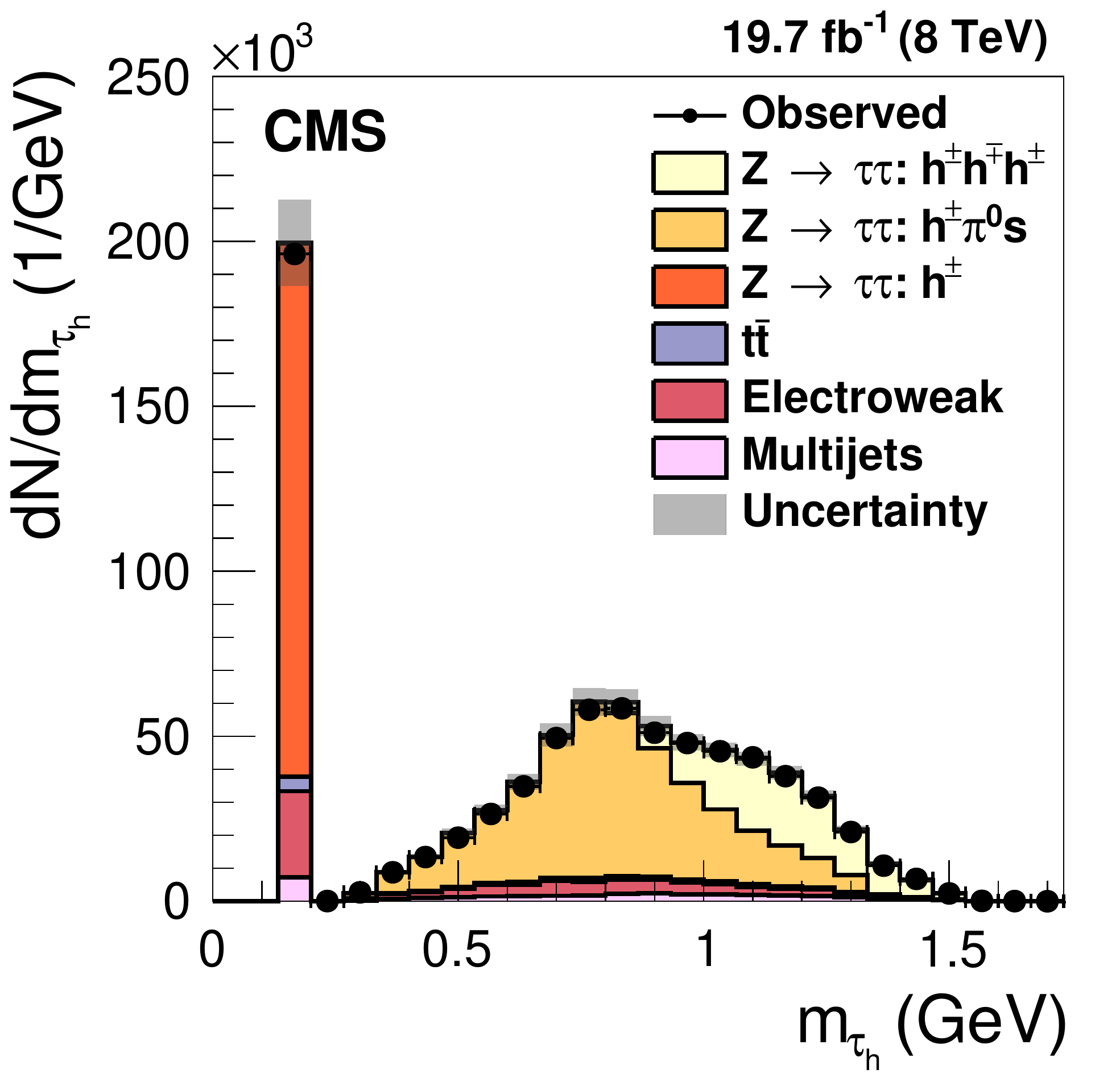}
\caption{
  Distribution of reconstructed $\Pgt$ decay mode (left) and of $\tauh$ mass (right)
  in $\cPZ/\Pggx \to \Pgt\Pgt$ events selected in data compared to the MC expectation.
  The $\cPZ/\Pggx \to \Pgt\Pgt$ events are selected in the decay channel with a muon and a $\tauh$.
The $\tauh$ is required to be reconstructed in one of the three allowed decay modes and to be isolated~\cite{TAU-14-001}.
}
\label{fig:tauIdAlgorithm_ZTT_dm_and_mTau}
\end{figure}

The $\tau$ decay mode and $\tauh$ energy reconstruction have been validated
with the same sample of $\cPZ/\Pggx \to \Pgt\Pgt$ events selected in the $\mu\tauh$ final state used for the tag-and-probe study described above.
In addition, the $\tauh$ candidates are required to be reconstructed by the HPS algorithm in one of the three possible decay modes and to be isolated.
Figure~\ref{fig:tauIdAlgorithm_ZTT_dm_and_mTau} compares the expected and observed distributions of the $\tau$ decay mode
and the $\tauh$ invariant mass, denoted $m_{\tauh}$.
The agreement in the $\tau$ decay mode distributions confirms that the simulation properly models
the identification of the individual $\tauh$ constituents through an accurate description of the tracking efficiency and of the photon reconstruction in the ECAL.
The $m_{\tauh}$ distribution is used to measure the $\tauh$ energy scale.
For $\tauh$ reconstructed in the $\Ph^{\pm}$ mode with a single charged hadron as a constituent,
$m_{\tauh}$ equals the pion mass.
For the other decay modes, however, the reconstructed $m_{\tauh}$ depends on the energy scale
at which each constituent is reconstructed.
In these two decay modes, a template fit is performed to the observed $m_{\tauh}$ distribution,
with the $\tauh$ energy scale as a nuisance parameter that coherently shifts all components of the $\tauh$ four-momentum.
The fit results in a small increase of the $\tauh$ energy scale, by about 0.5\% (1.5\%) for the $\Ph^{\pm \mp \pm}$ ($\Ph^{\pm}\pi^0\mathrm{s}$) decay mode, which leads to a slight shift of the $m_{\tauh}$ distribution.
For the figure, this correction of the $\tauh$ energy scale was applied to simulated $\tauh$.

\section{Summary and outlook}
\label{sec:summary}

The CMS detector was designed 20 years ago to  identify energetic
and isolated leptons and photons and measure their momenta with high
precision, to provide a calorimetric determination of jets and
missing transverse momentum, and to efficiently tag b quark jets.
The CMS detector turned out to feature properties well-suited
for particle-flow (PF) reconstruction. For the first time in a hadron
collider experiment, a PF algorithm aimed at identifying
and reconstructing all final-state particles was implemented.

The technical challenges posed by the complexity of proton-proton collisions
and the amount of material in the tracker were overcome with the development
of new, high-performance reconstruction algorithms in the different
subdetectors, and of discriminating particle identification algorithms
combining their information. The PF reconstruction computing time was kept
under control both for offline data processing and for triggering the data
acquisition, irrespective of the final state intricacy.  The resulting
global event description augmented the performance of all physics objects
(efficiency, purity, response bias, energy and angular resolutions, etc.),
thereby reducing the associated systematic biases and the need for a
posteriori corrections. Knowledge of the detailed particle content of these
physics objects enhanced the scope of many physics analyses.

Excellent agreement was obtained between the simulation and the data
recorded by CMS at a centre-of-mass energy of 8\TeV, thereby validating
the use of PF reconstruction in real data-taking conditions. The PF approach
also paved the way for particle-level pileup mitigation methods, the simplest
of which have been presented in this paper for an average of 20 and up to
35 concurrent pileup interactions. Machine learning algorithms based on
the detailed PF information were shown to preserve the improved physics
object performance even in the presence of a large number of background
particles produced in pileup interactions.

The future CMS detector upgrades have been planned to provide optimal conditions
for PF performance. In the first phase of the upgrade programme, a better and
lighter pixel detector~\cite{CMS:2012sda} will reduce the rate of
misreconstructed charged-particle tracks, and the readout of multiple layers
with low noise photodetectors in the hadron calorimeter~\cite{CMS:2012tda}
will improve the neutral-hadron identification, which currently limits the
jet energy resolution.
The second phase~\cite{CMS:2015tda} will include a lighter and extended tracker
(integrated into the level 1 trigger) and high-granularity endcap calorimeters,
enhancing the PF capabilities for online and offline reconstruction.
These detector evolutions, accompanied by the necessary PF software developments,
should help to respond to the new challenges posed by the 200 pileup interactions
per bunch crossing foreseen at the LHC by the end of the next decade.

\begin{acknowledgments}
\hyphenation{Bundes-ministerium Forschungs-gemeinschaft Forschungs-zentren Rachada-pisek} We congratulate our colleagues in the CERN accelerator departments for the excellent performance of the LHC and thank the technical and administrative staffs at CERN and at other CMS institutes for their contributions to the success of the CMS effort. In addition, we gratefully acknowledge the computing centres and personnel of the Worldwide LHC Computing Grid for delivering so effectively the computing infrastructure essential to our analyses. Finally, we acknowledge the enduring support for the construction and operation of the LHC and the CMS detector provided by the following funding agencies: the Austrian Federal Ministry of Science, Research and Economy and the Austrian Science Fund; the Belgian Fonds de la Recherche Scientifique, and Fonds voor Wetenschappelijk Onderzoek; the Brazilian Funding Agencies (CNPq, CAPES, FAPERJ, and FAPESP); the Bulgarian Ministry of Education and Science; CERN; the Chinese Academy of Sciences, Ministry of Science and Technology, and National Natural Science Foundation of China; the Colombian Funding Agency (COLCIENCIAS); the Croatian Ministry of Science, Education and Sport, and the Croatian Science Foundation; the Research Promotion Foundation, Cyprus; the Secretariat for Higher Education, Science, Technology and Innovation, Ecuador; the Ministry of Education and Research, Estonian Research Council via IUT23-4 and IUT23-6 and European Regional Development Fund, Estonia; the Academy of Finland, Finnish Ministry of Education and Culture, and Helsinki Institute of Physics; the Institut National de Physique Nucl\'eaire et de Physique des Particules~/~CNRS, and Commissariat \`a l'\'Energie Atomique et aux \'Energies Alternatives~/~CEA, France; the Bundesministerium f\"ur Bildung und Forschung, Deutsche Forschungsgemeinschaft, and Helmholtz-Gemeinschaft Deutscher Forschungszentren, Germany; the General Secretariat for Research and Technology, Greece; the National Scientific Research Foundation, and National Innovation Office, Hungary; the Department of Atomic Energy and the Department of Science and Technology, India; the Institute for Studies in Theoretical Physics and Mathematics, Iran; the Science Foundation, Ireland; the Istituto Nazionale di Fisica Nucleare, Italy; the Ministry of Science, ICT and Future Planning, and National Research Foundation (NRF), Republic of Korea; the Lithuanian Academy of Sciences; the Ministry of Education, and University of Malaya (Malaysia); the Mexican Funding Agencies (BUAP, CINVESTAV, CONACYT, LNS, SEP, and UASLP-FAI); the Ministry of Business, Innovation and Employment, New Zealand; the Pakistan Atomic Energy Commission; the Ministry of Science and Higher Education and the National Science Centre, Poland; the Funda\c{c}\~ao para a Ci\^encia e a Tecnologia, Portugal; JINR, Dubna; the Ministry of Education and Science of the Russian Federation, the Federal Agency of Atomic Energy of the Russian Federation, Russian Academy of Sciences, the Russian Foundation for Basic Research and the Russian Competitiveness Program of NRNU ``MEPhI"; the Ministry of Education, Science and Technological Development of Serbia; the Secretar\'{\i}a de Estado de Investigaci\'on, Desarrollo e Innovaci\'on, Programa Consolider-Ingenio 2010, Plan de Ciencia, Tecnolog\'{i}a e Innovaci\'on 2013-2017 del Principado de Asturias and Fondo Europeo de Desarrollo Regional, Spain; the Swiss Funding Agencies (ETH Board, ETH Zurich, PSI, SNF, UniZH, Canton Zurich, and SER); the Ministry of Science and Technology, Taipei; the Thailand Center of Excellence in Physics, the Institute for the Promotion of Teaching Science and Technology of Thailand, Special Task Force for Activating Research and the National Science and Technology Development Agency of Thailand; the Scientific and Technical Research Council of Turkey, and Turkish Atomic Energy Authority; the National Academy of Sciences of Ukraine, and State Fund for Fundamental Researches, Ukraine; the Science and Technology Facilities Council, UK; the US Department of Energy, and the US National Science Foundation.

{\tolerance=800
Individuals have received support from the Marie-Curie programme and the European Research Council and Horizon 2020 Grant, contract No. 675440 (European Union); the Leventis Foundation; the A. P. Sloan Foundation; the Alexander von Humboldt Foundation; the Belgian Federal Science Policy Office; the Fonds pour la Formation \`a la Recherche dans l'Industrie et dans l'Agriculture (FRIA-Belgium); the Agentschap voor Innovatie door Wetenschap en Technologie (IWT-Belgium); the Ministry of Education, Youth and Sports (MEYS) of the Czech Republic; the Council of Scientific and Industrial Research, India; the HOMING PLUS programme of the Foundation for Polish Science, cofinanced from European Union, Regional Development Fund, the Mobility Plus programme of the Ministry of Science and Higher Education, the National Science Center (Poland), contracts Harmonia 2014/14/M/ST2/00428, Opus 2014/13/B/ST2/02543, 2014/15/B/ST2/03998, and 2015/19/B/ST2/02861, Sonata-bis 2012/07/E/ST2/01406; the National Priorities Research Program by Qatar National Research Fund; the Programa Clar\'in-COFUND del Principado de Asturias; the Thalis and Aristeia programmes cofinanced by EU-ESF and the Greek NSRF; the Rachadapisek Sompot Fund for Postdoctoral Fellowship, Chulalongkorn University and the Chulalongkorn Academic into Its 2nd Century Project Advancement Project (Thailand); and the Welch Foundation, contract C-1845.
\par}
\end{acknowledgments}
\bibliography{auto_generated}
\cleardoublepage \appendix\section{The CMS Collaboration \label{app:collab}}\begin{sloppypar}\hyphenpenalty=5000\widowpenalty=500\clubpenalty=5000\textbf{Yerevan Physics Institute,  Yerevan,  Armenia}\\*[0pt]
A.M.~Sirunyan, A.~Tumasyan
\vskip\cmsinstskip
\textbf{Institut f\"{u}r Hochenergiephysik,  Wien,  Austria}\\*[0pt]
W.~Adam, E.~Asilar, T.~Bergauer, J.~Brandstetter, E.~Brondolin, M.~Dragicevic, J.~Er\"{o}, M.~Flechl, M.~Friedl, R.~Fr\"{u}hwirth\cmsAuthorMark{1}, V.M.~Ghete, C.~Hartl, N.~H\"{o}rmann, J.~Hrubec, M.~Jeitler\cmsAuthorMark{1}, A.~K\"{o}nig, I.~Kr\"{a}tschmer, D.~Liko, T.~Matsushita, I.~Mikulec, D.~Rabady, N.~Rad, B.~Rahbaran, H.~Rohringer, J.~Schieck\cmsAuthorMark{1}, J.~Strauss, W.~Waltenberger, C.-E.~Wulz\cmsAuthorMark{1}
% ...
\vskip\cmsinstskip
% ...
1:~~Also at Vienna University of Technology, Vienna, Austria\\
% ...

\end{sloppypar}
\end{document}